\documentclass{tcibook}
\usepackage{fancyhea}
\usepackage{work}
\usepackage{bm}       
\usepackage{graphicx}
\usepackage{hyperref}      
\usepackage{multirow}
\usepackage{slashed}
\usepackage{amsmath,amssymb}

\newcommand{\nc}{\newcommand}  

\def\ie{{\it i.e.}}

\def\Acknowledgements{\bigskip  \bigskip \begin{center} \begin{large}
             \bf ACKNOWLEDGEMENTS \end{large}\end{center}}

\def\beq{\begin{equation}}
\def\eeq#1{\label{#1}\end{equation}}
\def\eeqn{\end{equation}}

\newenvironment{Eqnarray}%
   {\arraycolsep 0.14em\begin{eqnarray}}{\end{eqnarray}}
\def\beqa{\begin{Eqnarray}}
\def\eeqa#1{\label{#1}\end{Eqnarray}}
\def\eeqan{\end{Eqnarray}}

\nc{\ra}{\rightarrow}  
\nc{\slsh}{\slash\hspace*{-0.22cm}}
\def\Re{{\cal R \mskip-4mu \lower.1ex \hbox{\it e}\,}}
\def\Im{{\cal I \mskip-5mu \lower.1ex \hbox{\it m}\,}}

\nc{\vev}[1]{ \left\langle {#1} \right\rangle }
\nc{\bra}[1]{ \langle {#1} | }
\nc{\ket}[1]{ | {#1} \rangle }
\nc{\fb}{\,{\rm fb}^{-1}}
\nc{\ev}{{\rm eV}}
\nc{\kev}{{\rm keV}}
\nc{\Mev}{{\rm MeV}}
\nc{\gev}{{\rm GeV}}
\nc{\tev}{{\rm TeV}}
\nc{\mev}{{\rm MeV}}

\def\L{{\cal L}}

\def\O{{\cal O}}

\def\del{\partial}
\def\Dslash{\not{\hbox{\kern-4pt $D$}}}
\def\dslash{\not{\hbox{\kern-2pt $\del$}}}
\def\pslash{\not{\hbox{\kern-2pt $p$}}}
\def\ETmiss{ \not{\hbox{\kern-4pt $E$}}_T }

\def\eff{{\mbox{\scriptsize eff}}}

\def\ee{e^+e^-}

\def\mw{m_W}
\def\mt{m_t}

\def\msb{{\bar{\ssstyle M \kern -1pt S}}}

\newcommand{\mste}{m_{\tilde{t}_1}}

\newcommand{\SM}{\mathrm {SM}}

\newcommand{\MW}{M_W}
\newcommand{\MZ}{M_Z}

\newcommand{\Mh}{M_h}
\newcommand{\MH}{M_H}
\newcommand{\MHSM}{M_H^{\SM}}

\newcommand{\tsf}{\theta\kern-.20em_{\tilde{f}}}
\newcommand{\tsfp}{\theta\kern-.20em_{\tilde{f}\prime}}
\newcommand{\tsq}{\theta\kern-.15em_{\tilde{q}}}
\newcommand{\sweff}{\sin^2{\theta^\ell}_{\mathrm{eff}}}

\newcommand{\mcha}[1]{m_{\tilde \chi^\pm_{#1}}}

\newcommand{\VL}{\left( \begin{array}{c}}
\newcommand{\VR}{\end{array} \right)}
\newcommand{\ML}{\left( \begin{array}{cc}}
\newcommand{\MLd}{\left( \begin{array}{ccc}}
\newcommand{\MLv}{\left( \begin{array}{cccc}}
\newcommand{\MR}{\end{array} \right)}

\newcommand{\sw}{s_\mathrm{w}}
\newcommand{\cw}{c_\mathrm{w}}

\newcommand{\lsim}
{\;\raisebox{-.3em}{$\stackrel{\displaystyle <}{\sim}$}\;}
\newcommand{\gsim}
{\;\raisebox{-.3em}{$\stackrel{\displaystyle >}{\sim}$}\;}

\newcommand{\VBFNLO}{\textsc{VBFNLO}}
\providecommand{\href}[2]{#2}

\newcommand{\ii}{\mathrm{i}}

\newcommand{\pd}{\partial}
\newcommand{\LL}{\mathcal{L}}
\newcommand{\vV}{\mathbf{V}}
\newcommand{\vT}{\mathbf{T}}
\newcommand{\vW}{\mathbf{W}}

\newcommand{\trs}[1]{\operatorname{tr}\left[#1\right]}
\newcommand{\pp}{{\prime\,2}}
\newcommand{\vB}{\mathbf{B}}
\newcommand{\vD}{\mathbf{D}}

\newcommand{\GeV}{\text{GeV}}

\newcommand{\bss}{\begin{tiny}}
\newcommand{\ess}{\end{tiny}}

\newcommand{\ppl}  {\mathcal{P}_{e^+}}
\newcommand{\pmi}  {\mathcal{P}_{e^-}}
\newcommand{\ppm}  {\mathcal{P}_{e^\pm}}
\newcommand{\ALR}{\mbox{$A_{\rm {LR}}$}}
\newcommand {\stl}  {\sin^2 \theta_{{\rm eff}}^\ell}
\newcommand\POWHEG{{\tt POWHEG}}
\newcommand\POWHEGBOX{{\tt POWHEG~BOX}}
\newcommand\PYTHIA{{\tt PYTHIA}}

\newcommand\HERWIG{{\tt HERWIG}}

\newcommand{\sintheff}{\ensuremath{\sin^2\!\theta^\ell_{\rm eff}}}

\newcommand{\abinv}{\ensuremath{\mathrm{ab}^{-1}}}
\newcommand{\fbinv}{\ensuremath{\mathrm{fb}^{-1}}}

\begin{document}

\def\bibname{References}
\bibliographystyle{plain}

\raggedbottom

\pagenumbering{roman}

\parindent=0pt
\parskip=8pt
\setlength{\evensidemargin}{0pt}
\setlength{\oddsidemargin}{0pt}
\setlength{\marginparsep}{0.0in}
\setlength{\marginparwidth}{0.0in}
\marginparpush=0pt


\pagenumbering{arabic}

\renewcommand{\chapname}{chap:intro_}
\renewcommand{\chapterdir}{.}
\renewcommand{\arraystretch}{1.25}
\addtolength{\arraycolsep}{-3pt}



\chapter{Study of Electroweak Interactions at the Energy Frontier}
\label{chap:electroweak}

\begin{center}\begin{boldmath}



\begin{center}

\begin{large} {\bf Conveners: A.~Kotwal and D.~Wackeroth} \end{large}

M.~Baak,
A.~Blondel,
A.~Bodek,
R.~Caputo,
T.~Corbett,
C.~Degrande,
O.~Eboli,
J.~Erler,
B.~Feigl,
A.~Freitas,
J.~Gonzalez Fraile,
M.C.~Gonzalez-Garcia,
J.~Haller,
J.~Han, 
S.~Heinemeyer,
A.~Hoecker,
J.~L.~Holzbauer,
S.-C.~Hsu,
B.~J\"ager,
P.~Janot,
W.~Kilian,
R.~Kogler, 
P.~Langacker,
S.~Li,
L.~Linssen,
M.~Marx,
O.~Mattelaer,
J.~Metcalfe,
K.~M\"onig,
G.~Moortgat-Pick,
M.-A.~Pleier,
C.~Pollard,
M.~Ramsey-Musolf,
M.~Rauch,
J.~Reuter,
J.~Rojo,
M.~Rominsky,
W.~Sakumoto,
M.~Schott, 
C.~Schwinn,
M.~Sekulla,
J.~Stelzer,
E.~Torrence,
A.~Vicini,
G.~Weiglein,
G.~Wilson,
L.~Zeune

\end{center}



\end{boldmath}\end{center}

\begin{center} {\bf Abstract} \end{center}

With the discovery of the Higgs boson, the spectrum of particles in
 the Standard Model (SM) is complete. It is more important than ever
 to perform precision measurements and to test for deviations from SM
 predictions in the electroweak sector. In this report, we investigate 
two themes in the arena of precision electroweak
measurements: the electroweak precision observables (EWPOs)
that test the particle content and couplings in the SM and the minimal supersymmetric SM,
and the measurements involving multiple gauge bosons in
the final state which provide unique probes of the basic tenets of
electroweak symmetry breaking.
Among the important EWPOs we focus our discussion on $M_W$ and $\sin^2\theta_\eff^\ell$ 
and on anomalous
 quartic gauge couplings probed by triboson production and vector boson
 scattering. We investigate the thresholds of precision that need to be
 achieved in order to be sensitive to new physics. We study the
 precision that can be achieved at various facilities on these
 observables. We discuss the calculational tools needed to predict SM
 rates and distributions in order to perform these measurements at the
 required precision. This report summarizes the work of the Energy Frontier Precision Study of Electroweak 
Interactions working group of the 2013
 Community Summer Study (Snowmass).


\section{Introduction}
\label{sec:electroweak-introduction}

Particle physics research at the energy frontier has entered an
exciting era: Experiments at the CERN Large Hadron Collider (LHC) are
exploring the fabric of matter at an unprecedented level of precision
and are expected to provide answers to some of the most fundamental
questions in science.  The recent discovery of a Higgs boson with
SM-like properties at the LHC marks the beginning of an exciting
journey with the goal to fully reveal the nature of the mechanism
responsible for the generation of mass and its messenger, the Higgs
boson.  Besides the study of the Higgs boson at the LHC and future
collider experiments, these experiments at the energy frontier strive
to discover new particles and to gain new insights in the fundamental
principles that govern all dynamics and properties of matter,
i.e. beyond what is described by the Standard Model (SM) of particle
physics.
   
The SM is a thoroughly tested framework for describing
electromagnetic, weak and strong interactions of the fundamental
constituents of matter, based on a symmetry principle and
mathematically formulated as a renormalizable Quantum Field Theory.
The SM successfully describes all presently observed electroweak and
strong interactions of matter particles (quarks and leptons) and of
the mediators of the fundamental forces (photon, $W$ and $Z$ bosons, and
the gluon).  Despite this enormous success of the SM, it is generally
accepted that the SM is merely a low-energy approximation to a more
fundamental theory, which is expected to reveal itself at the LHC or
at future high-energy experiments, in the form of the emergence of new,
non-SM particles and interactions. A promising candidate for a theory
beyond the SM, which also provides a dark matter candidate, is
Supersymmetry (SUSY), an additional symmetry connecting fermions and
bosons. The LHC is presently searching for signals of SUSY, and
already succeeded in excluding a  range of possible manifestations
of SUSY. While direct signals of new particles ({\it i.e.}, the
on-mass shell production of non-SM particles) may require collider
energies not yet accessible, it is possible that new physics manifests
itself first in form of small deviations between measurements and
equally precise predictions of properties of SM particles. The deviations
 can arise due to the  virtual presence of new particles in quantum loops and 
in  new amplitudes generated by their exchange at tree-level. 

This is the realm of precision electroweak physics, where well-defined
electroweak precision observables (EWPO) are being measured 
  in the interactions of $W$ and $Z$ bosons and
are equally well predicted by complex quantum-field theoretical
calculations of these quantum loop effects of SM and beyond-the-SM
(BSM) particles. The powerful concept of precision physics not only
tests the SM as a full-fledged Quantum Field Theory, but also provides
indirect access to currently unobserved sectors of the SM and
beyond. Examples of successful applications of precision physics in
the recent past include the test of the electroweak sector of the 
SM at the 0.1\% level at LEP
and the SLC~\cite{ALEPH:2010aa}, an indirect prediction of the mass of the
top quark and the SM Higgs boson prior to their discovery
respectively in $p\bar p$ collisions at the Tevatron and $pp$
collisions at the LHC, and exclusion of, or severe constraints on,
various extensions of the SM (e.~g. Technicolor). In this report, in Sec.~\ref{sec:electroweak-precision}, we
will study the potential of EWPOs measured at future high-energy
colliders for revealing signals of new physics, constraining the
parameter space of BSM models, or providing additional information
about the underlying model once a new particle is discovered.

Apart from UV-complete theories such as SUSY, 
an alternate way to indirectly search for signals of BSM physics 
is based on Effective Field Theories (EFT). 
If the new physics scale is well above the energies reached in
experiments, the new degrees of freedom cannot be produced directly and
the new physics appears only as new interactions between the known
particles. These new interactions are included in the Lagrangian as
higher-dimensional operators which are invariant under the SM symmetries and
suppressed by the new physics scale $\Lambda$,
\begin{equation}
{\cal L_{EFT}} = {\cal L}_{SM} + \sum_{d>4}\sum_i \frac{c_i}{\Lambda^{d-4}}{\cal O}_i
\label{eq:eft}
\end{equation}
where $d$ is the dimension of the operators. In the limit
$\Lambda\to\infty$, this EFT Lagrangian reduces to the SM one. Since the
$c_i$ are fixed by the complete high energy theory, any extension of
the SM can be parametrized by this Lagrangian where the coefficients
of the operators are kept as free parameters. Below the new physics
scale, only the operators with lowest dimensions can give a large
contribution and need to be kept (unless the coefficient of a higher-dimension operator is substantially larger). 
 Once truncated, the EFT Lagrangian becomes predictive even without
fixing the coefficients and parametrizes any heavy new physics
scenario. However, it should be kept in mind that this truncated
Lagrangian is only valid at energies below the new physics scale. 

EFT operators are a useful method for parameterizing the predictions of various
 strongly-interacting light Higgs (SILH) models~\cite{Giudice:2007fh}
 which describe 
 the Higgs boson  as a pseudo-Goldstone Boson arising from 
 the breaking of a larger symmetry. The lightness of the Higgs
 boson is the big question raised by the non-stability of the
  SM Higgs potential under the effect of quantum loops. While SUSY
 offers an elegant solution which is weakly-coupled and perturbative, 
 EFT operators provide a starting point for exploring strongly-coupled
 solutions to this important question. Some of these operators induce deviations
 in EWPOs and Higgs couplings as well as multi-boson production, while
 others are uniquely probed by multi-boson production. 
 
In this report, in Sec.~\ref{sec:electroweak-multiboson}, we will use the EFT approach to explore the potential of multi-boson processes at the LHC and other machines  
for providing information about the scale and the dynamics of the new physics.

\section{Electroweak precision physics}
\label{sec:electroweak-precision}

In this section we focus on two important observables, $M_W$ and
$\sin^2\theta_\eff^{\ell}$. We start with a discussion on the theoretical
predictions, followed by discussions of their measurements at various
machines.

\subsection{Uncertainties in predictions of $Z$ pole observables, $\sin^2\theta_\eff^{\ell}$ and $M_W$ }
\label{sec:electroweak-ewpo-theory}


At $e^+ e^-$ colliders, near the $Z$-peak the differential cross
section for $e^+ e^- \to f\bar{f}$ can be written as\footnote{For a
  review, see Ref.~\cite{zpole}.}
\begin{eqnarray}
\frac{d\sigma}{d\cos\theta}
 &=& {\cal R}_{\rm ini} \biggl [
  \frac{9}{2}\pi\,\frac{\Gamma_{ee}\Gamma_{ff}(1-{\cal P}_e{\cal A}_e)
   (1+\cos^2\theta) + 2({\cal A}_e-{\cal P}_e){\cal A}_f\cos\theta}%
  {(s-\MZ^2)^2-\MZ^2\Gamma_Z^2} + \sigma_{\rm non-res} \biggr ], 
  \label{eq:electroweak-xsecz} \\
{\rm where} \;
\Gamma_{ff} &=& {\cal R}^f_V g_{Vf}^2 + {\cal R}^f_A g_{Af}^2, \qquad
\Gamma_Z = \sum_f \Gamma_{ff}, \\
{\cal A}_f &=& 2\frac{g_{Vf}/g_{Af}}{1+(g_{Vf}/g_{Af})^2}
 = \frac{1-4|Q_f|\sin^2\theta^f_{\rm eff}}{1-4\sin^2\theta^f_{\rm eff} +
     8(\sin^2\theta^f_{\rm eff})^2}.
\end{eqnarray}
Here $\Gamma_Z$ is the total $Z$ decay width, $\Gamma_{ff}$ is the
partial width for the decay $Z\to f\bar{f}$, and $g_{Vf}$/$g_{Af}$ are
the effective vector/axial-vector coupling that mediate this
decay. These effective couplings include higher-order loop corrections
to the vertex, except for QED and QCD corrections to the external
$f\bar{f}$ system, which are captured by the radiator functions ${\cal
  R}^f_V$ and ${\cal R}^f_A$. The factor $\cal R_{\rm ini}$, on the
other hand, accounts for QED radiation in the
initial-state. (Specifically, as written in
Eq.~\ref{eq:electroweak-xsecz}, it describes these effects
\emph{relative} to the final-state radiation contribution for
$e^+e^-$.)

Equation~\ref{eq:electroweak-xsecz} explicitly spells out the leading
$Z$-pole contribution, while additional effects from photon exchange
and box corrections are included in the remainder $\sigma_{\rm
  non-res}$.

The ratio of the vector and axial vector couplings of fermions to the
$Z$ boson, $g_{Vf}$ and $g_{Af}$, is commonly parametrized through the
effective weak mixing angle $\sin^2\theta^f_{\rm eff}$.  At $e^+ e^-$
colliders it can be determined from the angular distribution of
fermions in $e^+ e^- \to f \bar f$ processes with respect to the
scattering angle $\cos\theta$ or from the dependence on the initial
electron polarization ${\cal P}_e$:
\begin{eqnarray}
A_{\rm FB} &\equiv& \frac{\sigma(\cos\theta > 0) - \sigma(\cos\theta < 0)}%
{\sigma(\cos\theta > 0) + \sigma(\cos\theta < 0)}
 = {\cal R}_{\rm FB} \, {\textstyle\frac{3}{4}} {\cal A}_e {\cal A}_f, \\
A_{\rm LR} &\equiv& \frac{\sigma({\cal P}_e > 0) - \sigma({\cal P}_e < 0)}%
{\sigma({\cal P}_e > 0) + \sigma({\cal P}_e < 0)} = {\cal A}_e.
\end{eqnarray}
${\cal R}_{\rm FB}$ accounts for QCD and QED corrections.
The total cross-section, decay width $\Gamma_Z$, and branching ratios
of the $Z$ boson are measured from the rates and lineshape of the
cross sections $\sigma_{e^+ e^- \to f\bar{f}}(s)$ on the $Z$ pole
($\sqrt{s} = \MZ$) and for at least one value of $\sqrt{s}$ each above
and below the pole ($\sqrt{s} = \MZ\pm\Delta E$).

At hadron colliders, the effective weak mixing angle can be determined
from the forward-backward asymmetry of the process $q\bar{q} \to
\ell^+\ell^-$ ($\ell=e,\mu$) near the $Z$ pole. However, one cannot
determine on a event-by-event basis from which side the quark and the
antiquark were coming. For a $p\bar{p}$ initial state, it is generally
assumed that the (anti)quark originated from the (anti)proton,
respectively, and the dilution effect from the opposite possibility is
evaluated based on Monte-Carlo simulations \cite{sintev}. For a $pp$
initial state, the boost direction of the $\ell^+\ell^-$ system is
defined as the quark direction \cite{sinlhc}, based on the observation
that the valence quarks from the proton tend to be more energetic than
the sea antiquarks. Again, dilution effects from the wrong
quark-antiquark assignment are studied with Monte-Carlo generators.

Due to the high precision of the experimental measurements for these
observables, much effort has gone into their theoretical calculation
within the Standard Model (SM).

The effective weak mixing angle can be written as
\begin{equation}
  \sin^2\theta^f_{\rm eff} = \frac{1}{4 |Q_f|}(1+{\cal R}e\frac{g_{Vf}}{g_{Af}})=(1-\frac{M_W^2}{M_Z^2})(1+\Delta\kappa),
\end{equation}
where $\Delta\kappa$ denotes the contribution from radiative
corrections. At tree level, $\Delta\kappa=0$ and the effective weak
mixing angle coincides with  the so-called {\em on-shell}
weak mixing angle.

For leptonic final states, the effective weak mixing angle
$\sin^2\theta^\ell_{\rm eff}$ has been calculated to the complete
two-loop order \cite{sineffl2,sineffl2a}, and 3- and 4-loop
corrections of order ${\cal O}(\alpha\alpha_s^2)$ \cite{sineffl3} and
${\cal O}(\alpha\alpha_{\rm s}^3)$ \cite{sineffl4} are also
known. Furthermore, the leading ${\cal O}(\alpha^3)$ and ${\cal
  O}(\alpha^2\alpha_s)$ contributions for large values of $m_t$
\cite{sineffmt} or $m_H$ \cite{sineffmh} have been computed.

The current uncertainty from unknown higher orders is estimated to
amount to about $4.5\times 10^{-5}$ \cite{sineffl2a}, which mainly
stems from missing ${\cal O}(\alpha^2\alpha_s)$ and ${\cal
  O}(N_f^2\alpha^3,\,N_f^3\alpha^3)$ contributions beyond the leading
$m_t^4$ and $m_t^6$ terms, respectively. (Here $N_f^n$ denotes
diagrams with $n$ closed fermion loops. Based on experience from lower
orders, the ${\cal O}(\alpha^3)$ diagrams with several closed fermion
loops are expected to be dominant.) The calculation of these
corrections requires three-loop vertex integrals with self-energy
sub-loops and general three-loop self-energy integrals, which
realistically can be expected to be worked out in the foreseeable
future. The remaining ${\cal O}(\alpha^3)$ and four-loop terms are
tentatively estimated to amount to $\approx 10^{-5}$. This estimate
should improve when the aforementioned calculations have been
completed.

When extracting $\sin^2\theta^\ell_{\rm eff}$ from $A_{\rm FB}$ and
$A_{\rm LR}$, the initial- and final-state QED radiator functions
${\cal R}_i$ must be taken into account. In general, the QED
corrections are known to ${\cal O}(\alpha)$ for the differential cross
section and to ${\cal O}(\alpha^2)$ for the integrated cross section
(see Ref.~\cite{zfitter} for a summary). However, for the LR asymmetry
these corrections completely cancel up to NNLO \cite{QEDLR}, while for the FB
asymmetry they cancel if hard photon contributions are excluded, i.e.\
they cancel up to terms of order $E_\gamma/\sqrt{s}$ \cite{QEDLR,
QEDFB, hollikee}.  Therefore, a sufficiently precise result for the
soft-photon contribution with $E_\gamma < E_\gamma^{\rm cut}$ can be
obtained using existing calculations for small enough $E_\gamma^{\rm
cut}$, while the hard-photon contribution ($E_\gamma > E_\gamma^{\rm
cut}$) can be evaluated with numerical Monte-Carlo methods.

Other important $Z$ pole observables are $R_b$ and the $Z$ width.  For
the branching fraction $R_b = \Gamma_b/\Gamma_{\rm had}$, two-loop
corrections of ${\cal O}(\alpha\alpha_s)$, ${\cal O}(N_f\alpha^2)$,
and ${\cal O}(N_f^2\alpha^2)$ are known \cite{Zqqaas,rb}. Assuming
geometric progression of the perturbative series, the remaining
higher-order contributions are estimated to contribute at the level of
$\sim 2\times 10^{-4}$. As before, the contribution from electroweak
two-loop diagrams without closed fermion loops is expected to be
small. The dominant missing contributions are the same as for
$\sin^2\theta^q_{\rm eff}$.

For the total width $\Gamma_Z$, only an approximate result for the
electroweak two-loop corrections in the limit of large $m_t$ is known
\cite{gzmt}.  The remaining ${\cal O}(N_f\alpha^2)$ may be relatively
large.  As deduced from the size of individual vertex factors in the
calculation of $R_b$ \cite{rb}, and assuming that there are no
relevant cancellations between them, the uncertainty of $\Gamma_Z$
associated with these corrections is estimated to be a few MeV, which
is by far dominant compared to missing three-loop contributions.
However, the ${\cal O}(N_f\alpha^2)$ correction can be computed with
existing methods without conceptual difficulties.

Besides $Z$-pole observables, the $W$-boson mass, $\MW$ plays an
important role for electroweak precision physics. Theoretically, it
can be predicted from the muon decay rate. After subtraction of QED
radiation effects \cite{muqed}, muon decay can be described by an
effective four-fermion interaction with the Fermi coupling constant
$G_\mu$, which in the SM is given by
\begin{equation}
\frac{G_\mu}{\sqrt{2}} = \frac{\pi\alpha\MZ^2}{2\MW^2(\MZ^2-\MW^2)}(1+\Delta r),
\end{equation}
where $\Delta r$ summarizes the electroweak (non-QED) higher-order
corrections. This equation can be solved numerically for $\MW$.

Within the SM, $\MW$ has been computed including full two-loop
corrections \cite{mw2,Awramik:2003rn} and leading 3- and 4-loop
corrections \cite{sineffl3,sineffl4,sineffmt}. The intrinsic
theoretical error is estimated to be about 4~MeV, mostly due from
missing ${\cal O}(\alpha^2\alpha_s)$ and ${\cal
  O}(N_f^2\alpha^3,\,N_f^3\alpha^3)$ contributions beyond the
leading-$\mt$ approximation. Inclusion of these effects, which would
require the computation of the 3-loop self-energies, would reduce the
perturbative error to less than 1~MeV.

The current status of the theoretical calculations and prospects for
the near future are summarized in Tab.~\ref{sumz}.  Note that
$\sigma_{\rm non-res}$ in Eq.~\ref{eq:electroweak-xsecz} is suppressed
by $\Gamma_Z/\MZ$ compared to the leading pole term, so that the known
one-loop corrections are sufficient to reach NNLO precision at the $Z$
pole.

\newcommand{\lesim}{\,\raisebox{-.1ex}{$_{\textstyle<}\atop^{\textstyle\sim}$}\,}

\begin{table}[t]
\renewcommand{\arraystretch}{1.3}
\centering
\begin{tabular}{lrlc}
\hline
Quantity & Current theory error & Leading missing terms & Est.\ future theory
error
\\
\hline\hline
$\sin^2\theta^\ell_{\rm eff}$ & $4.5\times 10^{-5}\qquad$ & 
 ${\cal O}(\alpha^2\alpha_s)$, ${\cal O}(N_f^{\ge 2}\alpha^3)$ & $1...1.5\times
 10^{-5}$ \\ 
\hline
$R_b$ & $\sim 2\times 10^{-4}\qquad$ & ${\cal O}(\alpha^2)$, 
 ${\cal O}(N_f^{\ge 2}\alpha^3)$ & $\sim 1\times 10^{-4}$ \\
\hline
$\Gamma_Z$ & few MeV$\qquad$ & ${\cal O}(\alpha^2)$, 
 ${\cal O}(N_f^{\ge 2}\alpha^3)$ & $<1$~MeV \\
\hline
$\MW$ & 4~MeV$\qquad$ & ${\cal O}(\alpha^2\alpha_s)$, ${\cal O}(N_f^{\ge 2}\alpha^3)$ & $\lesim 1$~MeV \\ 
\hline
\end{tabular}
\caption{Some of the most important precision observables for $Z$-boson
production and decay and the $W$ mass (first column), their present-day estimated theory error
(second column), the dominant missing higher-order corrections (third column),
and the estimated improvement when these corrections are available (fourth
column). In many cases, the leading parts in a large-mass expansion are already
known, in which case the third column refers to the remaining pieces at the
given order. The numbers in the last column are rough order-of-magnitude
guesses.}
\label{sumz}
\end{table}

The known corrections to the effective weak mixing angles and the
partial widths are implemented in programs such as {\tt
Zfitter} \cite{zfitter,zfitter2} and {\tt
Gfitter}~\cite{gfitter}. However, these programs are based on a
framework designed for NLO but not NNLO corrections. In particular,
there are mismatches between the electroweak NNLO corrections to the
$Zf\bar{f}$ vertices and QED/QCD corrections to the external legs due
to approximations and factorization assumptions. Another problem is
the separation of leading and sub-leading pole terms in
Eq.~\ref{eq:electroweak-xsecz} \cite{sineffl2a}. While these
discrepancies may be numerically small, it would be desirable to
construct a new framework that treats the radiative corrections to
$Z$-pole physics systematically and consistently at the NNLO level and
beyond. Such a framework can be established based on the pole
scheme \cite{pole}, where the amplitude is expanded about the complex
pole $s=\MZ^2-i \MZ\Gamma_Z$, with the power counting
$\Gamma_Z/\MZ \sim
\alpha$.

In addition to intrinsic theoretical error, the predictions of
$\sin^2\theta^\ell_{\rm eff}$, $\MW$, \emph{etc.} also depend on input
parameters and their experimental uncertainties.  The parametric
uncertainties in the currently best SM predictions for $M_W$ and
$\sin^2\theta_\eff^{l}$ due to uncertainties in $m_t,
\Delta\alpha_{had}^{(5)}$ and $M_Z$ of
Tables~\ref{tab:electroweak-mwsin2tpresent},\ref{tab:electroweak-mwsin2tfuture}
have been determined with the help of the parametrization formulae of
Ref.~\cite{Awramik:2003rn} for $\MW$ and of Ref.~\cite{sineffl2a} for
$\sin^2\theta^\ell_{\rm eff}$.

Two recent determinations of the five-quark hadronic contribution to
$\alpha(M_Z)$ find $\Delta \alpha_{\rm had}^{(5)}(M_Z)=(275.7 \pm 1.0)
\times 10^{-4}$~\cite{Davier:2010nc} and $\Delta\alpha_{\rm
  had}^{(5)}(M_Z)=(276.26 \pm 1.38) \times
10^{-4}$~\cite{Hagiwara:2011af}. The residual theory uncertainties due
to missing higher order corrections as listed in
Table~\ref{tab:electroweak-mwsin2tpresent} have been taken from
Refs.~\cite{Awramik:2003rn,sineffl2a}.  Using the following measured
values in the calculation of $M_W$ and $\sin^2\theta_{\eff}^{l}$:
$\mt=173.2 \pm 0.9$ GeV~\cite{mtop}, $\alpha_s(M_Z)=0.1184 \pm
0.0007$~\cite{pdg}, $\MZ=91.1876 \pm 0.0021$ GeV~\cite{zpole}, and
$\MH=125 \pm 1$~GeV, one finds $M_W=80.3603 \pm 0.0076$ GeV and
$\sin^2\theta_{\eff}^{l}=0.23127 \pm 0.00007$ for $\Delta \alpha_{\rm
  had}^{(5)}(M_Z)=(276.26 \pm 1.38)\times
10^{-4}$~\cite{Hagiwara:2011af} and $M_W=80.3614 \pm 0.0074$ GeV and
$\sin^2\theta_{\eff}^{l}=0.23129 \pm 0.00007$ for $\Delta \alpha_{\rm
  had}^{(5)}(M_Z)=(275.7 \pm 1.0) \times
10^{-4}$~\cite{Davier:2010nc}.
\begin{table}[h]
\begin{center}
\begin{tabular}{l|cp{2.5cm}ccc}\hline 
 & $\Delta m_t=0.9$ GeV  & 
$\Delta (\Delta \alpha_{\rm had}^{(5)})=1.38(1.0)  \times  10^{-4}$ & $\Delta M_Z=2.1$ MeV &  missing h.o. & total \\ 
\hline \hline
  $\Delta M_W$ [MeV] & 5.4  & 2.5 (1.8)  &  2.6  &  4.0 & 7.6 (7.4)\\
  $\Delta \sin^2\theta^\ell_\eff [10^{-5}]$ &  2.8  &  4.8 (3.5)   & 1.5  &  4.5  & 7.3 (6.5)\\ \hline
\end{tabular}
\caption{Current parametric and theory uncertainties of SM predictions of $M_W$ and $\sin^2\theta^\ell_{\eff}$.}
\label{tab:electroweak-mwsin2tpresent}
\end{center}
\end{table}

Table~\ref{tab:electroweak-mwsin2tfuture} clearly shows that
measurements of $M_W$ at the few MeV level, and
$\sin^2 \theta_\eff^\ell$ at the level of $10^{-5}$, requires
significant improvements in the parametric uncertainties from
$m_{t}, \Delta \alpha_{\rm had}^{(5)}$ and $M_Z$ as well as in higher
order calculations. Parametric uncertainties from $m_{top}$ and $\Delta
  \alpha_{had}$, if reduced by a factor of 2 compared to current
  uncertainties, will prevent them from exceeding the anticipated
  total precision on $M_W$ at the LHC. At the ILC and TLEP a factor of
  5 and 10 improvement, respectively, in the parametric uncertainties
  is needed, which is only achievable if the precision on $M_Z$ is
  considerably improved as well. TLEP can improve the $M_Z$ precision
  by a factor of at least 10 (for a discussion of prospects for an improvement
in $\Delta M_Z$ see Section~\ref{sec:otherewpos}). 
The LHC may be able to achieve $\Delta m_{t} \sim 0.5$ GeV but further progress at
the LHC will likely be limited by theoretical uncertainties in the
non-perturbative QCD effects associated with translating the
kinematically-reconstructed $m_{t}$ to the pole mass. We refer to the
summary report of the Snowmass working group on {\em Fully
understanding the top quark} for a detailed discussion of the
prospects for future precision $m_t$ measurements.
\begin{table}[h]
\begin{center}
\begin{tabular}{l|cp{2.0cm}ccc}\hline  
 & $\Delta m_t=0.5(0.1)$ GeV  & $\Delta (\Delta \alpha_{\rm had}^{(5)})=5 \times 10^{-5}$ & $\Delta M_Z=2.1$ MeV & missing h.o. & total \\ 
\hline \hline
 $\Delta M_W$ [MeV] & 3.0 (0.6)  & 1.0  &  2.6  &  1.0 & 4.2 (3.0) \\
 $\Delta \sin^2\theta^\ell_{\rm eff} [10^{-5}]$ &  1.6 (0.3)  &  1.8   & 1.5  &  1.0  &  3.0 (2.6)\\ \hline
\end{tabular}
\caption{Anticipated parametric and theory uncertainties of SM predictions of $M_W$ and $\sin^2\theta^\ell_{\eff}$.}
\label{tab:electroweak-mwsin2tfuture}
\end{center}
\end{table}

In many new physics models, the leading contributions beyond the SM to electroweak precision observables can be described by the {\em oblique} parameters $S,T,U$ \cite{Peskin:1991sw}:
\begin{eqnarray}
\Delta r &\approx& \Delta r^{\rm SM} + \frac{\alpha}{2s_W^2}\Delta S - \frac{\alpha c_W^2}{s_W^2}\Delta T + \frac{s_W^2-c_W^2}{4s_W^4} \Delta U, \\
\sin^2\theta^\ell_\eff &\approx& (\sin^2\theta^\ell_\eff)^{\rm SM}
 + \frac{\alpha}{4(c_W^2-s_W^2)}\Delta S - \frac{\alpha s_W^2c_W^2}{c_W^2-s_W^2}\Delta T,
\end{eqnarray}
where $s_W^2=1-c_W^2=1-M_W^2/M_Z^2$, and $S,T,U$ are given at 1-loop level in terms of the transverse parts of 1-PI gauge boson self-energies, $\Pi_{VV'}$:
\begin{eqnarray}
\alpha S &=& 4s_W^2c_W^2 \left [
 \Pi'_{ZZ}(0) - \frac{c_W^2-s_W^2}{s_Wc_W}\Pi'_{Z\gamma} -
 \Pi'_{\gamma\gamma}(0)\right], \\
\alpha T &=& \Delta\rho = \frac{\Pi_{WW}(0)}{\MW^2}-\frac{\Pi_{ZZ}(0)}{\MZ^2},\\
\alpha U &=& 4s_W^2 \left [
 \Pi'_{WW}(0) - c_W^2 \Pi'_{ZZ}(0) - 2s_Wc_W\Pi'_{Z\gamma} - s_W^2\Pi'_{\gamma\gamma}(0) \right].
\end{eqnarray}
Note that $\Delta T=0$ ($\Delta\rho=0$) and $\Delta U=0$ in case of
exact custodial $SU(2)$ symmetry.  A discussion of the current and
expected precision in the determination of the oblique parameters~$S$
and $T$, for a fixed value of $U=0$, can be found in
Section~\ref{sec:electroweak-ewpo-gfitter}.

\subsection{Uncertainties in measurements of $M_W$ and $\sin^2\theta_{\eff}^\ell$ at hadron colliders}
\label{sec:electroweak-had}

In this section we discuss the theoretical inputs and the experimental
aspects of measuring these EWPOs at hadron colliders.

\subsubsection{Theory and PDF aspects: $M_W$}
\label{sec:electroweak-had-theory}


In hadronic collisions, the $W$ boson mass can be determined from the
transverse mass distribution of the lepton pair, $m_T(l\nu)$,
originating from the $W$ boson decay, $W\to\ell\nu$, and the transverse
momentum distribution of the charged lepton or neutrino.  Both QCD and
electroweak (EW) corrections play an important role in the measurement
of $W$ boson observables at hadron colliders. For the anticipated
experimental precision in the measurement of $M_W$ at the Tevatron and
the LHC, as presented in Section~\ref{sec:electroweak-had-mw-exp}, it
is imperative to control predictions for the relevant observables at
the per-mille level.  For instance, the transverse momentum
distribution of the $W$ boson is an important ingredient in the
current $W$ mass measurement at the Tevatron (see,
e.~g., Ref.~\cite{Kotwal:2008zz} for a review).  In lowest order (LO) in
perturbation theory, the $W$ boson is produced without any transverse
momentum. Only when QCD corrections are taken into account does the
$W$ boson acquire a non-negligible transverse momentum, $p_T^W$. For a
detailed understanding of the $p_T^W$ distribution, it is necessary to
resum the soft gluon emission terms, and to model non-perturbative QCD
corrections. This has either be done using calculations targeted
specifically for resummation and parametrizing non-perturbative
effects (see, e.~g., Refs.~\cite{Balazs:1997xd}
and~\cite{Landry:1999an}), or interfacing a calculation of $W$ boson
production at next-to-leading order (NLO) in QCD with a parton-shower
Monte Carlo (MC) program and tuning the parameters used to describe
the non-perturbative effects. The latter approach has been pursued in
Refs.~\cite{Frixione:2002ik,Alioli:2008gx,Hamilton:2008pd}.  Fixed
higher-order QCD corrections to fully differential distributions in
$W$ boson production are known through next-to-next-to-leading
order~\cite{Anastasiou:2003ds,Melnikov:2006di,Catani:2009sm}.

While QCD corrections only indirectly affect the $W$ mass extracted
from the $m_T(l\nu)$ distribution, QED radiative corrections can
considerably distort the shape of this distribution in the region
sensitive to the $W$ mass. For instance, final-state photon radiation
is known to shift $M_W$ by ${\cal O}(100$~MeV)~
\cite{Aaltonen:2007ps,Abazov:2009cp,cdfwmass,d0wmass,unknown:2003sv,
  Abe:1994qn,Affolder:2000mt,Abazov:2002xj}.  In the last few years,
significant progress in providing predictions including EW corrections
to $W$ boson production in hadronic collisions has been made. The
complete ${\cal O}(\alpha)$ EW radiative corrections to
$p\,p\hskip-7pt\hbox{$^{^{(\!-\!)}}$} \to W^{\pm} \to\ell^{\pm} \nu$
($\ell=e,\,\mu$) were calculated by several
groups~\cite{Wackeroth:1996hz,Baur:1998kt,Dittmaier:2001ay,Baur:2004ig,
  Arbuzov:2005dd,CarloniCalame:2006zq,Zykunov:2008zz} and found to
agree ~\cite{Buttar:2006zd,Gerber:2007xk}. First steps towards going
beyond fixed order in QED radiative corrections in $W$ boson
production were taken in
Refs.~\cite{CarloniCalame:2003ux,Placzek:2003zg,Golonka:2005pn,Hamilton:2006xz,Brensing:2007qm}
by including the effects of final-state multiple photon radiation. For
a review of the state-of-the-art of predictions for $W$ and $Z$ boson
production at hadron colliders see, e.~g.,
Refs.~\cite{Buttar:2006zd,Gerber:2007xk,Laenen:2009zz}.  Given the
anticipated accuracy of the $W$ boson mass measurement at the Tevatron
and the LHC, it is necessary to not only fully understand and control
separately higher-order QCD and EW corrections, but also their
combined effects.  A first study of combined effects can be found in
Ref.~\cite{Cao:2004yy}, where final-state photon radiation was added
to a calculation of $W$ boson production which includes NLO and
resummed QCD corrections. This study showed that the difference in the
effects of EW corrections in the presence of QCD corrections and of
simply adding the two predictions may be not negligible in view of the
anticipated precision. Moreover, in the relevant kinematic region,
i.e.  around the Jacobean peak, the QCD corrections tend to compensate
some of the effects of the EW corrections.  In
Ref.~\cite{Balossini:2009sa} the full set of EW ${\cal O}(\alpha)$
corrections of {\tt HORACE}~\cite{CarloniCalame:2006zq} was combined
with the NLO QCD corrections to $W$ boson production simulated by the
generator {\tt MC@NLO} ~\cite{Frixione:2002ik}, which is matched with
the parton-shower MC program {\tt Herwig}~\cite{Corcella:2000bw}. The
results of a combination of the EW ${\cal O}(\alpha)$ corrections to
$W$ and $Z$ boson production as implemented in {\tt
  SANC}~\cite{Arbuzov:2005dd,Arbuzov:2007db} with {\tt
  Pythia}~\cite{Sjostrand:2006za} and {\tt Herwig} can be found in
Ref.~\cite{Richardson:2010gz}, without, however, performing a matching
of NLO QCD corrections to the parton shower.  Recently, the complete
EW ${\cal O}(\alpha)$ radiative corrections to $W$ and $Z$ boson
production became available in {\tt
  POWHEG}~\cite{Bernaciak:2012hj,Barze:2012tt,Barze:2013yca}, which
allows one to study the effects of NLO EW corrections in the presence of
QCD radiation and with both {\tt Pythia} and {\tt Herwig}. However,
this approach can only capture part of the two-loop mixed QCD-EW
corrections, and only a complete 2-loop calculation of ${\cal O}
(\alpha \alpha_s)$ corrections will provide a reliable estimate of the
theoretical uncertainty due to missing higher-order corrections. In
view of recent improvements in the calculation of two-loop
corrections~\cite{robertprivate}, it is reasonable to expect that
these calculations are available at the timescale of a final LHC
measurement of $M_W$~\footnote{See, e.~g., Ref.\cite{Kilgore:2011pa}
  for a calculation of the two-loop virtual mixed
  QCD$\times$QCD corrections to the Drell-Yan process, 
  and Ref.~\cite{stefanradcorr} for a recent discussion of factorizable contributions.}.

The  PDF uncertainty must be considerably reduced for a target
uncertainty of $\Delta M_W=9~(5)$~MeV at the Tevatron (LHC).
An analysis performed in Ref.~\cite{Bozzi:2011ww} has shown that the LHC
measurements of the $W$ charge asymmetry using the 2012 data has the
potential to reduce the $W$ mass uncertainty at the Tevatron by about
a factor of two (see also Section~\ref{sec:electroweak-had-mw-exp}).
But none of the PDF sets studied in Ref.~\cite{Bozzi:2011ww}
included the recent constraints from the LHC measurements of electroweak
boson production, which constrain PDFs for the same flavor combinations and
the same kinematic regions relevant for $M_W$ determinations.
In order to update this study taking this information into account, we
have repeated the determination of $\Delta M_W$ at NLO-QCD (with {\tt
  DYNNLO} for the theory modeling) using the same method as in
Ref.~\cite{Bozzi:2011ww} but using now the
NNPDF2.3~\cite{Ball:2012cx}.  This set is particularly suited for the
determination of the $W$ mass at the LHC since it already includes
constraints from $W$ and $Z/\gamma^*$ production data from ATLAS, CMS
and LHCb.
In order to reduce the statistical fluctuations, a dedicated set of
$N_{\rm rep}=1000$ replicas has been produced and used to compute the
theory predictions for the $W$ transverse mass distribution at
NLO-QCD~\cite{Rojo:2013nia}.
All of the results shown below correspond to $W^+$ boson production, but
we know from~\cite{Bozzi:2011ww} that they should apply as well for
$W^-$ boson production.

\begin{figure}[h]
\centering
\includegraphics[scale=0.42]{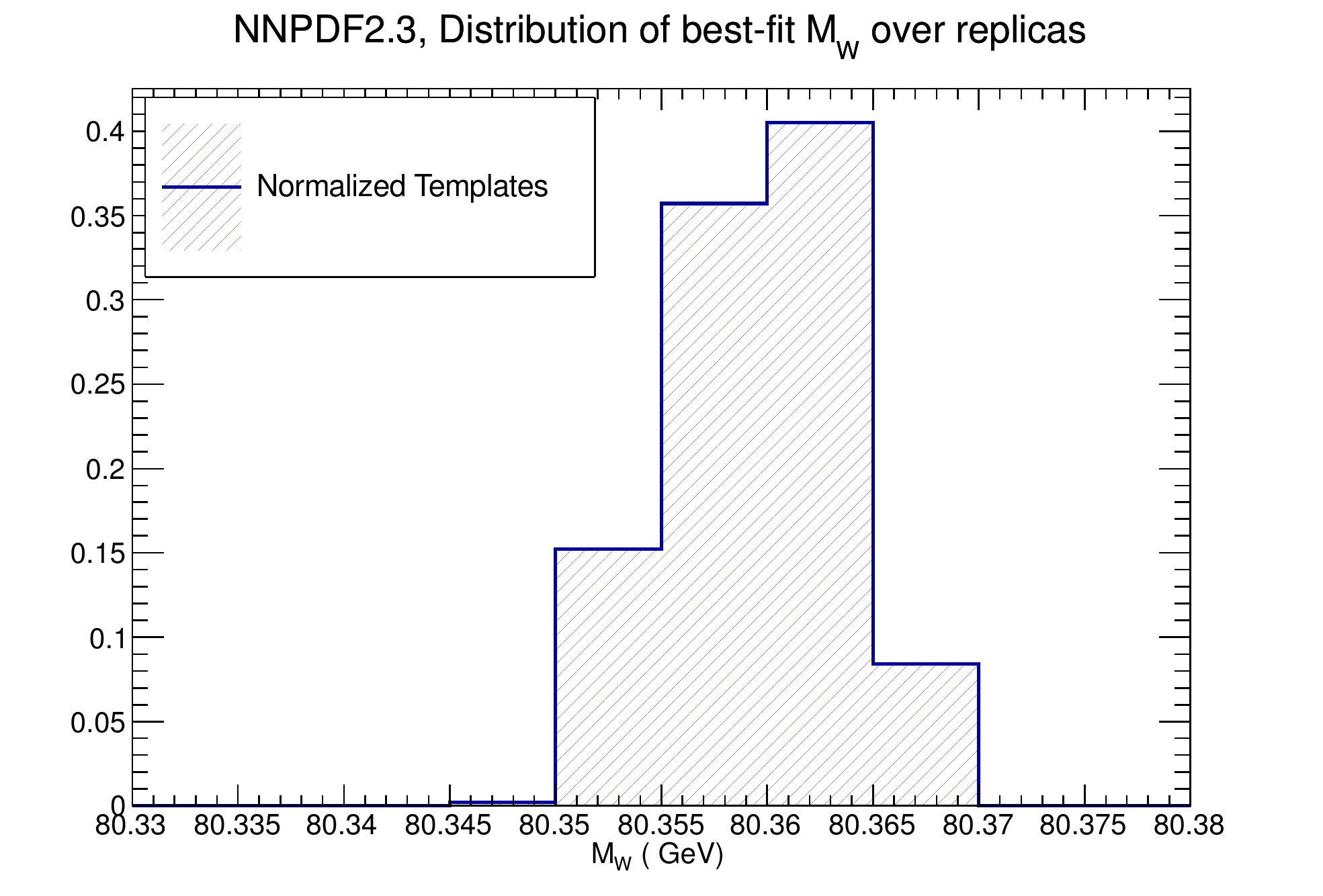}
\caption{\small The distribution of the best-fit $M_W$ value
obtained from the comparison of the 1000 replicas of NNPDF2.3
with the reference templates. The mean and the width value extracted from this histogram is $M_W\pm \delta_{\rm PDF}=80.3605 \pm 0.0051$~GeV.}
\label{fig:histo}
\end{figure}

%

For each of the $N_{\rm rep}=1000$ PDF replicas, the fit determines
which is the value of $M_W$ which maximizes the agreement with the
template distributions. The distribution of best-fit $M_W$ over the
NNPDF2.3 replicas for normalized
distributions is shown in Fig.~\ref{fig:histo}.
%
%
%
Note that our procedure entails a methodological uncertainty due to the finite
statistics used to generate the templates, which is not explicitly accounted for here. 
Note also that our estimate of $\delta_{\rm PDF}=5$ MeV applies to the
$m_T^W$ distribution fits only, and to derive the final estimate we
would need as well the results obtained with the MSTW~\cite{Martin:2009iq} 
and CT10~\cite{Lai:2010vv}
sets. Finally, the PDF uncertainty depends on the details of the
experimental distribution being fit, and therefore the effects of the
experimental resolution are also important.  Resolution effects tend
to increase the PDF uncertainty.

\begin{figure}[h]
\centering
\includegraphics[scale=0.37]{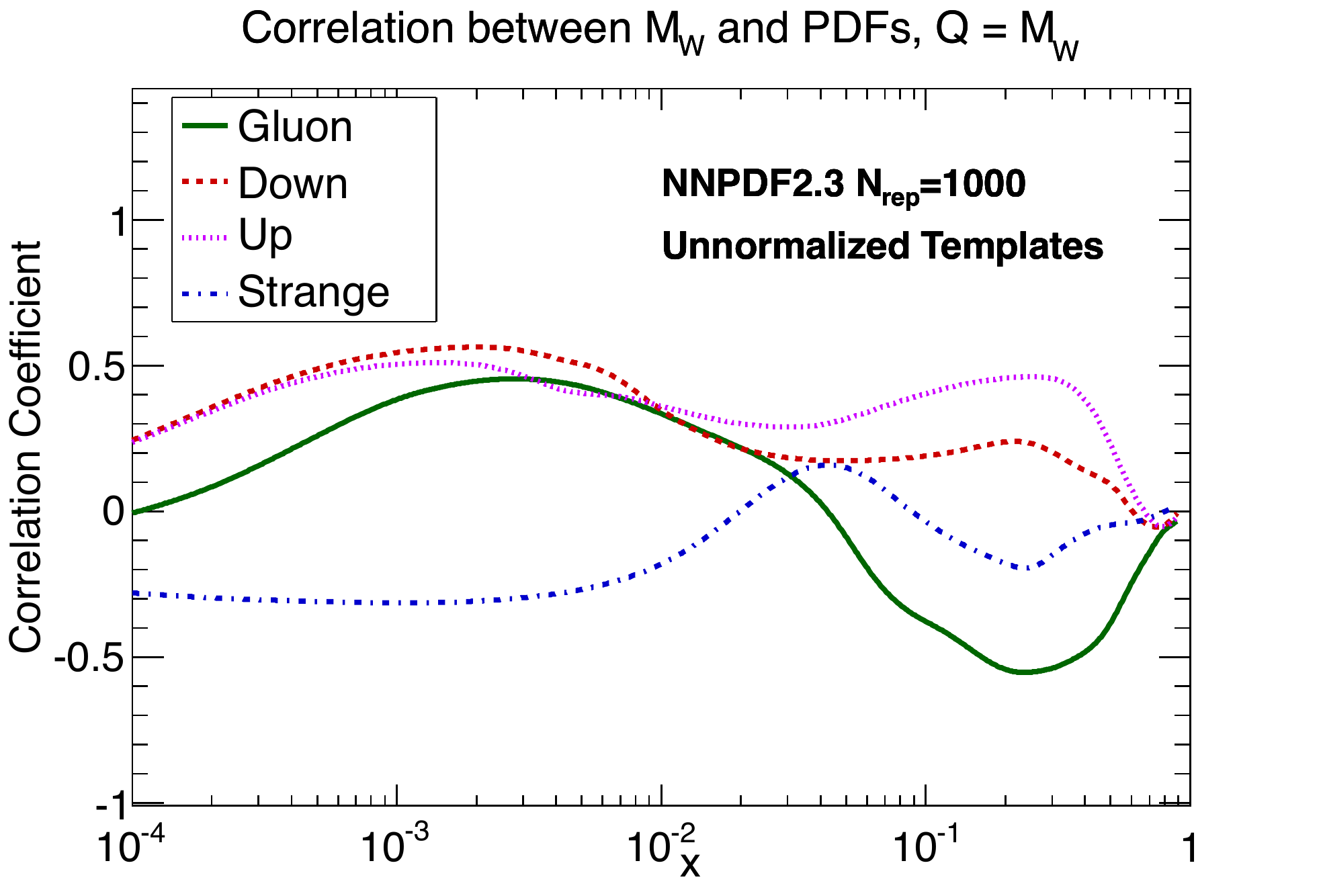}
\includegraphics[scale=0.37]{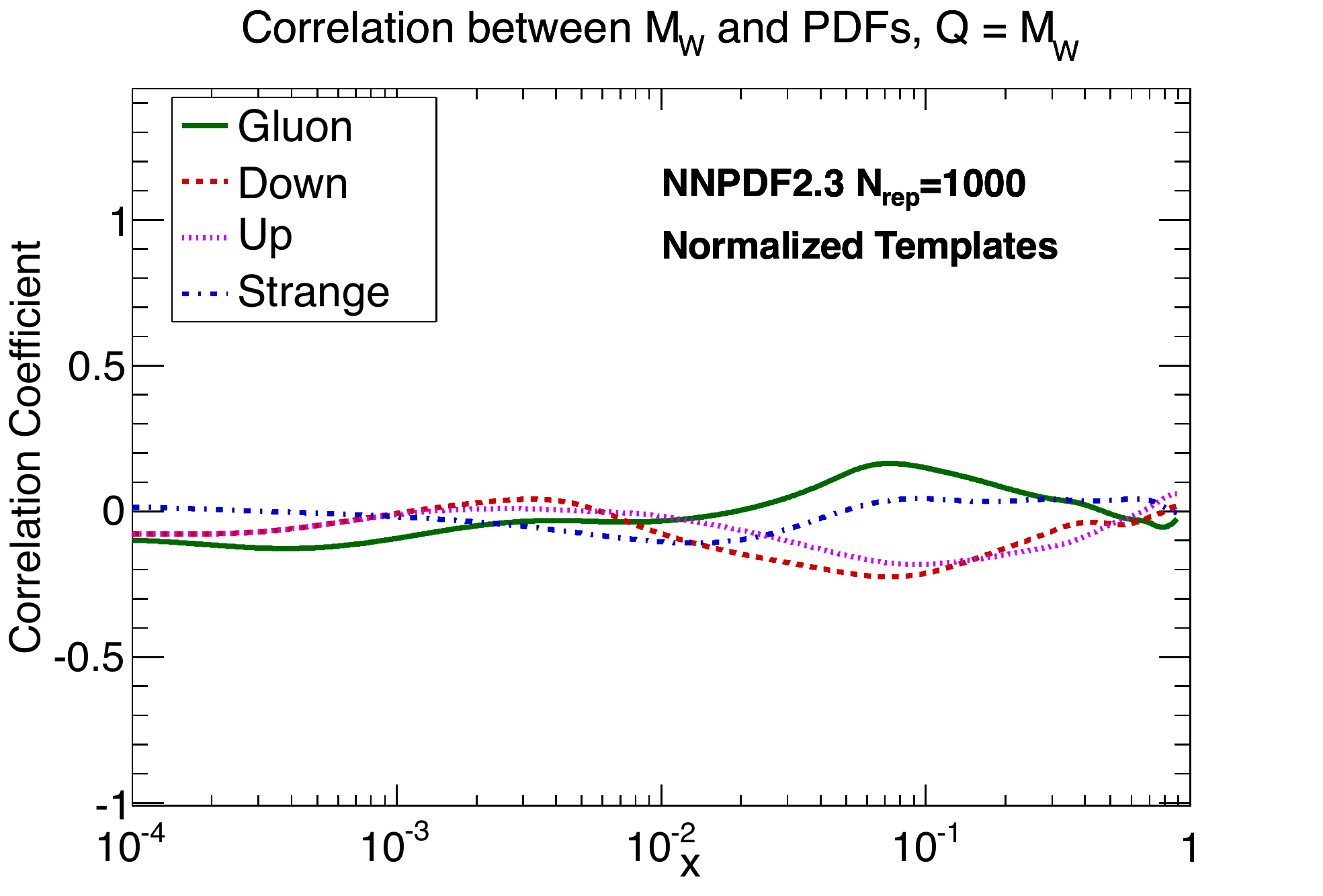}
\caption{\small  Correlations between different PDF flavours
and the $M_W$ determination at LHC 7 TeV, as a function of
Bjorken-$x$, for unnormalized (left plot) and normalized (right plot)
templates.
Note that experimental measurements always use normalized templates, and thus only the results obtained with these templates are of relevance. 
The predictions from the 1000 replicas of NNPDF2.3 have been used in
the computation.  }
\label{fig:correlations}
\end{figure}

In order to determine if a particular PDF combination is responsible
for the bulk of the PDF uncertainties in $M_W$, it is useful to compute the
correlations~\cite{Demartin:2010er} 
between the $N_{\rm rep}=1000$ PDF replicas of NNPDF2.3 and the 1000
determinations of $M_W$ obtained from the template fits for each replica.
The results are shown in Fig.~\ref{fig:correlations} for the unnormalized
templates (left plot) and for the normalized templates (right plot).
In the case of the unnormalized templates, the correlation between PDFs and $M_W$
is similar to the case of the inclusive $W^+$ cross section~\cite{Ball:2011mu}.
On the other hand, for the normalized templates the correlations are
much smaller, showing that the normalization effectively decorrelates
the $M_W$ fits with respect to the PDFs. Since experimental
measurements always use normalized templates, this is the result of
relevance.
It is clear that there is not a particular range of Bjorken-$x$ or a
particular quark flavor that dominates the $M_W$ measurement.
This implies that, in order to further reduce the PDF uncertainty in
the $M_W$ measurement from $m_T^W$, one needs new data constraining
all quark flavors and gluons in the broadest possible $x$ range.

The results that we have just discussed were based on the
determination of $M_W$ from the $W$ transverse mass distribution.
This distribution receives small higher-order QCD corrections,
but its accurate determination at the LHC will be challenging,
to achieve a competitive $M_W$ measurement.
Next we report on preliminary work towards the extension of the results
of~\cite{Bozzi:2011ww} to a template-fit analysis of the lepton
transverse momentum distribution, which has been successfully used at
the Tevatron to measure $M_W$.

As opposed to the transverse mass distribution, the
lepton transverse momentum, $p_T^l$ is substantially modified by
higher-order QCD corrections, given its strong correlation with the
$W$ boson transverse momentum, $p_T^W$, which vanishes at the Born
level.
For this distribution the use of resummed calculations for the $W$
boson $p_T^W$ is required, either using analytical $p_T^W$ resummation
or NLO-QCD calculations matched to QCD parton showers, with a
significant increase in the amount of CPU time needed to generate the
theory templates.

\begin{figure}[h]
\centering
\includegraphics[scale=0.15]{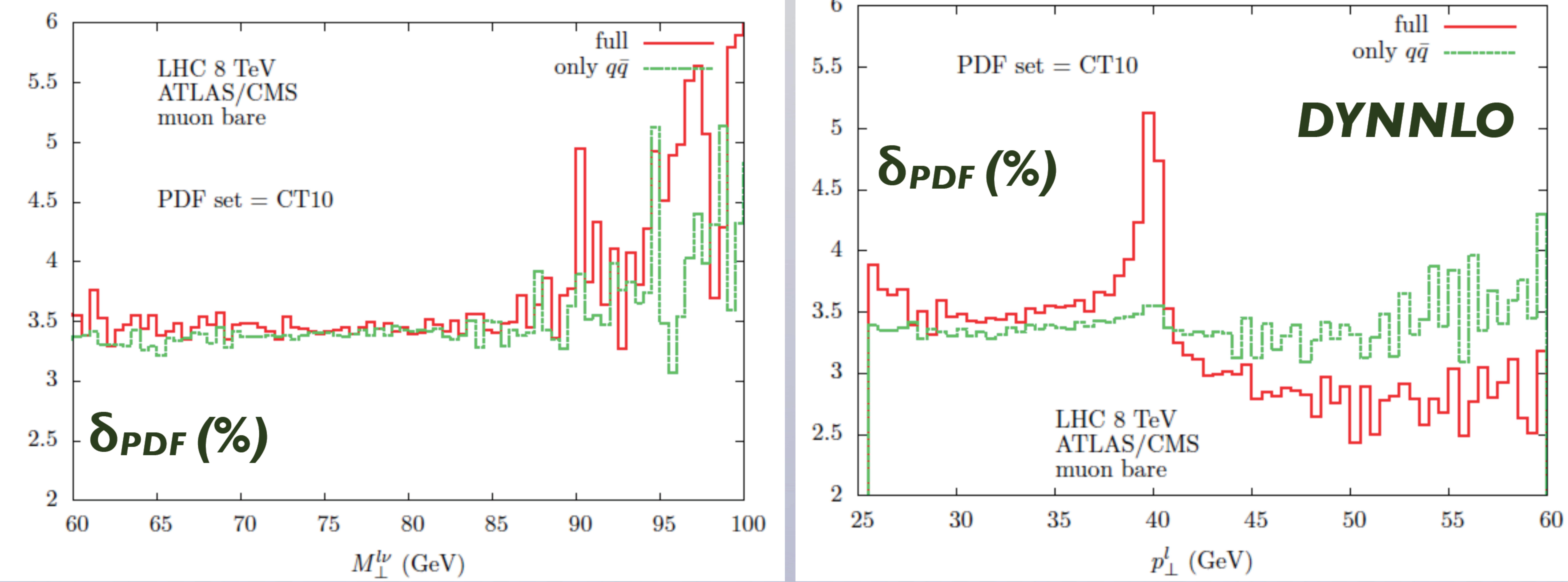}
\caption{\small  The total relative PDF
uncertainty and the separate contribution of quark-antiquark
diagrams  for the transverse
mass (left plot) and the lepton $p_T$ (right plot) distributions,
computed at NLO with {\tt DYNNLO}.
}
\label{fig:pt}
\end{figure}

The relevance of NLO-QCD corrections implies that the gluon PDF yields
also a more important contribution to the PDF uncertainty on $M_W$
than in the transverse mass case.  In order to confirm this, in
Fig.~\ref{fig:pt} we show the contribution of quark-antiquark terms to
the total PDF uncertainty in the transverse mass and lepton $p_T$
distributions, computed at NLO-QCD with {\tt DYNNLO}.
It is clear that for the lepton $p_T^l$ distribution the contribution of
the quark-gluon subprocess is substantial, in particular near the
Jacobean peak.
It should be stressed that the results presented in Fig.~\ref{fig:pt}
have been obtained at fixed order, whereas a fully resummed
calculation would be ideal at least in the lepton $p_T^l$ case.
Furthermore, the quark-antiquark contribution alone provides a correct
estimate of its PDF uncertainty, but only the results including all
the partonic subprocesses are sensible in terms of physical
distributions.  With these two caveats in mind, it is clear that a
dedicated analysis should be pursued in order to limit as much as
possible the contribution to $\Delta M_W$ due to the gluon PDF.  For
example, ratios of $W$ over $Z$ distributions provide a significant
cancellation of contributions which are common in the two cases, 
e.g. the quark-gluon initiated subprocesses, strongly reducing the
corresponding contribution to the PDF uncertainty.

As a final remark, let us mention that PDFs with QED contributions
included should be taken into account to consistently assess the
corrections to the $M_W$ fits induced by QED effects.
The recently released NNPDF2.3 QED set~\cite{Ball:2013hta} is
specially suitable for this purpose, since not only does it include
the combined NNLO QCD with LO QED corrections, but it also includes
the most recent constraints from electroweak gauge boson production
data at the LHC.
The implications of NNPDF2.3 QED for $M_W$ determinations should be
the topic of detailed studies in the near future.

\subsubsection{Experimental aspects: $M_W$}
\label{sec:electroweak-had-mw-exp}


The Tevatron experiments have made precise measurements of the $W$
boson mass~\cite{Aaltonen:2012bp, Abazov:2012bv}. The combined uncertainty on $M_W$ using CDF and D\O\
measurements is 16 MeV~\cite{Aaltonen:2013iut}, significantly surpassing the combined LEP
precision of 33 MeV. This is a noteworthy achievement for hadron
collider experiments. Furthermore, additional statistics are available
at the Tevatron (approximately a factor of 4 at CDF and a factor of 2
at D\O) and very large samples are available at the LHC (which will grow further in the coming years). The
Tevatron experiments have demonstrated that many systematic
uncertainties related to calibrations can be reduced as the statistics
of the calibration samples and other control samples increase~\cite{Kotwal:2008zz}. This is
a non-trivial demonstration  since consistency between multiple
calibration methods and channels, e.g. as shown in the CDF measurements~\cite{Aaltonen:2012bp,Aaltonen:2007ps}, is an essential component of a robust
analysis. 

Uncertainties due to parton distribution functions (PDFs) and
electroweak radiative corrections rely on external experimental and
theoretical input. Improvements in theoretical calculations have led
to reductions in the latter. Collider measurements of boson
distributions have provided constraints on PDFs and increased
statistics in the future should continue to do so.

Table~\ref{tab:electroweak-mwexptev} shows the projections from CDF and D\O\ on the $M_W$
uncertainty they expect to achieve with their respective final
datasets from the Tevatron.  These projections build mostly on the 4-5
measurements that these experiments  have each made over the last two
decades, which show that careful analysis of data has led to
the approximate scaling of many systematic uncertainties with
statistics. The datasets have grown by a factor of 200-500 over this
time period. The projections to the full dataset assume some
improvement in the understanding of the tracking and calorimetry,
modest improvement in the understanding of radiative corrections, and
a factor of two improvement in the PDF uncertainty over the next few
years.  The analysis in Ref.~\cite{Bozzi:2011ww} has shown that the LHC
measurements of the $W$ charge asymmetry using the 2012 data has the
potential to reduce the $W$ mass uncertainty at the Tevatron by about
a factor of two.  Thus, it is reasonable to assume that the final
Tevatron measurements will achieve a combined uncertainty of 9-10 MeV,
as projected.

\begin{table}[t]
\begin{center}
\begin{tabular}{l|cccccc}  
$\Delta M_W$ [MeV] & CDF & D0   & combined & final CDF & final D0 & combined  \\ \hline
$\L [\fb]$ & 2.2  & 4.3 (+1.1) &7.6  & 10 & 10 & 20 \\ \hline
PDF  & 10 & 11  & 10  & 5   & 5   &5  \\
QED rad.  & 4 & 7  & 4  & 4   & 3   & 3 \\
$p_T(W)$ model  &  5 & 2 & 2  & 2   & 2   & 2 \\
other systematics  &  10 & 18 & 9  & 4   & 11   & 4  \\
$W$ statistics  &  12 & 13 & 9  & 6   & 8   & 5 \\
Total  &  19 & 26 (23) & 16  &  10  &  15  & 9  \\
\end{tabular}
\caption{Current and projected uncertainties in the measurement of 
$M_W$ at the Tevatron.}
\label{tab:electroweak-mwexptev}
\end{center}
\end{table}

According to Ref.~\cite{Bozzi:2011ww}, the PDF uncertainty at the LHC is
about a factor of two larger than at the Tevatron. Thus, further
improvements in the PDF uncertainty will be required to produce higher
precision on the $W$ boson mass. This will require a program of
measurements of differential boson distributions such as (i) the $Z$
boson rapidity distribution, (ii) the charged lepton rapidity
distribution from $W$ boson decays, (iii) the $W$ charge asymmetry
distribution, and (iv) the $W$+charm production which constrains the strange
quark contribution. Combined with the increasing understanding of the LHC
detectors, we suggest that a PDF uncertainty below 5 MeV is a reasonable
target for the LHC (see also Sec.~\ref{sec:electroweak-had-theory} for a detailed discussion).
As shown in Table~\ref{tab:electroweak-mwexplhc}, we propose targets for
$m_W$ precision at the LHC, approaching 5 MeV in the long term. Note
that detailed detector studies and improved analysis techniques are
just as important in this endeavor as the growth of the data
statistics. We consider having a 5 MeV target for the total precision
as a reasonable ambition for the LHC.

\begin{table}[t]
\begin{center}
\begin{tabular}{l|ccc}  
$\Delta M_W$ [MeV] & \multicolumn{3}{c}{LHC}  \\ \hline 
$\sqrt{s}$ [TeV] & 8  & 14  & 14  \\
$\L [\fb]$ & 20   &300  & 3000  \\ \hline
PDF            & 10 & 5  & 3      \\
QED rad.       & 4 & 3  & 2    \\
$p_T(W)$ model &  2 & 1 & 1   \\
other systematics &  10 & 5 & 3    \\
$W$ statistics    &  1 & 0.2 & 0   \\
Total             &  15 & 8 & 5    \\
\end{tabular}
\caption{Current and target uncertainties in the measurement of 
$M_W$ at the LHC.}
\label{tab:electroweak-mwexplhc}
\end{center}
\end{table}

\subsubsection{Experimental aspects: $\sin^2\theta_\eff^\ell$}
\label{sec:electroweak-had-exp-sin2theta}


At hadron colliders, investigations around the $Z$ resonance in single neutral-current
vector-boson, $q\bar{q} \rightarrow \gamma,Z \rightarrow l^+l^-$, with
charged leptons $l$ in the final state, allow a precise measurement of
the electroweak mixing angle from the forward-backward asymmetry
$A_{\rm FB}$. The results of a measurement of
$\sin^2\theta_{eff}^l$ at the Tevatron by the CDF and D0
collaborations and at the LHC by the ATLAS and CMS collaborations are
presented in Table~\ref{tab:electroweak-s2thexptev} and
Table~\ref{tab:electroweak-s2thexplhc}, respectively.
 
At the Tevatron, because the quark direction is better defined for
$\bar pp$ than for $pp$ collisions, the measurement of
$\sin^2\theta_\eff^l$ is less sensitive to PDF uncertainties and
higher order QCD corrections. In addition, three significant
improvements have been recently introduced in the analysis at CDF.
The first is the introduction of the event weighting
technique~\cite{Bodek:2010}, which to first order results in the
cancellation of acceptance errors and also reduces the statistical
errors by 20$\%$. The second is the introduction of momentum scale
corrections~\cite{Bodek:2012}, which remove the bias in the
determination of muon momenta, and the third is the
consideration of electroweak radiative corrections using
Zfitter~\cite{Aaltonen:2013wcp}. Therefore, smaller error bars are
expected for the final analysis of the full Run II Tevatron data as
shown in Table~\ref{tab:electroweak-s2thexptev}. The errors in the
$e^+e^-$ channel are smaller than in the $\mu^+\mu^-$ channel, if
forward electrons (i.e. large $\cos\theta$) are included in the
analysis.
Based on the recent improvements in the CDF analysis, we expect
similar errors with the full Run II data set at D0. 
 The recent CDF measurement~\cite{Aaltonen:2013wcp}   with an $e^+e^-$ sample corresponding to 2.1 $\fb$ of integrated   luminosity 
   yields (statistical and systematic errors are added
   linearly): $\sin^2\theta_\eff^l= 0.2328 \pm 0.0011$. D0 measures $\sin^2\theta_\eff^l  = 0.2309 \pm 0.0008$~(stat) $\pm 0.0006$~(syst)~\cite{Abazov:2011ws} using  an $e^+e^-$ sample corresponding to 5.0
  $\fb$ of integrated   luminosity.

\begin{table}[h]
\begin{center}
\begin{tabular}{l|cc|ccc}
$\Delta \sin^2\theta_\eff^l$ [$10^{-5}$] & CDF & D0  &  final CDF & final CDF  & final CDF 
\\ 
 final state        & $e^+e^-$ & $e^+e^-$ & $\mu^+ \mu^-$& $e^+e^-$ &combined 
 \ \\ \hline
$\L [\fb]$         & 2.1 & 5.0 & 9.0  &9.0 & 9 $\mu\mu+9~e^+e^ -$
\\ \hline
PDF                &   12 & 48 &  12 & 12  &12 
\\
higher order corr. &  13   & 8  & 13   & 13  & 13
 \\
other systematics  &   5  & 38 &  5 &   5   & 5
\\
statistical        & 90  & 80  &  80 &  40   & 40  
 \\ \hline
total   $\Delta \sin^2\theta_\eff^l$           & $\bf{92}$  & $\bf{101}$  &$\bf{ 82}$    & $\bf{44}$  &  $\bf{41}$ 
\\\hline
\end{tabular}
\caption{Current and target uncertainties in the measurement of  $\sin^2\theta_\eff^l$ at the Tevatron.}
\label{tab:electroweak-s2thexptev}
\end{center}
\end{table}

At the LHC, the measurement of the forward-backward asymmetry $A_{FB}$
at the $Z$ boson pole is complicated by the fact that the $pp$ initial
state dilutes the $A_{FB}$ in the $q \bar{q}$ collision. As a result,
the measurement is sensitive to the
PDFs. Table~\ref{tab:electroweak-s2thexplhc} shows the uncertainties
from the current LHC analyses.  Systematic uncertainties due to
experimental effects will very likely reduce with higher statistics as
efficiencies and resolutions are better measured using control
samples. In order to exploit this potential, however, a significant
improvement in the understanding of PDFs will be required. We note
that the PDF uncertainty will need to reduce by a factor of $\sim 7$
for the LHC measurement of $\sin^2\theta_\eff^l$ to have
precision comparable to the LEP and SLC measurements. A factor of 2 reduction in
the systematic uncertainty due to missing higher order corrections
will also be required.  In the following we discuss in more detail the
challenges involved in reaching the target uncertainties shown in
Table~\ref{tab:electroweak-s2thexplhc} based on the experience from
the recent ATLAS analysis.

\begin{table}[t]
\begin{center}
\begin{tabular}{l|cc|ccc}  
$\Delta \sin^2\theta_\eff^l$ [$10^{-5}$] & ATLAS & CMS & \multicolumn{3}{c}{LHC/per experiment}   \\ \hline
$\sqrt{s}$ [TeV] & 7   & 7    & 8   & 14   & 14 \\
$\L [\fb]$       & 4.8 & 1.1  & 20  & 300  & 3000 \\ \hline
PDF              & 70  &  130 &  35      & 25  & 10\\
higher order corr. &20  & 110  & 20      & 15  & 10 \\
other systematics  & 70  & 181 & 60 (35)  & 20  & 15 \\
statistical        & 40  & 200 &  20     & 5   &  2  \\
Total              &108  & 319 &  75 (57) & 36  & 21\\
\end{tabular}
\caption{Current and target uncertainties in the measurement of 
$\sin^2\theta_\eff^l$ at the LHC. The target uncertainties are based on expected advancements in 
both theory and experiment as described in the text. A conservative and more optimistic (in parentheses) target uncertainty 
is provided for the measurement at 8 TeV.}
\label{tab:electroweak-s2thexplhc}
\end{center}
\end{table}

The main difficulty in measuring $\sin^2\theta_\eff^l$ at the LHC from
the forward backward asymmetry, $A_{FB}$, lies in the fact that it is a $pp$
collider. Since both beams have valence quarks (as opposed to
anti-quarks), there is an ambiguity in the incoming quark
direction. This ambiguity gives rise to a dilution, or reduction, in
the $A_{FB}$. The effect of dilution can be resolved in part by using
the momentum of the $Z$ boson along the longitudinal direction ($z$) to
determine the direction of the outgoing lepton with respect to the
quark.  However, for events produced in the central part of the
detector, there remains about a 50\% probability of misidentifying the
quark direction. Therefore the best region of phase space to make this
measurement is at large $p_{z}$, or equivalently rapidity, of the $Z$ boson.

The ATLAS $\sin^2\theta_\eff^l$ measurement~\cite{atlassin2theta} utilizes electrons and
muons not only in the central region of the detector, which are
standard for most measurements, but also electrons in the forward
region ($2.5 < |\eta|< 4.9$). 
However, there are some difficulties in using forward
electrons. The forward calorimeters are not as highly segmented
and there are no tracking detectors, so reconstruction relies on less
information. Also, distinguishing between photons and electrons is not
possible.  Finally, electrons in the forward region are more sensitive
to pile-up, which not only increases the background but also makes
background modeling more difficult.
These difficulties can be overcome
by requiring one central electron and one forward electron in the $Z$
reconstruction. 

This approach means that ATLAS has produced three
search channels in total in the 2011 analysis: a muon channel with two
central muons, an electron channel with two central electrons (CC
electron), and an electron channel with one central electron and one
forward electron (CF channel). 
To measure $\sin^2\theta_\eff^l$, the $A_{FB}$ spectra from data were
compared to the asymmetry spectra from Monte Carlo (MC) predictions
produced with varying initial values of the weak mixing angle. 
The results from the three channels are as follows

\[\mathrm{CC~electron:} \sin^2\theta_\eff^l = 0.2288 \pm0.0009(\mathrm{stat.})\pm0.0014(\mathrm{syst.})  \]
\[\mathrm{muon:} \sin^2\theta_\eff^l = 0.2294 \pm0.0009 \mathrm{(stat.)}\pm0.0014\mathrm{(syst.)}  \]
\[\mathrm{CF~electron:} \sin^2\theta_\eff^l = 0.2304 \pm0.0006\mathrm{(stat.)}\pm0.0010\mathrm{(syst.)}  \]

Despite having fewer events, the CF electron has the smallest total
uncertainty. This is due to the reduced effect of dilution in this
channel, which allows better discrimination between the MC templates.  
The uncertainties in the combined  ATLAS measurement of $\sin^2\theta_\eff^l = 0.2297 \pm0.0004\mathrm{(stat.)}\pm0.009\mathrm{(syst.)}$
are outlined in Table~\ref{tab:electroweak-s2thexplhc}. The systematic
uncertainty is dominated by the PDF uncertainty (0.0007) which is
correlated and therefore did not see the reduction that the other
uncertainties did. This uncertainty was estimated using the CT10 NLO
PDF set. 
The total uncertainty for this measurement matches the precision of the most recent results from the Tevatron
experiments shown in Table~\ref{tab:electroweak-s2thexptev}.

The $\sin^2\theta_\eff^l$ ATLAS measurement is limited by the PDF
uncertainty in the two central channels, and by the energy scale and
smearing in the CF electron channel. The future
of this measurement lies in reducing the PDF uncertainty for the
central channels. It is a factor of 2 larger than the next largest 
uncertainty. However, it is the CF electron channel that shows the
most potential for being competitive with the LEP and SLC
experiments. If the PDF and energy scale/smearing uncertainties were
both reduced by a factor of two, ATLAS would become
competitive. Although the energy scale/smearing will be increasingly
difficult with increased pile-up conditions as well as higher trigger
thresholds due to the increased luminosity, the increased statistics
as well as better knowledge of the detector (with more use and
simulation) will allow us to work toward a significant reduction.

For the 2012 projection with 8 TeV and 20 fb$^{-1}$ of data shown in
Table~\ref{tab:electroweak-s2thexplhc}, the PDF and energy scale/smearing
uncertainties will most likely not be reduced to the point where it
will be competitive with the world's best measurements. The
statistical uncertainty will be reduced by a factor of two (with the
4x increase in data). The MC statistical uncertainty can be reduced to
a negligible amount which reduces the ``other systematics" column to 6
instead of 7 in a conservative scenario (shown in parentheses is a
more optimistic scenario where it is assumed that energy
scale/smearing uncertainties can be considerably reduced).  Assuming
that the inclusion of LHC data in updated PDF fits yields a reduction
of the PDF error by a factor of 2 (see discussion in
Section~\ref{sec:electroweak-had-theory}), this would lead to a total
uncertainty of 75(57)$\times$10$^{-5}$.

The target uncertainties (for $\sqrt{s} = 14$~TeV and 300 fb$^{-1}$ and 3000 fb$^{-1}$ respectively)
  for a future ATLAS measurement of $\Delta
\sin^2\theta_\eff^l$ shown in Table~\ref{tab:electroweak-s2thexplhc} are
based on the following reasoning and expectations for advancements
in both theory and experiment.  We assume that
with advancements in MC generators the uncertainty due to higher order
corrections will decrease dramatically. Currently this is taken as a
systematic uncertainty in the ATLAS and CMS measurements, however in the future, it could
just be a shift or not required at all.  
It is difficult to reduce the energy scale/smearing uncertainty
in the forward channel, but this task  can be made the priority  if this became the limiting systematic uncertainty.  The last 
challenge will be triggering on CF electron events. Since there is
only one central electron, ATLAS relies on the single-electron trigger. For
the 2015 run, these thresholds will increase compared to the
2012 level, making it more difficult  to trigger on signal events. There
is currently work going on in ATLAS to overcome this challenge, so that forward-electron $Z$-boson events can still be used in the future.


\subsection{Uncertainties in measurements of $M_W$ and $\sin^2\theta_\eff^\ell$ at lepton clolliders}
\label{sec:electroweak-lep}

We now turn to the discussion of these EWPOs at lepton colliders,
starting with theoretical aspects and moving to experimental issues.

\subsubsection{Theory status of $W$-pair production at $\ee$ colliders}
\label{sec:electroweak-lep-theory}


The possibilities offered by future $\ee$ colliders make it mandatory
to improve the accuracy of the theoretical predictions of the $\ee\to
W^+W^-$ cross section beyond the level achieved for LEP2.  A precise
measurement of $\mw$ with an accuracy $\sim 5~\mev$ from a threshold
scan in the GigaZ option of an ILC or at TLEP requires a cross-section
calculation with a precision of a few per-mille in the threshold
region $\sqrt s\sim 2\mw$.  At high center of mass energies $\sqrt s
\gsim 800$~GeV in the second phase of an ILC or at CLIC, which are
particularly relevant for measurements of anomalous triple gauge
couplings, electroweak radiative corrections are enhanced so that NNLO
corrections can become relevant. In the following we review recent
calculations that improve the theoretical predictions in these regimes
and assess the remaining theoretical uncertainties.

\begin{figure}[h]
\begin{center}
\includegraphics[width=0.4\hsize]{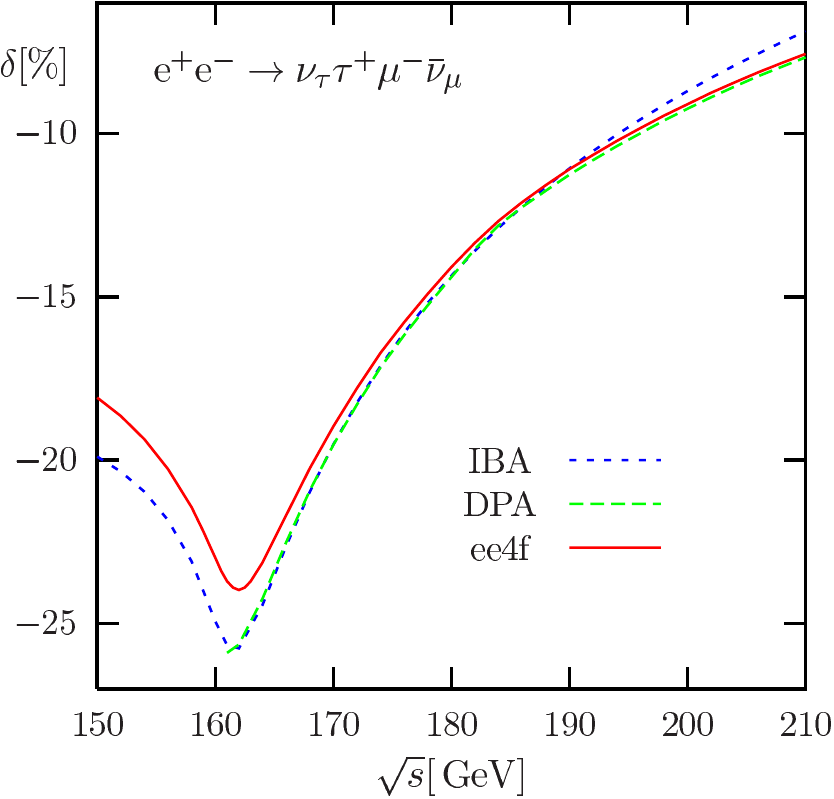}
\includegraphics[width=0.4\hsize]{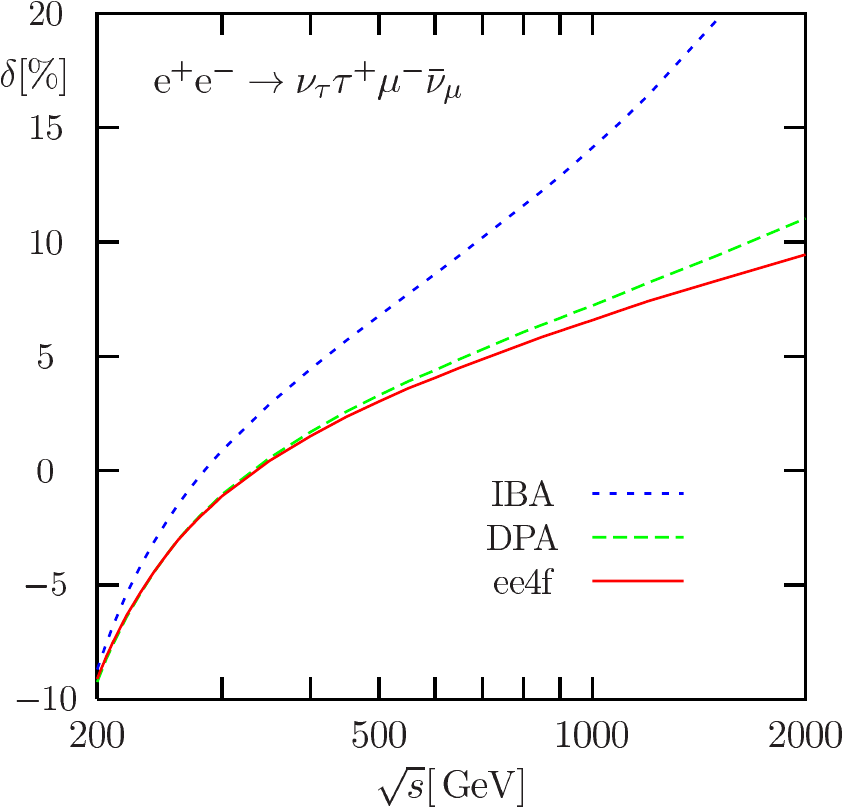}
\caption{Relative corrections to the total cross section for $\ee\to\nu_\tau\tau^+\mu^-\bar\nu_\mu$, normalized by the Born cross section without ISR improvement: improved Born approximation~(IBA, blue dashed), double-pole approximation~(DPA, green short-dashed) and the NLO calculation for the $4$-fermion final state (ee4f, red). Taken from Ref.~\cite{Denner:2005es}.}
\label{fig:electroweak-wwxsec}
\end{center}
\end{figure}

Precise theoretical predictions for $W$-pair production have to take
the $W$-boson decay into account and treat the full $4$-fermion final
state.  The state of the art during the LEP2 run consisted of the
so-called double-pole approximation~(DPA) utilized in the computer
programs RACOONWW~\cite{Denner:1999kn} and
YFSWW3~\cite{Jadach:2000kw}. In the DPA the quantum corrections to
four-fermion production are consistently decomposed into the
corrections to on-shell $W$-boson production and decay (factorizable
corrections), soft-photon corrections connecting $W$-production,
propagation and decay stages (non-factorizable corrections), and into
a non-resonant remainder.  More recently, a complete NLO calculation
of $4$-fermion production was performed~\cite{Denner:2005es},
including loop corrections to singly- and non-resonant diagrams and
treating unstable particles in the complex mass scheme.  As can be
seen in Figure~\ref{fig:electroweak-wwxsec} the results of the DPA for
the total cross section agree well with the full $\ee\to 4\mathrm{f}$
calculation for energies $200\,\gev \lsim\sqrt{s}\lsim 500\,\gev$.
But the full calculation is required in the threshold region
$160\,\gev\lsim \sqrt{s}\lsim 170\,\gev$ and at energies $\sqrt{s} > 500\,\gev$
where off-shell effects 
become important.  The description of differential distributions for
hadronic $W$-decays could be improved in the future since presently
only QCD corrections to the inclusive $W$-decay width are included.

At center of mass energies $s\gg m_W^2$ as relevant for the
measurement of anomalous triple gauge boson couplings, large radiative
corrections to $W$-pair production arise due to so-called Sudakov
logarithms $\log^2(s/\mw^2)$~\cite{Fadin:1999bq}.  The NNLO
corrections to on-shell $W$-pair production at NNLL accuracy
(\ie~corrections of the form $\alpha^2\log^m(s/\mw^2)$ with $m=2,3,4$)
have been computed in~\cite{Kuhn:2007ca}. They are of the order of
$5\%$ ($15\%)$ for $\sqrt{s}=1\,\tev$ ($\sqrt{s}=3\,\tev$) and therefore should be
taken into account in the second phase of an ILC or at CLIC.  The
remaining uncertainty due to the uncalculated single-logarithmic NNLO
terms has been estimated to be $1$--$2\%$~\cite{Kuhn:2007ca}.  It
might also be relevant to consider NNLO Sudakov logarithms for the
full $4$-fermion final state instead of using the approximation of
on-shell $W$-bosons.

\begin{figure}[h]
\begin{center}
\includegraphics[width=0.4\hsize]{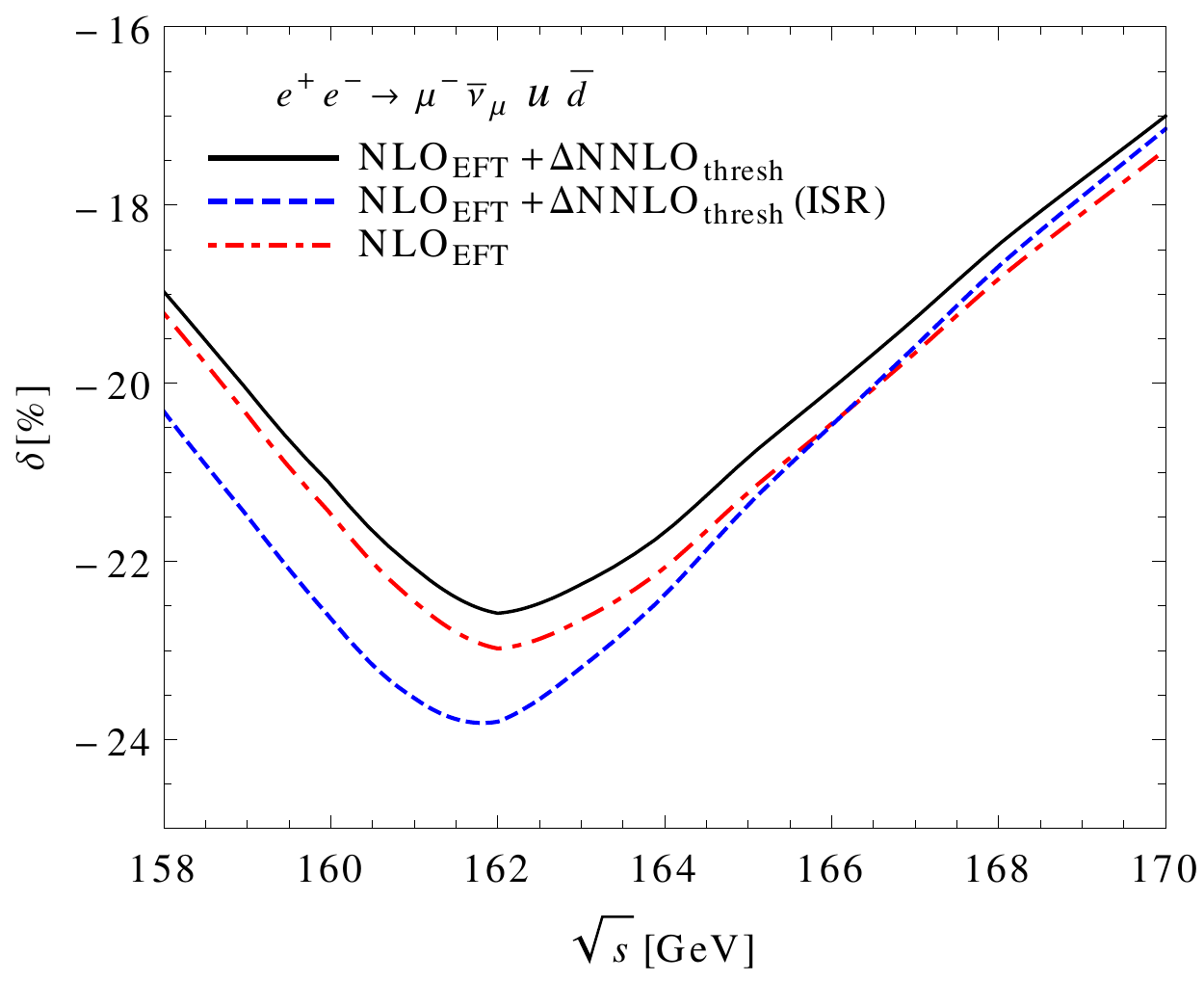}
\caption{Relative corrections to the total cross section for $\ee\to\mu^-\bar\nu_\mu u\bar d$ in the EFT approach, normalized by the Born cross section without ISR improvement: NLO~(red, dash-dotted), NLO with dominant NNLO corrections~(black, solid). In the blue, dashed curve the higher-order corrections are convoluted with ISR structure functions. 
Based on Ref.~\cite{Actis:2008rb}.
}
\label{fig:electroweak-ww-thresh}
\end{center}
\end{figure}

Near the $W$-pair production threshold, Coulomb corrections of the
form $(\alpha/\beta)^n$, with $\beta=\sqrt{1-4\mw^2/s}$, are enhanced
over the remaining corrections of the same order in $\alpha$.
Therefore the second-order Coulomb correction $\sim
(\alpha/\beta)^2$~\cite{Fadin:1995fp} and contributions of the form
$\alpha^2/\beta$ are expected to be the dominant NNLO corrections near
threshold. These have been calculated in~\cite{Actis:2008rb} using an
effective-field theory~(EFT) approach~\cite{Beneke:2007zg}.  As can be
seen in Figure~\ref{fig:electroweak-ww-thresh}, the effect of the
threshold-enhanced NNLO corrections is of the order of $0.5\%$. The
remaining uncertainty of the $\mw$-measurement from a threshold scan
due to uncalculated NNLO corrections has been estimated to be below
$\Delta\mw\approx\, 3~\mev$~\cite{Actis:2008rb}. The current best
prediction for the total cross section near threshold is obtained by
adding the dominant NNLO corrections to the NLO$_{\mathrm{ee4f}}$
result~\cite{Denner:2005es} that includes one-loop singly and
non-resonant diagrams beyond the NLO EFT calculation. For the future,
a calculation of the leading NNLO corrections for differential
distributions is desirable.  As a caveat, further uncertainties arise
since the cross section at threshold is very sensitive to
initial-state radiation~(ISR).  In the NLO$_{\mathrm{EFT}}$ and
$\Delta \mathrm{NNLO}_{\mathrm{thresh}}$ results in
Figure~\ref{fig:electroweak-ww-thresh} as well as in the DPA and ee4f
results in Figure~\ref{fig:electroweak-wwxsec}, a leading-logarithmic
resummation of ISR effects~\cite{Skrzypek:1992vk} is performed for
the Born cross section while higher-order corrections are added
without ISR improvement.  In the blue-dashed curve in
Figure~\ref{fig:electroweak-ww-thresh}, the higher-order
EFT corrections are improved by ISR resummation.\footnote{Note that
  the IBA approximation in Figure~\ref{fig:electroweak-wwxsec}
  also contains ISR improvement of the first Coulomb correction.}  The
numerical impact of this next-to-leading logarithmic~(NLL) effect is
as large as $2\%$ directly at threshold while it soon becomes
negligible at higher energies. Therefore a consistent NLL-treatment of
ISR might be required for sufficient theoretical control over the
cross section at threshold.


\subsubsection{Experimental aspects: $M_W$}
\label{sec:electroweak-lep-mw-exp}


The three most promising approaches to measuring the $W$ mass at an $e^{+} e^{-}$ collider are: 
\begin{itemize}
\item Polarized threshold scan of 
the $\mathrm{W}^+\mathrm{W}^-$ cross-section as discussed in~\cite{Wilson_2001aw}.
\item Kinematically-constrained reconstruction of $\mathrm{W}^+\mathrm{W}^-$ using 
constraints from four-momentum conservation and optionally mass-equality 
as was done at LEP2.
\item Direct measurement of the hadronic mass. This can be applied 
particularly to single-W events decaying hadronically 
or to the hadronic system in semi-leptonic $\mathrm{W}^+\mathrm{W}^-$ events.
\end{itemize}

The three different methods are summarized in the following
tables. There is one reasonably complete study related to a polarized
threshold scan at ILC~\cite{Wilson_2001aw} which has been updated for
this Snowmass workshop.  There is also a new,  much more precise method
for determining the beam energy {\em in situ} using di-muon events at ILC
which has been developed in more depth during this workshop and was
presented at~\cite{Wilson_LC2013}.  This gives the potential to reduce
the beam energy uncertainty on the W mass to 0.8 MeV (limited by
stand-alone momentum scale uncertainties estimated at 10 ppm). This
previously important systematic for the threshold method - and
dominant systematic for the kinematically-constrained reconstruction
method appears to be no longer such a critical issue.  The reported
tables should be taken as reasonable indications of the potential
performance.  W mass measurements were statistics limited for these
methods at LEP2. It is clear that large improvements in the
systematics are feasible at future machines like ILC. Exactly how much
better can be done is something that can not be predicted with
absolute certainty, given the orders of magnitude of improvement. In
practice it is something that typically can only be pinned down once a
detector is operating. 

\begin{table}[h]
\begin{center}
\begin{tabular}{l|c|c|c|c|c}
$\Delta M_W$ [MeV]               & LEP2   & ILC     &       ILC          & $e^+ e^-$      &   TLEP       \\ \hline
$\sqrt{s}$ [GeV]                 &  161   &  161    &      161           &     161   &   161        \\
$\cal{L}$ [fb$^{-1}$]            & 0.040  &  100    &      480           &     600   &   3000$\times$4       \\ 
$P(e^{-})$ [\%]                  &   0    &  90     &       90           &     0     &    0         \\
$P(e^{+})$ [\%]                  &   0    &  60     &       60           &     0     &    0         \\ \hline
systematics                      &  70    &         &                    &     ?     &   $<0.5$        \\
statistics                       &  200   &         &                    &    2.3?   &   0.5       \\
experimental total               &  210   &  3.9    &      1.9           &    $>$2.3 &   $<0.7$     \\ \hline
beam energy                      &  13    &  0.8-2.0 &   0.8-2.0         &    0.8-2.0  &  0.1     \\ 
radiative corrections            &   -    &  1.0    &      1.0           &    1.0    &   1.0        \\ \hline
total                            &  210   &  4.1-4.5    &  2.3-2.9       &    $>$2.6-3.2 &   $<1.2$     \\ \hline
\end{tabular}
\caption{Current and preliminary target uncertainties in the measurement of
$M_W$ at $e^+e^-$ colliders close to $WW$ threshold, including an estimate for a future 
theoretical uncertainty due to missing higher-order corrections.}
\label{tab:electroweak-mwexplep_THRESHOLD}
\end{center}
\end{table} 


Table~{\ref{tab:electroweak-mwexplep_THRESHOLD}} has projected results for running 
close to the $WW $ threshold. ILC can collide highly longitudinally polarized 
electrons and positrons - this is particularly advantageous for a threshold scan.
In the tables it is assumed that if ILC undertakes a dedicated scan near threshold 
that this would be done with the highest polarization levels achievable.
The estimated uncertainties assume that the beam energy scale 
can be established from collision data at the level of 1 part in $10^5$ leading 
to a corresponding experimental uncertainty on $M_W$ of 0.8~MeV. This has been shown to be statistically feasible 
using di-muon events provided that the momentum scale is determined to the same precision. 
This appears feasible using J/${\psi}$ events in $Z$ boson decays.
The ILC numbers are based on a detailed and updated study with realistic assumptions on detection efficiency, 
polarization determination, backgrounds, efficiency and normalization errors using a 6-point scan with four different 
beam helicity combinations. The ILC numbers include the (small) effects from beamstrahlung on the cross-section 
and take advantage of the 150~fb cross-section of multi-hadron production for determinining the beam polarizations 
from the data. 
In addition, the table includes an indicative estimate of the anticipated theoretical uncertainty 
associated with interpreting cross-section measurements near threshold in terms of $M_W$ of 1.0~MeV.
A discussion of the present status of predictions for $W$-pair production at threshold can be found in Section~\ref{sec:electroweak-lep-theory}.
A detailed assessment of the anticipated theoretical shape and normalization uncertainties on 
the cross-section behavior with center-of-mass energy 
and including the effects of realistic experimental acceptance for all the four-fermion final states would 
in principle be needed to report a firm theoretical error estimate.
In the table for the ILC, the systematics are essentially currently included in the overall error 
as the multi-parameter fit adjusts the systematics as nuisance parameters constrained within 
a priori uncertainties taken as 0.1\% for relative efficiency and absolute integrated luminosity. 
The beam polarizations and backgrounds are fitted simultaneously from the data. In the context of the polarized 
scan this measurement is essentially {\it statistics} dominated. 

\begin{table}[h]
\begin{center}
\begin{tabular}{l|c|c|c|c}
$\Delta M_W$ [MeV] & LEP2 & ILC & ILC & ILC   \\ \hline
$\sqrt{s}$ [GeV]                & 172-209  &    250     &   350     &    500     \\
$\cal{L}$ [fb$^{-1}$]           & 3.0      &    500     &   350     &   1000     \\ 
$P(e^{-})$ [\%]                 & 0        &     80     &    80     &     80     \\
$P(e^{+})$ [\%]                 & 0        &     30     &    30     &     30     \\ \hline
beam energy                     & 9        &    0.8     &   1.1     &    1.6     \\ 
luminosity spectrum             & N/A      &    1.0     &   1.4     &    2.0     \\ \hline
hadronization                   & 13       &    1.3     &   1.3     &    1.3     \\
radiative corrections           &  8       &    1.2     &   1.5     &    1.8     \\
detector effects                & 10       &    1.0     &   1.0     &    1.0     \\
other systematics               &  3       &    0.3     &   0.3     &    0.3     \\ \hline   
total systematics               & 21       &    2.4     &   2.9     &    3.5     \\
statistical                     & 30       &    1.5     &   2.1     &    1.8     \\
total                           & 36       &    2.8     &   3.6     &    3.9     \\ \hline
\end{tabular}
\caption{Current and preliminary estimated experimental uncertainties in the measurement of
$M_W$ at $e^{+}e^{-}$colliders from kinematic reconstruction in the $q\bar{q}\ell\nu_{\ell}$ channel with $\ell=e,\mu$}
\label{tab:electroweak-mwexplep_KR}
\end{center}
\end{table}

Table~{\ref{tab:electroweak-mwexplep_KR}} has projected results for kinematic reconstruction using the semi-leptonic channels 
as was used at LEP2. Details of this method are in the recently submitted LEP2 legacy paper~\cite{Schael:2013ita} 
and the systematics discussed there are used as the basis for this discussion.
At LEP2 the fully hadronic channel was also used. It is not expected to be competitive at the sub-10 MeV 
level because of final-state interaction effects and is so is neglected for these projections.
There have not been dedicated studies on the semi-leptonic channel for ILC, but the measurements 
at LEP2 can be used to estimate/bracket some of the primary uncertainties.
The beam energy uncertainty is taken again as a $10^{-5}$ uncertainty at 250~GeV leading to an error of 0.8~MeV. 
At higher energies this uncertainty is scaled linearly with center-of-mass energy reflecting in part 
less statistics for {\em in situ} checks. Systematic errors associated with knowledge of 
the luminosity spectrum $dL/dx_{1}dx_{2}$ are estimated to be at the 1~MeV level at 250 GeV and 
will increase with center-of-mass energy. The table assumes a linear dependence.  
Two of the primary systematics associated with the $W$ boson mass measurement at LEP2, namely from hadronization and detector effects 
will be controlled much better with the modern ILC detectors and a more than one hundred times larger dataset. 
In particular for example it is reasonable to expect that the 7 MeV error associated with a 0.3\% uncertainty on 
the muon energy scale in for example the OPAL analysis is reduced to negligible (naively 0.02 MeV). The hadronization 
errors which dominated the LEP2 systematic uncertainty were a result of 
several effects. The much larger statistics envisaged at ILC will allow the kaon and proton fractions in $W$ boson decays 
to be measured at least ten times better and the 
particle-flow based jet reconstruction should make it more feasible 
to use identified particles in reconstructing jets. Given the improvements in the detector and statistics, improvements in the 
leading experimental systematics by a factor of 10 can be envisaged. 
The radiative corrections systematic can presumably be improved with further work. 
The growing importance of ISR at higher center-of-mass 
energies suggests that this systematic will degrade as the center-of-mass energy increases.
The effective statistical error is not completely straightforward to estimate as it includes effects 
from ISR and beamstrahlung which often degrade the validity of the kinematic constraints both of which are substantially 
larger at higher center-of-mass energy. It has been shown that these effects can be 
ameliorated in the fully hadronic channel~\cite{Beckmann:2010ib} by allowing for such photon radiation. 
It is expected that similar methods will be useful to improve the effective resolution in the semi-leptonic channel too 
although this is not as highly constrained given the unobserved neutrino. 
This method is likely to be {\it systematics} dominated.

\begin{table}[h]
\begin{center}
\begin{tabular}{l|c|c|c|c}
$\Delta M_W$ [MeV] & ILC & ILC & ILC & ILC  \\ \hline
$\sqrt{s}$ [GeV]                 &   250      &     350     &   500     &   1000     \\
$\cal{L}$ [fb$^{-1}$]            &   500      &     350     &  1000     &   2000     \\
$P(e^{-})$ [\%]                  &    80      &     80      &    80     &     80     \\
$P(e^{+})$ [\%]                  &    30      &     30      &    30     &     30     \\ \hline
jet energy scale                 &   3.0      &     3.0     &   3.0     &    3.0     \\
hadronization                    &   1.5      &     1.5     &   1.5     &    1.5     \\
pileup                           &   0.5      &     0.7     &   1.0     &    2.0     \\ \hline
total systematics                &   3.4      &     3.4     &   3.5     &    3.9     \\ 
statistical                      &   1.5      &     1.5     &   1.0     &    0.5     \\
total                            &   3.7      &     3.7     &   3.6     &    3.9     \\ \hline
\end{tabular}
\caption{Preliminary estimated experimental uncertainties in the measurement of
$M_W$ at $e^{+}e^{-}$colliders from direct reconstruction of the hadronic mass in single-$W$ and $WW$ events 
where one $W$ boson decays hadronically. 
Does not include $WW$ with $q\bar{q}\ell\nu_{\ell}$ where $\ell=e,\mu$.}
\label{tab:electroweak-mwexplep_MH}
\end{center}
\end{table} 


Table~\ref{tab:electroweak-mwexplep_MH} has projected results from the
direct measurement of the hadronic mass. This measurement depends
primarily on how well the hadronic mass scale can be determined. It
essentially does not depend at all on measurements of the beam energy
or luminosity spectrum and so is very complementary to the previous
two methods.  In the particle-flow approach it is in principle
possible to cast this as primarily a ``bottom-up'' problem of
determining the tracker momentum scale, the electro-magnetic
calorimeter scale and the calorimeter energy scale for neutral hadrons
and it is these components that affect the jet energy scale.  Over the
course of the envisaged ILC program it is anticipated that the samples
of Z bosons decaying to hadrons where the Z boson mass is currently
known to 2.1 MeV should make it feasible to target a 3 MeV error
originating from the jet energy scale. The hadronization error is
anticipated to be dominated by knowledge of the $K^0_L$ and neutron
fractions. The pile-up entry refers to primarily $\gamma \gamma
\rightarrow \mathrm{hadrons}$ events coincident with $W$ boson
events. The contribution of such events to the measured hadronic mass
can be mitigated and is not expected to be a dominant systematic error
- but it will be more problematic at higher center-of-mass energies.
The statistical error depends on the jet energy resolution and the
consequent hadronic mass resolution.  The hadronic mass resolution for
a particular event varies substantially depending primarily on the
fractions of energy in charged particles, photons and neutral hadrons
in the event.  The effective hadronic mass resolution is therefore a
strong function of the analysis method. A full convolution fit with
more advanced reconstruction techniques like $\pi^0$ mass-constrained
fitting offers the potential to improve the $W$ boson mass statistical
error by a factor of 2.2 over that naively estimated from the observed
average jet energy resolution in full simulation studies.  In the
estimates presented, we have been conservative and have assumed that
the actual improvement factor of a realistic and mature analysis is
1.4. This method is likely to be {\it systematics} dominated.

\underline{Prospects at CLIC}\\

CLIC has yet to study the potential precision for a $W$ boson mass
measurement from direct reconstruction. However, it is anticipated
that with more than 50 million single $W$ boson events, a statistical
precision on the $W$ boson mass of a few MeV will be
achievable~\cite{Abramowicz:2013tzc}. CLIC does not foresee operation
at either 91 GeV or 161 GeV.

\underline{Prospects at TLEP}\\

Studies of the prospects at TLEP have just begun and will continue
over the next few years. As described in~\cite{TLEPwhitepaper}, the
statistical precision achievable at TLEP using the $WW$ threshold scan
is very high, with about 25 million $W$ boson pairs produced at
threshold.  A statistical precision of about 1 MeV per experiment,
leading to 0.5 (0.7) MeV combining four (two) experiments, should be
possible.

 The key experimental issue is the calibration of the beam energy
 using the resonant depolarization technique. Using a subset of the
 collider bunches {\em in situ} to perform this calibration, the
 uncertainty on the beam energy of about 100 keV has been
 projected. The main question is whether this technique can be made
 operational at a beam energy of 81 GeV, and it is motivated in~\cite{TLEPwhitepaper}
 that it can be.

 In Table~\ref{tab:electroweak-mwexplep_THRESHOLD}, a statistical
 uncertainty of 0.5 MeV is mentioned (to represent 4 experiments)
 and a placeholder for other experimental systematics (such as
 backgrounds) of $< 0.5$ MeV is also included. Finally, a placeholder for
 the uncertainty in QED radiative corrections of 1 MeV is included,
 partly to indicate that this theoretical uncertainty is a major
 challenge for TLEP; on the other hand, the TLEP potential may provide
 strong motivation to improve the radiative correction calculations
 further.  A target for the total $M_W$ uncertainty of 0.5 MeV is
 quoted in~\cite{TLEPwhitepaper}; including placeholders for other
 systematics in Table~\ref{tab:electroweak-mwexplep_THRESHOLD}, a projected
 total uncertainty of $<1.2$ MeV is obtained. Clearly, more studies are
 needed to check that the ultimate statistical precision of TLEP can
 be fully exploited.

\subsubsection{Experimental aspects: $\sin^2\theta_\eff^\ell$}
\label{sec:electroweak-lep-exp-sin2theta}



With polarized beams, the left-right asymmetry $\ALR$ provides the
most sensitive measurement of the effective weak mixing angle.
Details of this measurement at a polarized linear collider running at
the Z-pole are reported in \cite{kmogigaz,peter_mike,Erler:2002su}.
With $10^9$ Z bosons, an electron polarization of 80\% and no positron
polarization, a statistical error of $\Delta A_{\rm {LR}} = 4 \cdot
10^{-5}$ can be realized, although the systematic uncertainty
achievable on the absolute polarization measurement will be
significantly worse.  Extrapolating from the SLD experience, where
$\Delta \mathcal{P}/\mathcal{P}=0.5\%$ has been
achieved \cite{alrsld}, it is assumed that at a future facility a
polarization uncertainty of $0.25\%$ is
realistic \cite{Boogert:2009ir}, leading to $\Delta \ALR = 3.8 \cdot
10^{-4}$ or $\Delta \stl = 5 \cdot 10^{-5}$.

If positron polarization is available, a significantly more precise
measurement can be made using a modified Blondel
scheme~\cite{Blondel:1987wr} which removes the need for an absolute
polarization measurement.  The total cross section with both beams
being polarized can be written as $\sigma \, = \, \sigma_{\rm
unpol} \left[ 1 - \ppl \pmi + \ALR (\ppl - \pmi) \right]$. If all
four helicity combinations are measured, $\ALR$ can be determined as
\[
\ALR \, = \, \sqrt{\frac{
    ( \sigma_{++}+\sigma_{-+}-\sigma_{+-}-\sigma_{--})
    (-\sigma_{++}+\sigma_{-+}-\sigma_{+-}+\sigma_{--})}{
    ( \sigma_{++}+\sigma_{-+}+\sigma_{+-}+\sigma_{--})
    (-\sigma_{++}+\sigma_{-+}+\sigma_{+-}-\sigma_{--})}},
\]
which is independent of the absolute polarization.  Only 10\% of the
total luminosity needs to be delivered in the `wrong-helicity'
combinations (++, --). The statistical uncertainty which can be
achieved with $\pmi = 80\%$ and $\ppl > 30\%$ is $\Delta \ALR <
5 \cdot 10^{-5}$ or $\Delta \stl < 6 \cdot 10^{-6}$.  The statistical
uncertainty improvement for higher positron polarization values is
relatively mild, asymptotically approaching $\Delta \ALR = 2.5 \cdot
10^{-5}$.

Even though an absolute polarization measurement is not needed, a
precise measurement of the polarization difference between the beam
helicity states is required.  If the polarization in the two helicity
states is written as $\ppm = \pm |\ppm| + \delta \ppm$, the dependence
of the measured $\ALR$ value on the polarization difference is given
by $\rm{d} \ALR / \rm{d} \delta \ppm \approx 0.5$.  Extrapolating from
the SLC experience, it has been estimated that $\delta \ppm$ can be
measured to around $10^{-4}$~\cite{peter_mike}.  With
effort, this uncertainty might be reduced further, although the
difficulty of quickly reversing the positron helicity may limit how
precisely the relative positron polarization difference can be
measured.  Several other experimental systematic uncertainties on
$\ALR$ also need to be controlled at the $10^{-4}$ level, including
asymmetries in the luminosity delivered and backgrounds observed in
the different helicity combinations.

Due to $\gamma-Z$ interference, the dependence of $\ALR$ on the
collision energy $\sqrt{s}$ is given by $\mathrm{d} \ALR
/ \mathrm{d} \sqrt{s} = 2 \times 10^{-2}$/GeV.  The difference
$\sqrt{s} - \MZ$ thus needs to be known to $\sim 10$~MeV to match the
experimental precision achievable with electron polarization only, and
to $\sim 1$~MeV with polarized positrons.  A multi-point scan of the
$Z$ peak is foreseen to provide the relative calibration of the
collision energy $\sqrt{s}$ with respect to $m_Z$ at the $\sim 1$~MeV
level.  The collision energy $\sqrt{s}$ must be understood including
any beamstrahlung, which causes a significant $\approx 50$~MeV shift
to the luminosity-weighted mean $\sqrt{s}$, depending on the exact
collision parameters.  As long as the beamstrahlung distribution is
constant throughout the $Z$ scan, however, this effect will be
calibrated out to first order by the scan.  Direct measurements of
this beamstrahlung shape to a few percent should also be possible
using Bhabha acolinearity and the di-muon momentum spectrum.

Overall, an uncertainty of $\Delta \ALR = 10^{-4}$ can be achieved at
a polarized linear collider with $10^9$ Z bosons, corresponding to an
uncertainty on the effective weak mixing angle of $\Delta \stl < 1.3 \cdot
10^{-5}$~\cite{kmogigaz}.

Polarized beams can also be produced at circular $e^+e^-$ storage
rings, and transverse polarizations of $\sim 50\%$ were observed at
LEP~\cite{Assman:1995tb}. To exploit this polarization for a
measurement of $\ALR$ requires spin rotators to be installed to
provide longitudinal polarization at the interaction points.  With the
luminosity available at a machine like TLEP, very small statistical
uncertainties on $\ALR$ can be achieved.  The collision energy
$\sqrt{s}$ can be measured to a precision of 100~keV using resonant
depolarization as was used at LEP for the measurement of the $Z$ boson
mass \cite{Assmann:2004gc}. This uncertainty limits the achievable
precision on the effective weak mixing angle to $\Delta \stl = 1 \cdot 10^{-6}$~\cite{TLEPwhitepaper}. 

In a storage ring, it is difficult to reverse the helicity of the
colliding bunches on a short time scale, so the scheme used at a
linear collider is not applicable.  To avoid being limited by an
absolute polarization measurement, the original Blondel scheme has
been proposed where four helicity combinations are formed from
polarized and unpolarized electron and positron bunches circulating in
the same fill~\cite{Blondel:1987wr}.  This scheme again relies on
the ability to measure the difference in polarization between the
electron and positron bunches ($\delta \ppm$), and also to measure or
limit any residual polarization in the nominally unpolarized bunches.
An added complication is that the longitudinal polarization needs to
be known at the interaction point, while the polarization measurements
at a storage ring are most easily made on the transversely-polarized
beams away from the interaction points, so possible asymmetries in the
spin rotators must be carefully considered.  Detailed estimates of the
other experimental systematic uncertainties achievable at TLEP are not
currently available, although in general they are similar to the
experimental challenges faced at a linear $e^+e^-$ collider for the
same measurement. In Table~\ref{tab:electroweak-s2thexplep}, a placeholder for the systematics at TLEP is based on scaling with statistics; a better estimate is needed. 
 Higher order calculations needed for the extraction of $\stl$ also require further study and are absent in this table at the moment. A target of $\Delta \stl = 1 \cdot 10^{-6}$ is quoted
 in~\cite{TLEPwhitepaper}; in Table~\ref{tab:electroweak-s2thexplep} the systematics-dominated target of $\Delta \stl = 3 \cdot 10^{-6}$ is subject to further study of the TLEP potential. 

\begin{table}[t]
\begin{center}
\begin{tabular}{l|cc}  
$\Delta \sin^2\theta_\eff^l$ [$10^{-5}$] & ILC/GigaZ & TLEP(Z)   \\ \hline
$\sqrt{s}$ [GeV]   & 91  & 91 \\
$\L [\fb]$         & 30    & 3000$\times$4 \\ \hline
systematics        & 1.1 & 0.2 \\
statistical        & 0.5 & 0.1\\
higher order corr. & ?   & ? \\
beam energy        & 0.5 & 0.05 \\
total              & 1.3 & 0.3 \\
\end{tabular}
\caption{Projected target uncertainties in the measurement of 
$\sin^2\theta_\eff^l$ at $e^+ e^-$ colliders. Systematic uncertainties for TLEP have been scaled with statistics; whether this scaling can be achieved remains a question to study. 
 Higher-order calculations required for the measurement also need to be investigated. }
\label{tab:electroweak-s2thexplep}
\end{center}
\end{table}

\subsection{Summary of  experimental target accuracies and theory uncertainties for
 $\sin^2\theta_\eff^\ell$ and $M_W$}
\label{sec:electroweak-mwstconcl}

In Table~\ref{tab:electroweak-mwsin2t} we summarize the discussion of
the previous sections by showing the target accuracies for the
measurements of $M_W$ and $\sin^2\theta_{eff}^\ell$ at the LHC, ILC and
TLEP and the estimated future theory uncertainties of their SM
predictions. 

\begin{table}[h]
\begin{center}
\begin{tabular}{l|cc|c|ccc|c|c}  
                    & LHC & LHC  & ILC/GigaZ & ILC      & ILC     & ILC   & TLEP          & SM prediction \\ \hline \hline
$\sqrt{s}$ [TeV]    & 14  & 14   & 0.091     & 0.161    & 0.161   & 0.250 & 0.161         & - \\
$\L [\fb]$          & 300 & 3000 &          & 100      & 480     & 500   & 3000$\times$4 & -\\ \hline
$\Delta M_W$ [MeV]  & 8  & 5     & -         & 4.1-4.5  & 2.3-2.9 & 2.8   &  $<1.2$          & 4.2(3.0) \\
$\Delta 
\sin^2\theta_\eff^\ell$ 
[$10^{-5}$]         & 36 & 21     &  1.3      &   -     &  -       & -  &    0.3        & 3.0(2.6)\\ \hline
\end{tabular}
\caption{Target accuracies for the measurement of $M_W$ and $\sin^2\theta_{eff}^\ell$ at the LHC, ILC and TLEP, also including estimated future theoretical uncertainties due to missing higher-order corrections,
  and theory uncertainties of their SM predictions. The uncertainties on the SM predictions are provided for $\Delta m_t=0.5(0.1)$~GeV (see Table~\ref{tab:electroweak-mwsin2tfuture} for details). At present the measured values for $M_W$ and
$\sin^2\theta_\eff^\ell$ are:  
$M_W=80.385 \pm 0.015$ GeV~\cite{Group:2012gb} and $\sin^2\theta_{\rm eff}^\ell= (23153 \pm 16) \times 10^{-5}$~\cite{zpole}
compared to their current SM predictions of Section~\ref{sec:electroweak-ewpo-theory}:
$M_W=80.360 \pm 0.008$~GeV and $\sin^2\theta_{\rm eff}^\ell= (23127 \pm 7.3) \times 10^{-5}$.}
\label{tab:electroweak-mwsin2t}
\end{center}
\end{table}

\subsection{Other EWPOs at lepton colliders}
\label{sec:otherewpos}

Besides $M_W$ and $\stl$, other EWPOs of great interest include the
$Z$-pole observables such as the $Z$ boson mass and width, and its
hadronic and leptonic partial widths. Estimates of statistical
sensitivity and beam energy calibration uncertainty are presented for
TLEP in~\cite{TLEPwhitepaper}.  This machine is capable of producing
about a trillion $Z$ bosons. Pending further study, preliminary
targets of 0.1 MeV uncertainty have been stated for $M_Z$ and
$\Gamma_Z$.  A relative precision of $5 \times 10^{-5}$ is stated to
be a reasonable target for the ratio of the $Z$-boson's
hadronic-to-leptonic partial widths at TLEP, as well as for the ratios
of the $Z$-boson leptonic widths (as a test of lepton
universality). These estimates would represent a factor of $\approx
20$ improvement over LEP.

 The vertex correction to the $Z \to b \bar{b}$ partial width is also
 of interest, which is sensitively probed by $R_b = \Gamma_{Z \to b
   \bar{b}} / \Gamma_{\rm had}$.  A precision of $2-5 \times 10^{-5}$
 is stated as a reasonable goal for the measurement of $R_b$ at TLEP,
 a factor of $\approx 10$ improvement over LEP and SLC.

A discussion of present and future anticipated theory errors of
predictions for $R_b$ and $\Gamma_Z$ can be found in
Section~\ref{sec:electroweak-ewpo-theory}.

\subsection{Prospects for determinations of SM parameters from global fits with GFITTER }
\label{sec:electroweak-ewpo-gfitter}


Measurements at future colliders will increase the experimental
precision of key observables sensitive to electroweak loop
effects. Among these are the $W$ boson mass, the top quark mass, and
the effective weak mixing angle. Alongside the construction of these
machines, progress in the calculation of multi-loop corrections to
these observables, and also in the determination of $\Delta
\alpha_{\rm had}(M_Z^2)$, is expected. Taken together in the global
electroweak fit, these improvements will provide tests of the
consistency of the SM with unprecedented power.

This section presents a short summary of preliminary studies foreseen
to be published soon. To date results of the global electroweak fit
are compared with expectations for the Large Hadron Collider (LHC)
with $\int Ldt=300\,{\rm fb}^{-1}$ at $\sqrt{s}=14$\,TeV and the
International Linear Collider (ILC) with GigaZ option~\cite{Baer:2013cma}.
The left columns of Table~\ref{tab:results_best} summarize the current
and the projected experimental precisions for the observables used in
the fit.
 For the studies of fit prospects at the LHC and ILC
presented here, the central values of the assumed future measurements
have been adjusted to obtain a common fit value of
$M_H\simeq126\,$GeV.  We assume that the theoretical uncertainties in
the SM predictions of $M_W$ and \sintheff reduce from the current
$\delta_{\rm theo} M_W=4\;$MeV and $\delta_{\rm
  theo}\sintheff=4.7\cdot10^{-5}$ to $1\;$MeV and $10^{-5}$,
respectively. We refer to our past
publications~\cite{Flacher:2008zq,Baak:2011ze,Baak:2012kk} for details
about the theoretical calculations and the statistical methods used.
  
\begin{figure}[h]
\begin{center}
\includegraphics[width=0.48\textwidth]{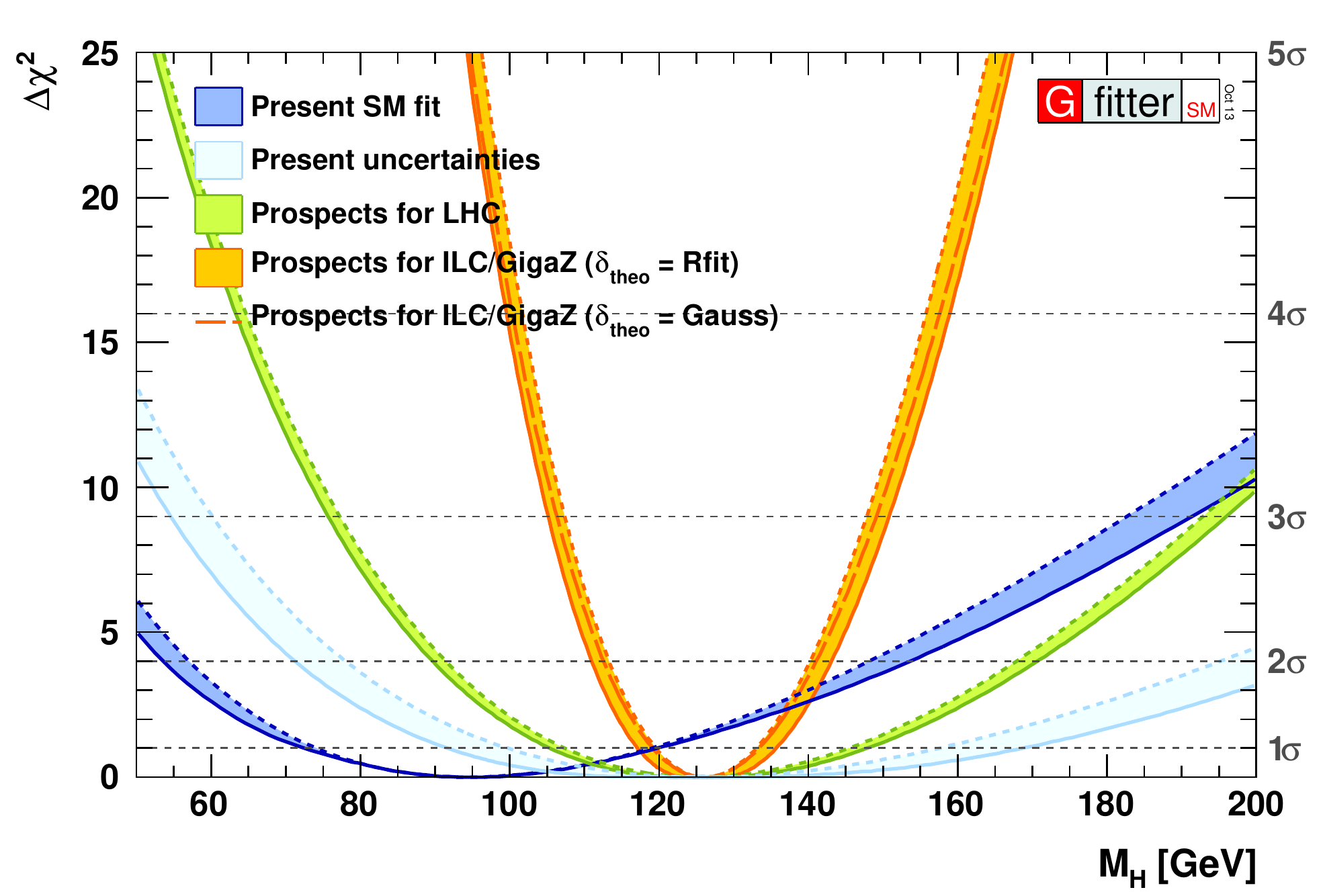}
\includegraphics[width=0.48\textwidth]{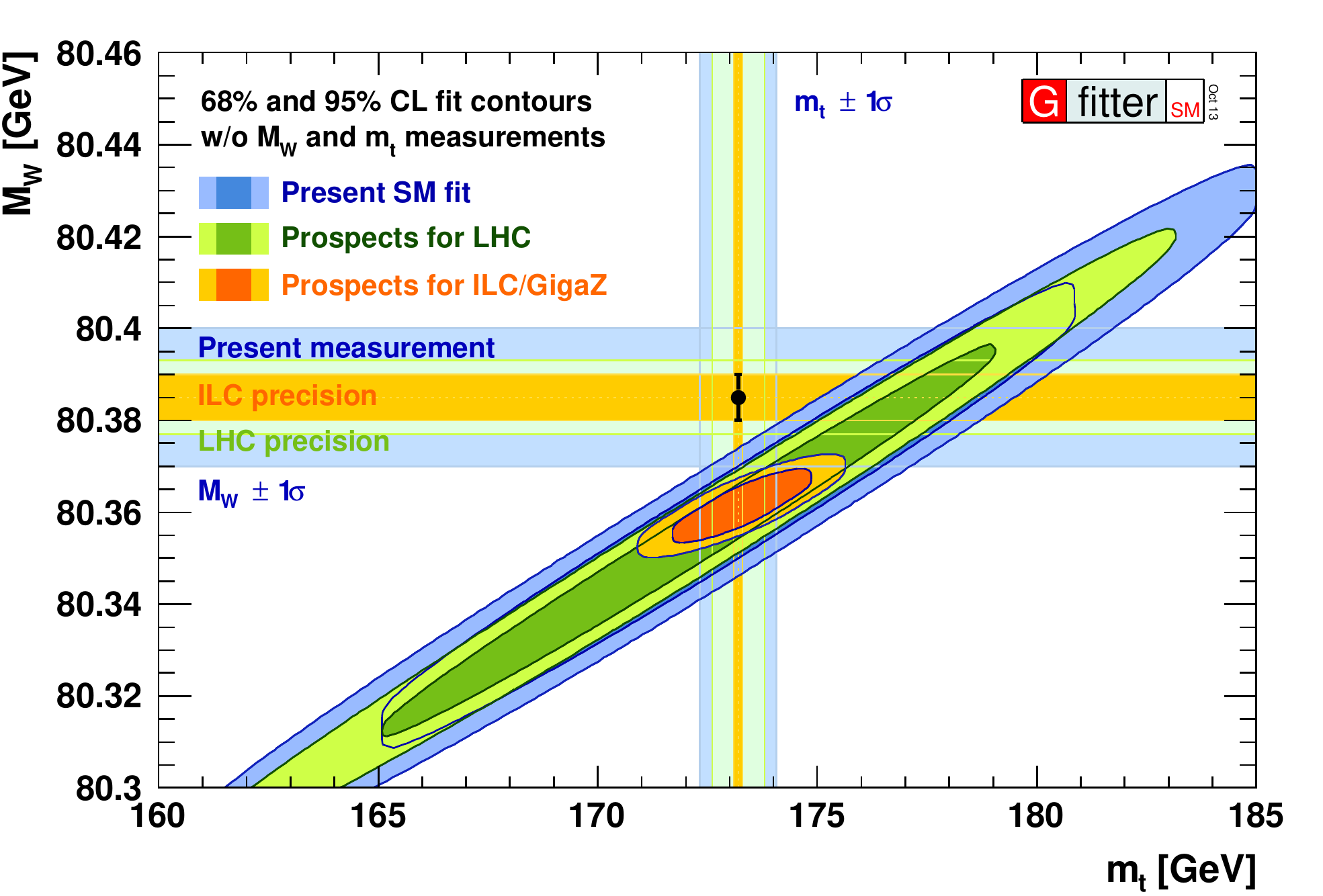}
\end{center}
\vspace{-0.2cm}
\caption{Fit results for the present and assumed future scenarios compared to 
  the direct measurements. 
  For the future scenarios the central values of the input 
  measurements are adjusted to reproduce the SM with $M_H\simeq126\;$GeV.
  Left: $\Delta \chi^2$ profiles versus $M_H$; in blue the present result,
  and in light blue, green and orange the present, LHC and ILC/GigaZ scenarios are shown, 
  respectively, all using the future fit setup with corresponding uncertainties.
  Right: $M_W$ versus $m_t$; the horizontal and
  vertical bands indicate in blue today's precision of the direct measurements, 
  and in light green and orange the extrapolated precisions for LHC and ILC/GigaZ,
  respectively.  
  }
\label{fig:sm1}
\end{figure}

Indirect determinations of the SM parameters and observables are
obtained by scanning the $\Delta\chi^2$ profile in fits where the
corresponding input constraint is ignored. Examples for such profiles
of the Higgs boson mass are shown in the left panel of
Fig.~\ref{fig:sm1}. The resulting one-sigma uncertainties are listed
in Table~\ref{tab:results_best}.
\begin{table}[b!]
\setlength{\tabcolsep}{0.0pc}
{\normalsize                                                                                   
\begin{tabular*}{\textwidth}{@{\extracolsep{\fill}}lcccccc}                                                                                                                                 
\hline\noalign{\smallskip}
                                   & \multicolumn{3}{c}{Experimental input [{\footnotesize $\pm 1 \sigma$]}   } &  \multicolumn{3}{c}{Indirect determination [{\footnotesize $\pm 1\sigma_{\rm exp}\pm 1\sigma_{\rm theo}$}]} \\
Parameter               &   Present &   LHC  &   ILC/GigaZ                   &  Present      & LHC   & ILC/GigaZ \\
\noalign{\smallskip}\hline\noalign{\smallskip}
$ M_{H}$        {\footnotesize[GeV]}              & $\pm0.4$ & $<\pm0.1$ & $<\pm0.1$ & $^{+32}_{-26}\;^{+12}_{-8}$   & $^{+20}_{-18}\;^{+3}_{-2}$  & $^{+6.9}_{-6.6}\;^{+2.7}_{-2.4}$ \\
$ M_{W}$        {\footnotesize [MeV]}              & $\pm15 $ & $ \pm8$  & $\pm5  $ &  $\pm6.3\; \pm\!4.2$        & $\pm4.8\; \pm\!1.0$     & $\pm1.9\; \pm\!1.7$        \\
$ M_{Z}$        {\footnotesize [MeV]}              & $\pm2.1$ & $\pm2.1$ & $\pm2.1$ &  $\pm11.6\; \pm\!3.8$      & $\pm6.9\; \pm\!0.8$       & $ \pm2.6\; \pm\!1.1$       \\
$ m_{t}$        {\footnotesize [GeV]}              & $\pm0.9$ & $\pm0.6$ & $\pm0.1$ & $\pm2.3\; \pm\!0.8$        & $ \pm1.4\; \pm\!0.2$  & $\pm0.7\;^{+0.3}_{-0.2}$ \\
$\sintheff$   {\footnotesize $[\cdot 10^{-5}]$} & $\pm16$  & $\pm16$  & $\pm1.3$ & $\pm4.8\; \pm\!5.0 $       & $\pm2.7\; \pm\!1.1$      & $\pm2.0\;\pm\!1.2$        \\
$ \Delta \alpha^{5}_{\rm had}M_Z^2$ {\footnotesize $[\cdot 10^{-5}]$} & $\pm10$  & $\pm4.7$ & $\pm4.7$ & $\pm43\; \pm\!15$   & $\pm36\; \pm\!4$  & $ \pm5.7\;\pm\!2.9$   \\
$ R_l^0             $ {\footnotesize $[\cdot 10^{-3}]$} & $\pm25$        & $\pm25$        & $\pm4$         & --                & --               & --                \\
$ \alpha_S(M_Z)$    {\footnotesize $[\cdot 10^{-4}]$} & --         & --         & --         & $\pm27\; \pm\!1$         & $\pm27\; \pm\!1$      & $^{+6.8}_{-6.3}\;^{+0.3}_{-0.2}$           \\
\noalign{\smallskip}\hline\noalign{\smallskip}
$ S|_{U=0}$           &  --     &   --  &   --                    & $ \pm0.09$ &  $ \pm0.09$   & $ \pm0.02$     \\
$ T|_{U=0}$           &  --     &   --  &   --                    & $ \pm0.07$ &  $ \pm0.06$   & $ \pm0.02$     \\
\noalign{\smallskip}\hline
\noalign{\smallskip}
\end{tabular*} \\
}
\caption{Current and estimated future uncertainties in the input observables 
  (left), and the precision obtained from the fit without using a given observable
  as input (right). The experimental and theoretical 
  uncertainties are separately given, where in the Rfit scheme the total 
  error corresponds to the linear sum of both contributions. The value 
  of $\alpha_s(M_Z^2)$ is not used directly as input in the fit. 
  The exact value of the uncertainties of the future $M_H$ measurements is not relevant for the fit. 
  For all indirect determinations shown in this table (including the present $M_H$ determination) the assumed central values of the input measurements have been adjusted to obtain a common fit value of $M_H\simeq126\;$GeV.}
\label{tab:results_best} 
\end{table}

The assumed improvements in the experimental precision of $M_W$ and
$m_t$ from the LHC lead to a reduction of the uncertainty in the
indirect determination of $M_H$ (present:$\;^{+44}_{-34}\,$GeV,
LHC:$\;^{+23}_{-20}\,$GeV). Substantial gain is achieved with the
ILC/GigaZ with an expected uncertainty of$\;^{+10}_{-9}$\,GeV. In all
scenarios the uncertainties on the indirect $M_H$ determination are
quoted for a common fit value of $M_H\simeq126$\,GeV.\footnote{A fit
  using the uncertainties and the central values of the present
  measurements yields $M_H=94^{+25}_{-22}$\,GeV~\cite{Baak:2012kk}.}

\begin{figure}[h]
\begin{center}
\includegraphics[width=0.48\textwidth]{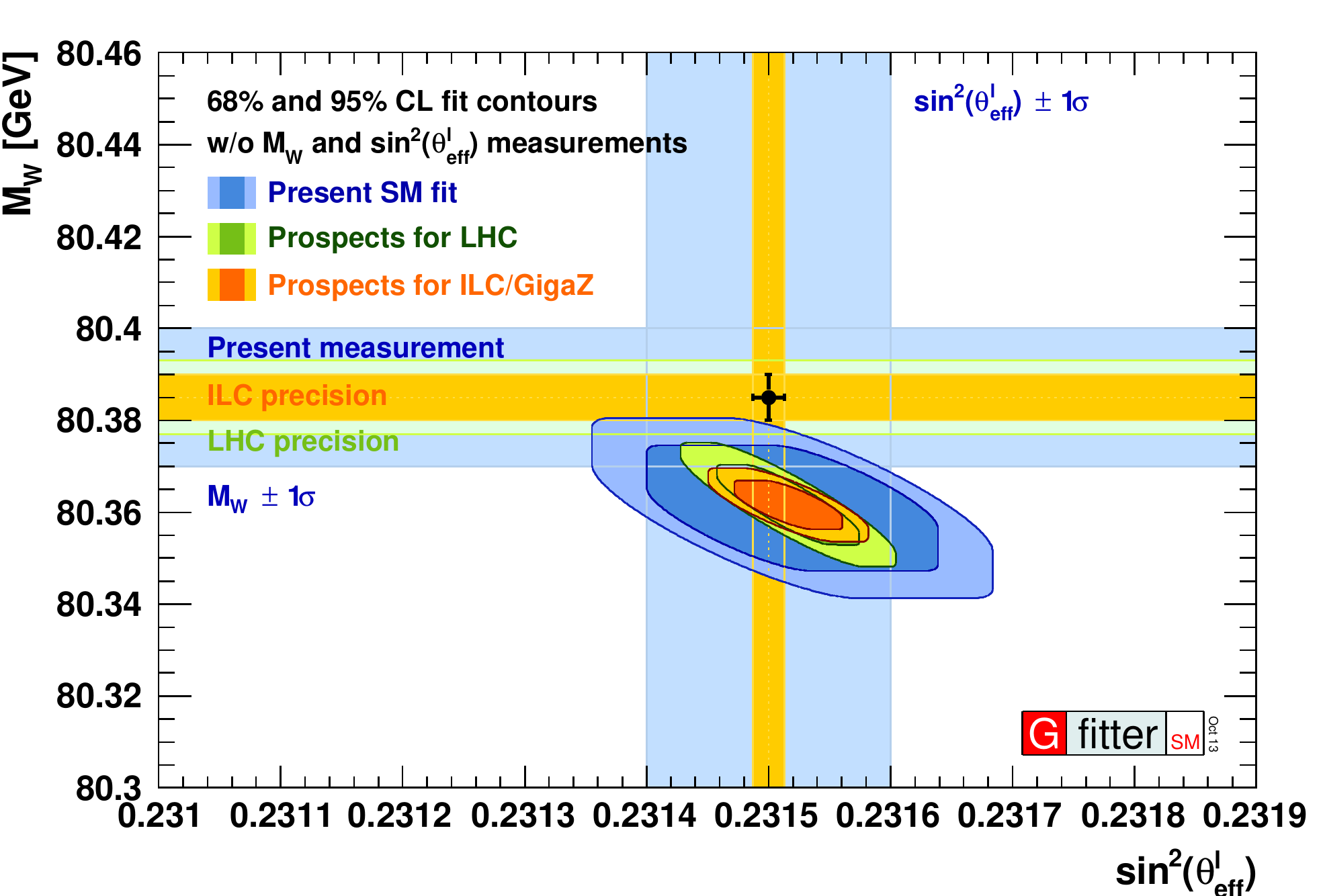}
\includegraphics[width=0.48\textwidth]{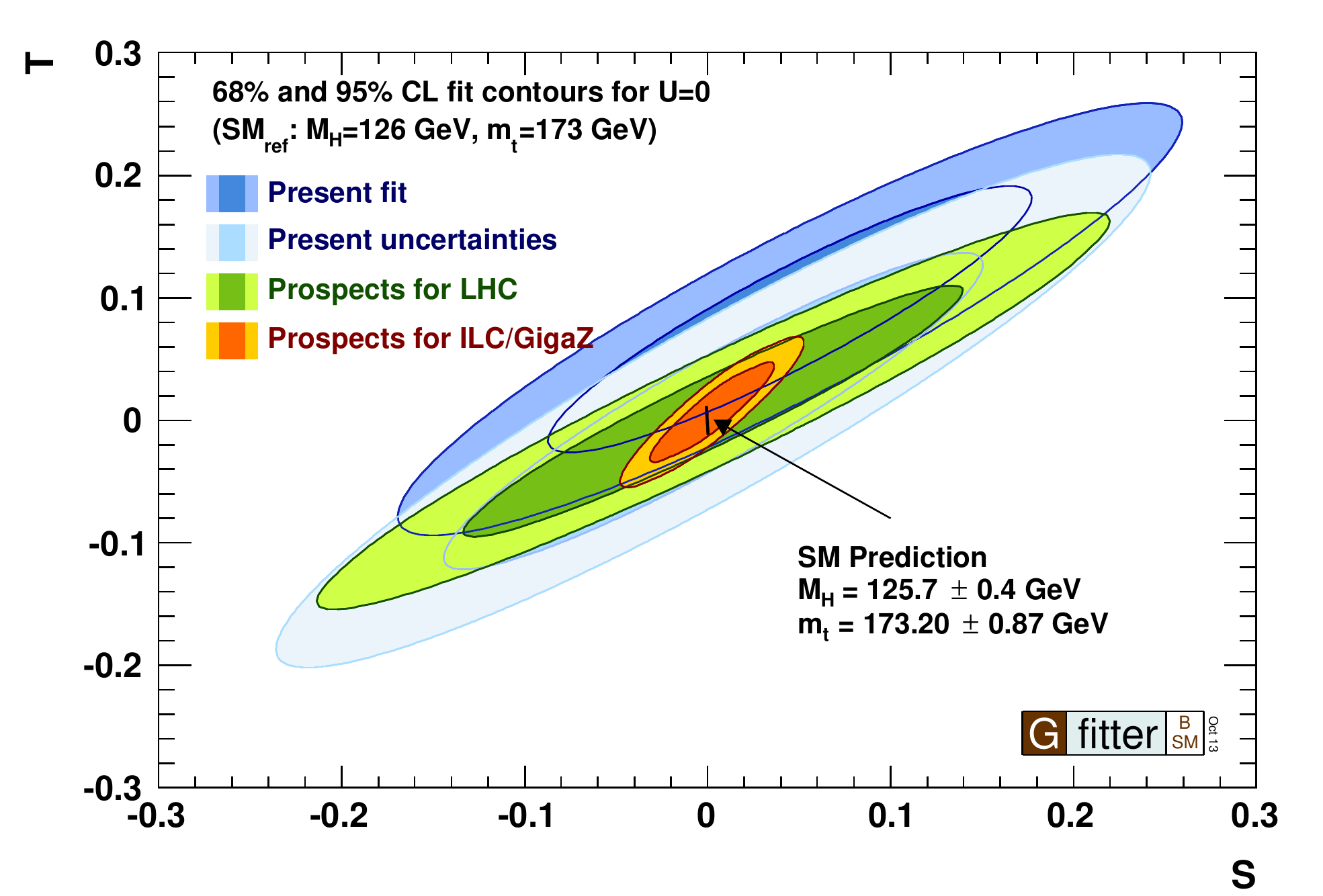}
\end{center}
\vspace{-0.2cm}
\caption{Fit results for the present and assumed future scenarios.
  Left: fit contours for $M_W$ versus \sintheff compared to the direct measurements;
  in blue, orange and green the present, LHC and ILC/GigaZ scenarios are shown, 
  respectively. For the future scenarios the central values of the input 
  measurements are adjusted to reproduce the SM with $M_H\simeq126\;$GeV;
  horizontal and vertical bands indicate today's and the expected future precision 
  of the direct measurements.   
  Right: constraints of the oblique parameters $S$ and $T$, with $U=0$ fixed, 
  for the present data (blue), the present uncertainties with central values adjusted to 
  obtain $M_H\simeq126$\,GeV (light blue), the LHC (green) and ILC/GigaZ prospects (orange). }
\label{fig:sm2}
\end{figure}
The current uncertainty on the indirect $M_W$ determination from the
fit of $11\;$GeV will improve to $5.8\;$GeV with the LHC and to
$3.6\;$MeV at the ILC/GigaZ. A comparison of the direct $M_W$
measurement at the ILC ($5\;$MeV expected precision) with the more
precise indirect determination will provide a stringent test of the
SM.

The right panel of Fig.~\ref{fig:sm1} shows the allowed areas for fits
with fixed variable pairs $M_W$ versus $m_t$ in the three scenarios
(current fit, LHC prospects, ILC prospects).  Also shown by the
horizontal and vertical bands are the one-sigma ranges of the current
direct measurements (blue), as well as the LHC (green) and ILC/GigaZ
expectation (yellow). A significant rise in precision is found for the
future scenarios.

The left panel of Fig.~\ref{fig:sm2} shows the corresponding results
in the $M_W$ versus \sintheff plane. The high precision of the
expected \sintheff measurement at the ILC will enable a stringent test
of the internal SM consistency in this area.

The expected constraints on the oblique
parameters~\cite{Peskin:1990zt,Peskin:1991sw} $S$ and $T$,
parametrizing possible new physics contributions to the neutral and to
the difference between neutral and charged weak current, respectively,
for a fixed value of $U=0$, are shown in the right panel of
Fig.~\ref{fig:sm2}. The blue ellipses indicate the 68\% and 95\%
contours of the current electroweak fit. The light blue ellipses show
the results using the current uncertainties with adjusted central
values to obtain $M_H\simeq126$\,GeV, the green ellipses that of the
LHC, and the orange ellipses show the precision expected for the
ILC/GigaZ.  The corresponding numerical values are given in
Table~\ref{tab:results_best}.  The sensitivity to new physics is
improved over a factor of three compared with that of today.

\subsection{EWPOs in the MSSM}
\label{sec:electroweak-ewpo-mssm}


Precision measurements of SM observables have proven to be a powerful
probe of BSM physics via virtual effects of the additional BSM
particles.  In general, precision observables (such as particle
masses, mixing angles, asymmetries etc.)  constitute a test of the
model at the quantum-loop level, since they can be calculated within a
certain model beyond leading order in perturbation theory, depending
sensitively on the other model parameters, and can be measured with
equally high precision.  Various models predict different values of
the same observable due to their different particle content and
interactions. This permits to distinguish between, e.g., the SM and a
BSM model, via precision observables. Naturally, this requires a very
high precision of both the experimental results and the theoretical
predictions.  (It should be kept in mind that the extraction of
precision data often assumes the SM.)
Important EWPOs are the $W$~boson mass, $\MW$, and the effective
leptonic weak mixing angle, $\sweff$, where the top quark mass plays a
crucial role as input parameter.  As an example for BSM physics the
Minimal Supersymmetric Standard Model (MSSM) is a prominent showcase and will be used here for illustration.


\bigskip
The first analysis concerns the $W$~boson mass. 
The prediction of $\MW$ in the MSSM depends on the masses, mixing
angles and couplings of all MSSM particles.  Sfermions, charginos,
neutralinos and the MSSM Higgs bosons enter already at one-loop level
and can give substantial contributions to $\MW$. The evaluation used
here consists of the complete available SM calculation, a full MSSM
one-loop calculations and all available MSSM two-loop
corrections~\cite{Heinemeyer:2006px,MWlisa}. 
Due to the strong MSSM parameter dependencies, it is
expected to obtain restrictions on the MSSM parameter space in the
comparison of the $\MW$ prediction and the experimental value.

The results for the general MSSM can be obtained in an extensive parameter
scan~\cite{MWlisa}. The ranges of the various SUSY parameters are given
in Table~\ref{tab:electroweak-scanparam}. $\mu$ is the Higgsino mixing
parameter, 
$M_{\tilde {F}_i}$ denotes the soft SUSY-breaking parameter for sfermions of
the $i$th family for left-handed squarks ($F = Q$), right-handed up- and
down-type squarks ($F = U, D$), left-handed sleptons ($F = L$) and
right-handed sleptons ($F = E$). $A_f$ denotes the trilinear sfermion-Higgs
couplings, $M_3$ the gluino mass parameter and $M_2$ the SU(2) gaugino mass
parameter, where the U(1) parameter is fixed as $M_1 = 5/3 s_w^2/c_w^2
M_2$. $M_A$ is the CP-odd Higgs boson mass and $\tan\beta$ the ratio of the two
Higgs vacuum expectation values. 

\begin{table}[htb!]
\centering
\begin{tabular}{cccl}
\hline
Parameter &  Minimum &  Maximum \\
\hline
$\mu$ & -2000       & 2000 \\
$M_{\tilde{E}_{1,2,3}}=M_{\tilde{L}_{1,2,3}}$ & 100       & 2000 \\
$M_{\tilde{Q}_{1,2}}=M_{\tilde{U}_{1,2}}=M_{\tilde{D}_{1,2}}$ & 500       & 2000 \\
$M_{\tilde{Q}_{3}}$     & 100       & 2000 \\
$M_{\tilde{U}_{3}}$     & 100       & 2000 \\
$M_{\tilde{D}_{3}}$     & 100       & 2000 \\
$A_e=A_{\mu}=A_{\tau}$   & -3$\,M_{\tilde{E}}$      & 3$\,M_{\tilde{E}}$     \\
$A_{u}=A_{d}=A_{c}=A_{s}$& -3$\,M_{\tilde{Q}_{12}}$  & 3$\,M_{\tilde{Q}_{12}}$ \\
$A_b$ & -3$\,$max($M_{\tilde{Q}_{3}},M_{\tilde{D}_{3}}$)  &
3$\,$max($M_{\tilde{Q}_{3}},M_{\tilde{D}_{3}}$) \\ 
$A_t$ & -3$\,$max($M_{\tilde{Q}_{3}},M_{\tilde{U}_{3}}$)  &
3$\,$max($M_{\tilde{Q}_{3}},M_{\tilde{U}_{3}}$) \\ 
$\tan\beta$     & 1       & 60 \\
$M_3$     & 500     & 2000 \\
$M_A$     & 90      & 1000\\
$M_2$     &100      &1000\\
\hline
\end{tabular}
\caption{MSSM parameter ranges. All parameters with mass dimension are given
  in GeV.}
\label{tab:electroweak-scanparam}
\end{table}

All MSSM points included in the results have the neutralino as LSP and
the sparticle masses pass the lower mass limits from direct searches
at LEP.  The Higgs and SUSY masses are calculated using 
{\tt FeynHiggs} (version 
2.9.4)~\cite{Frank:2006yh,Degrassi:2002fi,Heinemeyer:1998np,Heinemeyer:1998yj,Hahn:2009zz}.
For every point, it was tested whether it is allowed by direct Higgs
searches using the code {\tt HiggsBounds} (version 3.8.0)
\cite{Bechtle:2008jh,Bechtle:2011sb}.  This code tests the MSSM points
against the limits from LEP, Tevatron and  the LHC. 

\begin{figure}[htb]
\begin{center}
\includegraphics[width=0.5\hsize]{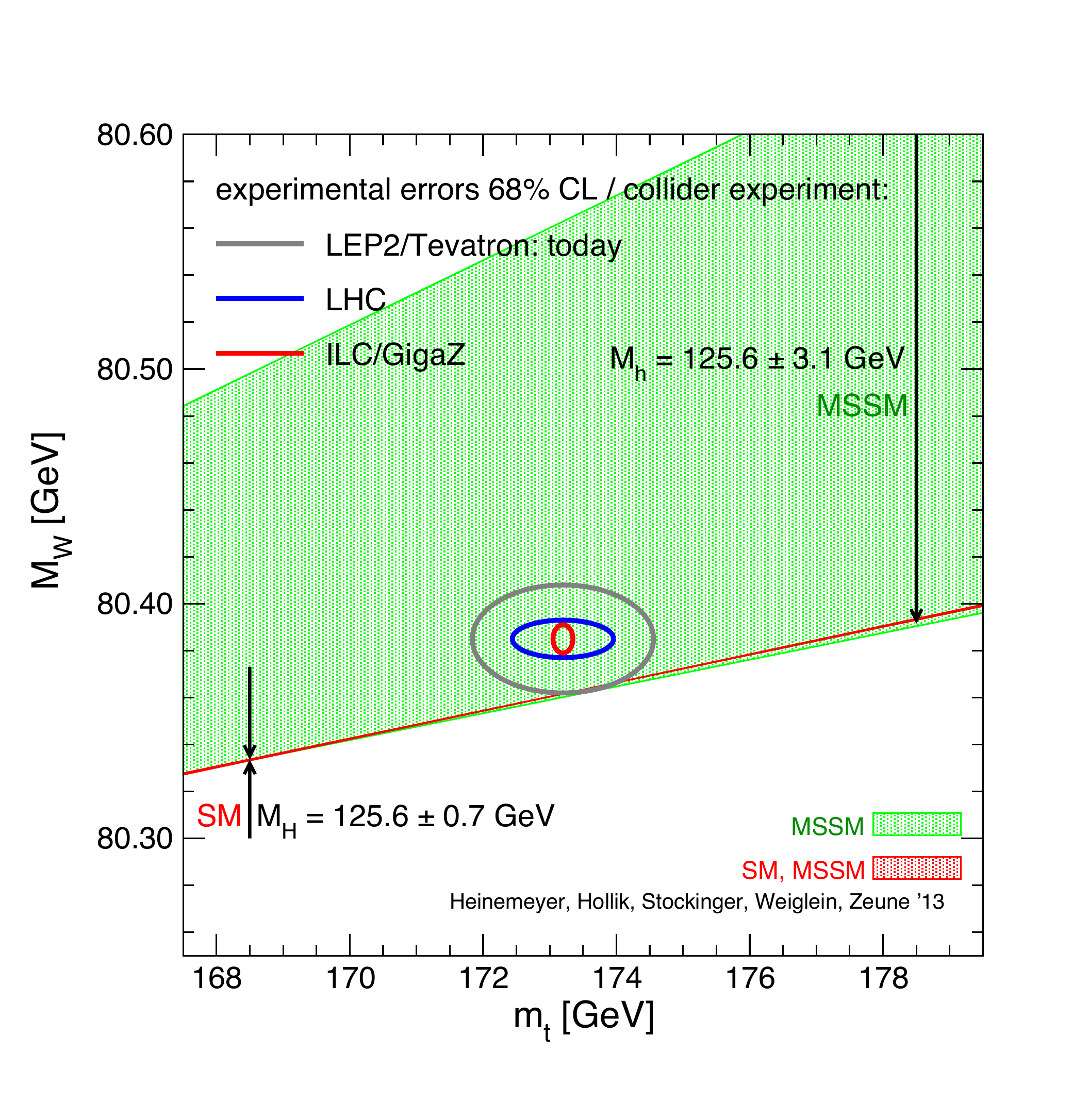}
\caption{Predictions for $\MW$ as a function of $m_t$ in the SM and
  MSSM (see text). The gray, blue and red ellipses denote the current,
  and the target LHC and ILC/GigaZ precision, respectively, as
  provided in Table~\ref{tab:electroweak-mwsin2t}.  }
\label{fig:electroweak-mtmwmh125}
\end{center}
\end{figure}

The results for $\MW$ are shown in Fig.~\ref{fig:electroweak-mtmwmh125}
as a function of $m_t$, assuming the light $CP$-even Higgs $h$ in the region
$125.6 \pm 0.7 (3.1)$~$\gev$ in the SM (MSSM) case.  
The red band indicates the overlap region of the
SM and the MSSM.  The leading one-loop SUSY contributions arise from
the stop sbottom doublet. However requiring $M_h$ in the region 
$125.6 \pm 3.1$~$ \gev$ restricts the parameters in the stop
sector~\cite{Heinemeyer:2011aa} and with it the possible $\MW$
contribution.  Large $\MW$ contributions from the other MSSM sectors
are possible, if either charginos, neutralinos or sleptons are light.

The gray ellipse indicates the current experimental uncertainty,
whereas the blue and red ellipses shows the anticipated future LHC and
ILC/GigaZ precisions, respectively (for each collider experiment
separately) of Table~\ref{tab:electroweak-mwsin2t}, along with $m_t = 172.3 \pm 0.9 ~ (0.5, 0.1)$~GeV for 
the current (LHC, ILC) measurement of the top quark mass.  While, at the
current level of precision, SUSY might be considered as slightly
favored over the SM by the $\MW$-$m_t$ measurement, no clear
conclusion can be drawn. The smaller blue and red ellipses, on the
other hand, indicate the discrimination power of the future LHC and
ILC/GigaZ measurements. With the improved precision a small part of
the MSSM parameter space could be singled out.

\begin{figure}[htb]
\begin{center}
\includegraphics[width=0.48\hsize]{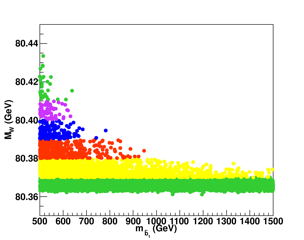}
\includegraphics[width=0.48\hsize]{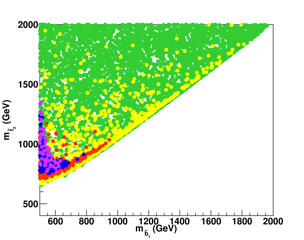}
\caption{Results of a MSSM parameter
  scan to illustrate what can be learned from an improved $M_W$
  measurement under the assumption a light stop is found with
  $m_{\tilde t_1}=400 \pm 40$ GeV: green points: all points in
  the scan with $M_h =125.6 \pm 3.1$ GeV and $m_{\tilde t_1}=400 \pm 40$ GeV,
  yellow, red, blue, purple points: 
  $M_W=80.375 \pm 0.005$ GeV (yellow), 
 $M_W=80.385 \pm 0.005$  GeV (red), 
  $M_W=80.395 \pm 0.005$ GeV (blue), 
 and $M_W=80.405 \pm 0.005$ GeV (purple)
.}
\label{fig:electroweak-fig1lisa}
\end{center}
\end{figure}

In a second step we apply the precise ILC measurement of $\MW$ to
investigate its potential to determine unknown model parameters. Within
the MSSM we assume the hypothetical future situation that a light scalar
top has been discovered with $\mste = 400 \pm 40$~$ \gev$ at the LHC, but
that no other new particle has been observed. We set lower limits of 
$100$~$ \gev$ on sleptons, $300$~$ \gev$ on charginos, $500$~$ \gev$ on 
squarks of the third generation and $1200$~$ \gev$ on the
remaining colored particles. The neutralino mass is constrained by the GUT relation $M_1 \approx M_2 / 2$ so that 
 setting a lower limit of 300~GeV on charginos also sets the lower mass limits on neutralinos of $\sim  150$~GeV.
 We have selected the points from our scan
accordingly. Any additional particle observation would lead to an even
more restricted set of points and thus strengthen the parameter
determination. In Fig.~\ref{fig:electroweak-fig1lisa},
we show the ``surviving'' points from our scan.
All points fulfil $\Mh = 125.6 \pm 3.1$~$ \gev$ and $\mste = 400 \pm 40$~$ \gev$. Orange, red, blue and purple 
points denote in addition $W$~boson mass values of 
$\MW = 80.375, 80.385, 80.395, 80.405 \pm 0.005$~$ \gev$, respectively. 
In the right-hand figure we show the results as a function of the masses of the 
heavy scalar top and the light scalar bottom. It can be seen that these
unknown mass scales are restricted to small intervals if $80.385$~$ \gev$ or higher $M_W$ values  are assumed as central experimental values
(i.e.\ sufficiently different from the SM prediction). In this situation
the precise $\MW$ measurement could give clear indications of where to
search for these new particles (or how to rule out the simple MSSM
picture). For instance, a measurement of $M_W=80.405 \pm 0.005$ GeV (purple MSSM scenarios) would indicate that 
$m_{\tilde b_1} <800$~GeV. 

The MSSM parameter space
could also be constrained by a precise measurement of $\sweff$. 
The evaluation of the latter is performed at the same level of accuracy
as for $\MW$~\cite{Heinemeyer:2007bw}.

\begin{figure}[htb!]
\begin{center}
\includegraphics[width=0.58\textwidth]{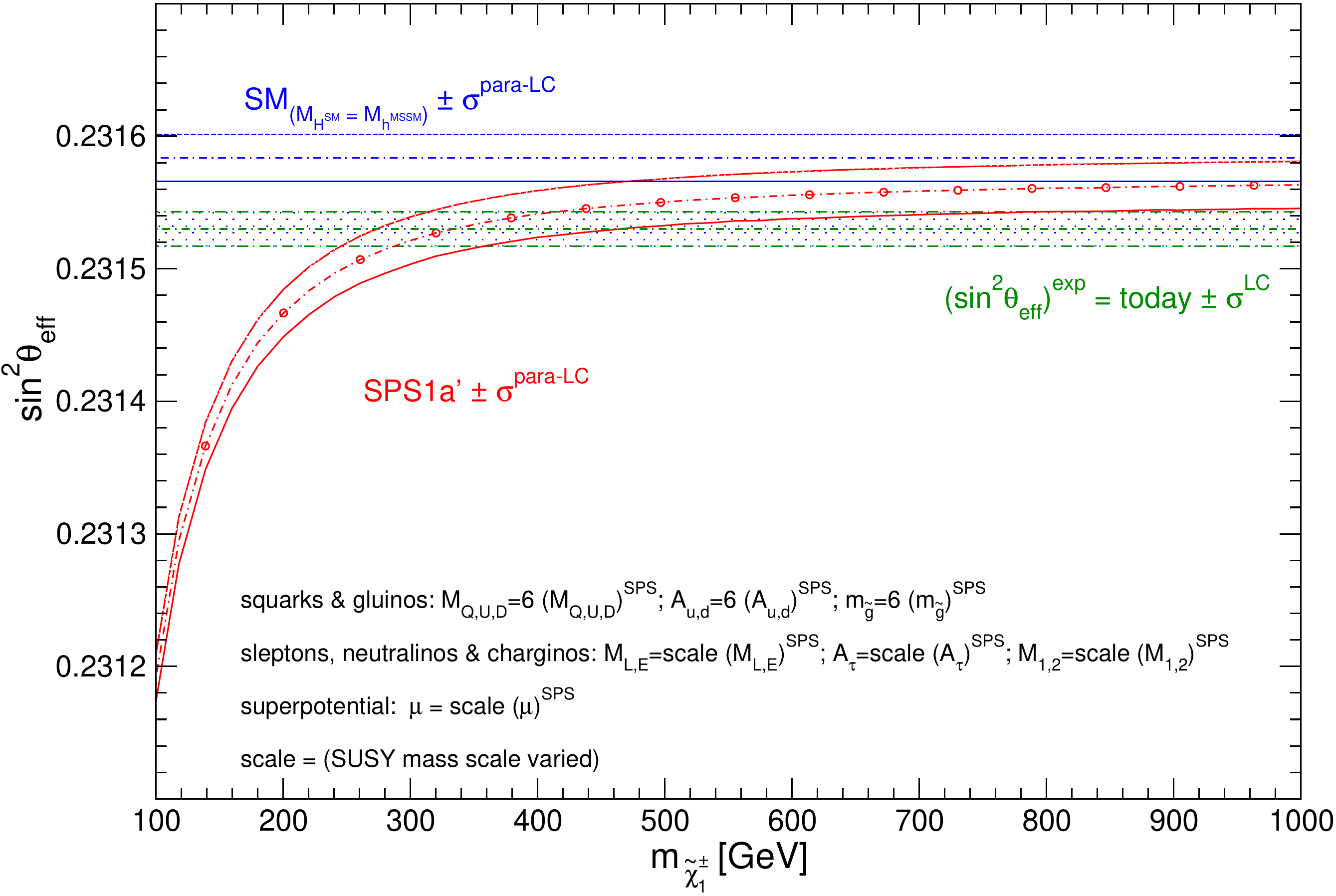}
\caption{
Theoretical prediction for $\sweff$ in the SM and the MSSM (including
prospective parametric theoretical uncertainties) compared to
the experimental precision at the ILC with GigaZ option.  
An SPS~1a$'$ inspired scenario is used, where the squark and gluino
mass parameters
are fixed to 6~times their SPS~1a$'$ values. The other mass 
parameters are varied with a common scale factor (see text). 
}
\label{fig:electroweak-sw2eff-theo} 
\end{center}
\end{figure}

In the first example it is investigated
whether the high accuracy achievable at
the GigaZ option of the ILC would provide sensitivity to indirect effects of
SUSY particles even in a scenario where the 
superpartners are so heavy that they escape detection at the
LHC~\cite{Heinemeyer:2007bw}. 
We consider in this context a scenario with very heavy squarks and a 
very heavy gluino. It is based on the values of the SPS~1a$'$ benchmark
scenario~\cite{Allanach:2002nj}, but the squark and gluino
mass parameters
are fixed to 6~times their SPS~1a$'$ values. The other masses are 
scaled with a common scale factor given by the light chargino mass, 
except $M_A$ which we keep fixed at its SPS~1a$'$ value.
In this scenario 
the strongly interacting particles are too heavy to be detected at the
LHC, while, depending on the scale-factor, some colour-neutral particles
may be in the ILC reach. In Fig.~\ref{fig:electroweak-sw2eff-theo} we
show the prediction for $\sweff$ in
this scenario as a function of the lighter chargino
mass, $\mcha1$. The prediction includes the parametric
uncertainty, $\sigma^{\rm para-LC}$, induced by the ILC measurement of $\mt$, 
$\Delta\mt = 100$~$ \mev$, and the numerically more
relevant prospective future uncertainty on $\Delta\alpha^{(5)}_{\textup{had}}$,
$\delta(\Delta\alpha^{(5)}_{\textup{had}})=5\times10^{-5}$. 
The MSSM prediction for $\sweff$
is compared with the experimental resolution with GigaZ precision,
$\sigma^{\rm LC} = 0.000013$, using for simplicity the current
experimental central value. The SM prediction (with 
$\MHSM = \Mh^{\rm MSSM}$) is also shown, applying again the parametric 
uncertainty $\sigma^{\rm para-LC}$.
Despite the fact that no coloured SUSY particles would be observed at
the LHC in this scenario, the ILC with its high-precision measurement
of $\sweff$ in the GigaZ mode could resolve indirect effects of SUSY
up to $\mcha1 \lsim 500$~$ \gev$. This means that the
high-precision measurements at the LC with GigaZ option could be
sensitive to indirect effects of SUSY even in a scenario where SUSY
particles have {\em neither \/} been directly detected at the LHC nor
the first phase of the ILC with a centre of mass energy of up to 
$500$~$ \gev$.

\bigskip
We now analyse the sensitivity of $\sweff$ together with $\MW$ 
to higher-order effects in the MSSM by investigating a broad parameter
scan range similar as in Tab.~\ref{tab:electroweak-scanparam}.
Only the constraints on the MSSM parameter space
from the LEP Higgs searches~\cite{Barate:2003sz,Schael:2006cr} and the lower
bounds on the SUSY particle masses previous to the LHC SUSY searches were
taken into account. However, the SUSY particles strongly affected by the LHC
searches are the squarks of the first and second generation and the
gluino. Exactly these particles, however, have a very small effect on
the prediction of $\MW$ and $\sweff$ and thus a negligible effect on
this analysis. 

\begin{figure}[htb!]
\begin{center}
\includegraphics[width=0.58\textwidth]{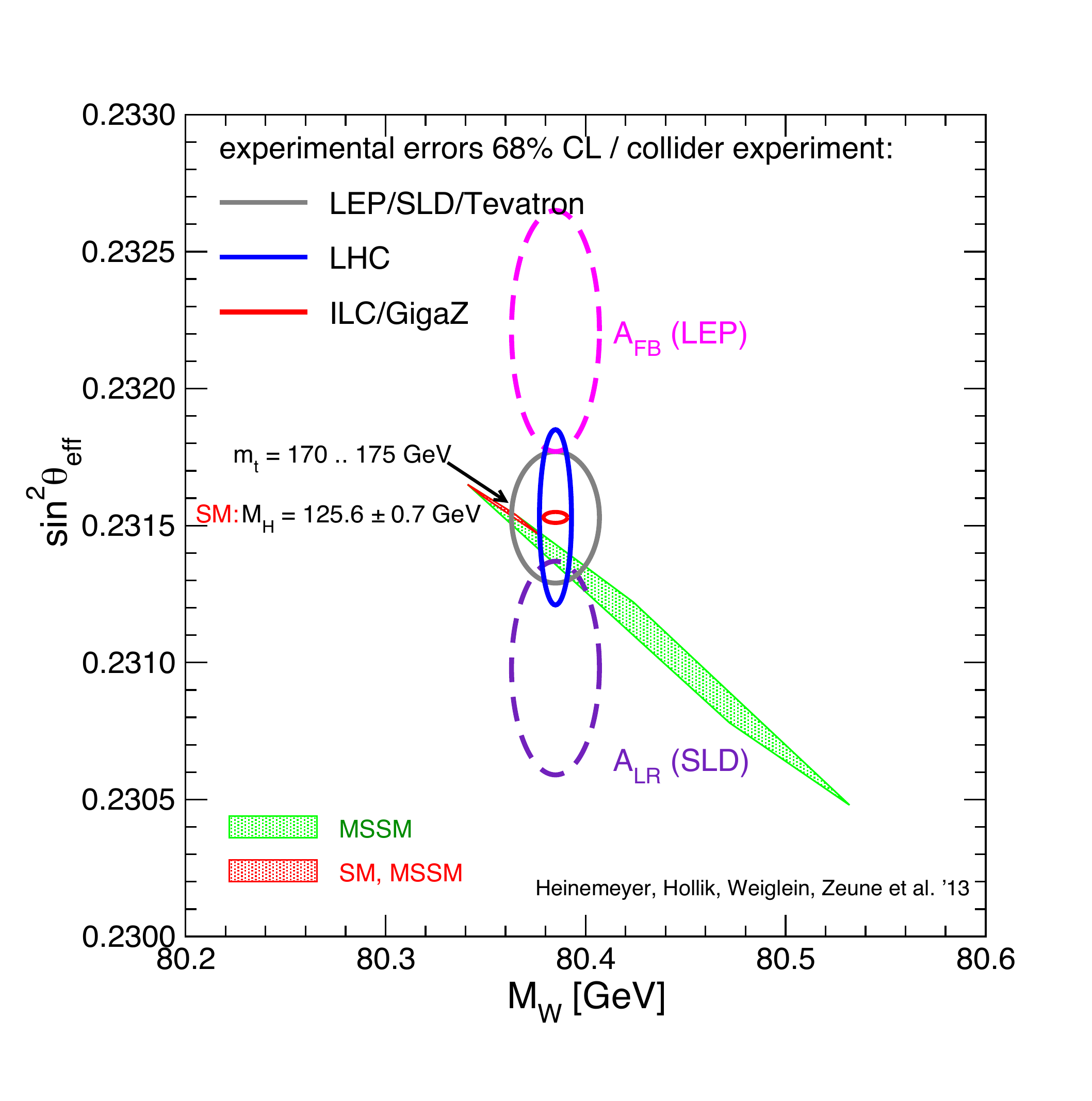}
\end{center}
\vspace{-1em}
\caption{MSSM parameter scan for $\MW$ and $\sweff$ (see text). 
  Today's 68\%~C.L.\ ellipses (from 
  $A_{\rm FB}^b({\rm LEP})$, $A_{\rm LR}^e({\rm SLD})$ and the world
  average) are shown 
  as well as the anticipated LHC and ILC/GigaZ precisions, drawn around today's
  central value. 
}
\label{fig:electroweak-sw2efftheo-Scans2} 
\end{figure}

In Fig.~\ref{fig:electroweak-sw2efftheo-Scans2} we
compare the SM and the MSSM predictions for $\MW$ and $\sweff$
as obtained from the scatter plot data. 
The predictions within the two models 
give rise to two regions in the $\MW$--$\sweff$ plane, red for the SM
and green for the MSSM. The SM region also forms part of
the MSSM-allowed intervals in the decoupling regime. 
For the SM $\MHSM = 125.6 \pm 0.7$~$ \gev$ has been required, whereas for
the MSSM the Higgs mass measurement is met with a larger uncertainty due
to the still large theory uncertainties in the $\Mh$
calculation~\cite{Degrassi:2002fi}. 
The variation with $\mt$ from $170$ to $175$~$ \gev$ is indicated.
The 68\%~C.L.\ experimental results
for $\MW$ and $\sweff$ are indicated in the plot, given for the current
precision and for the target LHC and ILC/GigaZ accuracies, see
Tab.~\ref{tab:electroweak-mwsin2t}. 
The center ellipse corresponds to the current world average. 
Also shown are the error ellipses corresponding to the two individual
most precise measurements of $\sweff$, based on $A_{\rm LR}^e$ by SLD
and $A_{\rm FB}^b$ by LEP, corresponding to
\begin{eqnarray}
\label{eq:electroweak-afb}
A_{\rm FB}^b({\rm LEP}) &:& \sweff^{\rm exp,LEP} = 0.23221 \pm 0.00029~, \\[.1em]
\label{alr}
A_{\rm LR}^e({\rm SLD}) &:& \sweff^{\rm exp,SLD} = 0.23098 \pm 0.00026~, \\[.1em]
\label{eq:electroweak-sweff-exp}
                       &~~& \sweff^{\rm exp,aver.} = 0.23153 \pm 0.00016~,
\end{eqnarray}
where the latter one represents the average~\cite{zpole}.  The first
(second) value prefers a value of $\MHSM \sim 32$~$ (437)$~$ \gev$.
The two measurements differ by about $3\,\sigma$.  The averaged value
of $\sweff$, as given in Eq.~\ref{eq:electroweak-sweff-exp}, prefers
$\MHSM \sim 110$~$ \gev$.
One can see that the current averaged value is compatible with the SM
with $\MHSM \sim 125.6$~$ \gev$
and with the MSSM. The value of $\sweff$ obtained from $A_{\rm LR}^e$(SLD)
clearly favors the MSSM over the SM.
On the other hand, the value of $\sweff$ obtained from $A^b_{\rm FB}$(LEP) 
together with the $\MW$ data from LEP and the Tevatron would correspond to an
experimentally preferred region that deviates from the predictions of both
models. 
This unsatisfactory solution can only be resolved by new measurements.
The anticipated LHC accuracy for $\sweff$ would have only a limited
potential to resolve this discrepancy, as it is larger than the current
uncertainty obtained from the LEP/SLD average. 
On the other hand, a $Z$~factory, i.e.\ the GigaZ option, would be an
ideal solution, as is indicated by the red ellipse.
The anticipated
ILC/GigaZ precision of the combined $\MW$--$\sweff$ measurement could
put severe constraints on each of the models and resolve the discrepancy
between the $A_{\rm FB}^b$(LEP) and $A_{\rm LR}^e$(SLD) measurements. 
If the central value of an improved measurement with higher precision should
turn out to be close to the central value favored by the current measurement
of $A_{\rm FB}^b({\rm LEP})$, this would mean that the electroweak
precision observables $\MW$ and $\sweff$ could rule out both the SM and the
most general version of the MSSM.

\subsection{EWPOs and $Z'$ bosons}
\label{sec:electroweak-ewpo-zprime}


EWPOs also constrain possible new physics scenarios such as $U(1)'$
gauge extensions of the SM.  Current constraints~\cite{Erler:2009jh}
on the associated $Z'$ boson masses, $M_{Z'}$, are generally
comparable and in some cases stronger than the direct lower search
limits from LEP and the Tevatron.  The 8~TeV LHC data have extended
the lower limits to roughly 2.5~TeV (depending on the model).
However, the LHC dilepton and dijet resonance searches are insensitive
to the $Z$-$Z'$ mass mixing angle, $\theta_{ZZ'}$.  Current EWPOs
constrain $\theta_{ZZ'}$ to the $10^{-2}$ level and very often well
below this.  The EWPOs projected for the ILC including the GigaZ option as
shown in Table~\ref{tab:electroweak-mwsin2t} (most importantly the measurements of $M_W$ to 2.3-2.9~MeV,
the effective weak mixing angle to $1.3 \times 10^{-5}$, and $m_t$ to
0.1~GeV) would improve the $\theta_{ZZ'}$ limits by almost another
order of magnitude.  This is important, since in specific models,
$M_{Z'}$ and $\theta_{ZZ'}$ are not independent.  As an example,
consider the popular benchmark case of the $Z_\chi$ boson (appearing
in $SO(10)$ GUT models) with a $U(1)'$ breaking Higgs sector
compatible with Supersymmetry.  In this case, the projected EWPOs
would experience noticeable shifts for $M_{Z'}$ values of up to 6~TeV,
without assuming any improvement in $\Delta \alpha_{\rm had}$.  The
EWPOs are also important for leptophobic $Z'$ bosons where the LHC
sets weaker mass limits.

In the case of a $Z'$ discovery at the LHC, it becomes mandatory to
achieve the highest possible accuracy in the EWPOs.  As an
illustration, suppose a future LHC run discovers a dilepton resonance
with an invariant mass of 3~TeV.  Even if one would succeed to
determine the spin of the resonance, it would not be possible to
simultaneously obtain meaningful information on the coupling strength
and on $\theta_{ZZ'}$, by using LHC data alone.  But the EWPOs would
determine the size and the sign of $\theta_{ZZ'}$ which would give
valuable information on the $U(1)'$ breaking sector and {\em
simultaneously\/} constrain the $T$ parameter to the level of $\pm
0.01$, thereby constraining possible additional non-degenerate fermion
(or scalar) multiplets that may be necessary to cancel gauge anomalies
related to the $U(1)'$.

\section{New interactions in vector boson scattering and tri-boson processes}
\label{sec:electroweak-multiboson}

Multi-boson production in various topologies provides a unique way to
probe new physics. Assuming that the 125 GeV boson discovered at the
LHC is the SM Higgs boson, it is natural to assume that electroweak
symmetry breaking occurs according to the SM Higgs
mechanism. Therefore, deviations from the SM in multi-boson production
can be parameterized by SU(2)$_L \times $U(1)$_Y$ gauge-invariant operators which do not
introduce any new sources of EWSB. If the new physics associated with
these operators occurs at a high mass scale, one is motivated to use
the formulation of Effective Field Theory (EFT) to organize the
operators in order of increasing dimensionality.
Here we will consider an EFT, which includes
dimension-6 and dimension-8 operators that modify the
interactions among electroweak gauge bosons, described by the following Lagrangian:
\begin{equation}
{\cal L}_{EFT} = {\cal L}_{SM} + \sum_{i=WWW,W,B, \\\Phi W,\Phi B} \frac{c_i}{\Lambda^{2}}{\cal O}_i+\sum_{j=0,1} 
\frac{f_{S,j}}{\Lambda^{4}}{\cal O}_{S,j}+\sum_{j=0,\ldots,9}  \frac{f_{T,j}}{\Lambda^{4}}{\cal O}_{T,j}+\sum_{j=0,\ldots,7}\frac{f_{M,j}}{\Lambda^{4}}{\cal O}_{M,j}
\label{eq:eft2}
\end{equation}
A detailed discussion of these operators is provided in Section~\ref{sec:electroweak-eft}.

As an example of new physics in the Higgs sector, let us consider the
interaction of the Higgs doublet field $\Phi$ with a new scalar field $S$ of
the form $\Phi^\dagger \Phi S$. 
This operator can mediate $\Phi \Phi
\to \Phi \Phi$ scattering via $s$ and $t$ channel exchange of the $S$
boson. In the limit of the mass of $S$ being much larger than the
energy of this scattering process, the lowest dimension effective
operator induced is the dimension-4 operator $(\Phi^\dagger
\Phi)^2$, which mimics the quartic Higgs potential in the SM. At the next order in the momentum flowing along the $S$
propagator, the effective operator induced is $$O_{\phi d} = \frac{c_{\phi d}
}{M_S^2} \partial_\mu (\Phi^\dagger \Phi) \partial^\mu (\Phi^\dagger
\Phi) $$ where the coefficient is enhanced by the coupling of the
$\Phi$ to the $S$ field and is suppressed by the squared mass of the
$S$ boson. This example illustrates a tree-level contribution to a
higher-dimension operator due to a new interaction with a massive
scalar field.  After the Higgs field $\Phi$ acquires a vev, the
operator $O_{\phi d}$ changes the normalization of the Higgs field and
therefore changes its coupling to the electroweak gauge bosons. As a
result, the unitarization of the vector boson scattering (VBS) amplitudes is
altered and we would expect anomalous contributions.
 
An example of a dimension-8 operator is provided by the analogue of
the QED light-by-light scattering mediated by the electron box
loop. In the limit that the electromagnetic field is weak and slowly
varying, this process is described by the Euler-Heisenberg Lagrangian
$$ {\cal L}_{EH} = \frac{1}{2} (E^2 - B^2) + 2 \frac{\alpha^2}{ 45 m^4} [ (E^2 - B^2)^2 + 7 (E.B)^2 ]$$
where $E$ and $B$ are the electric and magnetic field strengths,
$\alpha$ is the electromagnetic coupling and $m$ is the electron
mass. The second term represents the $\gamma \gamma \to \gamma \gamma$
scattering EFT operator induced by the electron box diagram when the
photon energies are much smaller than the electron mass. This term can
be re-written as a linear combination of the operators $(F_{\mu \nu} F^{\mu \nu})^2$
and $(F_{\mu \rho} F^{\mu \sigma}) (F^{\nu \rho} F_{\nu \sigma})$, where
$F$ is the electromagnetic field strength tensor.
 
Similarly, one may imagine a new heavy fermion coupling to the
electroweak gauge bosons and inducing  a four-boson contact
interaction via a box loop.  Such an interaction can mediate anomalous
triple gauge boson production and anomalous vector boson
scattering. The operator would be suppressed by four powers of the
heavy fermion mass and enhanced by potentially strong coupling between
the new fermion and the longitudinal vector bosons.  These dimension-8
operators are described by the operators ${\cal O}_{T,i}, i=0,1,2$ of Eqs.~\ref{eq:t0},~\ref{eq:t1} and~\ref{eq:t2}.

In the following, a review of studies using VBS and triboson channels is presented. The main purpose of these studies is to estimate the improvement
 of sensitivity to these operator coefficients as a function of integrated luminosity and collider energy.

\subsection{Theory of non-standard EW gauge boson interactions}
\label{sec:electroweak-eft}


While the translation between simplified new physics models in the EW
sector to an EFT is described in~\cite{Reuter:2013gla} and also
presented in Section~\ref{sec:electroweak-resonances}, there are also
ambiguities for the low-energy EFT. This results from the choice of
operator bases. In the following sections, which are taken from
Ref.~\cite{Degrande:2013rea}, we give a brief overview of dimension-6
and dimension-8 operators, discuss the EFTs in different operator
bases, and provide translations from one basis to another.  This
should simplify the comparison between many different studies that
have been performed for several past, present and future collider
experiments. We also address unitarization and discuss the role of
higher-order corrections in studies of non-standard EW interactions
using \VBFNLO{} and a \POWHEGBOX{} implementation of higher-order QCD
corrections to $WWjj$ production.

\subsubsection{Dimension-six operators for electroweak vector boson pair and triple production and scattering}
\label{sec:dim6}

If baryon and lepton numbers are conserved, only operators with even
dimension can appear in the EFT. Consequently, the largest new physics
contribution is expected from dimension-six operators.  Three CP
conserving dimension-six operators,
\begin{equation}
 \begin{aligned}
   {\cal O}_{WWW}&=\mbox{Tr}[W_{\mu\nu}W^{\nu\rho}W_{\rho}^{\mu}]\\
{\cal O}_W&=(D_\mu\Phi)^\dagger W^{\mu\nu}(D_\nu\Phi)\\
{\cal O}_B&=(D_\mu\Phi)^\dagger B^{\mu\nu}(D_\nu\Phi),
\end{aligned}
\label{opTGC}
\end{equation}

and two CP violating dimension-six operators,
\begin{equation}
 \begin{aligned}
{\cal O}_{\tilde WWW}&=\mbox{Tr}[{\tilde W}_{\mu\nu}W^{\nu\rho}W_{\rho}^{\mu}]\\
{\cal O}_{\tilde W}&=(D_\mu\Phi)^\dagger {\tilde W}^{\mu\nu}(D_\nu\Phi),
\end{aligned}
\label{opTGCCP}
\end{equation}
affect the triple and quartic gauge couplings. Here $\Phi$ denotes the Higgs doublet field.
The covariant derivative for such a field with hypercharge $Y=1/2$ is given by
\begin{equation}
 D_\mu \equiv \partial_\mu + i \frac{g'}{2} B_\mu  + i g W_\mu^i \frac{\tau^i}{2} 
\label{eq:covariant}
\end{equation}
where $\tau^i, i=1,2,3$ are the $SU(2)_I$ generators  
with $\mbox{Tr}[ \tau^i \tau^j ] = 2\delta^{ij}$. 
The field strength tensors of the $SU(2)_I$ ($W^i_\mu$) and $U(1)_Y$ ($B_\mu$)  gauge fields read
\begin{equation}
 \begin{aligned}
W_{\mu\nu} & =  \frac{i}{2} g\tau^i (\partial_\mu W^i_\nu - \partial_\nu W^i_\mu
	+ g \epsilon_{ijk} W^j_\mu W^k_\nu ) \\ 
B_{\mu \nu} & = \frac{i}{2} g' (\partial_\mu B_\nu - \partial_\nu B_\mu)  \; .
\end{aligned}
\label{eq:fieldtensors}
\end{equation}
As in the SM, trilinear gauge couplings (TGCs) and quartic gauge
couplings (QGCs) induced by dimension-six operators are completely
related by the requirement to guarantee gauge invariance.  In
addition, three CP-conserving operators
\begin{eqnarray}
\mathcal{O}_{\Phi d} &=& \partial_\mu\left(\Phi^\dagger \Phi\right)\partial^\mu\left(\Phi^\dagger \Phi\right)\label{opphid} \\
\mathcal{O}_{\Phi W} &=& \left(\Phi^\dagger\Phi\right) \mbox{Tr}[W^{\mu\nu}W_{\mu\nu}] 
\label{opphiw} \\
\mathcal{O}_{\Phi B} &=& \left(\Phi^\dagger\Phi\right)  B^{\mu\nu}B_{\mu\nu}  
\label{opphib} 
\end{eqnarray}
and two CP-violating dimension-six operators
\begin{equation}
 \begin{aligned}
\mathcal{O}_{\tilde{W}W} &= \Phi^{\dagger} {\tilde{W}}_{\mu\nu} {W}^{\mu\nu} \Phi \\
\mathcal{O}_{\tilde{B}B} &= \Phi^{\dagger} {\tilde{B}}_{\mu\nu} {B}^{\mu\nu} \Phi 
\end{aligned}
\label{eq:cpodd} 
\end{equation}
modify the coupling of the Higgs boson to the weak gauge bosons and therefore
the four-gauge-boson amplitudes.
The list of vertices relevant to three- and four-gauge-boson amplitudes 
of each operator is displayed in Table~\ref{tab:vert}.
\begin{table}[ht]
\centering
{\small 
\begin{tabular}{|l|c|c|c|c|c|c|c|c|c|c|} \hline
   					&ZWW	&AWW	&HWW	&HZZ	&HZA	&HAA	&WWWW	&ZZWW	&ZAWW	&AAWW	\\
   \hline
   ${\cal O}_{WWW}$		&X		&X		&		&		&		&		&X		&X		&X		&X\\
   ${\cal O}_{W}$		&X		&X		&X		&X		&X		&		&X		&X		&X		&\\
   ${\cal O}_{B}$		&X		&X		&		&X		&X		&		&		&		&		&\\
   ${\cal O}_{\Phi d}$		&		&		&X		&X		&		&		&		&		&		&\\
   ${\cal O}_{\Phi W}$		&		&		&X		&X		&X		&X		&		&		&		&\\
   ${\cal O}_{\Phi B}$ 		&		&		&		&X		&X		&X		&		&		&		&\\
   \hline
   ${\cal O}_{\tilde WWW}$	&X		&X		&		&		&		&		&X		&X		&X		&X\\
   ${\cal O}_{\tilde W}$	&X		&X		&X		&X		&X		&		&		&		&		& \\
${\cal O}_{\tilde WW}$	&		&		&X		&X		&X		&X		&		&		&		&\\
${\cal O}_{\tilde BB}$	&		&		&		&X		&X		&X		&		&		&		& \\ \hline
\end{tabular}
}
\caption{The vertices induced by each operator are marked with X in the corresponding column. The vertices that are not relevant for three- and four-gauge-boson amplitudes have been omitted.}
\label{tab:vert}
\end{table}
We have neglected the operators affecting the couplings of the bosons
to fermions as they can be measured in other processes such as $Z$
boson decay.  This is a minimal set of independent dimension-six
operators relevant to amplitudes involving vertices of three and four
electroweak gauge bosons.  Additional dimension-six operators
invariant under SM symmetries can be constructed but they can be shown
to be equivalent to a linear combination of the previous operators by
using equations of motion. Consequently, the choice of the basis of
operators is not unique. Other choices than the one presented here
can be found in the literature. For example, the operators $Q_{\Phi
  D}$ and $Q_{\Phi WB}$ in Ref.~\cite{Grzadkowski:2010es} have been
replaced in this paper by ${\cal O}_W$ and ${\cal O}_B$. Our basis
avoids the otherwise necessary redefinition of the masses of the gauge
bosons and the mixing of the neutral vector bosons.  The operator
$\mathcal{O}_{\Phi d}$ does not contain any gauge boson since
$\Phi^\dagger \Phi$ is a singlet under all the SM gauge
groups. However, it contributes to the Higgs field's kinetic term
after $\Phi$ has been replaced by its value in the unitary gauge,
i.e. with
\begin{equation}
\Phi=\left(0,\frac{v+h}{\sqrt2}\right)^T
\label{eq:unitary}
\end{equation}
one finds
\begin{equation}
\label{eq:phid}
\mathcal{O}_{\Phi d} \ni v^2 \partial_\mu h \partial^\mu h,
\end{equation}
and it requires a renormalization of the Higgs field,
\begin{equation}
h \to h(1-\frac{c_{\Phi d}}{\Lambda^2} v^2),
\end{equation}
in the full Lagrangian. The Higgs couplings to all particles including
the electroweak gauge bosons are consequently multiplied by the same
factor. $\mathcal{O}_{\Phi W}$ and $\mathcal{O}_{\Phi B}$ modify the
kinetic term of the gauge bosons after the Higgs doublet has been
replaced by its vacuum expectation value ($v$). Those two operators
require a renormalization of the gauge fields and the gauge
couplings. As a matter of fact, their part proportional to $v^2$ is
entirely absorbed by those redefinitions and can therefore be removed
directly in the definition of the operators, i.e.
\begin{equation}
 \begin{aligned}
\mathcal{O}_{\Phi W} &= \left(\Phi^\dagger\Phi-v^2\right) \mbox{Tr}[W^{\mu\nu}W_{\mu\nu}] 
\label{eq:phiw}\\
\mathcal{O}_{\Phi B} &= \left(\Phi^\dagger\Phi-v^2\right)  B^{\mu\nu}B_{\mu\nu}  
 \end{aligned}
\end{equation}
It is now clear that those operators affect only the vertices with one
or two Higgs bosons and not the TGCs or the QGCs.

\subsubsection{Dimension-eight operators for genuine QGCs}
\label{sec:dim8}

As can be seen in Table~\ref{tab:vert}, the dimension--six operators
giving rise to QGCs also exhibit TGCs. In order to separate the
effects of the QGCs we shall consider effective operators that lead to
QGCs without a TGC associated to them. Moreover, not all possible QGCs
are generated by dimension--six operators, for instance, these
operators do not give rise to quartic couplings among the neutral
gauge bosons \footnote{Notice that the lowest order operators leading
  to neutral TGCs are also of dimension eight. }.  The lowest
dimension operator that leads to quartic interactions but does not
exhibit two or three weak gauge boson vertices is of dimension
eight\footnote{Effective operators possessing QCGs but no TGCs can be
  generated at tree level by new physics at a higher
  scale~\cite{Arzt:1994gp}, in contrast with operators containing TGCs
  that are generated at loop level.  }. The counting is straightforward: we can get a weak boson field either from the covariant
derivative ($D_\mu$ of Eq.~\ref{eq:covariant}) of $\Phi$ or from the
field strength tensor of Eq.~\ref{eq:fieldtensors}. In either case,
the vector field is accompanied by $v$ (after using
Eq.~\ref{eq:unitary}) or a derivative $\partial_\mu$. Therefore,
genuine quartic vertices are of dimension 8 or higher.

The idea behind using dimension--eight operators for QGCs is that the
anomalous QGCs are to be considered as a straw man to evaluate the LHC
potential to study these couplings, without having any theoretical
prejudice about their size.  There are three classes of genuine QGC
operators~\cite{Eboli:2006wa}:

\newpage
\underline{Operators containing only  $D_\mu\Phi$}

This class contains two independent operators, {\em i.e.}
\begin{align}
  {\cal O}_{S,0} &= \left [ \left ( D_\mu \Phi \right)^\dagger
 D_\nu \Phi \right ] \times
\left [ \left ( D^\mu \Phi \right)^\dagger
D^\nu \Phi \right ] \; ,
\label{eq:s0}\\
  {\cal O}_{S,1} &= \left [ \left ( D_\mu \Phi \right)^\dagger
 D^\mu \Phi  \right ] \times
\left [ \left ( D_\nu \Phi \right)^\dagger
D^\nu \Phi \right ] \; ,
\end{align}
where 
the
Higgs covariant derivative is given by the expression  in
Eq.~\ref{eq:covariant}. These operators can be generated when we
integrate out a spin--one resonance that couples to gauge--boson pairs
as discussed in Section~\ref{sec:electroweak-resonances}.

The operators ${\cal O}_{S,0}$ and ${\cal O}_{S,1}$ contain quartic
$W^+W^-W^+W^-$, $W^+W^-ZZ$ and $ZZZZ$ interactions that do not depend
on the gauge boson momenta; for a comparative table showing all QGCs
induced by dimension--eight operators see Table~\ref{tab:vertices}.
In our framework, the QGCs are accompanied by vertices with more than
4 particles due to gauge invariance.  In order to simply rescale the
SM quartic couplings containing $W^\pm$ and $Z$ it is enough to choose
$f_{S,0} = - f_{S,1} = f$ which leads to SM quartic couplings modified by
a factor $(1+f v^4/8\Lambda^4)$, where $v$ is the Higgs vacuum expectation
value ($v \simeq 246$~GeV).

\underline{Operators containing $D_\mu\Phi$ and two field strength tensors}

QGCs are also generated by considering two electroweak field strength
tensors and two covariant derivatives of the Higgs
doublet~\cite{Eboli:2006wa}:
%
\begin{equation}
\begin{aligned}
 {\cal O}_{M,0} &=   \hbox{Tr}\left [ {W}_{\mu\nu} {W}^{\mu\nu} \right ]
\times  \left [ \left ( D_\beta \Phi \right)^\dagger
D^\beta \Phi \right ] \; ,
\\
 {\cal O}_{M,1} &=   \hbox{Tr}\left [ {W}_{\mu\nu} {W}^{\nu\beta} \right ]
\times  \left [ \left ( D_\beta \Phi \right)^\dagger
D^\mu \Phi \right ] \; ,
\\
 {\cal O}_{M,2} &=   \left [ B_{\mu\nu} B^{\mu\nu} \right ]
\times  \left [ \left ( D_\beta \Phi \right)^\dagger
D^\beta \Phi \right ] \; ,
\\
 {\cal O}_{M,3} &=   \left [ B_{\mu\nu} B^{\nu\beta} \right ]
\times  \left [ \left ( D_\beta \Phi \right)^\dagger
D^\mu \Phi \right ] \; ,
\\
  {\cal O}_{M,4} &= \left [ \left ( D_\mu \Phi \right)^\dagger {W}_{\beta\nu}
 D^\mu \Phi  \right ] \times B^{\beta\nu} \; ,
\\
  {\cal O}_{M,5} &= \left [ \left ( D_\mu \Phi \right)^\dagger {W}_{\beta\nu}
 D^\nu \Phi  \right ] \times B^{\beta\mu} \; ,
\\
  {\cal O}_{M,6} &= \left [ \left ( D_\mu \Phi \right)^\dagger {W}_{\beta\nu}
{W}^{\beta\nu} D^\mu \Phi  \right ] \; ,
\\
  {\cal O}_{M,7} &= \left [ \left ( D_\mu \Phi \right)^\dagger {W}_{\beta\nu}
{W}^{\beta\mu} D^\nu \Phi  \right ] \; ,
\end{aligned}
\label{opM0to7}
\end{equation}
where the field strengths $W_{\mu\nu}$ and $B_{\mu\nu}$ have been
defined above in Eq.~\eqref{eq:fieldtensors}. In this class of effective operators the
quartic gauge-boson interactions depend upon the momenta of the vector
bosons due to the presence of the field strength in their
definitions. Therefore, the Lorentz structure of these operators can
not be reduced to the SM one. The complete list of quartic vertices
modified by these operators can be found in Table~\ref{tab:vertices}.

\newpage
\underline{Operators containing only field strength tensors}

The following operators containing four field strength tensors
also lead to quartic anomalous couplings:
%
\begin{align}
 {\cal O}_{T,0} &=   \hbox{Tr}\left [ {W}_{\mu\nu} {W}^{\mu\nu} \right ]
\times   \hbox{Tr}\left [ {W}_{\alpha\beta} {W}^{\alpha\beta} \right ] \; ,
\label{eq:t0} \\
 {\cal O}_{T,1} &=   \hbox{Tr}\left [ {W}_{\alpha\nu} {W}^{\mu\beta} \right ] 
\times   \hbox{Tr}\left [ {W}_{\mu\beta} {W}^{\alpha\nu} \right ] \; ,
\label{eq:t1} \\
 {\cal O}_{T,2} &=   \hbox{Tr}\left [ {W}_{\alpha\mu} {W}^{\mu\beta} \right ]
\times   \hbox{Tr}\left [ {W}_{\beta\nu} {W}^{\nu\alpha} \right ]  \; ,
\label{eq:t2} \\
%
%
 {\cal O}_{T,5} &=   \hbox{Tr}\left [ {W}_{\mu\nu} {W}^{\mu\nu} \right ]
\times   B_{\alpha\beta} B^{\alpha\beta}  \; ,
\\
 {\cal O}_{T,6} &=   \hbox{Tr}\left [ {W}_{\alpha\nu} {W}^{\mu\beta} \right ]
\times   B_{\mu\beta} B^{\alpha\nu}  \; ,
\\
 {\cal O}_{T,7} &=   \hbox{Tr}\left [ {W}_{\alpha\mu} {W}^{\mu\beta} \right ]
\times   B_{\beta\nu} B^{\nu\alpha}  \; ,
\\
 {\cal O}_{T,8} &=   B_{\mu\nu} B^{\mu\nu}  B_{\alpha\beta} B^{\alpha\beta}
\label{eq:t8} \\
 {\cal O}_{T,9} &=  B_{\alpha\mu} B^{\mu\beta}   B_{\beta\nu} B^{\nu\alpha}  \; .
\label{eq:t9} 
\end{align}
It is interesting to note that the two last operators $ {\cal
  O}_{T,8}$ and $ {\cal O}_{T,9}$ give rise to QGCs containing only
the neutral electroweak gauge bosons.

Previous analyses~\cite{Belyaev:1998ih,Eboli:2000ad,Eboli:2003nq} of the LHC potential to study
QGCs were based on the non--linear realization of the gauge symmetry,
{\em i.e.} using chiral Lagrangians as for instance implemented in {\sc whizard}. 
The relation between the above framework
and chiral Lagrangians can be found in Section~\ref{sec:whizard-dim8}.

\begin{table}[h]
\centering
{\small
\begin{tabular}{|c|c|c|c|c|c|c|c|c|c|} \hline
   & WWWW & WWZZ & ZZZZ & WWAZ& WWAA& ZZZA & ZZAA& ZAAA& AAAA\\
\hline
${\cal O}_{S,0}$, ${\cal O}_{S,1}$ &  X & X & X &   &   &   &   &   &  \\
\hline 
${\cal O}_{M,0}$, ${\cal O}_{M,1}$,${\cal O}_{M,6}$ ,${\cal O}_{M,7}$ 
 &  X & X & X & X & X & X & X &   &   \\
\hline 
${\cal O}_{M,2}$ ,${\cal O}_{M,3}$, 
${\cal O}_{M,4}$ ,${\cal O}_{M,5}$ 
&    & X & X & X & X & X & X &   &   \\ \hline
${\cal O}_{T,0}$ ,${\cal O}_{T,1}$ ,${\cal O}_{T,2}$ 
& X   & X & X & X & X & X & X & X  &X   \\ \hline
${\cal O}_{T,5}$ ,${\cal O}_{T,6}$ ,${\cal O}_{T,7}$ 
&    & X & X & X & X & X & X & X  &X   \\ \hline
${\cal O}_{T,8}$ ,${\cal O}_{T,9}$ 
&    &  & X &  &  & X & X & X  &X   \\ \hline
\end{tabular}
}
\caption{Quartic vertices modified by each dimension-8 
operator are marked with $X$.}
\label{tab:vertices}
\end{table}

\subsubsection{Comparison with the anomalous coupling approach and the LEP convention for aQGCs}
\label{sec:anomalous}

The anomalous couplings approach is based on the Lagrangian
\cite{Hagiwara:1986vm}
\begin{equation}
 \begin{aligned}
  {\cal L}=&ig_{WWV}\left(g_1^V(W_{\mu\nu}^+W^{-\mu}-W^{+\mu}W_{\mu\nu}^-)V^\nu
+\kappa_VW_\mu^+W_\nu^-V^{\mu\nu}
+\frac{\lambda_V}{M_W^2}W_\mu^{\nu+}W_\nu^{-\rho}V_\rho^{\mu}
\right.\\&\left.
+ig_4^VW_\mu^+W^-_\nu(\partial^\mu V^\nu+\partial^\nu V^\mu)
-ig_5^V\epsilon^{\mu\nu\rho\sigma}(W_\mu^+\partial_\rho W^-_\nu-\partial_\rho W_\mu^+W^-_\nu)V_\sigma
\right.\\&\left.
+\tilde{\kappa}_VW_\mu^+W_\nu^-\tilde{V}^{\mu\nu}
+\frac{\tilde{\lambda}_V}{m_W^2}W_\mu^{\nu+}W_\nu^{-\rho}\tilde{V}_\rho^{\mu}
\right) \, ,
 \end{aligned}
 \label{eq:L}
\end{equation}
where $V=\gamma,Z$; $W_{\mu\nu}^\pm = \partial_\mu W_\nu^\pm -
\partial_\nu W_\mu^\pm$, $V_{\mu\nu} = \partial_\mu V_\nu -
\partial_\nu V_\mu$, $g_{WW\gamma}=-e$ and $g_{WWZ}=-e\cot\theta_W$.
The first three terms of Eq.~\ref{eq:L} are $C$ and $P$ invariant
while the remaining four terms violate $C$ and/or $P$.
Electromagnetic gauge invariance requires that $g_1^\gamma =1$ and
$g_4^\gamma=g_5^\gamma = 0$.  Finally there are five independent $C$-
and $P$-conserving parameters: $g_1^Z, \kappa_\gamma, \kappa_Z,
\lambda_\gamma, \lambda_Z$; and six $C$ and/or $P$ violating
parameters: $g_4^Z, g_5^Z, \tilde{\kappa}_\gamma, \tilde{\kappa_Z},
\tilde{\lambda}_\gamma, \tilde{\lambda_Z}$. This Lagrangian is not the
most generic one, since extra derivatives can be added in all of the
operators. Furthermore, there is no reason to remove those extra terms,
since they are not suppressed by $\Lambda$ but by $M_W$.

The effective field theory approach described in the previous section
allows one to calculate those parameters in terms of the coefficients
of the five dimension-six operators relevant for TGCs, i.~e. in terms
of the EFT coefficients $c_{WWW}, c_W, c_B, c_{\tilde{W}WW}$ and $
c_{\tilde{W}}$.  One finds for the anomalous TGC
parameters~\cite{Hagiwara:1993ck,Wudka:1994ny}:
\begin{align}
g_1^Z &= 1+c_W\frac{m_Z^2}{2\Lambda^2}\\
\kappa_\gamma &= 1+(c_W+c_B)\frac{m_W^2}{2\Lambda^2}\\
\kappa_Z &= 1+(c_W-c_B\tan^2\theta_W)\frac{m_W^2}{2\Lambda^2}\\
\lambda_\gamma &= \lambda_Z = c_{WWW}\frac{3g^2m_W^2}{2\Lambda^2}\\
g_4^V &= g_5^V=0\\
\tilde{\kappa}_\gamma &=
c_{\tilde{W}}\frac{m_W^2}{2\Lambda^2}\\
\tilde{\kappa}_Z &=
-c_{\tilde{W}}\tan^2\theta_W\frac{m_W^2}{2\Lambda^2}\\
\tilde{\lambda}_\gamma &= \tilde{\lambda}_Z = c_{\tilde{W}WW}\frac{3g^2m_W^2}{2\Lambda^2}
\end{align}
Defining $\Delta g_1^Z = g_1^Z - 1$, $\Delta \kappa_{\gamma,Z} = \kappa_{\gamma,Z} - 1$, the relation \cite{Hagiwara:1993ck}
\begin{equation}
\Delta g_1^Z=\Delta \kappa_Z + \tan^2\theta_W \Delta \kappa_\gamma
\end{equation}
and the relation $\lambda_\gamma = \lambda_Z$ reduce the five $C$ and $P$ conserving parameters down to three.  For the $C$ and/or $P$ violating parameters, the relation
\begin{equation}
0=\tilde \kappa_Z + \tan^2\theta_W \tilde \kappa_\gamma
\end{equation}
and the relations $\tilde\lambda_\gamma = \tilde\lambda_Z$ and
$g_4^Z=g_5^Z=0$ reduce the six $C$ and/or $P$ violating parameters
down to just two.

The Lagrangian of Eq.~\ref{eq:L} is not $SU(2)_L$ gauge invariant under linear transformations even
after imposing those relations because the quartic and higher
multiplicity couplings are not included. Furthermore, gauge invariance
requires also several relations between vertices with different number
of particles. 
The quartic couplings involving two photons have been parametrized in
a similar way. However, the parametrization is not generic enough and
does not include the contributions from the dimension-six operators.

The LEP2 constraints on the vertices $\gamma\gamma W^+ W^-$ and
$\gamma\gamma ZZ$~\cite{quarticATlep} described in terms of anomalous couplings
$a_0/\Lambda^2$ and $a_c/\Lambda^2$ can be translated into bounds on
%
%
%
%
%
%
%
%
%
%
$f_{M,0}$ -- $f_{M,7}$. 
%
%
%
In Ref.~\cite{Stirling:1999ek} (see also Refs~\cite{Belanger:1992qh,Eboli:1993wg}), genuine
anomalous quartic couplings involving two photons have been introduced as follows:
\begin{equation}
\begin{aligned}
{\cal L}_0 & =  - \frac{e^2}{16 \pi \Lambda^2} a_0 F_{\mu \nu} F^{\mu \nu} \vec{W}^\alpha \vec{W}_\alpha \\
{\cal L}_c & =  - \frac{e^2}{16 \pi \Lambda^2} a_c F_{\mu \alpha} F^{\mu \beta} \vec{W}^\alpha \vec{W}_\beta
\end{aligned}
\label{eq:aoac}
\end{equation}
with
\begin{equation}
\begin{aligned}
F^{\mu \nu} & =\partial^\mu A^\nu - \partial^\nu A^\mu  \\
\vec{W}_{\mu} & = 
\left(
\begin{array}{c}
\frac{1}{\sqrt{2}} (W_\mu^+ + W_\mu^-) \\
\frac{i}{\sqrt{2}} (W_\mu^+ - W_\mu^-) \\
\frac{Z_\mu}{\cos\theta_w} 
\end{array}
\right)
\end{aligned}
\label{eq:LEPfields}
\end{equation}
where $A_\mu$ and $W_\mu^\pm,Z_\mu$ denote the photon and weak fields, respectively.
%
%
%
%
%
%
%
%
%
%
Thus, using the conventions of Eq.~\eqref{eq:fieldtensors} for the fields in the operators $\mathcal{O}_{M,i}$,
and Eq.~\eqref{eq:LEPfields} for the fields in the operators ${\cal L}_0$ and ${\cal L}_c$, 
the following relations for the $WW\gamma\gamma$ (upper sign) and $ZZ\gamma\gamma$ (lower sign) vertices can be derived:
%
   \begin{alignat}{3}
    \frac{f_{M,0}}{\Lambda^4} &= \hspace{2ex} \frac{a_0}{\Lambda^2} \frac{1}{g^2 v^2}
    &&\qquad\text{and}\qquad&
   \frac{f_{M,1}}{\Lambda^4} &= -\frac{a_c}{\Lambda^2} \frac{1}{g^2 v^2}  \\
    \frac{f_{M,2}}{\Lambda^4} &=\hspace{2ex} \frac{a_0}{\Lambda^2} \frac{2}{g^2 v^2}
    &&\qquad\text{and}\qquad&
   \frac{f_{M,3}}{\Lambda^4} &= -\frac{a_c}{\Lambda^2} \frac{2}{g^2 v^2}  \\
    \frac{f_{M,4}}{\Lambda^4} &= \pm\frac{a_0}{\Lambda^2} \frac{1}{g^2 v^2}
    &&\qquad\text{and}\qquad&
   \frac{f_{M,5}}{\Lambda^4} &= \pm\frac{a_c}{\Lambda^2} \frac{2}{g^2 v^2}  \\
    \frac{f_{M,6}}{\Lambda^4} &= \hspace{2ex}\frac{a_0}{\Lambda^2} \frac{2}{g^2 v^2}
    &&\qquad\text{and}\qquad&
   \frac{f_{M,7}}{\Lambda^4} &= \hspace{2ex}\frac{a_c}{\Lambda^2} \frac{2}{g^2 v^2}  \, .
  \end{alignat}

\subsubsection{Conventions for non-standard electroweak gauge boson interactions in different MC programs}
\label{sec:conventions}

\underline{Dimension-8 operators: VBFNLO and MadGraph5}
\label{sec:mg-dim8}

The convention for the dimension-8-operators in VBFNLO is
the same as described in Section~\ref{sec:dim8}, and the
coefficients $f_i/{\Lambda^4}$ set in the input file are the
ones that multiply the operators of Section~\ref{sec:dim8}.
However, the MadGraph5 implementation by means of a UFO file~\cite{Eboli:anom4}
uses expressions for the field strengths which are slightly
different than the ones from Eq.\ref{eq:fieldtensors}:
\begin{equation}
\begin{aligned}
\widehat{W}_{\mu\nu} & \;=\;  \frac{1}{2} \tau^i (\partial_\mu W^i_\nu - \partial_\nu W^i_\mu
	+ g \epsilon_{ijk} W^j_\mu W^k_\nu ) = \frac{1}{i g}  W_{\mu\nu} \\ 
\widehat{B}_{\mu \nu} & \;=\;  (\partial_\mu B_\nu - \partial_\nu B_\mu) =  \frac{2}{i g'}   B_{\mu\nu}
\end{aligned}
\end{equation}
The resulting changes can be absorbed in a redefinition of the operator coefficients:
\begin{equation}
\begin{aligned}
 f_{S,0,1}   &= f_{S,0,1}^\text{VBFNLO}   \;=\;                            f_{S,0,1}^\text{MG5}   
\\
 f_{M,0,1}   &= f_{M,0,1}^\text{VBFNLO}   \;=\; - \frac{1}{g^2}      \cdot f_{M,0,1}^\text{MG5}   \\
 f_{M,2,3}   &= f_{M,2,3}^\text{VBFNLO}   \;=\; - \frac{4}{g'^2}     \cdot f_{M,2,3}^\text{MG5}  
\\
 f_{M,4,5}   &= f_{M,4,5}^\text{VBFNLO}   \;=\; - \frac{2}{g g'}     \cdot f_{M,4,5}^\text{MG5}   \\
 f_{M,6,7}   &= f_{M,6,7}^\text{VBFNLO}   \;=\; - \frac{1}{g^2}      \cdot f_{M,6,7}^\text{MG5}  \\
 f_{T,0,1,2} &= f_{T,0,1,2}^\text{VBFNLO} \;=\;   \frac{1}{g^4}      \cdot f_{T,0,1,2}^\text{MG5} \\
 f_{T,5,6,7} &= f_{T,5,6,7}^\text{VBFNLO} \;=\;   \frac{4}{g^2 g'^2} \cdot f_{T,5,6,7}^\text{MG5} \\
 f_{T,8,9}   &= f_{T,8,9}^\text{VBFNLO}   \;=\;   \frac{16}{g'^4}    \cdot f_{T,8,9}^\text{MG5} 
\end{aligned}
\label{eq:fsmtrelations}
\end{equation}

\underline{Dimension-8 operators: {\sc whizard}}
\label{sec:whizard-dim8}

{\sc whizard} uses different anomalous coupling operators than the ones described
in Section~\ref{sec:dim8}, assuming
a different symmetry group~\cite{Alboteanu:2008my}, and so a conversion is in general not possible.
However, a vertex-specific conversion exists for the operators ${\cal O}_{S,0}$
and ${\cal O}_{S,1}$ to their corresponding operators
\begin{align}
   {\cal L}^{(4)}_4 &= \alpha_4 \left[ \textrm{Tr} \left( V_\mu V_\nu \right) \right] ^2  \nonumber \\
   {\cal L}^{(4)}_5 &= \alpha_5 \left[ \textrm{Tr} \left( V_\mu V^\mu \right) \right] ^2 , \quad \textrm{with}\; V_\mu = \left( D_\mu \Sigma \right) \Sigma^\dagger \; .
\label{eqn:alpha4}
\end{align}
The conversion reads:
\begin{itemize}
 \item for the WWWW-Vertex: 
  \begin{align}
     \alpha_4 &=  \frac{f_{S,0}}{\Lambda^4} \frac{v^4}{8} \label{alpha4Tofs0}\\
     \alpha_4 + 2 \cdot \alpha_5 &=  \frac{f_{S,1}}{\Lambda^4} \frac{v^4}{8}
  \end{align}
 \item for the WWZZ-Vertex:
  \begin{align}
    \alpha_4 &=  \frac{f_{S,0}}{\Lambda^4} \frac{v^4}{16}\\
    \alpha_5 &=  \frac{f_{S,1}}{\Lambda^4} \frac{v^4}{16}
  \end{align}
 \item for the ZZZZ-Vertex:
  \begin{align}
    \alpha_4 + \alpha_5 =  \left(\frac{f_{S,0}}{\Lambda^4} + \frac{f_{S,1}}{\Lambda^4} \right) \frac{v^4}{16}
  \end{align}
\end{itemize}

\newpage
\underline{Dimension-6 operators: VBFNLO and MadGraph5}
\label{sec:mg-dim6}

The MadGraph model EWdim6 has been generated from FeynRules and
contains the operators from Eqs.~\ref{opTGC}, \ref{opTGCCP}, \ref{opphid}, \ref{opphiw} and \ref{opphib}, with the exception of ${\cal O}_{\tilde WW}$, ${\cal
O}_{\tilde BB}$ of Eq.~\ref{eq:cpodd} and ${\cal O}_{D\tilde W}$ of
Eq.~\ref{eq:cpodd2}~\footnote{We have neglected the CP violating
operators with the dual strength tensors affecting only the gauge
boson Higgs couplings, since measuring CP violation in the
four-weak-boson amplitude would be very challenging.}. The names of
the coefficients is displayed in Table~\ref{tab:cname}.  All the
coefficients include the $1/\Lambda^2$ as reminded by the "L2" at the
end of the names and are in TeV$^{-2}$. The model also has a new
coupling order $NP$ counting the power of $1/\Lambda$. Consequently,
each vertex from the dimension-six operators has NP=2.

\begin{table}[ht]
\centering
 \begin{tabular}{l|l}
 $c_{WWW}/\Lambda^2$ & CWWWL2\\\hline
 $c_{W}/\Lambda^2$ & CWL2\\\hline
 $c_{B}/\Lambda^2$ & CBL2\\\hline
 $c_{\tilde{W}WW}/\Lambda^2$ & CPWWWL2\\\hline
 $c_{\tilde{W}}/\Lambda^2$ & CPWL2\\\hline
 $c_{\Phi d}/\Lambda^2$ & CphidL2\\\hline
 $c_{\Phi W}/\Lambda^2$ & CphiWL2\\\hline
 $c_{\Phi B}/\Lambda^2$ & CphiBL2\\
 \end{tabular}
 \caption{Names of the couplings of the dimension-six operators present in the EWdim6 model of MadGraph5.}
 \label{tab:cname}
\end{table}


The operators from Eqs.~\ref{opTGC} and \ref{opTGCCP} in
Section~\ref{sec:dim6} are directly available in VBFNLO. The operators
$\mathcal{O}_{\tilde{W}W}$ and $\mathcal{O}_{\tilde{B}B}$ of Eq.~\ref{eq:cpodd}and $\mathcal{O}_{\Phi B}$ of Eq.~\ref{opphib} are available as well (${\cal O}_{\Phi B}$ is
called ${\cal O}_{BB}$ within VBFNLO).  Additionally, the operator
\begin{equation}
 \mathcal{O}_{WW} = \Phi^{\dagger} {W}_{\mu\nu} {W}^{\mu\nu} \Phi
\end{equation}
from VBFNLO can be related to the operator
${\cal O}_{\Phi W}$ of Eq.~\ref{opphiw} by choosing the coefficient as
\begin{align}
 c_{WW} &= 2 \cdot c_{\Phi W}
\end{align}
In addition to those operators, VBFNLO also provides the following CP-odd operators:
\begin{equation}\label{eq:cpodd2}
 \begin{aligned}
\mathcal{O}_{\tilde{B}} &= (D_{\mu} \Phi)^{\dagger} {\tilde{B}}^{\mu \nu} (D_{\nu} \Phi)  \\
\mathcal{O}_{B\tilde{W}} &= \Phi^{\dagger} {B}_{\mu\nu} {\tilde{W}}^{\mu\nu} \Phi  \\
\mathcal{O}_{D\tilde{W}} &= \textrm{Tr} \left( [D_\mu, {\tilde{W}}_{\nu\rho} ] [D^\mu, {W}^{\nu\rho}] \right) \, .
\end{aligned}
\end{equation}
However, only 4 of the 7 CP-odd operators are linearly independent, so 
the additional operators can be expressed in terms of the operators of Eqs.~\ref{opTGCCP} and \ref{eq:cpodd} as follows:
\begin{equation}
 \begin{aligned}
\mathcal{O}_{\tilde{B}} &= \mathcal{O}_{\tilde{W}} + \frac{1}{2} \mathcal{O}_{\tilde{W}W} - \frac{1}{2} \mathcal{O}_{\tilde{B}B}   \\
\mathcal{O}_{B\tilde{W}} &= -2 \, \mathcal{O}_{\tilde{W}} - \mathcal{O}_{\tilde{W}W} \\\
\mathcal{O}_{D\tilde{W}} &= -4 \, \mathcal{O}_{\tilde{W}WW} \, . 
 \end{aligned}
\end{equation}
The CP-conserving anomalous couplings implementation is also available in  VBFNLO with the
parameters $\Delta g_1^Z$, $\Delta \kappa_Z$, $\Delta \kappa_\gamma$, and $\lambda_\gamma$,
defined in Section~\ref{sec:anomalous}.

\subsubsection{Discussion of unitarity bounds and usage of form factors}
\label{sec:unitarity}

\begin{figure}[h]
\centering
\includegraphics[width=0.48\textwidth]{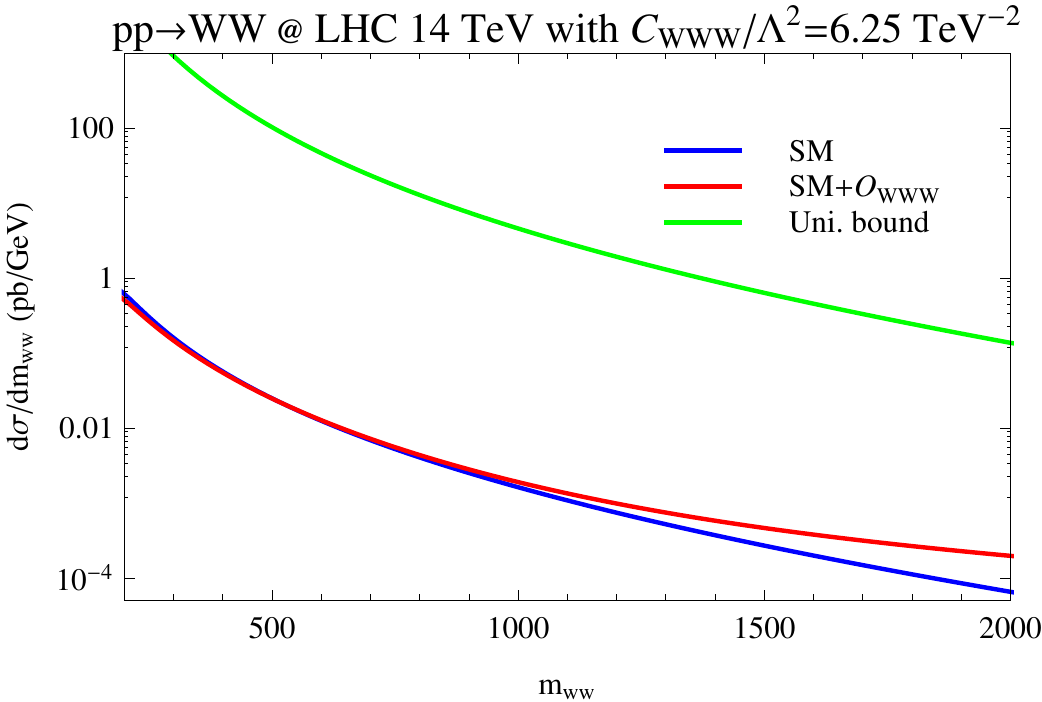}
\caption{$m_{WW}$ distributions in $W$-pair production at the 14 TeV LHC are displayed for the SM (in blue) and for the SM plus
  the dimension six operator $\mathcal{O}_{WWW}$ with $c_{WWW}/\Lambda^2=6.25$ TeV (in red). Also shown is the
  unitarity bound~\cite{Degrande:2012wf} (in
  green).}  
\label{fig:unid61}
\end{figure}

\begin{figure}[h]
\centering
\includegraphics[width=0.48\textwidth]{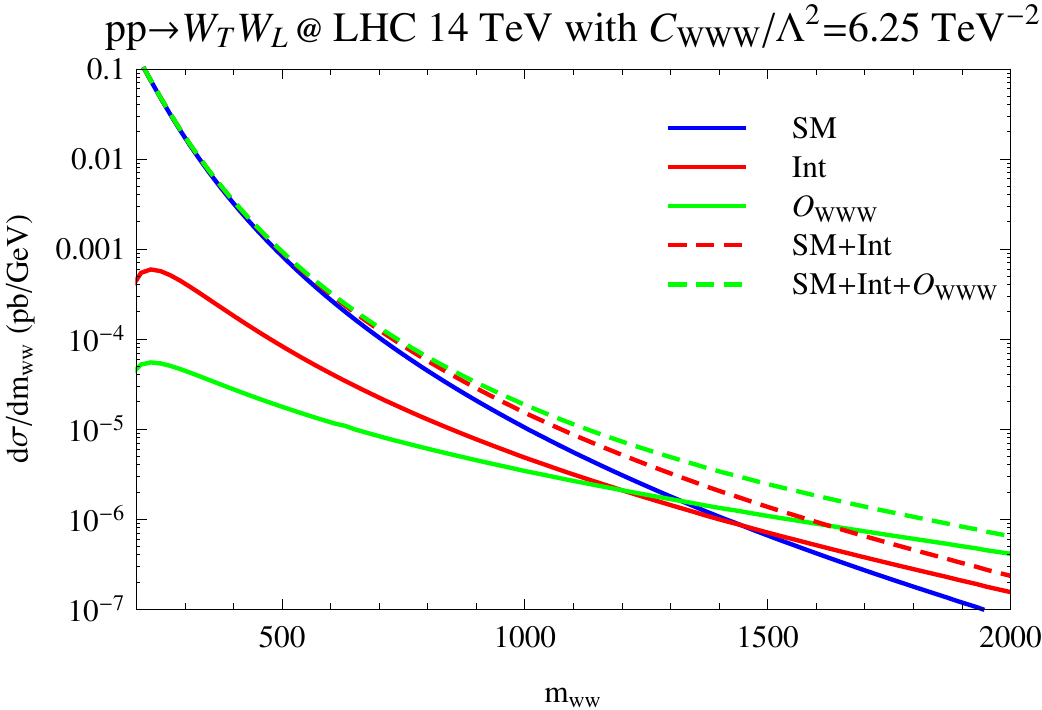}
\caption{$m_{WW}$ distributions in $W$-pair production at the 14 TeV LHC are displayed for the
  production of one longitudinally and one transversally polarized $W$
  boson, when considering the SM (solid blue line), only the interference between
  the SM and the dimension-six operator (solid red line), the sum
  of the two (dashed red line), only the square of the new physics
  amplitude (solid green line), and finally the total contribution from
  the SM and the dimension-six operator (dashed green line).}
\label{fig:unid62}
\end{figure}

The effective field theory is valid only below the new physics scale $\Lambda$ and no
violation of unitarity occurs in this regime.
In the regime where EFT is valid, the new physics contributions to a SM process, i.e. 
the interference of the SM amplitude with the higher-dimensional operators and the square
of the new physics amplitudes, are suppressed by increasing powers of
$1/\Lambda$,
\begin{equation}
\left| {\cal M}_{SM} + {\cal M}_{dim6}+{\cal M}_{dim8}+\ldots\right|^2 = \underbrace{\left| {\cal M}_{SM}\right|^2}_{\Lambda^0} + \underbrace{2 \Re\left({\cal M}_{SM}{\cal M}_{dim6}\right)}_{\Lambda^{-2}} +\underbrace{ \left|{\cal M}_{dim6}\right|^2 + 2 \Re\left({\cal M}_{SM}{\cal M}_{dim8}\right)}_{\Lambda^{-4}} +\ldots 
\end{equation}
For illustration we show in Figs.~\ref{fig:unid61},~\ref{fig:unid62} the invariant mass
distribution of the $W$-pair, $m_{WW}$, produced at the 14 TeV LHC,
with and without the contribution of the dimension six operator ${\cal
  O}_{WWW}$ of Eq.~\ref{opTGC}. As can be seen in Figs.~\ref{fig:unid61}, the
prediction for $m_{WW}$ including ${\cal O}_{WWW}$ is well below the
unitarity bound ~\cite{Degrande:2012wf} for this process in the
relevant energy regime.  However, as illustrated in Fig.~\ref{fig:unid62}, the
contributions of this operator to the amplitude squared for $W_LW_T$
production reach similar magnitude at $m_{WW}\approx1.3$ TeV and above
this energy the $1/\Lambda^4$ suppressed term overtakes the
$1/\Lambda^2$ suppressed contribution. Clearly, the $1/\Lambda$
expansion is only valid below this energy.   In typical analyses for anomalous couplings, the EFT does not break down for $m_{VV}$ as low as 1 TeV.


For dimension eight operators, the effect from unitarity violation
typically sets in earlier due to the higher exponent in $\Lambda$ in the
denominator. Hence, the task to avoid unphysical contributions from
regions where unitarity is violated becomes more important. In these
regions the EFT expansion in terms of suppressed additional
contributions to the SM part, our starting point, is no longer valid, as
each order becomes similarly important.

In experimental searches one has to ensure that the sensitivity to
anomalous gauge couplings is not driven by parameter regions where
unitarity is violated. As nature will ensure unitarity conservation in
the full model, such results would not be meaningful. Thereby, one can
take advantage of the fact that only energies up to the center-of-mass
energy of the collider are probed. For hadron colliders like the LHC,
the steep fall-off of the parton distribution functions means that the
effective probed energy range is even smaller, as the expected number of
signal events will be smaller than one above a certain energy and
therefore this region will not contribute.
However, if the bound for unitarity violation is lower than that, some
method to ensure that no sensitivity comes from this energy range needs
to be employed. One possibility is to use appropriate experimental cuts.
However, often processes will contain neutrinos and so the full
reconstruction of the partonic energy is not possible. Another option is to use
form factors. These are introduced to model an energy-dependent
cutoff, which in the full theory would be accomplished by new-physics
states at the scale $\Lambda$, which have been integrated out in the EFT
description. Various options are possible, for example a sharp cut-off
of the higher-dimensional contributions at a fixed energy scale, or a
dipole-like form factor as used in \VBFNLO{}, that gives a smoother
cut-off. The exact choice depends on the full model, so for an effective
theory description all choices are equally well motivated from the
theory side. The last possibility to ensure no unitarity violation
happens is a unitarity projection, like the $K$-matrix method
implemented in {\sc whizard} 
(see also Section~\ref{sec:electroweak-resonances}). There the amplitude $A$ is moved onto the
unitarity circle along a line connecting $\Re(A)$ and the imaginary unit
$i$. Physically, this corresponds to introducing an infinitely heavy and
wide resonance.  This scheme maximizes the contributions from anomalous
couplings while ensuring unitarity for all energies.

\subsubsection{The role of higher order corrections in multi-boson processes at the LHC}
\label{sec:ho}

Higher-order corrections play an important role for accurate predictions
at the LHC. In this section we study the impact of NLO QCD corrections
in vector-boson fusion and triboson processes and how they impact the
extraction of anomalous quartic gauge couplings. As example of these two
process classes we take the processes $W^+W^+jj$ and $W^+\gamma\gamma$,
respectively. The NLO results including anomalous QGCs presented in Sections~\ref{sec:wwjjvbfnlo} and ~\ref{sec:wggvbfnlo} have been obtained with \VBFNLO{}.
We discuss the impact of a parton shower on the example of $W^+W^+jj$ production with {\tt POWHEG+PYTHIA}~\cite{Jager:2011ms} in Section~\ref{sec:wwjjpowheg}.
Finally, in Section~\ref{sec:tribosonew} we discuss the impact of NLO
electroweak corrections in triboson processes.

\underline{Vector-boson-fusion process $W^+W^+jj$ with VBFNLO}
\label{sec:wwjjvbfnlo}

The production of a vector-boson pair via vector-boson
fusion~\cite{Jager:2006zc, Jager:2006cp, Bozzi:2007ur, Jager:2009xx,
Denner:2012dz} has a characteristic signature of two high-energetic,
so-called tagging jets in the forward region of the detector, which are
defined as the two jets with the largest transverse momentum. This can
be exploited experimentally by requiring that there is a large rapidity
separation ($\Delta\eta_{jj}>4$) between the tagging jets, they are in
opposite detector hemispheres ($\eta_{j_1} \times \eta_{j_2} < 0$) and
they possess a large invariant mass ($M_{jj} > 600$ GeV). Additional
central jet radiation at higher orders is strongly suppressed due to the
exchange of a color-singlet in the t-channel, in contrast to typical
QCD-induced backgrounds. Higher-order corrections are typically small,
below the 10\% level, and reduce the residual scale uncertainty to about
2.5\%. Choosing the momentum transfer between an incoming and an
outgoing parton along a fermion line proves to be particularly
advantageous, as then also corrections to important distributions are
small and flat over the whole range. 

\begin{figure}
\begin{center}
\includegraphics[height=0.3\textwidth]{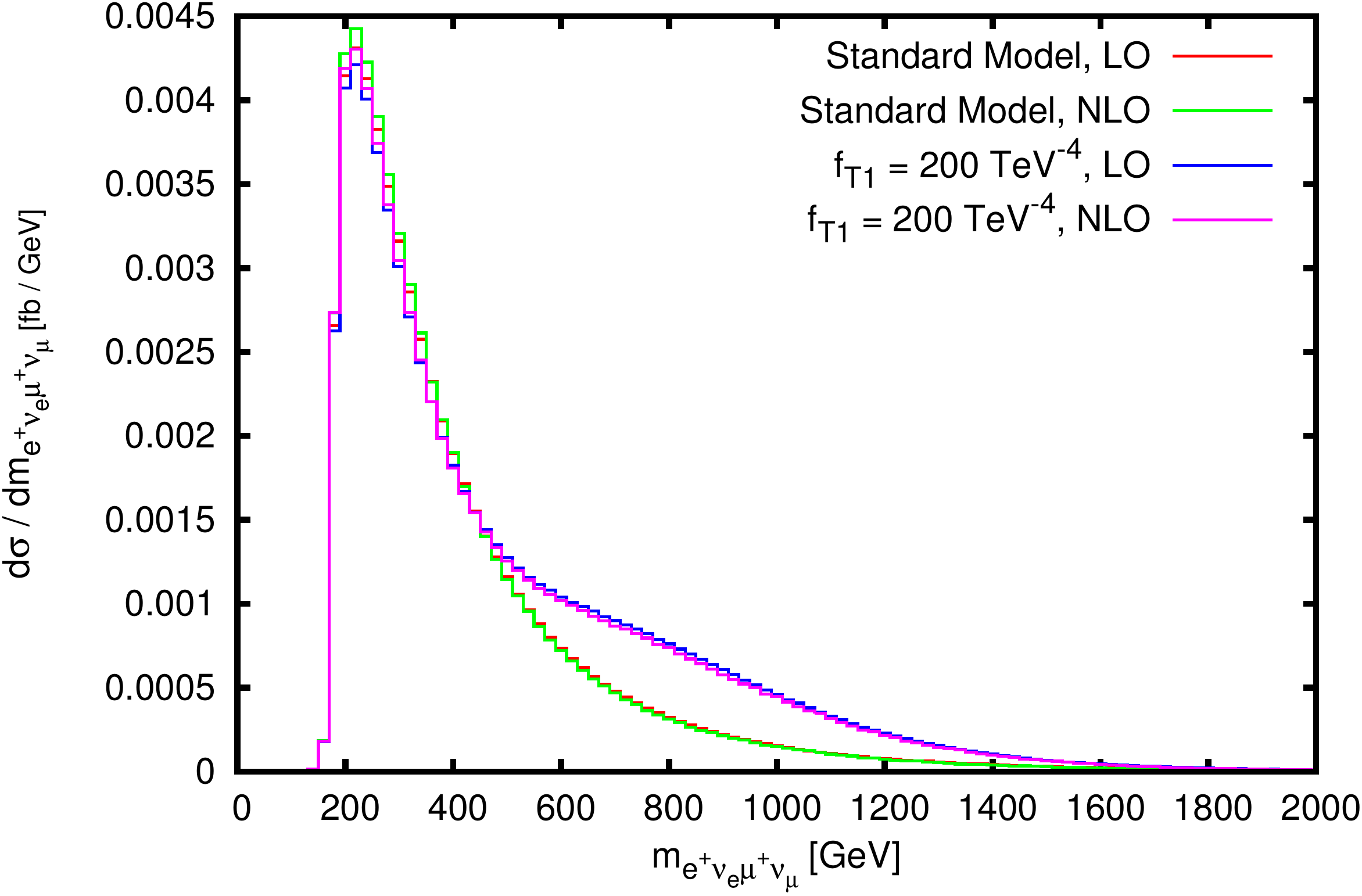} \quad
\includegraphics[height=0.3\textwidth]{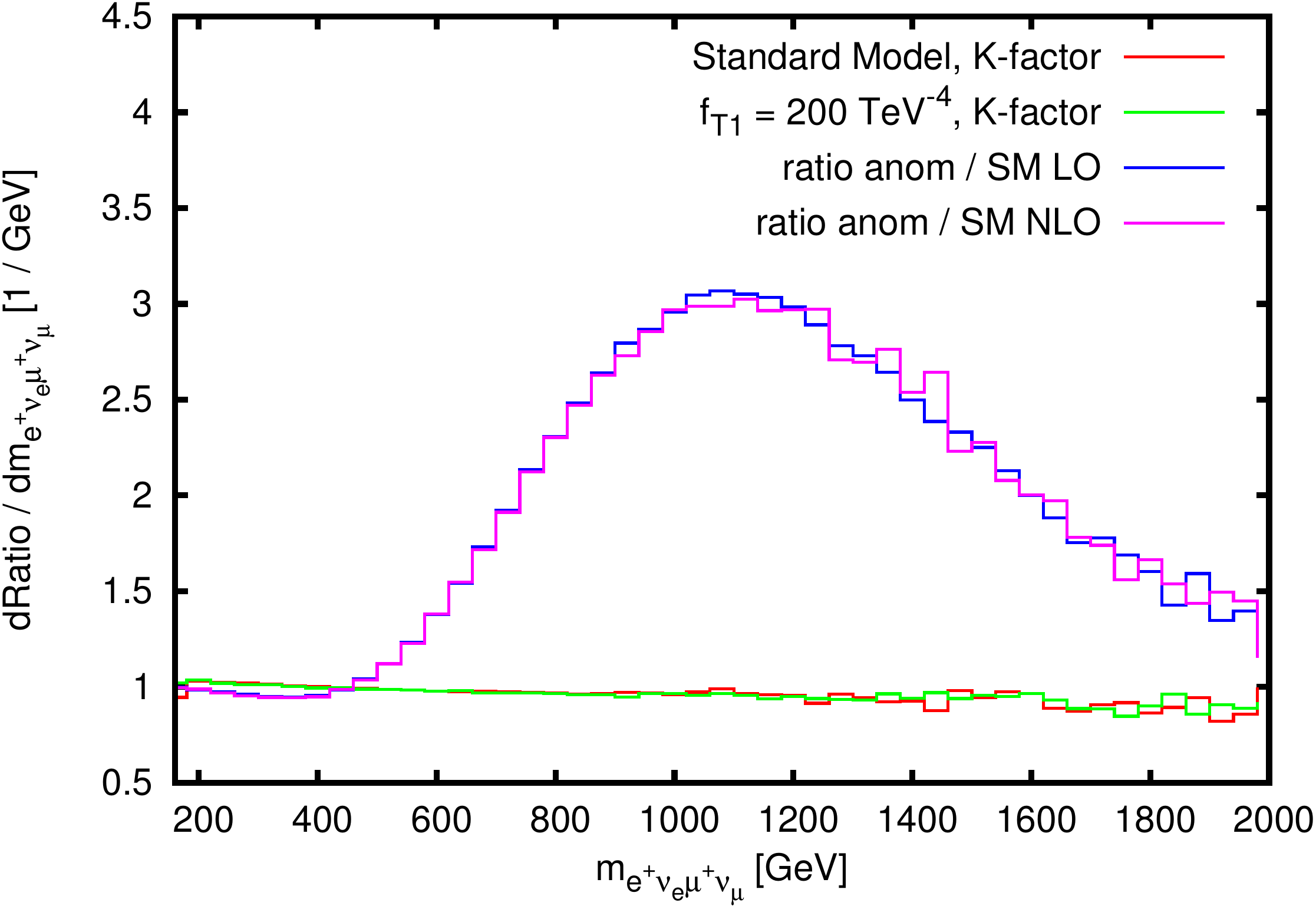}
\end{center}
\caption{Invariant-mass distribution of the two lepton, two neutrino
system. \textit{Left:} Differential cross section for the SM and with
anomalous coupling $T_1$ at LO and NLO. \textit{Right:} Differential
K-factors for the SM and with anomalous coupling as well as the
cross-section ratio between anomalous coupling and SM for LO and NLO.}
\label{NLO:VBF:mVV}
\end{figure}

As example we take the process $pp \rightarrow e^+ \nu_e \mu^+ \nu_\mu
jj$ with anomalous coupling $\frac{f_{T,1}}{\Lambda^4} = 200
\text{ TeV}^{-4}$ and form factor scale $\Lambda=1188$ GeV and exponent
$p=4$. The results for the total cross sections at LO and NLO are shown
in Table~\ref{NLO:VBF:cs}.
\begin{table}
\begin{center}
\begin{tabular}{l|cc}
& $\sigma_{\text{LO}}$ & $\sigma_{\text{NLO}}$ \\\hline
SM          & 1.169 fb & 1.176 fb \\
anom.coupl. & 1.399 fb & 1.388 fb \\
\end{tabular}
\caption{Total cross sections at LO and NLO for the process $pp
\rightarrow e^+ \nu_e \mu^+ \nu_\mu jj$ in the SM and with anomalous
coupling $\frac{f_{T,1}}{\Lambda^4} = 200 \text{ TeV}^{-4}$. Statistical
errors from Monte Carlo integration are below the per mille level.}
\label{NLO:VBF:cs}
\end{center}
\end{table}
Switching on the anomalous couplings increases the cross section by just
under 20\%, and NLO QCD corrections hardly change this number. This can
also be seen in Fig.~\ref{NLO:VBF:mVV} where we show the differential
distribution with respect to the invariant mass of the two leptons and
the two neutrinos.
In the left-hand plot we present the differential cross section in the
SM and with anomalous couplings switched on both at LO and NLO. Similar
to the integrated cross section, the difference between LO and NLO is
small in both cases. In contrast the anomalous couplings yield a
positive contribution to the cross section over the SM, which starts at
an invariant mass of about 500 GeV, before the formfactor, introduced to
preserve unitarity, damps the contributions again at higher invariant
masses. On the right-hand side we present two groups of ratios. The
differential K factor is flat and close to one both for the SM and the
anomalous coupling scenario. The second set shows the ratio of
differential anomalous-coupling over SM cross section both at LO and
NLO. The two curves agree well and show enhancements of the cross
section up to a factor of three. Hence, in this process higher-order
corrections do not influence the extraction of anomalous couplings.

\newpage
\underline{Vector-boson-fusion process $W^+W^+jj$ in the \POWHEGBOX}
\label{sec:wwjjpowheg}

NLO-QCD calculations are a crucial prerequisite for precision analyses
at the LHC, reducing theoretical uncertainties associated with hard
scattering processes significantly. On the other hand, a realistic
description of the additional hadronic activity that occurs in any
collider environment crucially relies on parton-shower Monte Carlo
generators such as \HERWIG{}~\cite{Corcella:2000bw} or
\PYTHIA{}~\cite{Sjostrand:2006za}. The perturbative accuracy of these
programs is, however, limited to leading logarithmic accuracy.
The most realistic yet accurate predictions available to date for
processes with many particles in the final state are thus obtained by
combining NLO-QCD calculations for the hard scattering with parton
shower programs, for example in the framework of the \POWHEG{}
formalism~\cite{Nason:2004rx,Frixione:2007vw}.  Such a matching can be
performed with the help of the \POWHEGBOX{}~\cite{Alioli:2010xd}, a
repository that provides all process-independent building blocks of
the matching procedure, while process-specific elements have to be
provided by the user.

Building on existing NLO-QCD calculations
\cite{Figy:2003nv,Oleari:2003tc,Jager:2009xx,Jager:2006zc}, recently
various VBF processes have been implemented in the
\POWHEGBOX{}~\cite{Nason:2009ai,Jager:2012xk,Schissler:2013nga,Jager:2011ms,Jager:2013mu}. The
code developed is publicly available from the project webpage, {\tt
  http://powhegbox.mib.infn.it/}, and can be tailored to the user's
needs for any dedicated study.
In order to assess the impact of parton-shower effects on NLO-QCD
predictions for VBF-induced $W^+W^+jj$ production at the LHC,
numerical analyses for a representative setup have been performed for
the $e^+\nu_e\mu^+\nu_\mu jj$ final state~\cite{Jager:2011ms}.  At a
collision energy of $\sqrt{s}=7$~TeV, the MSTW2008 parton distribution
functions~\cite{Martin:2009iq} are used for incoming protons and the
{\tt FASTJET} package~\cite{Cacciari:2005hq} for the reconstruction of
jets via the $k_T$~algorithm with a resolution parameter of
$R=0.4$. Events are showered with {\PYTHIA~6.4.21}, including
hadronization corrections and underlying event with the Perugia~0
tune.
At least two hard jets are required with $p_{T,j}\geq 20$~GeV and
$|y_j|\leq 4.5$, well-separated from each other such that
$|y_{j_1}-y_{j_2}|>4$, $y_{j_1}\times y_{j_2}<0$, and
$M_{j_1j_2}>600$~GeV.  In addition, an $e^+$ and a $\mu^+$ with
$p_{T,\ell}\ge 20$~GeV, $|y_\ell|\le 2.5$, $\Delta R_{j\ell}\ge
0.4$, $\Delta R_{\ell\ell}\ge 0.1$, located between the two tagging
jets, are requested. For the renormalization and factorization scales
dynamical choices bound to the kinematics of the underlying Born
configuration are made.

\begin{figure}[htp]
\begin{center}
\includegraphics[height=0.3\textwidth]{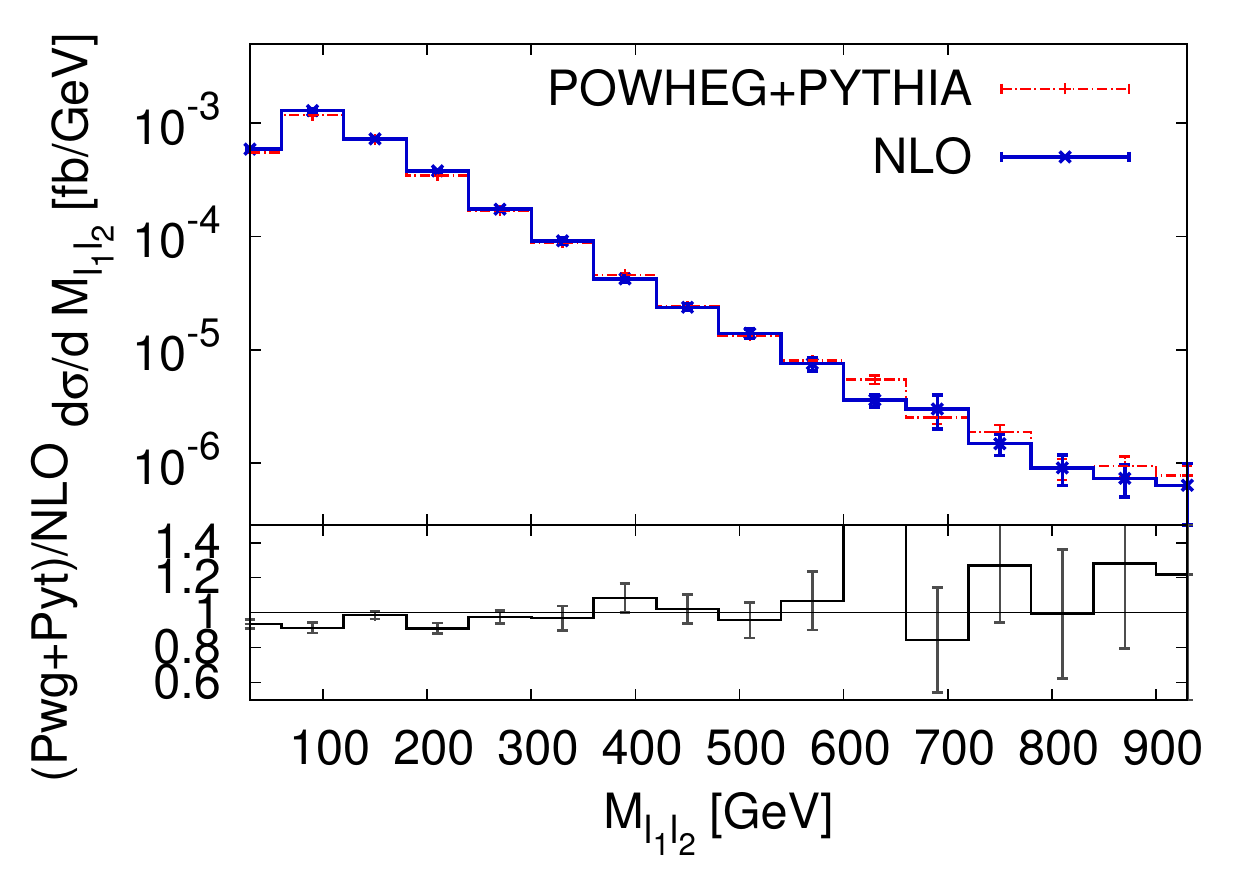} \quad
\includegraphics[height=0.3\textwidth]{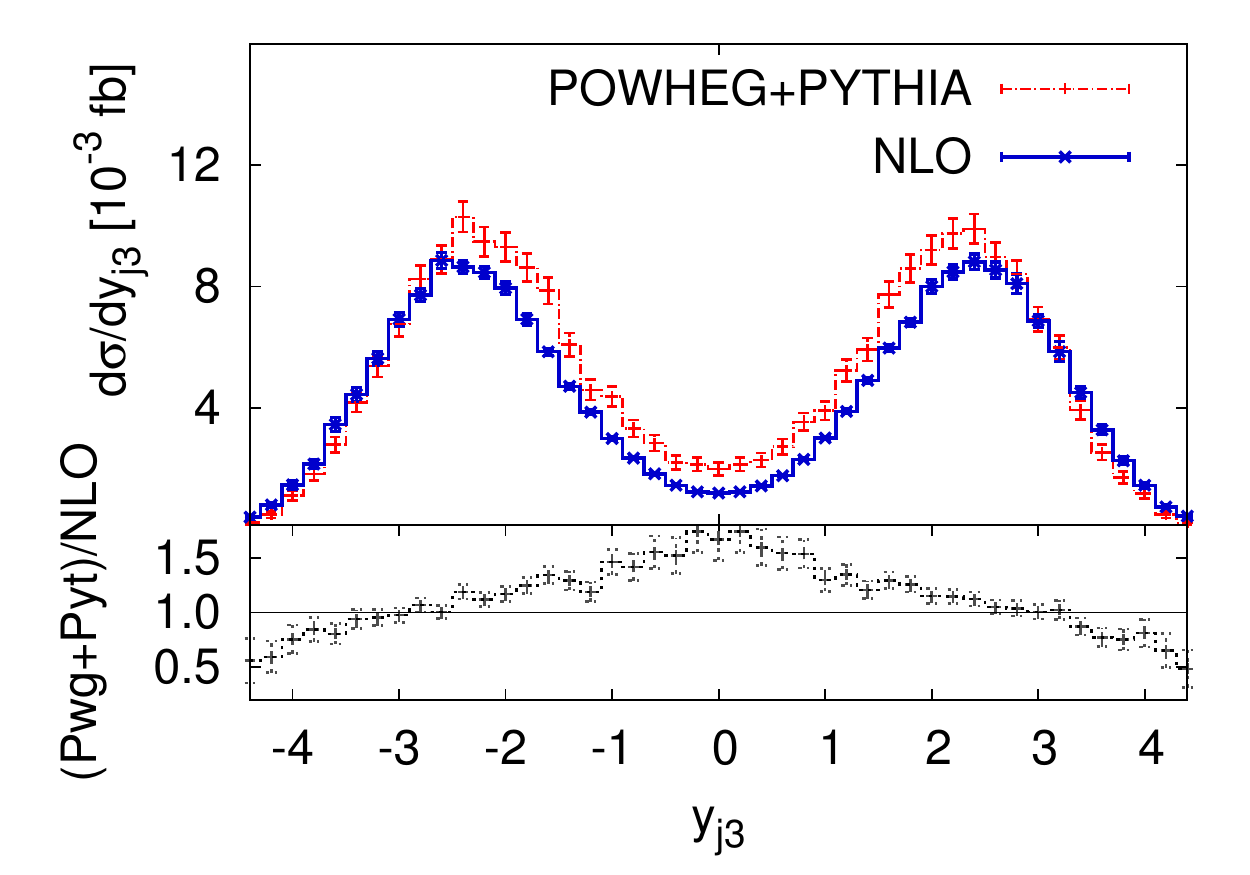}
\label{fig:vbf-wpp-powheg}
\caption{Invariant mass distribution of the charged lepton pair (left)
  and rapidity distribution of the third jet (right) in VBF-induced
  $e^+\nu_e\mu^+\nu_\mu jj$ production at the LHC with
  $\sqrt{s}=7$~TeV and the selection cuts described in the text.  The
  lower panels show the respective ratios of the {\tt POWHEG+PYTHIA}
  and the NLO-QCD results. Horizontal bars indicate statistical errors
  in each case. }
\end{center}
\end{figure}

In this setup distributions related to the tagging jets or the hard
leptons turn out to be rather insensitive to parton-shower effects. As
illustrated by Fig.~\ref{fig:vbf-wpp-powheg}~(left panel) for the
invariant mass distribution of the charged-lepton pair, the NLO-QCD
and the {\tt POWHEG+PYTHIA} results are very similar, both in
normalization and shape.
More pronounced effects of the parton shower occur in observables
related to the emission of an extra hard jet,
c.f.~Fig.~\ref{fig:vbf-wpp-powheg}~(right panel) for
$d\sigma/dy_{j_3}$. When the rapidity distribution of a third jet is
used in order to estimate central-jet veto efficiencies, this effect
should be carefully taken into account.

\underline{Triboson process $W^+\gamma\gamma$ with VBFNLO}
\label{sec:wggvbfnlo}

The second group of process where anomalous quartic gauge couplings can
be tested are the triboson processes~\cite{Lazopoulos:2007ix,
Hankele:2007sb, Campanario:2008yg, Binoth:2008kt, Bozzi:2009ig,
Bozzi:2010sj, Baur:2010zf, Bozzi:2011wwa, Bozzi:2011en, Bozzi:2012mh,
Campbell:2012ft}. The quartic vertex enters via an $s$-channel vector
boson, which decays into three vector bosons, while diagrams with two or
three bosons attached to the quark line as well as non-resonant
contributions form an irreducible background. These processes have been
shown to possess quite large K factors, typically between 1.5 and 1.8,
mostly due to the additional quark-gluon--induced production processes
first entering in the real-emission process. They also have a
considerable scale dependence.  While the dependence on the
factorization scale can be reduced by NLO QCD corrections, the strong
coupling constant first enters in the real emission part and therefore
shows a large variation with the scale.  

The example process we are considering here is $pp \rightarrow e^+ \nu_e
\gamma \gamma$~\cite{Baur:2010zf, Bozzi:2011wwa}. In this process the K
factor with a numerical value of about 3 is particularly large. This is
due to the fact that the SM amplitude vanishes when the two photons are
collinear and $\cos{\theta_W} = \frac13$, where $\theta_W$ is the angle
between the $W$ boson and the incoming quark in the partonic center-of-mass
frame.  This so-called radiation zero~\cite{Brown:1982xx, Baur:1993ir,
Baur:1997bn} is spoiled by the extra jet emission at NLO, therefore
giving huge K factors in these phase-space regions. The numerical values
for the integrated cross section are tabulated in
Table~\ref{NLO:VVV:cs}.
\begin{table}
\begin{center}
\begin{tabular}{l|cc}
& $\sigma_{\text{LO}}$ & $\sigma_{\text{NLO}}$ \\\hline
SM          & 1.124 fb & 3.674 fb \\
anom.coupl. & 1.216 fb & 3.787 fb \\
\end{tabular}
\caption{Total cross sections at LO and NLO for the process $pp
\rightarrow e^+ \nu_e \gamma \gamma$ in the SM and with anomalous
coupling $\frac{f_{T,6}}{\Lambda^4} = 2000 \text{ TeV}^{-4}$.
Statistical errors from Monte Carlo integration are below the per mille
level.}
\label{NLO:VVV:cs}
\end{center}
\end{table}
As anomalous coupling we choose the operator $T_6$ with
$\frac{f_{T,6}}{\Lambda^4} = 2000 \text{ TeV}^{-4}$, form factor scale
$\Lambda=1606$ GeV and exponent $p=4$. 

\begin{figure}
\begin{center}
\includegraphics[height=0.3\textwidth]{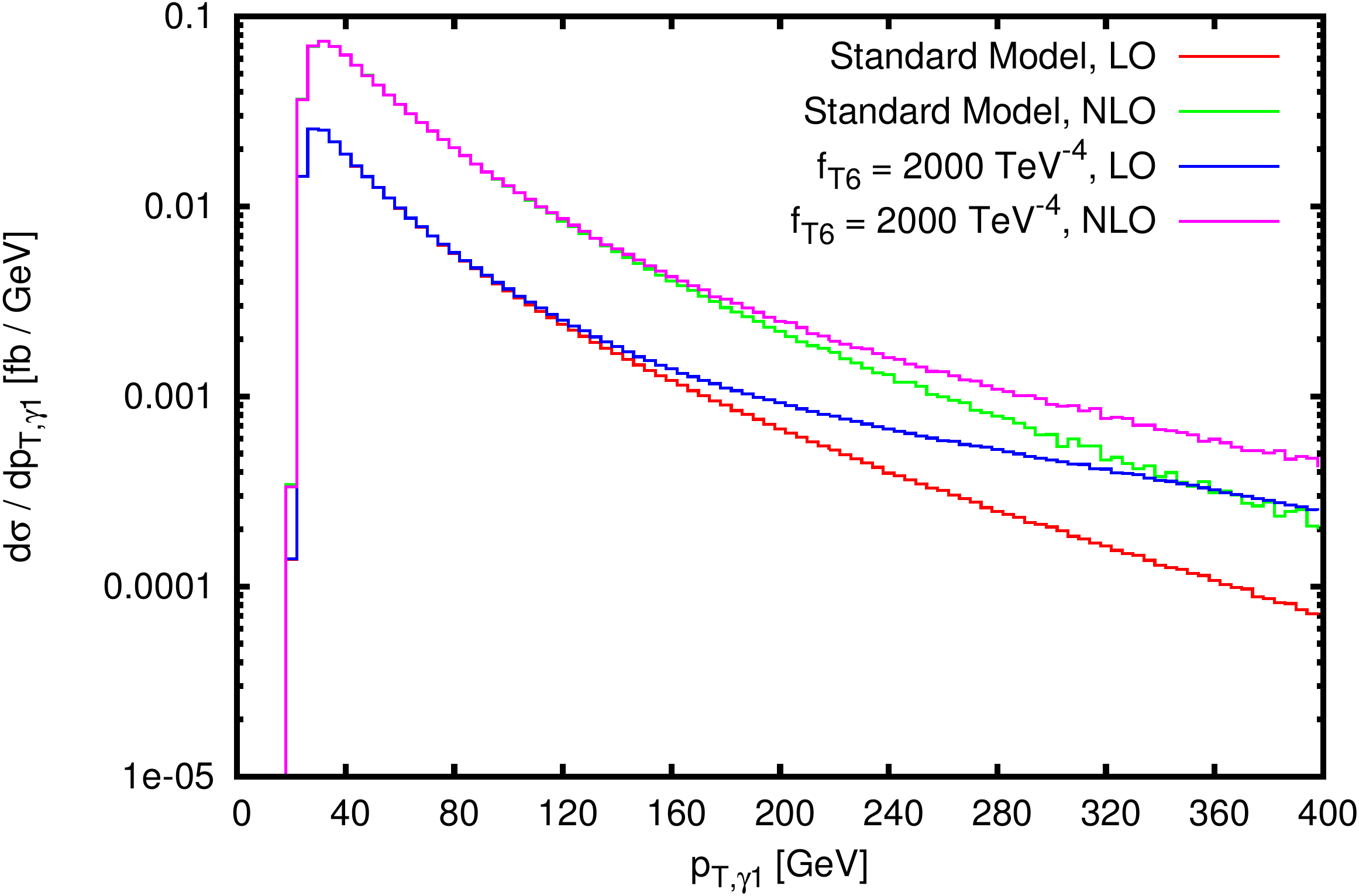} \quad
\includegraphics[height=0.3\textwidth]{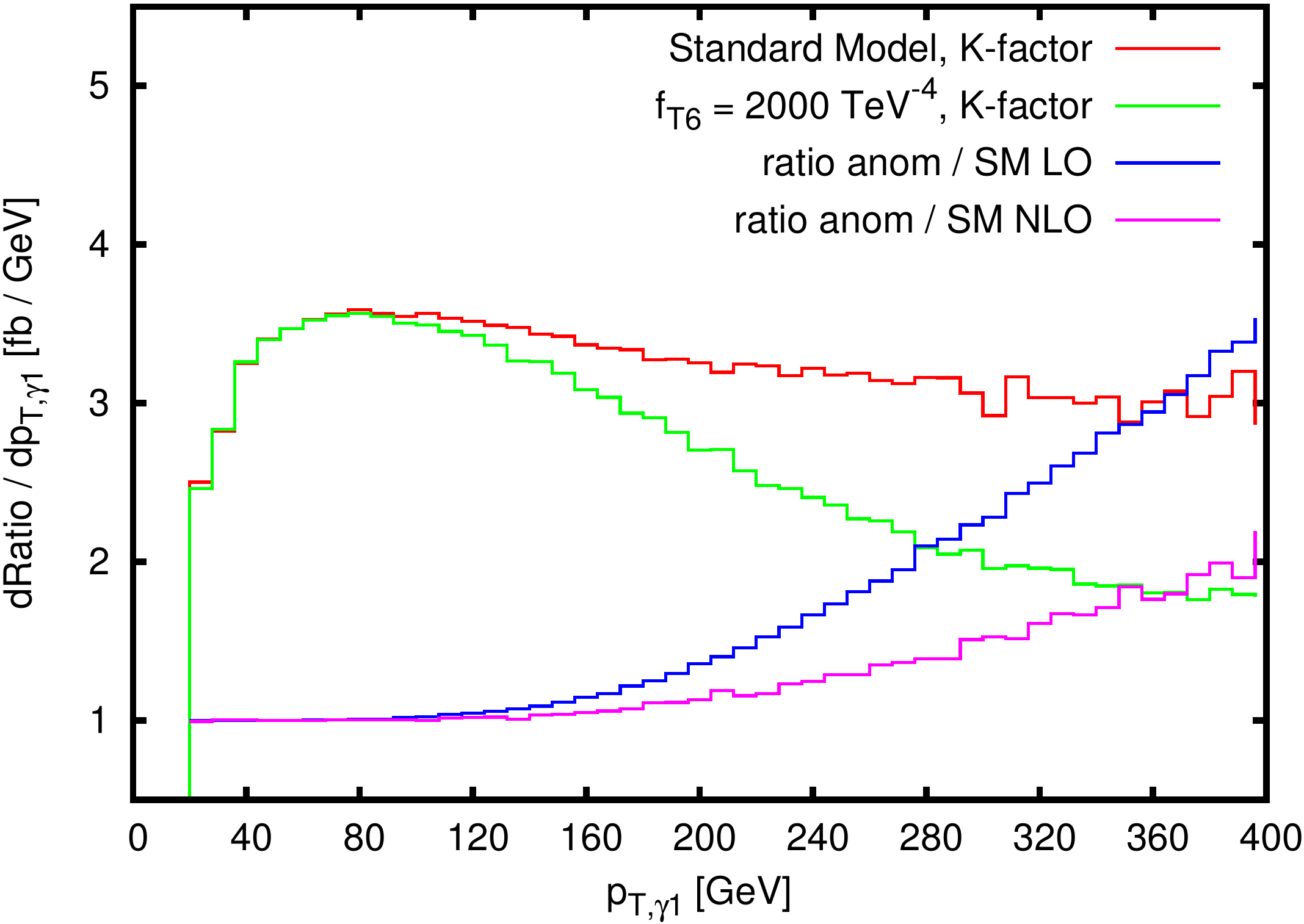}
\end{center}
\caption{Transverse-momentum distribution of the harder photon.
\textit{Left:} Differential cross section for the SM and with anomalous
coupling $T_6$ at LO and NLO. \textit{Right:} Differential K-factors for
the SM and with anomalous coupling as well as the cross-section ratio
between anomalous coupling and SM for LO and NLO.} 
\label{NLO:VVV:pTA1}
\end{figure}

Turning to differential distributions, we show the transverse momentum
distribution of the harder photon in Figure~\ref{NLO:VVV:pTA1}.
The left-hand side shows again the differential integrated cross
section. Both the SM and the anomalous-coupling scenario show
differential NLO cross sections which are significantly larger than
their LO counterpart. Contributions from anomalous couplings start to
contribute for transverse photon momenta above 100 GeV and their
relative size becomes gradually larger when going to higher momenta as
expected.

On the right-hand side one can see that the K-factor behavior differs
for the SM and the anomalous coupling scenario. While, in the SM, the K
factor is almost constant and only slightly decreases when going to
larger transverse momenta, there is a much stronger decrease when 
anomalous couplings are switched on. At the high end of the shown range,
the K factor has reached a value of around 1.8, which is the number
typically observed in other triboson processes involving $W$ bosons. As the
effect of the anomalous coupling increases, the cancellation between
different amplitudes gets gradually destroyed and the radiation zero
filled up. Only the effects from additional jet radiation remain,
yielding the smaller K factor. 

\begin{figure}
\begin{center}
\includegraphics[height=0.3\textwidth]{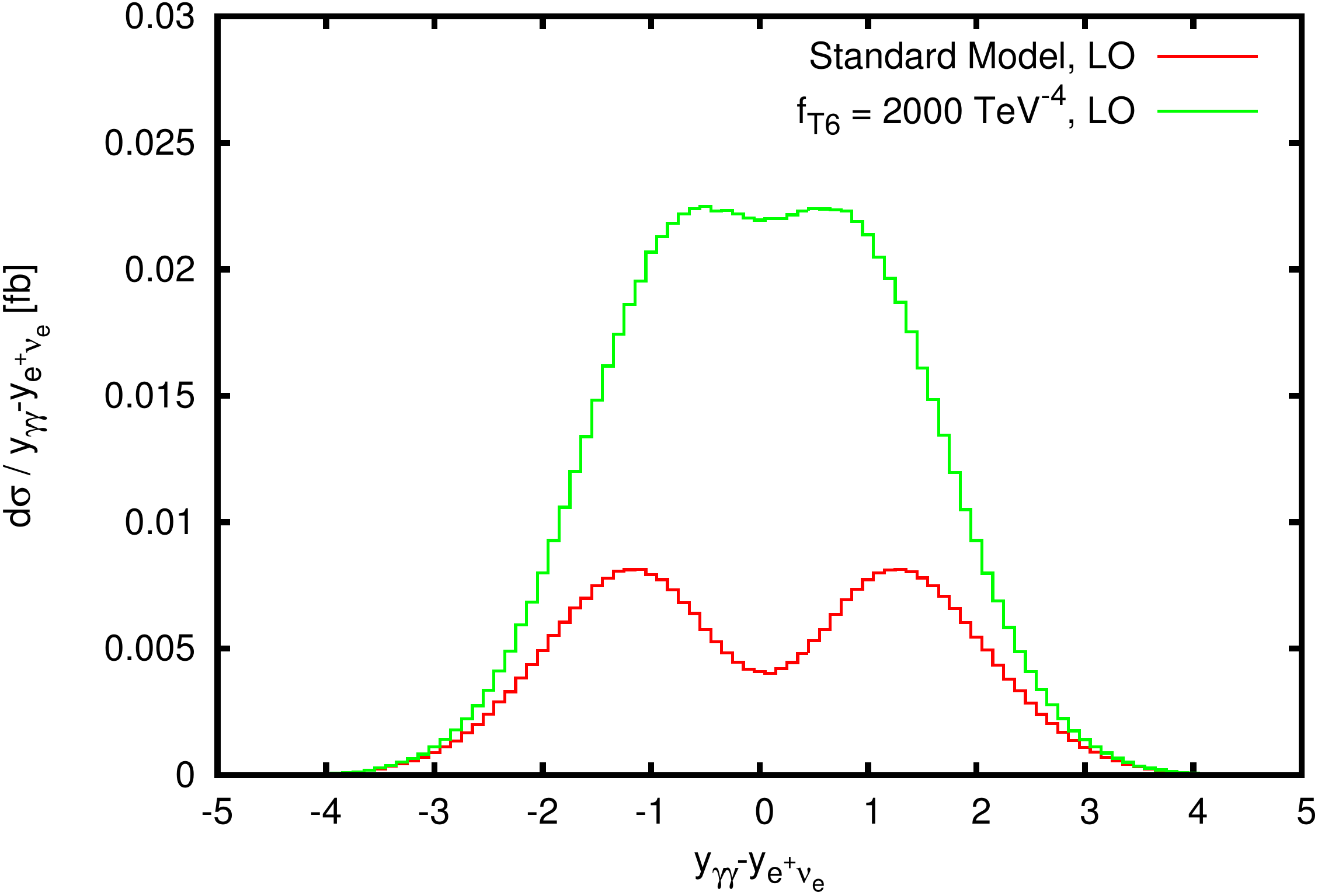} \quad
\includegraphics[height=0.3\textwidth]{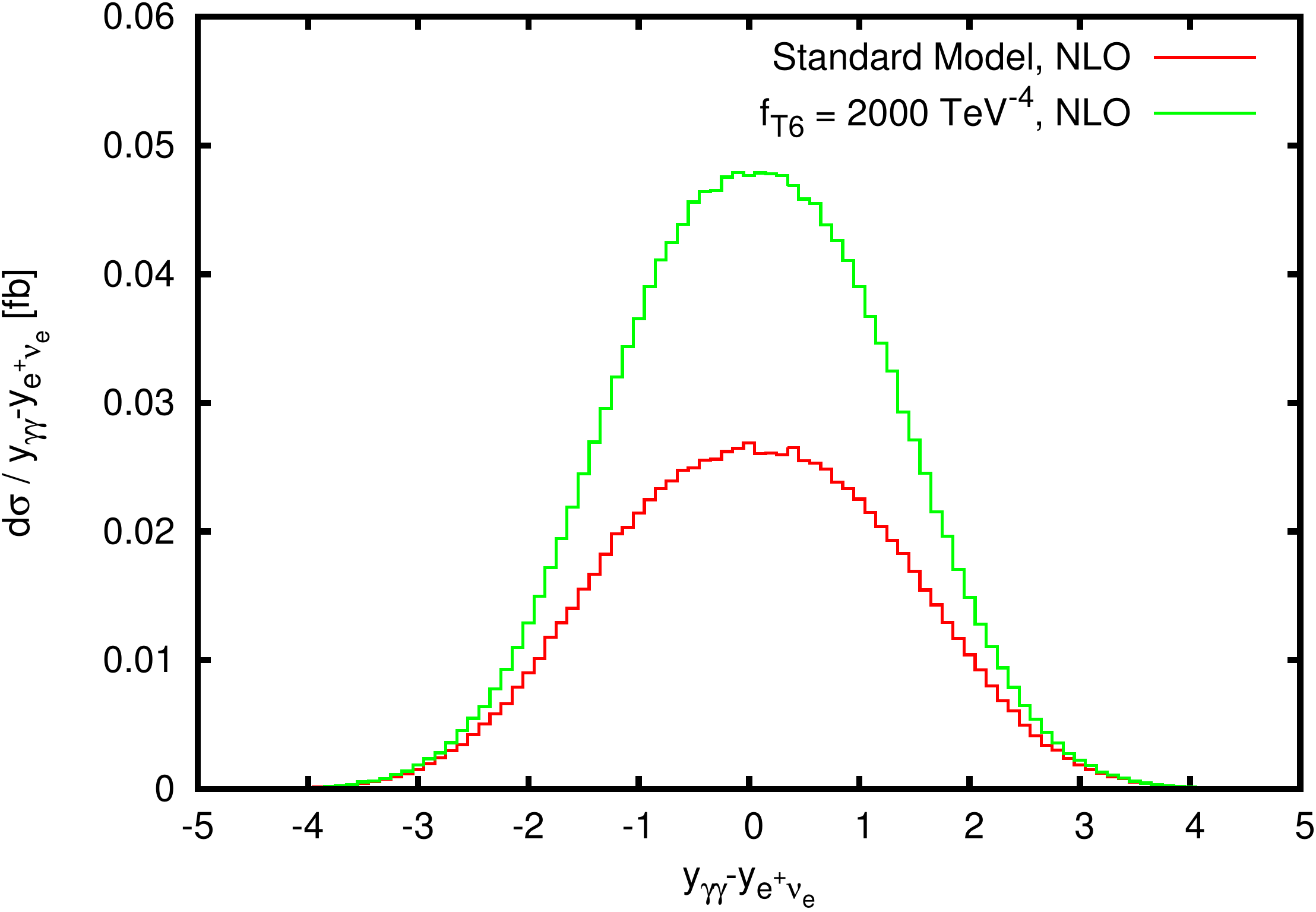}
\end{center}
\caption{Rapidity difference of the diphoton system and the
lepton-neutrino system for the SM and the anomalous coupling scenario.
\textit{Left:} LO distributions \textit{Right:} NLO distributions} 
\label{NLO:VVV:radzero}
\end{figure}

That this is indeed the case can be seen in Fig.~\ref{NLO:VVV:radzero}. 
Here we require additionally that the
transverse momentum of the harder photon exceeds 200 GeV and the
invariant mass of the lepton-neutrino system exceeds 75 GeV to suppress
radiation off the final-state lepton. The effect of the radiation zero
should be visible as a dip at zero in the rapidity difference between
the diphoton system and the lepton-neutrino system, which can be indeed
observed for the LO SM curve. In contrast the anomalous-coupling curve
shows no such behavior even at LO, and at NLO the dip is filled in both
cases.

Turning back to the right-hand plot of Fig.~\ref{NLO:VVV:pTA1}, the
ratio between anomalous-coupling and SM prediction decreases when going
from LO to NLO. This is due to the same effect, as part of the
additional contribution is caused by filling up the radiation zero,
which is no longer present at NLO because there already QCD effects have
caused this. Hence, for this process group, higher-order corrections
play an important role and cannot be neglected when determining the size
of or limits on anomalous quartic gauge couplings.

\underline{Electroweak corrections to triboson processes}
\label{sec:tribosonew}

The first calculation of electroweak NLO corrections for a triboson
processes at hadron colliders has appeared only very recently. Hence, no
publicly available Monte Carlo implementation is available at the
present stage. For gauge boson pair production via vector-boson fusion
electroweak corrections no results exist in the literature at the
current stage.

In Ref.~\cite{Nhung:2013jta} the full NLO corrections to on-shell $WWZ$
production have been considered. Besides the QCD corrections already
calculated in Refs.~\cite{Hankele:2007sb,Binoth:2008kt}, additional
virtual electroweak diagrams with loops up to the pentagon level appear
as well as real-emission processes with an additional external photon.
There, processes with both photon radiation and initial-state photons
are taken into account. The latter appear when using PDFs with
photons~\cite{Martin:2004dh,Ball:2013hta}. Additionally, in this case
the photon-initiated contribution of $\gamma\gamma \rightarrow WWZ$ is
added at tree-level.
The electroweak corrections are typically quite small for integrated
cross sections, of about -2\%. They can, however, get significant in
differential distributions. For example, looking at the
transverse-momentum distribution of the $Z$ boson, at the 14 TeV LHC, one
observes corrections of up to -30\% for transverse momenta of 1 TeV.
Thereby, the photon-initiated processes play an important role to partly
cancel large Sudakov virtual corrections.

\subsection{Current bounds on triple and quartic gauge boson couplings}
\label{sec:electroweak-bounds}

\begin{figure}[htp]
\begin{center}
\includegraphics[width=0.65\hsize]{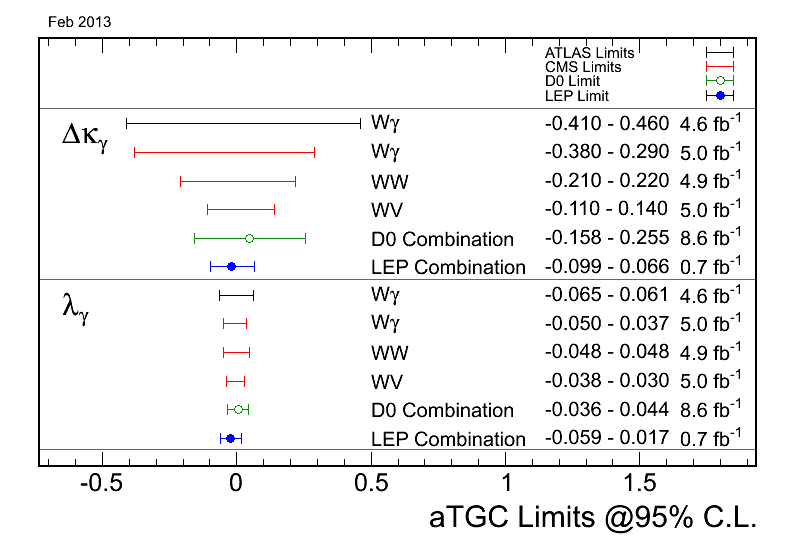}
\caption{Limits on anomalous $WW\gamma$ couplings. Tevatron limits use a form factor 
with the cut-off parameter $\Lambda=2$ TeV. Taken from Ref.~\cite{Lombardo:2013daa}.}
\label{fig:electroweak-tgclimits121}
\end{center}
\end{figure}
\begin{figure}[htp]
\begin{center}
\includegraphics[width=0.65\hsize]{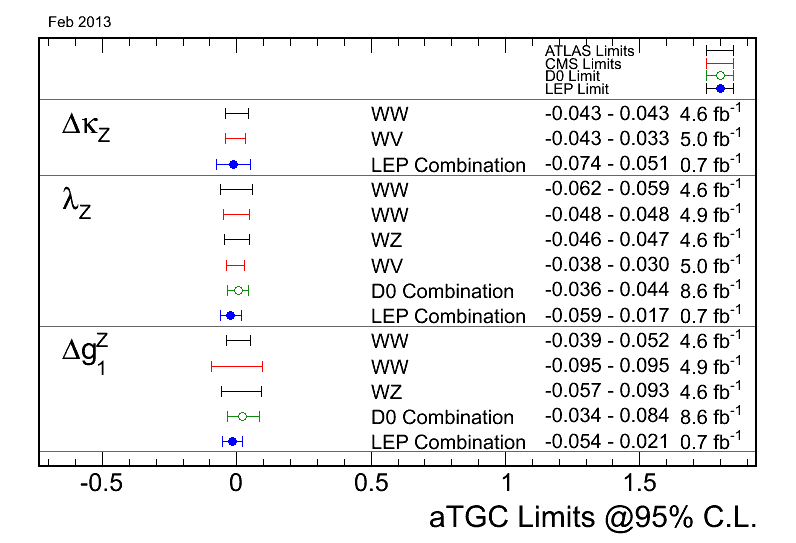}
\caption{Limits on anomalous $WWZ$ couplings. Tevatron limits use a form factor 
with the cut-off parameter $\Lambda=2$ TeV. Taken from Ref.~\cite{Lombardo:2013daa}.}
\label{fig:electroweak-tgclimits122}
\end{center}
\end{figure}
\begin{figure}[htp]
\begin{center}
\includegraphics[width=0.65\hsize]{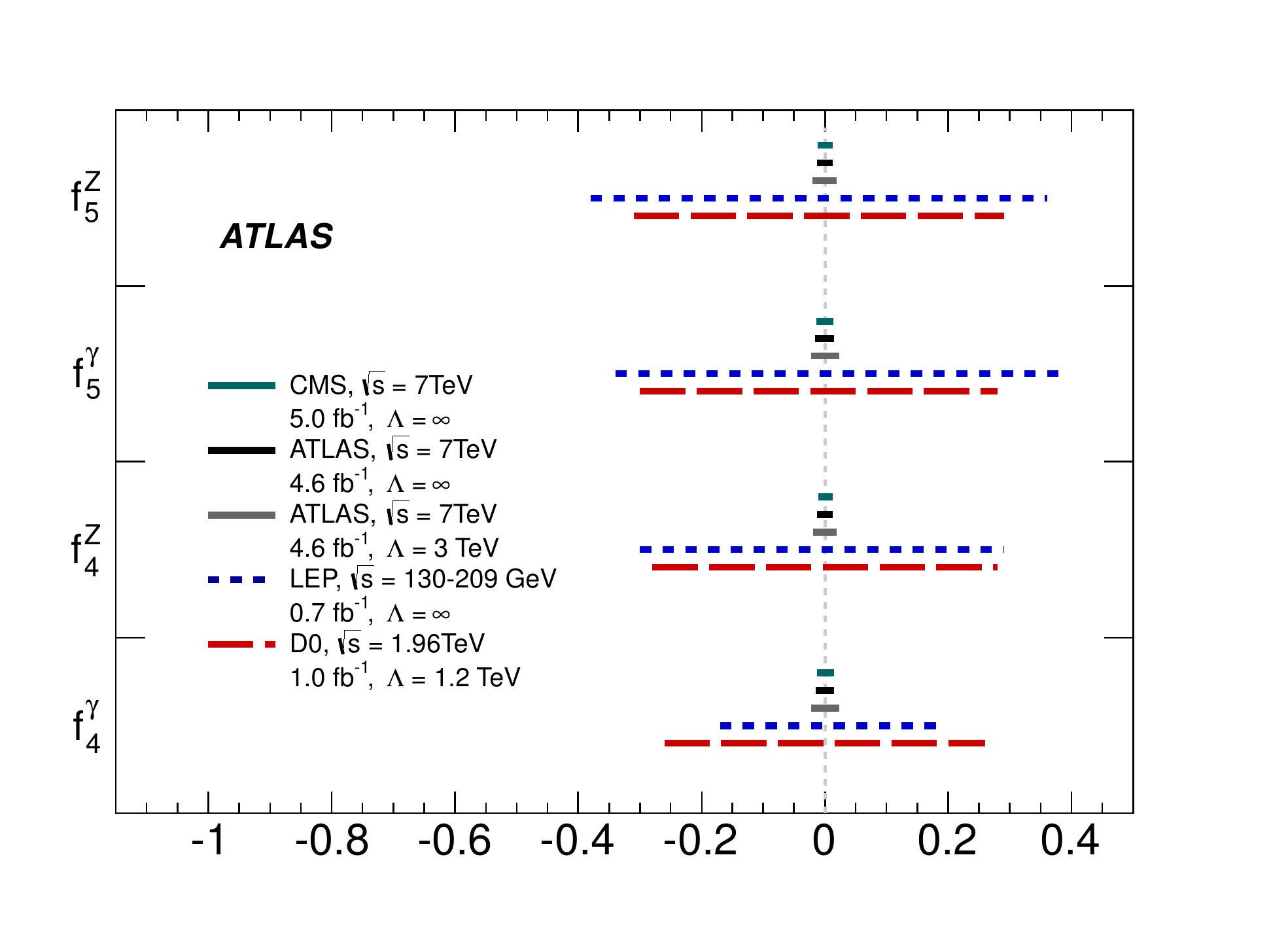}
\caption{Anomalous $ZZ\gamma$ and $ZZZ$ 95\% CL intervals from $ZZ$ production
  at ATLAS, CMS, LEP and the Tevatron experiments. If a form factor
  is used, the cut-off parameter $\Lambda$ is also shown. Taken from Ref.~\cite{Aad:2012awa}.}
\label{fig:electroweak-tgclimits34}
\end{center}
\end{figure}
\begin{figure}[htp]
\begin{center}
\includegraphics[width=0.48\hsize]{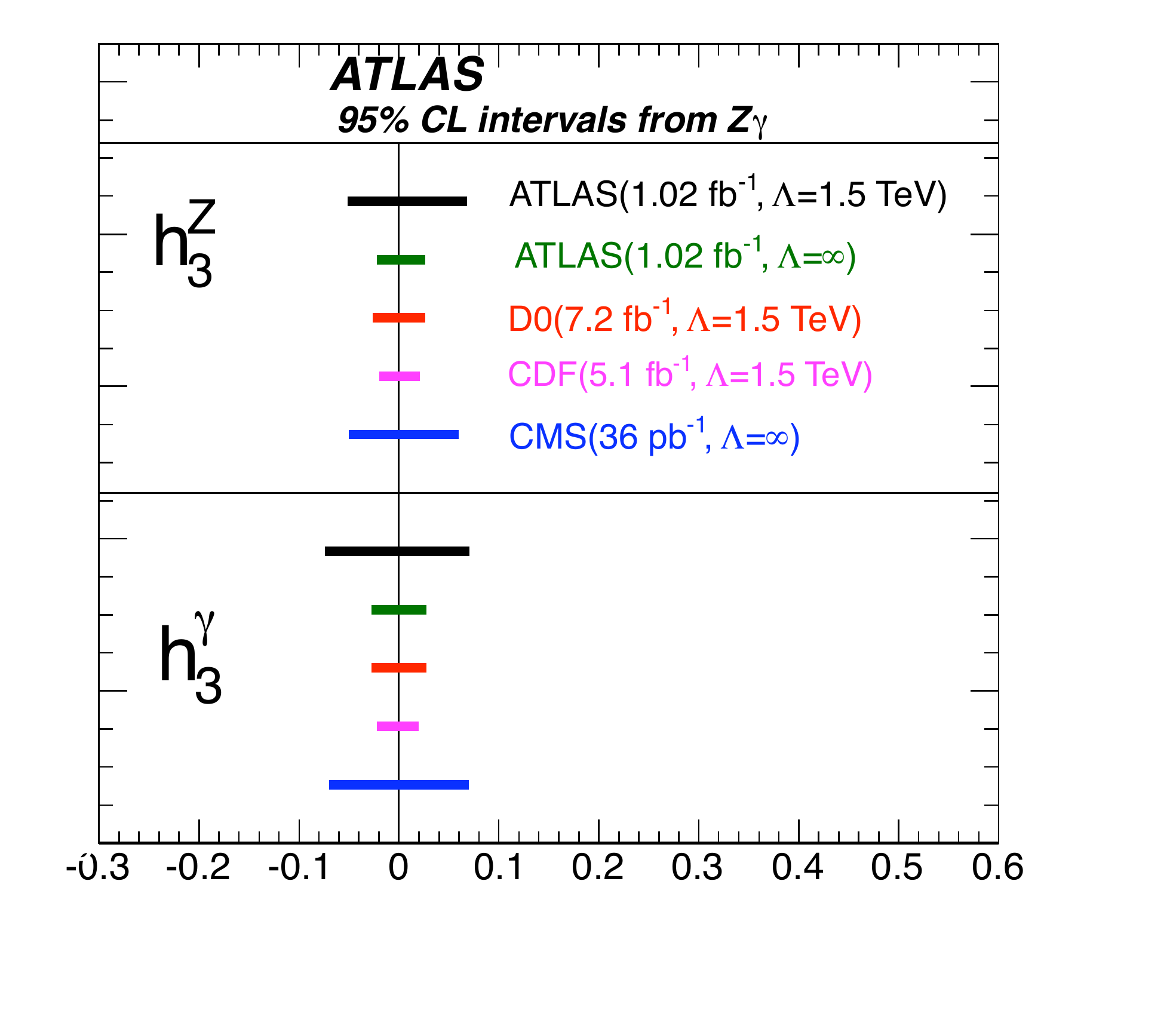}
\includegraphics[width=0.48\hsize]{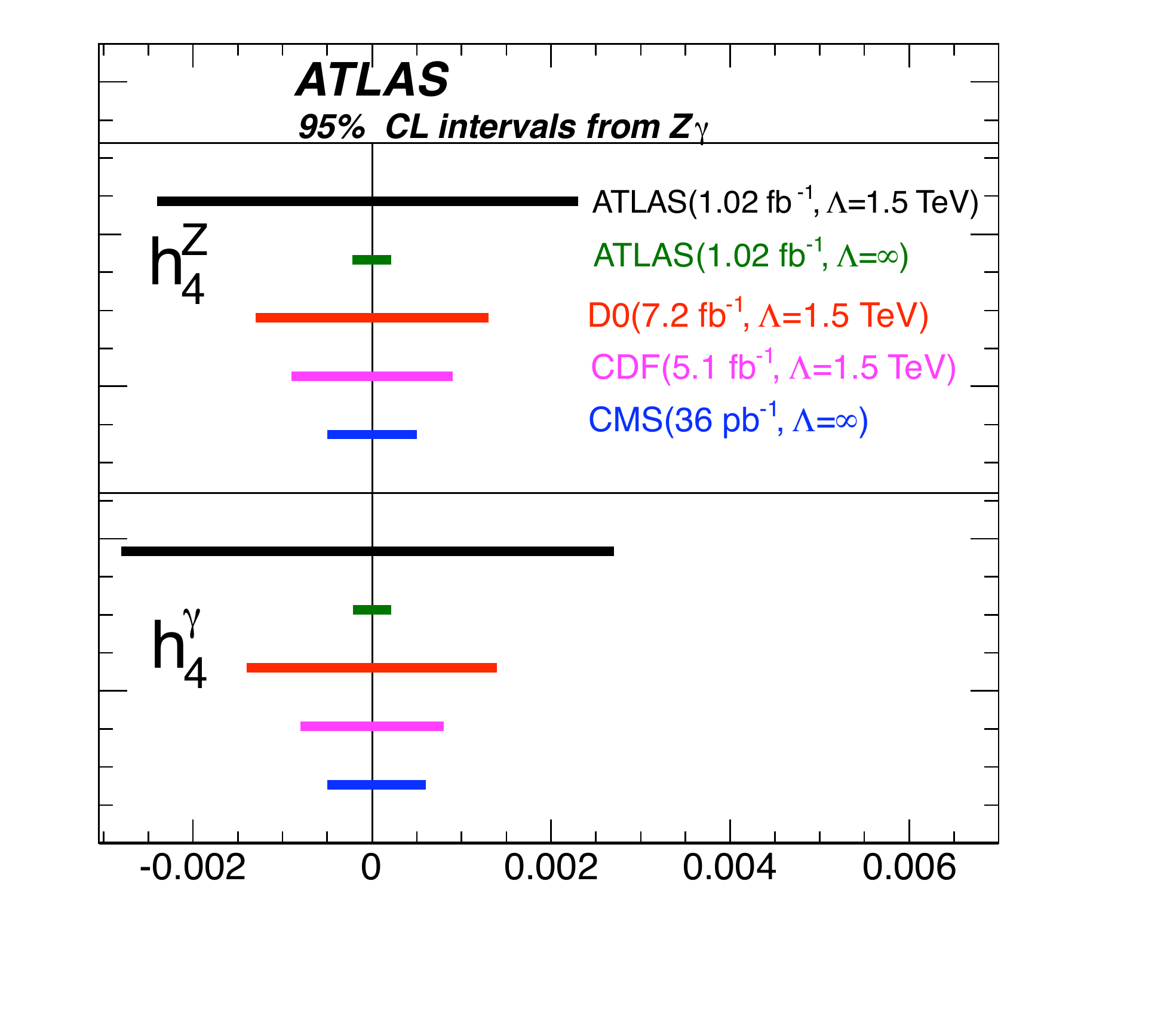}
\caption{The 95\% CL intervals for neutral aTGCs from ATLAS, CMS, LEP
  and the Tevatron experiments as obtained from $Z\gamma$ events. If a
  form factor is used, the cut-off parameter $\Lambda$ is also
  shown. Taken from Ref.~\cite{Aad:2012mr}.}
\label{fig:electroweak-tgclimits56}
\end{center}
\end{figure}

Current bounds on the aTGCs of Eq.~\ref{eq:L} from LEP, Tevatron and LHC
searches in $WW\gamma, WWZ, ZZ$ and $Z\gamma$ events are summarized in
Figs.~\ref{fig:electroweak-tgclimits121}, \ref{fig:electroweak-tgclimits122},
\ref{fig:electroweak-tgclimits34}, and
\ref{fig:electroweak-tgclimits56}. Constraints from a combined analysis of 
EWPOs and LEP data on aTGCs can be found in Ref.~\cite{Aihara:1995iq}, for instance. 

A review of current bounds on quartic gauge couplings from LEP can be
found in Ref.~\cite{quarticATlep}. Recently, stringent bounds on
anomalous quartic gauge couplings involving two photons, $a_0$ and
$a_c$ of Eq.~\ref{eq:aoac}, have been obtained from two-photon
production of a $W^+W^-$ pair at the LHC, as reported by
CMS~\cite{Chatrchyan:2013foa} (95\% CL intervals):
$|a_0/\Lambda^2|<0.00015~{\rm GeV}^{-2}, |a_c/\Lambda^2|<0.0005~{\rm
  GeV}^{-2}$ with a dipole form factor and $\Lambda_{cutoff}=500$~GeV,
and $|a_0/\Lambda^2|<4.0\times 10^{-6}~{\rm GeV}^{-2},
|a_c/\Lambda^2|<1.5 \times 10^{-5}~{\rm GeV}^{-2}$ without using a
form factor.

\subsection{Multi-boson processes at the 14 TeV LHC}
\label{sec:electroweak-multibosonhad14}


Studies on vector boson scattering (VBS) and triboson production have
been presented by ATLAS collaboration for $\sqrt {s} = 14$ TeV and
integrated luminosities of 300~fb$^{-1}$ and 3~ab$^{-1}$
respectively~\cite{ATLAS-Collaboration:1496527, ATLAS-Collaboration:1558703}.
  These studies showcase the greatly increased sensitivity
for new physics in these channels.

Studies of vector boson scattering in the $W^+ W^- jj \to \ell^+ \nu
\ell^- \nu jj$ have been presented based on the comparison of the
$m_{lljj}$ distribution from backgrounds (including $t \bar{t}$
production, diboson production with ISR jets, and SM VBS) and
anomalous VBS signal.  The statistical sensitivity has been
parameterized using the electroweak chiral lagrangian operator with
coefficient $\alpha_4$ (see Eq.~\ref{eqn:alpha4}).  In this formulation of new physics, this
particular operator is one of the least constrained since it preserves
the CP symmetry and the electoweak SU(2) custodial symmetry, does not
induce oblique corrections in the gauge boson propagators, and only
induces anomalous quartic couplings which have not been constrained by
past studies on trilinear guage couplings.  Unitarity is maintained by
using the inverse amplitude method~\cite{Truong:1988zp, Ballestrero:2011pe}.  Table~\ref{tab:dileptona4} shows
the results of this ATLAS study, as reproduced from their report. The
sensitivity to the $\alpha_4$ (referred to as $a_4$ in~\cite{ATLAS-Collaboration:1496527}) coefficient is increased by more than a
factor of 4 in the high-luminosity upgrade of the LHC. Using a dimension-8 operator formulation and considering simplified models of electroweak resonances, these $\alpha_4$ limits are converted
 to corresponding masses of heavy resonances in Table~\ref{tab:mreslima4}.
\begin{table}
\centering
\begin{tabular}{lccc}
\hline
parameter               & 300 fb$^{-1}$  & 1 ab$^{-1}$     & 3 ab$^{-1}$  \\
\hline
$\alpha_4$ &  0.066 & 0.025  & 0.016  \\
\hline
\hline
\end{tabular}
  \caption{Summary of expected upper limits for $\alpha_4$ at the 95\% confidence level using the  $pp \to W^+W^- +2j \to e \mu + 2\nu + 2j$ VBS search 
    at $pp$ collision center-of-mass energy of
    14 TeV at ATLAS~\cite{ATLAS-Collaboration:1496527}. See Table~\ref{tab:mreslima4} for corresponding resonance mass limits in a dimension-8 operator formulation and considering simplified models of electroweak resonances. }
\label{tab:dileptona4}
\end{table}

ATLAS has also presented a study of VBS in the $ZZ \to 4 \ell$ channel
which has a clean and fully reconstructible final state. In this
study, the K-matrix unitarization approach is used to model anomalous
quartic couplings and unitarization is achieved by including TeV-scale
resonances. Such resonances would be clearly visible in the $4 \ell$
invariant mass distribution.  Table~\ref{tab:zzSignificance},
reproduced from the ATLAS report, shows the statistical significance
of potential resonant signals given the background-only hypothesis,
for a number of resonance masses and couplings. The comparison of the
two scenarios with integrated luminosities of $300 \, \fb$ and $3000
\, \fb$ respectively showcases the discovery potential of the
high-luminosity upgrade.

ATLAS has estimated  the precision on the measurement of
the integrated cross section for the purely-electroweak SM process $pp
\to ZZ + 2j \to 4 \ell + 2j$.  In the kinematic region where the
tagging forward jets have $m_{jj} > 1$~TeV and the 4-lepton invariant
mass $m_{4 \ell} > 200 (500)$~GeV, a statistical precision of 10(15)\%
is achievable with $3000 \, \fb$, compared to 30(45)\% with $300 \,
\fb$.  Since a key prediction of the SM is that the Higgs boson
unitarizes longitudinal VBS, it is important to make the definitive
measurements of this cross section, which is only possible with the
high-luminosity upgrade in this clean and robust channel.
 
\begin{table}[h] 
\centering 
\begin{tabular}{lcc} \hline  model              & $300\,\fb$  & $3000\,\fb$  \\ 
\hline $m_{\rm resonance} = 500$~GeV, $g = 1.0$ &  $ 2.4 \sigma$  & $7.5 \sigma$  \\ 
\hline $m_{\rm resonance} = 1$~TeV, $g = 1.75$ &  $ 1.7 \sigma$  & $5.5 \sigma$  \\ 
\hline $m_{\rm resonance} = 1$~TeV, $g = 2.5$ &  $ 3.0 \sigma$  & $9.4 \sigma$  \\ 
\hline \hline \end{tabular} 
\caption{Summary of the expected sensitivity to
    anomalous VBS signal, quoted in terms of the background-only
    $p_0$-value expected for signal+background. The $p_0$-value has
    been converted to the corresponding number of Gaussian $\sigma$ in
    significance. These results are given for the $ pp \to ZZ + 2j \to
    \ell \ell \ell \ell +2j$ channel at $\sqrt{s} = 14 \,$TeV. The
    increase in significance with integrated luminosity is shown for
    different resonance masses and couplings $g$.\newline} \label{tab:zzSignificance}
\end{table}

ATLAS has also shown sensitivity studies using the fully-leptonic
decay modes of $W^\pm W^\pm$, $WZ$ and $ZZ$ channels in the VBS mode
as well as triboson results in the $Z \gamma \gamma$ channel.  These
results are quoted in the language of EFT higher-dimension
operators. The studies are performed in the kinematic regions where
unitarity is perserved.  In this context, ATLAS has studied one
dimension-6 operator, ${\cal O}$$_{\Phi W}$ of Eq.~\ref{opphiw} and four dimension-8 operators,
${\cal O}$$_{S,0}$ of Eq.~\ref{eq:s0} and
${\cal O}$$_{T,i}, i=1,8,9$ of Eqs.~\ref{eq:t1},~\ref{eq:t8}, and~\ref{eq:t9}.
Their values for $5\sigma$-significance discovery are summarised in
Table~\ref{tab:5sigsummary}, reproduced from the ATLAS report.  The
high-luminosity upgrade increases the discovery potential for the
operator coefficients by factors of 2-3, with further increases
possible using analysis optimizations.  If an anomaly is discovered
with 300~fb$^{-1}$, the corresponding operator coefficient can be
measured with a precision of 5\% or better with 3000~fb$^{-1}$ of
integrated luminosity, allowing detailed studies of the underlying
physics in this arena.
\begin{table}[h]
\centering
\begin{tabular}{c|c|c|c|c|c|c|c}
\hline\hline                
\multirow{2}{*}{Parameter} & \multirow{2}{*}{dimension} & \multirow{2}{*}{channel} & \multirow{2}{*}{$\Lambda_{UV}$ [TeV]} & \multicolumn{2}{|c|}{300 fb$^{-1}$} & \multicolumn{2}{|c}{3000 fb$^{-1}$}  \\
\cline{5-8}
                           &                            &                          &                                       & $5 \sigma$ & 95\% CL              & $5 \sigma$ & 95\% CL       \\
\hline
        $c_{\Phi W}/\Lambda^2$ & 6 & $ZZ$ & 1.9 & 34 TeV$^{-2}$ &  20 TeV$^{-2}$  & 16 TeV$^{-2}$  & 9.3 TeV$^{-2}$ \\
        $f_{S,0}/\Lambda^{4}$ & 8 & $W^\pm W^\pm$ & 2.0 & 10 TeV$^{-4}$ & 6.8 TeV$^{-4}$  &  4.5 TeV$^{-4}$  & 0.8 TeV$^{-4}$ \\
        $f_{T,1}/\Lambda^{4}$ & 8 & $WZ$ & 3.7 & 1.3 TeV$^{-4}$ & 0.7 TeV$^{-4}$ & 0.6 TeV$^{-4}$  & 0.3 TeV$^{-4}$ \\
        $f_{T,8}/\Lambda^{4}$ & 8 & $Z\gamma\gamma$ & 12 & 0.9 TeV$^{-4}$ & 0.5 TeV$^{-4}$ & 0.4 TeV$^{-4}$  & 0.2 TeV$^{-4}$ \\
        $f_{T,9}/\Lambda^{4}$ & 8 & $Z\gamma\gamma$ & 13 & 2.0 TeV$^{-4}$ & 0.9 TeV$^{-4}$ & 0.7 TeV$^{-4}$  & 0.3 TeV$^{-4}$ \\        
\hline\hline
\end{tabular}
\caption{$5 \sigma$-significance discovery values and 95\% CL limits for coefficients of higher-dimension operators. Madgraph5 is used for the event generation. $\Lambda_{UV}$ is the unitarity violation bound corresponding to the sensitivity with 3000 fb$^{-1}$ of integrated luminosity. See Table~\ref{tab:mreslim} for resonance mass limits corresponding to the $f_{S,0}/\Lambda^{4}$ sensitivity above, in a dimension-8 operator formulation and considering simplified models of electroweak resonances. 
}
\label{tab:5sigsummary}
\end{table}

The substantially improved sensitivity to these higher dimensional
operators highlights the potential of the LHC to probe one of the most
important aspects of the electroweak sector of the SM, namely, the
unitarization of the vector boson scattering amplitudes by the Higgs
mechanism. Since the "Mexican hat" Higgs potential is essentially just
a parameterization, a more "dynamical" explanation of this potential
in terms of the Higgs' interaction with new scalar, vector or fermion
fields involving strong dynamics can easily induce higher-dimension
operators as precursors to the more complete theory of the Higgs
sector.

Another example of the impact of the HL-LHC in studying the
unitarization mechanism is provided by the improved sensitivity to the
$\mathcal{O}_{\Phi d}$ operator of Eq.~\ref{opphid} shown in
Table~\ref{tab:wzSignificance_noUVCutOff}.  The threshold of interest in the magnitude of
this operator is provided by $v^{-2}$ where $v$ is the Higgs field's
vacuum expectation value, thus $v^{-2} = 16$ TeV$^{-2}$. As the
sensitivity to the magnitude of this operator falls below 16
TeV$^{-2}$, we obtain a direct test of the SM unitarization
mechanism. Table~\ref{tab:wzSignificance_noUVCutOff} shows that this threshold is crossed
by increasing the LHC integrated luminosity from 300 fb$^{-1}$ to 3000
fb$^{-1}$.
  
Vector boson scattering and triboson production are unique probes of
the possible high-energy dynamics underlying the Higgs
potential. Furthermore, the different operators reflect directly in
different energy dependencies of VBS and triboson production, and the
study of these processes can not only detect the presence of new
underlying dynamics but also distinguish between the operators through
the differences in the kinematic shapes.

\subsection{Multi-boson processes at HE pp colliders}
\label{sec:electroweak-multibosonhadHE}


Additional sensitivity studies have been performed using the
Snowmass-DELPHES detector
simulation~\cite{snowmass-multiboson-whitepaper}. These studies extend
the ATLAS investigations to higher energy $pp$ colliders, and also
include additional final states and higher-dimension operators. Madgraph5 has been used for the event generation and the relations of Eqn.~\ref{eq:fsmtrelations} apply. The
results presented in~\cite{snowmass-multiboson-whitepaper} are
summarized here and tables of numerical results have been reproduced
from this reference.  The main area of interest is anomalous quartic
gauge couplings which are only probed by vector boson scattering and
triboson production. Access to these processes has only opened up with
the availability of LHC data, and therefore they are promising new
avenues for discovery, for instance, of the composite nature of the
Higgs sector.

In the following, a review of studies using VBS and triboson channels is presented. A number of different combinations of operators and final states are investigated. 
 The main purpose of these studies is to estimate the improvement
 of sensitivity   to coefficients of higher-dimension operators as a function of     integrated luminosity and collider energy. 
 In cases where the same operator is used with different final states, we also learn which of these final states are more sensitive. 
 
Using the VBS $ZZ \to 4 \ell$ final state, the sensitivity to the
following operators was quantified for 14 TeV and 33 TeV $pp$
colliders; the dimension-6 operator ${\cal O}_{\Phi W}$ of Eqn.~\ref{opphiw}
and  the  dimension-8 operators ${\cal O}_{T,8}$ of Eqn.~\ref{eq:t8} and ${\cal O}_{T,9}$ of Eqn.~\ref{eq:t9}, 
with the results shown in Table~\ref{tab:zzSignificance_noUVCutOff}. As the sensitivity to these operators is probed with higher-energy colliders, the regime where the amplitude violates
 unitarity is probed more deeply. This is an indication of the colliders' ability to directly produce the particle excitations of the ultraviolet-complete field theory underlying the EFT. 
 The results in~\cite{snowmass-multiboson-whitepaper} are quoted both with and without the application of a unitarity-violating (UV) upper bound on the invariant mass of the multi-boson system. 
 When the bound is applied, the surviving events are restricted to the low-mass region where the corresponding beyond-SM amplitude does not violate unitarity. 
\begin{table}[h]
\centering
\begin{tabular}{c|c|c|c|c|c}
\hline\hline
\multirow{2}{*}{Parameter}                           & Luminosity & \multicolumn{2}{|c|}{14 TeV} & \multicolumn{2}{|c}{33 TeV}  \\
\cline{3-6}
                                                     & [fb$^{-1}$]             & $5 \sigma$ & 95\% CL              & $5 \sigma$ & 95\% CL                \\
\hline
\multirow{2}{*}{$c_{\Phi W}/\Lambda^2$ [TeV$^{-2}$]} & 3000                                   & 16.2 (16.2) &  9.7 (9.7)  & 13.2 (13.2)  & 8.2 (8.2) \\
\cline{3-6}
                                                     & 300                                    & 31.3 (31.5) & 18.2 (18.3)  & 23.8 (23.8)  & 14.7 (14.7) \\
\hline
\multirow{2}{*}{$f_{T,8}/\Lambda^{4}$ [TeV$^{-4}$]}   & 3000                                   & 2.9 (4.7)  & 1.7 (2.4)   & 1.6 (1.7)   & 1.0 (1.3) \\
\cline{3-6}
                                                     & 300                                    & 5.5 (8.4)  & 3.2 (5.3)   & 2.8 (2.3)   & 1.8 (1.8) \\
\hline
\multirow{2}{*}{$f_{T,9}/\Lambda^{4}$ [TeV$^{-4}$]}   & 3000                                   & 5.7 (6.3)  & 3.9 (4.6)   & 3.8 (6.6)   & 2.5 (3.5) \\
\cline{3-6}
                                                     & 300                                    & 8.7 (9.0)  & 6.2 (6.7)   & 6.3 (10.1)   & 4.2 (8.2) \\
\hline\hline
\end{tabular}
\caption{In $pp \to ZZ + 2j \to 4 \ell  + 2j$ processes, $5 \sigma$-significance discovery values and 95\% CL limits are shown
 for coefficients of high-dimension operators with 300 fb$^{-1}$/3000 fb$^{-1}$ of integrated luminosity.
 To show the impact of the UV bound, the corresponding results are shown in parentheses.}
\label{tab:zzSignificance_noUVCutOff}
\end{table}

Similarly, the studies of the VBS $WZ \to 3 \ell \nu$ were extended to 14 TeV and 33 TeV $pp$ colliders using the following operators; the dimension-8 operator ${\cal O}_{T,1}$ of Eqn.~\ref{eq:t1}
and the dimension-6 operator ${\cal O}_{\Phi d}$ of Eqn.~\ref{opphid}.
The results are shown in Table~\ref{tab:wzSignificance_noUVCutOff}. 
\begin{table}[h]
\centering
\begin{tabular}{c|c|c|c|c|c}
\hline\hline
\multirow{2}{*}{Parameter}                           & Luminosity  & \multicolumn{2}{|c|}{14 TeV} & \multicolumn{2}{|c}{33 TeV}  \\
\cline{3-6}
                                                     & [fb$^{-1}$]   & $5 \sigma$      & 95\% CL          & $5 \sigma$       & 95\% CL                \\
\hline
\multirow{2}{*}{$c_{\Phi d}/\Lambda^2$ [TeV$^{-2}$]} & 3000                                   & 15.2 (15.2) &  9.1 (9.1)  & 12.6 (12.7)  & 7.7 (7.7) \\
\cline{2-6}
                                                     & 300                                    & 28.5 (28.7) &  17.1 (17.1) & 23.1 (23.3)  & 14.1 (14.2) \\
\hline
\multirow{2}{*}{$f_{T,1}/\Lambda^{4}$ [TeV$^{-4}$]}   & 3000                                   & 0.6 (0.9)  & 0.4 (0.5)   & 0.3 (0.6)   & 0.2 (0.3) \\
\cline{2-6}
                                                     & 300                                    & 1.1 (1.6)  & 0.7 (1.0)   & 0.6 (0.9)   & 0.3 (0.6) \\
\hline\hline
\end{tabular}
\caption{In $pp \to WZ + 2j \to \ell \nu \ell \ell + 2j$ processes, $5 \sigma$-significance discovery values and 95\% CL limits are shown 
 for coefficients of higher-dimension operators with 300 fb$^{-1}$/3000 fb$^{-1}$ of integrated luminosity at
 14 TeV and 33 TeV. The results obtained after applying the UV bounds are shown in parentheses.}
\label{tab:wzSignificance_noUVCutOff}
\end{table}

Another sensitive channel is the VBS production of same-sign $W$ bosons, which unlike VBS $W^+ W^-$ is not dominated by $t \bar{t}$ production. Sensitivity studies to the
 ${\cal O}_{T,1}$ operator (Eqn.~\ref{eq:t1}) produce the results shown in Table~\ref{tab:ssWWSignificance_scenarios}. As part of this exercise, 
 different pileup configurations were investigated and found not to have a significant effect, as shown in Table~\ref{tab:ssWWSignificance_scenarios}. The implication is
 that none of the fully-leptonic channels considered in these studies are very sensitive to pileup. In this channel, the studies are extended to a 100 TeV $pp$ collider. 
\begin{table}[h]
\centering
\begin{tabular}{c|c|c|c|c|c}
\hline
\hline
{Parameter} & $\sqrt{s}$ & Luminosity & pileup & $5 \sigma$  & {95\% CL }  \\
                      & [TeV] & [fb$^{-1}$]&  & [TeV$^{-4}$] & [TeV$^{-4}$]  \\
\hline
        $f_{T,1}/\Lambda^{4}$ & 14  & 300  &  50 &  0.2 (0.4)  & 0.1 (0.2)    \\
        $f_{T,1}/\Lambda^{4}$ & 14  & 3000  &  140 &  0.1 (0.2)  & 0.06 (0.1)    \\
        $f_{T,1}/\Lambda^{4}$ & 14  & 3000  &  0 &  0.1 (0.2)  & 0.06 (0.1)    \\
        $f_{T,1}/\Lambda^{4}$ & 100  & 1000  &  40 &  0.001 (0.001) & 0.0004 (0.0004) \\
        $f_{T,1}/\Lambda^{4}$ & 100  & 3000 &  263 & 0.001 (0.001) & 0.0008 (0.0008) \\
        $f_{T,1}/\Lambda^{4}$ & 100  & 3000 &  0 &   0.001 (0.001) & 0.0008 (0.0008) \\
\hline
\hline
\end{tabular}
\caption{In $pp \to W^{\pm} W^{\pm} + 2j \to \ell \nu \ell \nu + 2j$ processes, $5 \sigma$-significance discovery values and 95\% CL limits are shown for coefficients the higher-dimension operator, $f_{T,1}/\Lambda^{4}$, for different
 machine scenarios without the UV cut and with the UV cut in parenthesis.
 Pileup refers to the number of $pp$ interactions per crossing. }
\label{tab:ssWWSignificance_scenarios}
\end{table}

Good sensitivity to higher-dimension operators is also obtainable from triboson production. Using the $WWW \to 3 \ell + 3 \nu$ final state, the following operators were investigated: the dimension-8 operator ${\cal O}_{T,0}$ of Eqn.~\ref{eq:t0}
and the dimension-6 operator ${\cal O}_{WWW}$ of Eqn.~\ref{opTGC}.
The sensitivity to these operators is presented in Table~\ref{tab:WWW5sigma}. 
\begin{table}[h]
\centering
\begin{tabular}{c|c|c|c|c|c}
\hline\hline
Parameter & dim. & Luminosity [fb$^{-1}$] & 14 TeV & 33 TeV & 100 TeV \\
\hline 
\multirow{3}{*}{$c_{WWW}/\Lambda^2$  [TeV$^{-2}$] }& \multirow{3}{*}{6} & 300 & 4.8 (8)  & - & - \\
\cline{3-6}
 &   & 1000 & - & - & 1.3 (1.5)  \\
 \cline{3-6}
  & &  3000 & 2.3 (2.5)  & 1.7 (2.0) & 0.9 (1.0) \\
  \hline
\multirow{3}{*}{  $f_{T,0} / \Lambda^4$ [TeV$^{-4}$] }&\multirow{3}{*}{ 8} & 300 & 1.2  & - & - \\
\cline{3-6}
   &  & 1000 & - & - & 0.004 \\
   \cline{3-6}
   & & 3000 & 0.6  & 0.05 & 0.002 \\   
  \hline\hline
  \end{tabular}
  \caption{In the $pp \to WWW \to 3\ell + 3\nu$ process, the $5\sigma$-significance discovery values are shown for the coefficients of higher order operators. The values in parentheses are obtained with the UV bound applied. 
 $pp$ colliders at $\sqrt{s} = 14, \; 33$ and 100~TeV are studied.  }
  \label{tab:WWW5sigma}
  \end{table}  

Finally, the final state $Z \gamma \gamma \to \ell \ell \gamma \gamma$ is used to investigate the sensitivity for the dimension-8 operators ${\cal O}_{M,i}, i=0,1,2,3$ of Eqn.~\ref{opM0to7},
and the results are presented in Table~\ref{tab:zaaSignificance}. 
\begin{table}[h]
\centering
\begin{tabular}{c|c|c|c|c|c}
\hline\hline
\multirow{2}{*}{Parameter}     & $\sqrt{s}$           & 14 TeV & 14 TeV & 33 TeV & 100 TeV \\
\cline{2-6}
                               & Lum.                 & 300 fb$^{-1}$ & \multicolumn{3}{c}{3000 fb$^{-1}$}  \\
\hline
\multirow{2}{*}{$f_{M,0}/\Lambda^4$ [TeV$^{-4}$]} & 5$\sigma$  & 7300 (830) & 3600 (310) & 1900 (190) & 750 (120) \\
\cline{2-6}
                                                 & 95\% CL  & 4200 (360) & 1200 (160) & 660 (120)  & 71 (59) \\
\hline
\multirow{2}{*}{$f_{M,1}/\Lambda^4$ [TeV$^{-4}$]} & 5$\sigma$  &  7600 (1600) & 3600 (680)  & 2100 (340) & 1000 (220) \\
\cline{2-6}
                                                 & 95\% CL  &  4500 (800)  & 1200 (290) & 770 (160) & 240 (126)  \\
\hline
\multirow{2}{*}{$f_{M,2}/\Lambda^4$ [TeV$^{-4}$]} & 5$\sigma$  & 3300 (130) & 510 (48) & 310 (26) & 120 (16) \\
\cline{2-6}
                                                 & 95\% CL  & 670 (56) & 160 (21) & 110 (13) & 25 (10) \\
\hline
\multirow{2}{*}{$f_{M,3}/\Lambda^4$ [TeV$^{-4}$]} & 5$\sigma$  & 2400 (250) & 720 (120)  & 320 (66) & 180 (34) \\
\cline{2-6}
                                                 & 95\% CL  & 820 (133) & 210 (52) & 130 (23) & 38 (15)\\
\hline\hline
\end{tabular}
\caption{In $pp \to Z\gamma\gamma \to l^{+}l^{-}\gamma\gamma$ processes, $5 \sigma$-significance discovery values and 95\% CL limits are shown 
 for coefficients of dimension-8 operators with integrated luminosity of 300 fb$^{-1}$ at $\sqrt{s} = 14$~TeV and 3000 fb$^{-1}$  at $\sqrt{s} = 14$~TeV, 33 TeV and 100 TeV, respectively.  
 To show the impact without the UV bound, the corresponding results are shown in parentheses.}
\label{tab:zaaSignificance}
\end{table}

The conclusions that can draw from these numerical results, have been stated in~\cite{snowmass-multiboson-whitepaper} and reproduced verbatim in this report for convenience, below;
\begin{itemize}
\item The VBS $ZZ$ final state, when used to probe the ${\cal O}_{\Phi W}$ dimension-6 operator, increases in
 sensitivity to the operator coefficient by a factor of $\approx 1.9$ when the luminosity is increased by a 
 factor of 10 from 300 fb$^{-1}$ to 3000 fb$^{-1}$, and by a factor of $\approx 1.2$ when the collider 
 energy 
 is increased from 14~TeV to 33~TeV. When considering the dimension-8 operators ${\cal O}_{T,8}$ (${\cal O}_{T,9}$),
 the sensitivity increases by a factor of 1.9 (1.5) due to the same luminosity increase and by a factor of
 $\approx 1.8$ (1.5) due to the energy increase. The sensitivity to the dimension-6 operator is not affected by
 imposing a UV bound, while the sensitivity to the dimension-8 operator is reduced by a factor of about 1.8
 when the bound is applied. 

\item The VBS $WZ$ final state, when used to probe the ${\cal O}_{\Phi d}$ dimension-6 operator, increases in
 sensitivity to the operator coefficient by a factor of $\approx 1.9$ when the luminosity is increased 
 from 300 fb$^{-1}$ to 3000 fb$^{-1}$, and by a factor of $\approx 1.2$ when the collider 
 energy 
 is increased from 14~TeV to 33~TeV. When considering the dimension-8 operator ${\cal O}_{T,1}$,
 the sensitivity increases by a factor of $\approx 1.8$ due to the same luminosity increase and by a factor of
 $\approx 2$ due to the energy increase. The sensitivity to the dimension-6 operator is not affected by
 imposing a UV bound, while the sensitivity to the dimension-8 operator is reduced by a factor of about 1.8
 when the bound is applied. 

\item The VBS same-sign $WW$ final state, when used to probe the ${\cal O}_{T,1}$ dimension-8 operator, increases in
 sensitivity to the operator coefficient by a factor of $\approx 2$ when the luminosity is increased 
 from 300 fb$^{-1}$ to 3000 fb$^{-1}$ at $\sqrt{s} = 14$~TeV. An increase in  collider  energy 
 from 14 TeV to 100~TeV increases the sensitivity by a factor of 100. 
 The sensitivity is not affected at $\sqrt{s} = 100$~TeV by
 imposing a UV bound because the bound is very high for the value of the coefficient probed. 
 The sensitivity at $\sqrt{s} = 14$~TeV is reduced by a factor of about 2 
 when the bound is applied. 

\item The triboson $WWW$ final state, when used to probe the ${\cal O}_{WWW}$ dimension-6
  operator, increases in
 sensitivity to the operator coefficient by a factor of $\approx 2$ when the luminosity
 is increased 
 from 300 fb$^{-1}$ to 3000 fb$^{-1}$ at $\sqrt{s} = 14$~TeV. An increase in  collider 
  energy 
 from 14 TeV to 33~TeV (100~TeV) increases the sensitivity by a factor of 1.3 (2.5). 
 These results are affected at the 10\% level by the application of the UV bound. 
 When probing the dimension-8 operator ${\cal O}_{T,0}$, the sensitivity to the operator
 coefficient increases by a factor of $\approx 2$ when the luminosity is increased
 from 300 fb$^{-1}$ to 3000 fb$^{-1}$ at $\sqrt{s} = 14$~TeV. An increase 
 in  collider  energy 
 from 14 TeV to 33~TeV (100~TeV) increases the sensitivity by a factor of 12 (300).
 This dramatic increase is tamed by the UV bound; we take this as an indication that
 $WWW$ triboson production is a sensitive channel for direct production of new particles
 as the collider energy is raised. 

\item The triboson $Z \gamma \gamma$ final state, when used to probe the ${\cal O}_{M,i}$ dimension-8
  operators, increases in
 sensitivity to the operator coefficient by a factor of $2-6$ (depending the operator considered) when the luminosity
 is increased 
 from 300 fb$^{-1}$ to 3000 fb$^{-1}$ at $\sqrt{s} = 14$~TeV. An increase in  collider 
  energy 
 from 14 TeV to 33~TeV (100~TeV) increases the sensitivity by a factor of $\approx 2$ (4 to 5). 
 These results are strongly affected by the application of the UV bound. 
 \end{itemize}

It is important to note that the sensitivities for the 33~TeV and 100~TeV colliders
 are based on 
 analyses that have not been re-optimized for higher energy colliders; the analyses were
  optimized for
  14 TeV only. Optimization of the analyses for higher collider energies is important and should be revisited in the future, as it will 
  lead to further improvements of the
  sensitivity to new physics at  those  machines.  The leptonic channels studied in~\cite{snowmass-multiboson-whitepaper}
 have been shown to be relatively insensitive to pileup effects.

\subsection{Multi-boson processes at lepton colliders}
\label{sec:electroweak-multibosonlep}

\providecommand{\Cdgz}{\ensuremath{\Delta g^\mathrm{Z}_1}}
\providecommand{\Cdgg}{\ensuremath{\Delta g^\mathrm{\gamma}_1}}
\providecommand{\Cdkz}{\ensuremath{\Delta \kappa_\mathrm{Z}}}
\providecommand{\Cdkg}{\ensuremath{\Delta \kappa_{\gamma}}}
\providecommand{\Ckg}{\ensuremath{\kappa_{\gamma}}}
\providecommand{\Ckz}{\ensuremath{\kappa_{\mathrm{Z}}}}
\providecommand{\Clg}{\ensuremath{\lambda_{\gamma}}}
\providecommand{\Clz}{\ensuremath{\lambda_{\mathrm{Z}}}}
\providecommand{\Cgv}[1]{\ensuremath{g^V_{#1}}}
\providecommand{\Cgz}[1]{\ensuremath{g^Z_{#1}}}
\providecommand{\Cgg}[1]{\ensuremath{g^{\gamma}_{#1}}}
\providecommand{\Ckzt}{\ensuremath{\tilde{\kappa}_\mathrm{Z}}}
\providecommand{\Clzt}{\ensuremath{\tilde{\lambda}_\mathrm{Z}}}
\providecommand{\Ckgt}{\ensuremath{\tilde{\kappa}_{\gamma}}}
\providecommand{\Clgt}{\ensuremath{\tilde{\lambda}_{\gamma}}}

\begin{figure}[htb!]
  \centering
  \includegraphics[width=0.4\linewidth,bb=33 17 492 468]{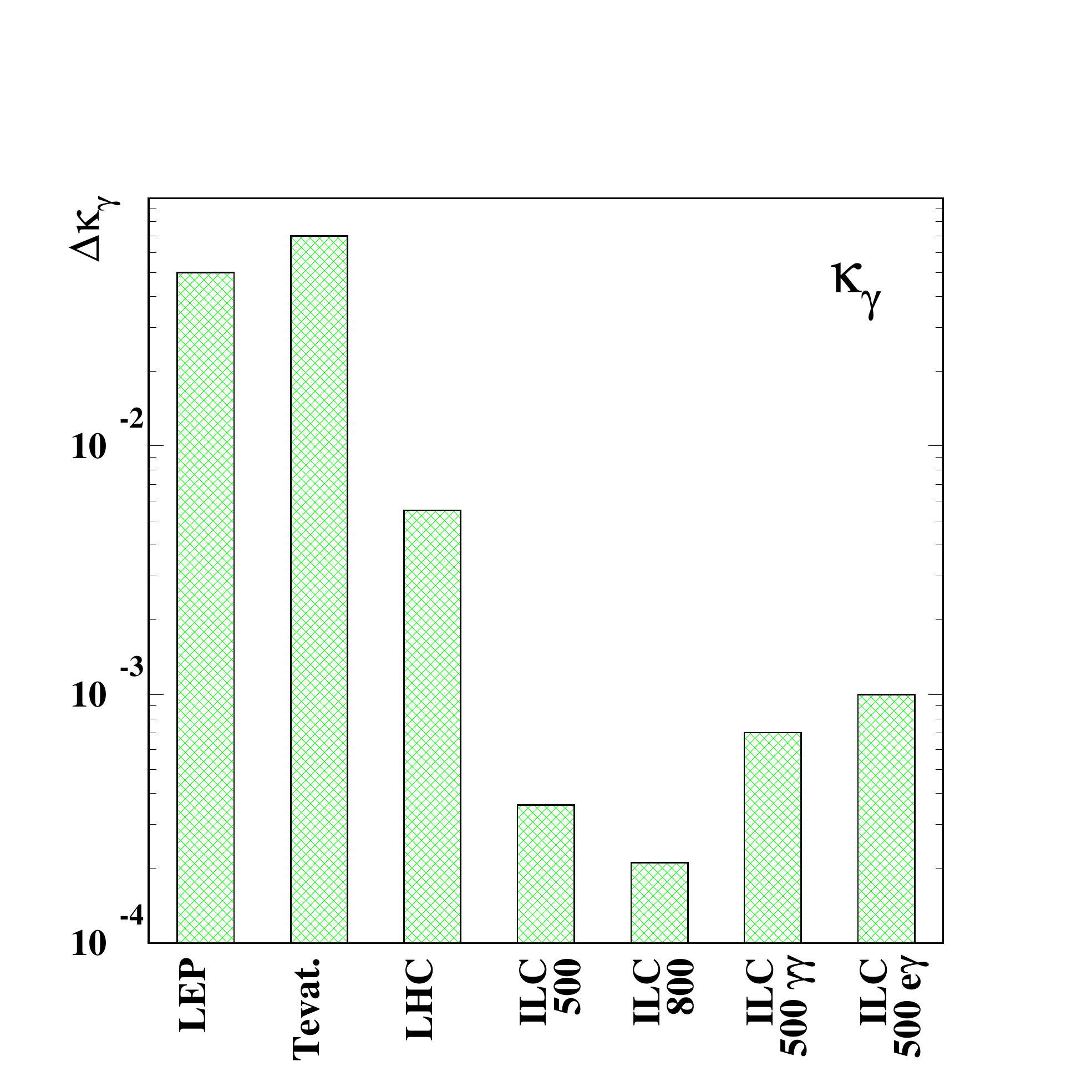}
  \hspace*{5mm}
  \includegraphics[width=0.4\linewidth,bb=33 17 492 468]{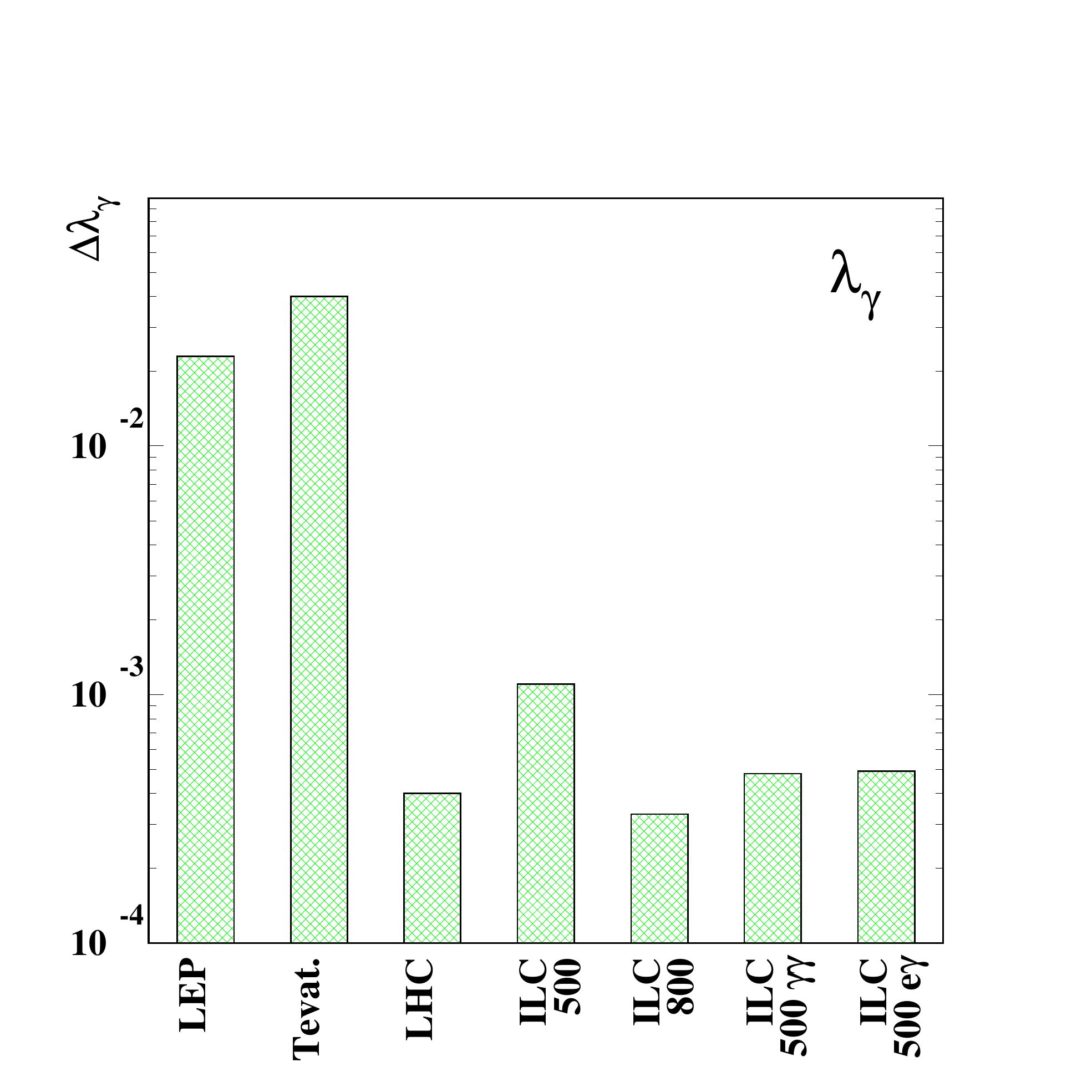}
  \caption{Comparison of $\Delta \kappa_\gamma$ and $\Delta
    \lambda_\gamma$ at different machines. For LHC and ILC three years
    of running are assumed (LHC: $300~\fbinv$, ILC
    $\sqrt{s}=500~\gev$: $500~ \fbinv$, ILC $\sqrt{s}=800~\gev$: $1000
    ~ \fbinv$). If available the results from multi-parameter fits
    have been used. Taken from Ref.~\cite{ilc:rdr,Freitas:2013xga}.}
 \label{fig:tgcilclhc}
\end{figure}

Di-boson processes, in particular $e^+ e^-\to W^+ W^-$, were used very
successfully at LEP to probe TGCs, and multi-boson processes have also
been studied at ILC and CLIC to estimate the sensitivity to TGCs and
QGCs. Table~\ref{tab:tgcilc} provides projected sensitivities to the
TGCs of Eq.~\ref{eq:L} at a 500 GeV and 800 GeV ILC with polarized
beams.  As shown in Fig.~\ref{fig:tgcilclhc} (see
Ref.~\cite{ilc:rdr}), most TGCs will be better constrained at ILC than
at the LHC, though the LHC will improve significantly upon LEP and the
Tevatron. For one specific anomalous coupling parameter, $\Delta
\lambda_\gamma$, the LHC and especially the HL-LHC (with 3000
fb$^{-1}$) is competitive with ILC800.

\begin{table}
\centering
\renewcommand{\arraystretch}{1.2}
\begin{tabular}[c]{|c|c|c|}
\hline
coupling & \multicolumn{2}{|c|}{error $\times 10^{-4}$} \\
\cline{2-3}
         & $\sqrt{s}=500~\gev$ & $\sqrt{s}=800~\gev$ \\
\hline
\hline
  \Cdgz  &$ 15.5 \phantom{0} $&$ 12.6 \phantom{0} $\\
  \Cdkg  &$  3.3 $&$  1.9 $\\
  \Clg   &$  5.9 $&$  3.3 $\\
  \Cdkz  &$  3.2 $&$  1.9 $\\
  \Clz   &$  6.7 $&$  3.0 $\\
\hline
\hline         
  \Cgz{5}&$ 16.5 \phantom{0} $&$ 14.4 \phantom{0} $\\
  \Cgz{4}&$ 45.9 \phantom{0} $&$ 18.3 \phantom{0} $\\
  \Ckzt  &$ 39.0 \phantom{0} $&$ 14.3 \phantom{0} $\\
  \Clzt  &$  7.5 $&$  3.0 $\\
  \hline
\end{tabular}
\caption{
  Results of the single parameter fits ($1 \sigma$) to the different 
  triple gauge couplings at the ILC for $\sqrt{s}=500 ~\gev$ with ${\cal L}=
  500~\fbinv$ and $\sqrt{s}=800 ~\gev$ with ${\cal L}=1000~\fbinv$;
  ${\cal P}_{e^-} = 80\%$ and ${\cal P}_{e^+} = 60\%$ has been used. Taken from Ref.~\cite{Freitas:2013xga}.}
\label{tab:tgcilc} 
\end{table}

Estimates of sensitivities to CP-even aTGCs in $W$-boson pair production at
CLIC have been provided in Ref.~\cite{Accomando:2004sz} and reproduced
for the Snowmass study in Ref.~\cite{Abramowicz:2013tzc}, as shown in
Table~\ref{tab:precSens}.  The TGCs $g_i^{L,R},\lambda_{L,R}$ and
$\Delta \kappa_{L,R}$ are defined in terms of the TGCs of
Eq.~\ref{eq:L}. The superscripts $\mathrm{L}$ ($\mathrm{R}$) refer to
the values obtained for amplitudes with left (right) handed electrons
and right (left) handed positrons.  The definitions for $g_1$, for
example, are
 \begin{eqnarray}
 g_1^\mathrm{L}&=& 4\sin^2\theta_\mathrm{W} g^\gamma_1+(2-4\sin^2\theta_\mathrm{W})g_1^Z\xi \\
 g_2^\mathrm{R}&=& 4\sin^2\theta_\mathrm{W} g_1^\gamma-4\sin^2\theta_\mathrm{W} g_1^Z\xi
\end{eqnarray}
where $\xi=s/(s-m_Z^2)$. For more details, see~\cite{Accomando:2004sz}.

\begin{table}[t]
  \centering
  \begin{tabular}{ccccccccc}
    \hline
    $\sqrt{s}~[\GeV]$ & $\mathrm{Re}(\Delta g_1^L)$ & $\mathrm{Re}(\Delta \kappa_L) $ & $\mathrm{Re}(\lambda_L)$ & $\mathrm{Re}(g_5^L)$ & $\mathrm{Re}(g_1^R)$ & $\mathrm{Re}(\Delta \kappa_R)$ & $\mathrm{Re}(\lambda_R)$ & $\mathrm{Re}(g_5^R)$ \\\hline
    500             & 2.6                         & 0.85                            & 0.59                     & 2.0                  & 10                   & 2.4                            & 3.6                      & 6.7                  \\
    800             & 1.6                         & 0.35                            & 0.24                     & 1.4                  & 6.2                  & 0.92                           & 1.8                      & 4.8                  \\
    3000            & 0.93                        & 0.051                           & 0.036                    & 0.88                 & 3.1                  & 0.12                           & 0.36                     & 3.2                  \\
    \hline
  \end{tabular}
  \caption{Sensitivity of CLIC to the real parts of CP-even TGCs in units of
    $10^{-3}$, defined in~\cite{Accomando:2004sz}. The integrated
    luminosities for the $500~\GeV$, $800~\GeV$ and $3000~\GeV$ stages are assumed
    here to be $500~\fbinv$, $1~\abinv$ and $3~\abinv$ respectively. Taken from Ref.~\cite{Abramowicz:2013tzc}}\label{tab:precSens}
\end{table}

Prospects for precision studies of anomalous quartic couplings at the
ILC have been studied in $e^+ e^- \to WWZ$ and $e^+ e^- \to ZZZ$
and VBS processes with and without beam polarization in
Ref.~\cite{Beyer:2006hx} (see also Ref.~\cite{Baer:2013cma} for a
review), and the results for $WWZ,ZZZ$ production are presented in Table~\ref{tab:a4a5ilc}. The
aQGCs have been parametrized in the effective chiral Lagrangian
approach as described in Section~\ref{sec:electroweak-resonances}.

In Section~\ref{sec:whizard-dim8} we provide a conversion of the
aQGCs coupling parameters $\alpha_4,\alpha_5$ of this approach to the
EFT coefficients $f_{S,0},f_{S,1}$ for $WWWW, WWZZ$ and $ZZZZ$
vertices. Using this translation, we can convert the
$\alpha_4,\alpha_5$ sensitivity from ILC in Table~\ref{tab:a4a5ilc} to
the $f_{S,0}, f_{S,1}$ basis used by the ATLAS studies in
Section~\ref{sec:electroweak-multibosonhad14} and the LHC and
higher-energy $pp$ collider studies in
Setion.~\ref{sec:electroweak-multibosonhadHE}. Using
Eq.~\ref{alpha4Tofs0}, we find that the ILC sensitivity translates to
90\% CL limits on $f_{S,0}/\Lambda^4 \approx 20$~TeV$^{-4}$, which are
about a factor of 20 weaker than the 95\% CL limit estimated by the
ATLAS study using 3000 fb$^{-1}$, as shown in
Table~\ref{tab:5sigsummary}. The sensitivity to the other dimension-8
operators, which induce purely quartic anomalous couplings, is even
higher at the LHC, as shown in
Sec.~\ref{sec:electroweak-multibosonhad14} and
Sec.~\ref{sec:electroweak-multibosonhadHE}. This is likely because the
${\cal O}$$_{S,0 (1)}$ operators do not contain derivatives of the
gauge boson fields, while such derivatives in other dimension-8
operators enhance the anomalous production at the higher LHC energy.

We arrive at the preliminary conclusion that aTGC's, which are induced
by dimension-6 operators, are significantly better probed by the
high-energy ILC options compared to the LHC.  On the other hand,
aQGC's, which are induced by dimension-8 operators, are significantly
better probed (by 1-2 orders of magnitude) by the LHC, due to the
stronger growth of the anomalous cross section with energy. This conclusion
 is supported by the comparison of 95\% CL limits in Table~\ref{tab:mreslim} for the
 LHC and Table~\ref{tab:mreslima4ilc} for the ILC. The LHC is sensitive to resonance
 masses that are higher by more than a factor of two, as compared to ILC1000. Note further
 that this comparison is based on the ${\cal O}$$_{S,0}$ operator of Eqn.~\ref{eq:s0}, for which the LHC sensitivity is not as strong
 as for the other operators (such as ${\cal O}$$_{T,i}, i=1,8,9$ of Eqn.~\ref{eq:t1},~\ref{eq:t8}, and~\ref{eq:t9})
  due to the difference in the number of derivatives of gauge fields. 

\begin{table}[t]
\[
\begin{array}{c|c||r|r|r||r||r}
\hline
\multicolumn{2}{c||}{} &
\multicolumn{3}{c||}{\mbox{WWZ}}&\mbox{ZZZ}&\mbox{best}\\
\multicolumn{2}{c||}{} &
\mbox{no pol.}& e^-\mbox{ pol.}& \mbox{both pol.}&\mbox{no pol.}\\
\hline
16\pi^2\Delta\alpha_4 &\sigma^+&9.79  &4.21 &1.90 &3.94&1.78 \\
                   &\sigma^-&-4.40  &-3.34 &-1.71 &-3.53&-1.48 \\
\hline
16\pi^2\Delta\alpha_5 &\sigma^+&3.05 &2.69  &1.17 &3.94&1.14 \\
                   &\sigma^-&-7.10  &-6.40  &-2.19 &-3.53 &-1.64 \\
\hline
\end{array}
\]
\caption{Sensitivity of the 1 TeV ILC (with 1~$\abinv$) to the aQGC coupling parameters $\alpha_4$ and $\alpha_5$ in $WWZ$ and $ZZZ$ production, expressed as
    $1\sigma$ errors. WWZ: two-parameter fit; ZZZ: one-parameter fit; best:
    best combination of both. Taken from Ref.~\cite{Beyer:2006hx}.}\label{tab:a4a5ilc}
\end{table}
%

\subsection{ Simplified Models for New Physics in Vector Boson Scattering }
\label{sec:electroweak-resonances}


Here we briefly describe the translation of very heavy resonances that
arise in simplified models and leave only traces in the form of
deviations in the SM couplings into higher-dimensional operators
leading to such deviations. A detailed discussion can be found in
Ref.~\cite{Reuter:2013gla}.

The simplified models discussed here contain the SM supplemented by all
possible resonances that could couple to the sector of EW gauge bosons
according to their spin and isospin quantum numbers. Such simplified models
cover cases ranging from Two- or Multi-Higgs doublet models, extended scalar
sectors, Technicolor models, models of complete or partial compositeness,
Little Higgs models, Twin Higgs models and many more.  Cases where there is
only a single resonance present could be described along these lines as well
as cases where there are more resonances (but maybe only one of them
accessible to LHC). The resonances are just parametrized by their mass,
possibly their width, as well as their couplings to the electroweak sector. As
simplified models are like any effective field theory not UV-complete,
perturbative unitarity of tree-level amplitudes in that setup are not
guaranteed (see also Section~\ref{sec:electroweak-eft}). To give a
prescription that can be used by the experiments in a model-independent setup
and does not yield overly optimistic results due to unphysical amplitude
contributions within exclusion limits, a unitarization formalism has been
introduced in Ref.~\cite{Reuter:2013gla} that projects back on amplitudes that
are genuinely unitary. This is insured by giving additive corrections to the
SM vector boson scattering augmented by the BSM resonances. A simple
implementation has been performed in the event generator {\sc
  Whizard}~\cite{Kilian:2007gr}.

Here we define a simplified model that is able to describe the essence
of a new physics sector that couples to the EW sector in an approach
as model-independent as possible. We refrain from discussing fermionic
resonances here as those would contribute only at the 1-loop order to
vector boson scattering, and concentrate on new bosonic resonances. To
do so, one needs to supplement the Lagrangian of the EW SM (accounting
for the discovery of the 125~GeV state as the SM Higgs boson but maybe
allowing its couplings to deviate within the limits of the EW
perturbation theory (EWPT) from their SM values). As the main
signatures to study the EW sector of the SM are diboson, triboson and
generically multi-boson production as well as vector-boson scattering
(VBS), and here particularly scattering of longitudinal gauge bosons,
it is convenient to use an operator basis containing explicitly the
longitudinal degrees of freedom (DOFs) of the EW gauge bosons. This
effective Lagrangian is basically identical to the chiral EW
Lagrangian~\cite{EWLag}, except that we linearize the Lagrangian by
adding the Higgs particle, and all higher-dimensional operators stem
from BSM contributions. So we implement $SU(2)_L \times U(1)_Y$ gauge
invariance, where the building blocks are the SM fermions, $\psi$, the
EW (transversal) gauge boson fields $W_\mu^a \; (a=1,2,3)$ and $B_\mu$
as well as the longitudinal DOFs, $\Sigma = \exp \left[ \frac{- i}{v}
  w^a \tau^a \right]$.

The effective Lagrangian is obtained by adding to the minimal (SM)
Lagrangian of Ref.~\cite{Reuter:2013gla} deviations from that
Lagrangian in the form of higher-dimensional operators allowed by
gauge symmetry as well as CP as follows:
\begin{equation}
  \LL_{\rm eff} = \LL_{\text{min}}  + \beta_1 \LL'_0 + \sum_i \alpha_i \LL_i + 
  \frac{1}{\Lambda} \sum_i \alpha_i^{(5)} \LL_i^{(5)} +  
  \frac{1}{\Lambda^2} \sum_i \alpha_i^{(6)} \LL_i^{(6)} + \ldots 
\end{equation}
where $\Lambda$ is (up to $\mathcal{O}(1)$ constants) the scale where
BSM physics potentially enters. The operators are 
\begin{xalignat}{2}
  \label{operators}
  \LL'_0 &=\; \frac{v^2}{4} \trs{\vT \vV_\mu} \trs{\vT \vV^\mu} & &
  \\
  \LL_1 &=\; \trs{\vB_{\mu\nu} \vW^{\mu\nu}} 
  &
  \LL_6 &=\; \trs{\vV_\mu \vV_\nu} \trs{\vT \vV^\mu} \trs{\vT
    \vV^\nu}
  \\
  \LL_2 &=\; \ii \trs{\vB_{\mu\nu} \lbrack \vV^\mu , \vV^\nu
    \rbrack}
  & 
  \LL_7 &=\; \trs{\vV_\mu \vV^\mu} \trs{\vT \vV_\nu} \trs{\vT
    \vV^\nu}
  \\
  \LL_3 &=\; \ii \trs{\vW_{\mu\nu} \lbrack \vV^\mu , \vV^\nu
    \rbrack} 
  &
  \LL_8 &=\; \tfrac14 \trs{\vT \vW_{\mu\nu}} \trs{\vT \vW^{\mu\nu}} 
  \\
  \LL_4 &=\; \trs{\vV_\mu \vV_\nu} \trs{\vV^\mu \vV^\nu}
  &
  \LL_9 &=\; \tfrac{\ii}{2} \trs{\vT \vW_{\mu\nu}} \trs{\vT \lbrack
    \vV^\mu , \vV^\nu \rbrack}
  \\
  \LL_5 &=\; \trs{\vV_\mu \vV^\mu} \trs{\vV_\nu \vV^\nu}
  &
  \LL_{10} &=\; \tfrac12 \left( \trs{\vT \vV_\mu} \trs{\vT \vV^\mu}
  \right)^2 
\end{xalignat}
Here, the field strength tensors are defined in terms of $W_\mu^a$ $(a=1,2,3)$ and $B_\mu$ as
\begin{align}
 \vW_{\mu\nu}&=\partial_\mu \vW_\nu - \partial_\nu \vW_\mu + i g \left [\vW_\mu, \vW_\nu \right ]\\
  \vB_{\mu\nu}&=\partial_\mu \vB_\nu - \partial_\nu \vB_\mu 
\end{align}
with
\begin{equation}
 \vW_{\mu}=W_\mu^a \frac{\tau^a}{2} \qquad \text{and}\qquad
  \vB_{\mu}=B_\mu \frac{\tau^3}{2}.
\end{equation}
Using the covariant derivative
\begin{equation}
 \vD_\mu \Sigma = \partial_\mu + i g \vW_\mu \Sigma -i g^\prime \Sigma \vB_\mu
\end{equation}
one defines $\vV$ as a field representing longitudinal vectors, $\vV =
\Sigma (\vD \Sigma)^\dagger$ that will be used shortly to write down
operators giving rise to modified couplings.  In order to write down
operators projecting out the neutral component, one uses $\vT = \Sigma
\tau^3 \Sigma^\dagger$.  For more technical details about this
formalism interpreted in that context of simplified models for
extended EW symmetry breaking,
cf.~\cite{Alboteanu:2008my,KRS}. Indirect information on new physics
is encoded in the operator coefficients $\beta_1$, $\alpha_i$.  From
EWPT (SLC/LEP/Tevatron measurements), one knows that $\alpha_i \ll 1$,
while on the other hand from naive dimensional analysis one would
assume $\alpha_i \sim 1/16\pi^2 \approx 0.006$ as they have to
renormalize divergences in an effective field theoretic simplified
model of a UV-complete BSM model. Using such a bottom-up approach, it
is notoriously difficult, as the usual setup as a ratio of the EW and
the BSM scale $\alpha_i = v^2/\Lambda^2$ is only valid up to unknown
operator normalization coefficients (that are in general coupling
constants of the UV-complete model), furthermore the power counting
can be highly nontrivial, producing unexpected scaling behavior of
operators.

\begin{figure}
  \centering
  \includegraphics[width=.54\textwidth]{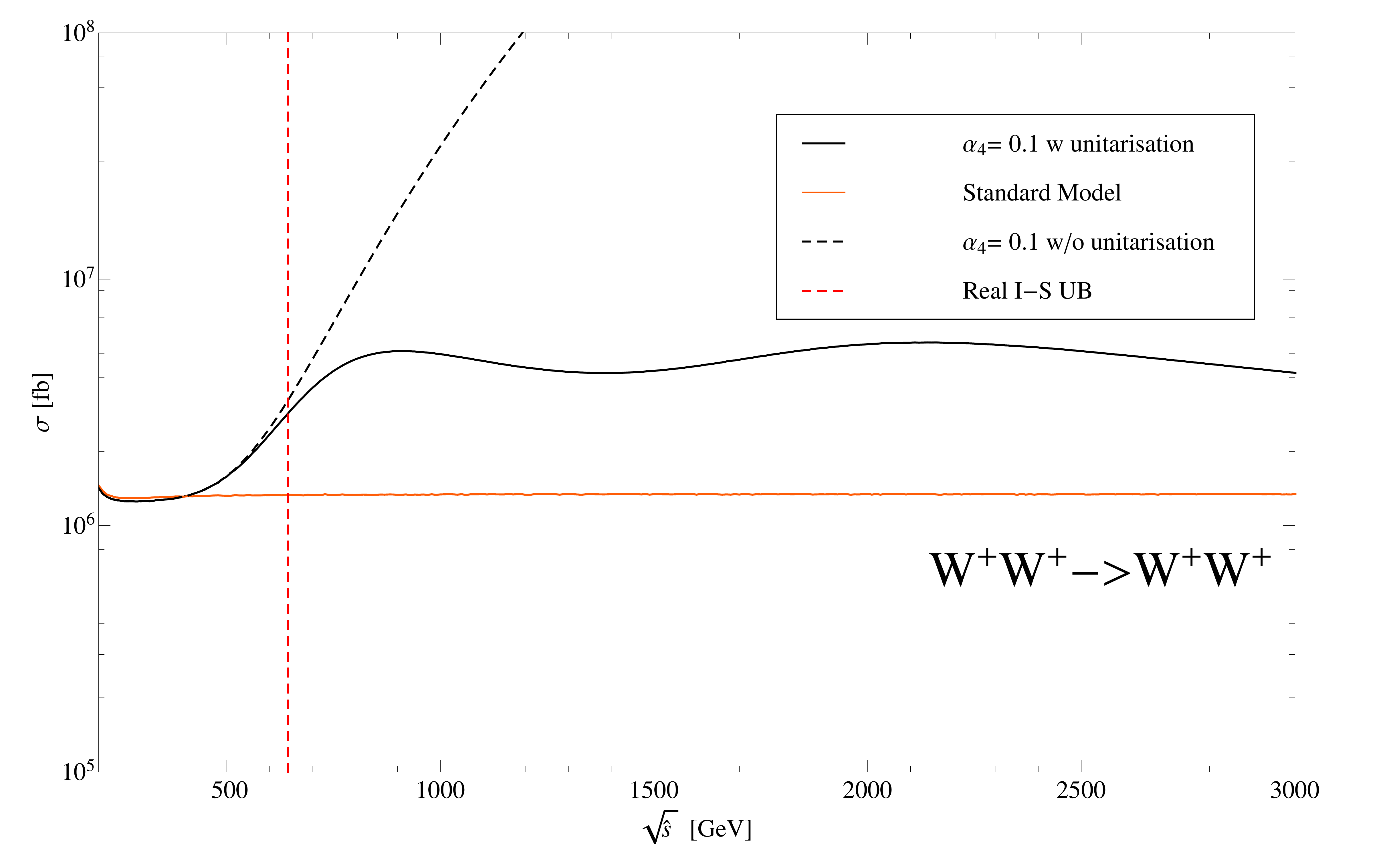}    
  \caption{Cross section of VBS with anomalous quartic gauge coupling
    $\alpha_4$ in $W^+W^+ \rightarrow W^+W^+$.  The orange line
    represents the Standard Model. Above the unitarity bound (red
    dashed line) one can clearly see that the cross section without
    unitarization scheme (black dashed line) violates unitarity. The
    unitarized cross section (black solid line) is going to saturate
    above the unitarity bound.}
  \label{fig:unitarity}
\end{figure}

Higher-dimensional operators do lead to deviations of the triple and
quartic gauge couplings from their SM values.  For completeness, we
repeat the formulae for triple and quartic gauge couplings, and how
they depend on the SM parameters as well as on the operator
coefficients of the effective Lagrangian above:
  \begin{align}
    \LL_{TGC} &=  \ii e\left[
      g_1^\gamma A_\mu \left(W^-_\nu W^{+\mu\nu} - W^+_\nu
        W^{-\mu\nu}\right) 
      + \kappa^\gamma W^-_\mu W^+_\nu A^{\mu\nu}
      + \frac{\lambda^\gamma}{M_W^2}W^-_\mu{}^\nu
      W^+_{\nu\rho} A^{\rho\mu} 
    \right]
    \nonumber\\ &\quad + \ii e\frac{\cw}{\sw}\left[
      g_1^Z Z_\mu \left(W^-_\nu W^{+\mu\nu} - W^+_\nu
        W^{-\mu\nu}\right) 
      + \kappa^Z W^-_\mu W^+_\nu Z^{\mu\nu}
      + \frac{\lambda^Z}{M_W^2}W^-_\mu{}^\nu W^+_{\nu\rho}
      Z^{\rho\mu} 
    \right] \\ 
  \LL_{QGC} &=
  e^2\left[ g_1^{\gamma\gamma} A^\mu A^\nu W^-_\mu W^+_\nu
    -g_2^{\gamma\gamma} A^\mu A_\mu W^{-\nu} W^+_\nu\right]
  \nonumber\\ &\quad
  + e^2\frac{\cw}{\sw}\left[ g_1^{\gamma Z} A^\mu Z^\nu
    \left(W^-_\mu W^+_\nu + W^+_\mu W^-_\nu\right)
    -2g_2^{\gamma Z} A^\mu Z_\mu W^{-\nu} W^+_\nu \right]
  \nonumber\\ &\quad
  + e^2\frac{\cw^2}{\sw^2}\left[ g_1^{ZZ} Z^\mu Z^\nu W^-_\mu W^+_\nu
    -g_2^{ZZ} Z^\mu Z_\mu W^{-\nu} W^+_\nu\right]
  \nonumber\\ &\quad
  + \frac{e^2}{2\sw^2}\left[ g_1^{WW} W^{-\mu} W^{+\nu} W^-_\mu W^+_\nu
    -g_2^{WW}\left(W^{-\mu} W^+_\mu\right)^2\right]
  + \frac{e^2}{4\sw^2\cw^4} h^{ZZ} (Z^\mu Z_\mu)^2
\end{align}
In these equations, the SM values are
$g_1^{\gamma,Z}=\kappa^{\gamma,Z}=1$, $\lambda^{\gamma,Z}=0$, and
$g_{1/2}^{VV'}=1$, $h^{ZZ} = 0$. The quantity $\delta_Z =
\frac{\beta_1+g^\pp\alpha_1}{\cw^2-\sw^2}$ takes into account the
definition of the EW scheme as well as the oblique corrections through
the $\rho$ parameter. In the presence of the operators
Eq.~\ref{operators}, one gets the following shifts: 
\begin{align}
  \Delta g_1^\gamma &= 0
  &
  \Delta\kappa^\gamma &= g^2(\alpha_2-\alpha_1) + g^2\alpha_3 
  + g^2(\alpha_9-\alpha_8)
  \\
  \Delta g_1^Z &= \delta_Z + \tfrac{g^2}{\cw^2}\alpha_3
  &
  \Delta\kappa^Z &= \delta_Z - g^\pp(\alpha_2-\alpha_1) + g^2\alpha_3 
  + g^2(\alpha_9-\alpha_8)
\end{align}
as well as 
\begin{align}
  \Delta g_1^{\gamma\gamma} &= \Delta g_2^{\gamma\gamma} = 0
  &
  \Delta g_2^{ZZ} &= 2\Delta g_1^{\gamma Z}
  - \tfrac{g^2}{\cw^4}(\alpha_5 + \alpha_7)
  \\
  \Delta g_1^{\gamma Z} &= \Delta g_2^{\gamma Z}
  = \delta_Z + \tfrac{g^2}{\cw^2}\alpha_3
  &
  \Delta g_1^{WW} &= 2\cw^2\Delta g_1^{\gamma Z} +
  2g^2(\alpha_9-\alpha_8) 
  + g^2\alpha_4
  \\
  \Delta g_1^{ZZ} &= 2\Delta g_1^{\gamma Z}
  + \tfrac{g^2}{\cw^4}(\alpha_4 + \alpha_6)
  &
  \Delta g_2^{WW} &= 2\cw^2\Delta g_1^{\gamma Z} +
  2g^2(\alpha_9-\alpha_8) 
  - g^2\left(\alpha_4 + 2\alpha_5\right)
  \vphantom{\tfrac{g^2}{\cw^2}}
  \\
  h^{ZZ} &= g^2\left[\alpha_4 + \alpha_5 
    + 2\left(\alpha_6+\alpha_7 + \alpha_{10}\right)\right]
\end{align}
The energy range of testing these anomalous couplings is bound by
unitarity.  Using $\alpha_4$ as example the invariant mass of VBS is
constrained by $ \sqrt{s} \leq \sqrt[4]{\frac{6}{7 \alpha_4}} v$.
Beyond this energy bound Monte-Carlo generators would generate too
many unphysical events and a useful statement about the tested
anomalous coupling is not possible.  Therefore dependent on the
anomalous coupling only a small part of the energy range of LHC can be
used.  A possible formalism to utilize nevertheless events in the
complete energy range is the K-Matrix scheme, which is described in
\cite{Reuter:2013gla}.  Following this scheme the isospin-spin
eigenamplitude will saturate for energies above the unitarity bound (
cf figure \ref{fig:unitarity}).  Therefore Monte Carlo generators
using K-Matrix generate the maximal number of events allowed by
unitarity.

We now make the connection to how BSM models in their
incarnation as EW resonances, coupling to the SM EW gauge boson sector
(particularly the longitudinal DOFs), generate such anomalous
couplings. To be as general as possible, we include weakly interacting
cases (e.g. Little Higgs models) where the new resonances are narrow
(proper particles), as well as strongly interacting cases
(e.g. compositeness or Technicolor models) where the new resonances
are rather wide and could even approach the case of a continuum
(e.g. unparticles or conformal sectors). As we know from EWPT,
$\beta_1 \ll 1$, so the $SU(2)_c$ custodial symmetry of weak isospin
(that in the SM is only broken by hypercharge $g' \neq 0$ and the
fermion masses) is valid to a very good approximation. From the spin
and isospin quantum numbers, only the resonances in the following
table can couple to system of two EW vector bosons,

\vspace{1mm}

\begin{center}
\begin{tabular}{|r|ccc|}\hline
  & 
  $J = 0$ & 
  $J = 1$ & 
  $J = 2$ 
  \\\hline
  $I = 0$ & 
  $\sigma^0 \;\text{(``Higgs'')}$ &
  $[\omega^0] \; (\gamma'/Z')$ &
  $f^0 \; \text{(KK graviton)}$ 
  \\          
  $I = 1$ & 
  $[\pi^\pm, \pi^0] \; \text{(2HDM)}$ & 
  $\rho^\pm, \rho^0 \; (W'/Z')$ & 
  $[a^\pm, a^0]$ 
  \\ 
  $I = 2$ & $\phi^{\pm\pm}, \phi^\pm, \phi^0 \;
  \text{(Higgs triplet)}$ & 
  $\text{---}$ &
  $t^{\pm\pm}, t^\pm, t^0$ 
  \\\hline
\end{tabular}
\end{center}

\vspace{1mm}

So only the scalars, vector or tensors can couple, and only the weak
isospins $I=0,1,2$ are allowed. The table shows prime examples for the
corresponding combinations where a specific choice for the hypercharge
has been made. The entries in brackets are combinations that are only
possible with $SU(2)_c$-violating couplings, and are not further
discussed here. The scalar isoscalar has the same quantum numbers as
the SM Higgs boson. The scalar isovector appears in Technicolor
models, while the isotensor can be found in the Littlest Higgs model,
e.g. Vector resonances appear in extra-dimensional models,
Technicolor, Little Higgs models and many more. Tensor resonances
without EW quantum numbers can be thought of as a recurrence of the
graviton, while the isovector and -tensor are quite exotic and appear
only, e.~g., in extended compositeness models.

As a next step, we relate these resonances from our simplified models
to anomalous couplings. Consider any kind of heavy resonance with
generic Lagrangian $ \LL_\Phi = z \Bigl[ \Phi \left( M_\Phi^2 + D D
\right)\Phi + 2 \Phi J \Bigr]$. Here, $z$ is a (wavefunction
re)normalization constant of the Lagrangian, and $D$ is the
gauge-covariant derivative. $J$ is the EW current to which that
particular resonance couples. Integrating out the resonance leads to
$\LL_\Phi^{\rm eff} = - \frac{z}{M^2} JJ + \frac{z}{M^4} J (D D) J +
\mathcal{O} (M^{-6})$. We now specialize to a scalar isoscalar
resonance $\sigma$, whose Lagrangian is given by $\LL_\sigma = -
\frac12 \Bigl[ \sigma (M_\sigma^2 + \pd^2) \sigma - g_\sigma v
\trs{\vV_\mu \vV^\mu} - h_\sigma \trs{\vT \vV_\mu} \trs{\vT \vV^\mu}
\Bigr]$. Integrating out the scalar, leads to the effective Lagrangian
\begin{equation*}
  \LL_\sigma^{\rm eff} = \frac{v^2}{8 M_\sigma^2} \biggl[
    g_\sigma \trs{\vV_\mu \vV^\mu} + h_\sigma \trs{\vT
      \vV_\mu} \trs{\vT \vV^\mu} \biggr]^2
\end{equation*}
From this one can read off that integrating out a scalar isoscalar
generates the following anomalous quartic couplings
\def\fact{\left(\frac{v^2}{8M_\sigma^2}\right)}
\begin{equation}
  \alpha_5 = g_\sigma^2 \fact \qquad
  \alpha_7 = 2 g_\sigma h_\sigma \fact \qquad
  \alpha_{10} = 2 h_\sigma^2 \fact
\end{equation}
One sees immediately, that a heavy SM Higgs would have fit into that
scheme, using the special couplings $g_\sigma = 1$ and $h_\sigma = 0$.

When one tries to turn constraints on anomalous couplings into direct
constraints on new physics, one faces the problem that there are too
many free parameters to over-constrain the system. There is however one
limiting case where one can do that which has been applied in the
context of studies of the possible search power of a 1 TeV ILC on
anomalous quartic couplings and their interpretation in terms of
resonances~\cite{Beyer:2006hx}: In the limit of a very broad resonance
(that couples rather strongly to the EW sector), the resonance is
close to a broad continuum: $\Gamma \sim M \gg
\Gamma(\text{non}-WW,ZZ) \sim 0$. Also, in that case the decays of
such a particular resonance into non-$W/Z$s can be ignored. From the
functional relation between the resonance width, its couplings and its
mass (again in the case of a scalar isoscalar)
\begin{equation}
  \Gamma_\sigma = \frac{g_\sigma^2 + \frac12 (g_\sigma^2 + 2
    h_\sigma^2)^2}{16 \pi} \left( \frac{M_\sigma^3}{v^2} \right) + 
  \Gamma (\text{non}-WW,ZZ)
\end{equation} 
one can then translate bounds for anomalous couplings directly into
those of the effective Lagrangian:
\begin{equation}
  \alpha_5 \leq \frac{4 \pi}{3} \left( \frac{v^4}{M_\sigma^4}
  \right) \approx \frac{0.015}{(M_\sigma \;\text{in TeV})^4}
  \quad \Rightarrow \quad 16 \pi^2 \alpha_5 \leq
  \frac{2.42}{(M_\sigma \;\text{in TeV})^4} 
\end{equation} 
Note that because of the different dependence of scalar and tensor
widths compared to vector widths, the limits behave differently
depending on the spin of the resonance:
\begin{center}
  \fbox{
    \begin{tabular}{rll}
      Scalar:& $\Gamma \sim g^2 M^3$, $\alpha
      \sim g^2/M^2$ & $\Rightarrow \quad 
        \alpha_{\text{max}}  
        \sim 1 / M^4$   \\
      Vector:& $\Gamma \sim g^2 M$, $\alpha
      \sim g^2/M^2$ & $\Rightarrow \quad 
        \alpha_{\text{max}}  
        \sim 1 / M^2$   \\
      Tensor:& $\Gamma \sim g^2 M^3$,
      $\alpha  
      \sim g^2/M^2$ & $\Rightarrow 
        \alpha_{\text{max}}  
        \sim 1 / M^4$   
    \end{tabular}}
\end{center}

\begin{table}[h]
\centering
\begin{tabular}{l|ccccc}
    \hline
    \text{Resonance}
    &
    $\sigma$
    &
    $\phi$
    &
    $\rho$
    &
    f
    &
    t
    \\
    \hline
    $\Gamma  [g^2 M^2/(64\pi v^2)] $
    &
    6
    &
    1
    &
    $\frac{4}{3}(\frac{v^2}{M^2})$
    &
    $\frac{1}{5}$
    &
    $\frac{1}{30}$
    \\
    \hline
    \\[-9pt]
    $\Delta\alpha_4  [(16\pi\Gamma/M)(v^4/M^4)]$  &
    0 &
    $\frac14$ &
    $\frac34$ &
    $\frac52$ &
    $-\frac58$
    \\[6pt]
  $\Delta\alpha_5  [(16\pi\Gamma/M)(v^4/M^4)]$  &
  $\frac{1}{12}$ &
  $-\frac{1}{12}$ &
  $-\frac{3}{4}$ &
  $-\frac{5}{8}$ &
  $\frac{35}{8}$
  \\[3pt]
  \hline
  \end{tabular}
  \caption{Width $\Gamma$ of the five different possible non-$SU(2)_c$ violating
    resonances for their decays into longitudinal EW gauge bosons, as
    well as their contributions to the anomalous quartic couplings
    parameters $\alpha_4$ and $\alpha_5$.}\label{tab:restype}
\end{table}
Table~\ref{tab:restype} shows the width of the five different possible
non-$SU(2)_c$ violating resonances for their decays into longitudinal
EW gauge bosons, as well as their contributions to the anomalous
quartic couplings parameters $\alpha_4$ and $\alpha_5$. In
Table~\ref{tab:mreslima4} we provide limits on $M$ based on the ATLAS limits
on $\alpha_4$ presented in Table~\ref{tab:dileptona4} (assuming $\Gamma \sim
M$, $v=0.246$~TeV).
\begin{table}[h]
\centering
\begin{tabular}{c|c|c}
\hline\hline                
Type of resonance  & LHC 300 fb$^{-1}$ & LHC 3000 fb$^{-1}$        \\
\hline
scalar $\phi$ & 0.9 TeV & 1.3 TeV \\
vector $\rho$ & 1.2 TeV & 1.7 TeV  \\
tensor $f$ & 1.6 TeV & 2.3 TeV \\
\hline\hline
\end{tabular}
\caption{
  95\% CL limits for the mass $M$ of a broad 
  resonance in simplified models obtained from limits 
  on $\alpha_4$ of Table~\ref{tab:dileptona4} and using the widths of Table~\ref{tab:restype} with $\Gamma \sim M$.}  
\label{tab:mreslima4}
\end{table}
The ATLAS limits on $f_{S,0}/\Lambda^4$ (see Table~\ref{tab:5sigsummary}) can also be translated into limits on
the mass $M$ of a broad EW resonance ($\Gamma \sim M$) as follows
(using Eq.~\ref{alpha4Tofs0}):
\begin{equation}\label{eq:mreslim}
M=\left(\frac{n c_R 16 \pi}{f_{S,0}/\Lambda^4}\right)^{\frac{1}{4}}
\end{equation}
where $c_R$ are the contributions to $\Delta \alpha_4$ of
Table~\ref{tab:restype} and $n=8,16$ for the $WWWW$ and $ZZWW$ case,
respectively. These equivalent resonance mass sensitivities are shown in Table~\ref{tab:mreslim}. Note that the sensitivities shown in Table~\ref{tab:mreslim} are better than
 those in Table~\ref{tab:mreslima4} because the latter were derived from the $W^+ W^-$ VBS channel which has a significant $t \bar{t}$ background, whereas the channels
 used for Table~\ref{tab:mreslim} do not suffer from this large background. 
\begin{table}[h]
\centering
\begin{tabular}{c|c|c|c|c}
\hline\hline                
\multirow{2}{*}{Type of resonance}  & \multicolumn{2}{|c|}{LHC 300 fb$^{-1}$} & \multicolumn{2}{|c}{LHC 3000 fb$^{-1}$}  \\
\cline{2-5} & $5 \sigma$ & 95\% CL              & $5 \sigma$ & 95\% CL       \\
\hline
scalar $\phi$ & 1.8 TeV & 2.0 TeV & 2.2 TeV & 3.3 TeV\\
vector $\rho$ & 2.3 TeV & 2.6 TeV & 2.9 TeV & 4.4 TeV \\
tensor $f$ & 3.2 TeV & 3.5 TeV & 3.9 TeV & 6.0 TeV \\
\hline\hline
\end{tabular}
\caption{$5 \sigma$-significance discovery values and 95\% CL limits for the mass $M$ of a broad resonance in simplified models. These values are obtained from the  $f_{S,0} / \Lambda^4$ values 
 of Table~\ref{tab:5sigsummary} and using Eq.~\ref{eq:mreslim} for the $WWWW$ case ($n=8$). These studies are more sensitive than those in Table~\ref{tab:mreslima4} because of the absence of the large $t \bar{t}$ 
 background in the $W^\pm W^\pm$ VBS channel used for these studies, while there is a large $t \bar{t}$ background in the  $W^+ W^-$ VBS channel used for Table~\ref{tab:mreslima4}. }  
\label{tab:mreslim}
\end{table}

In Ref.~\cite{Beyer:2006hx}, 1~$\sigma$ sensitivities on the anomalous
couplings $\alpha_i$ from VBS studies at a 1 TeV ILC (with 1~$\abinv$)
have been translated into 1~$\sigma$ limits on masses of
pure EW resonances ($M$). Table~\ref{tab:mreslima4ilc} shows the 
corresponding 95\% CL exclusion limits on $M$ in the $SU(2)_c$ conserving case and
assuming $\Gamma/M=1$.

\begin{table}[h]
\centering
\begin{tabular}{c|c}
\hline\hline                
Type of resonance & 95\% CL \\ \hline
scalar $\phi$ & 1.64 TeV \\
vector $\rho$ & 2.09 TeV \\
tensor $f$ & 2.76 TeV \\
\hline\hline
\end{tabular}
\caption{95\% CL exclusion limits for the mass
  $M$ of a pure EW resonance from $\alpha_4$ sensitivity studies in VBS 
  at a 1 TeV ILC (with 1~$\abinv$) in the $SU(2)_c$ conserving case and
assuming a single resonance with $\Gamma/M=1$. The 95\% CL limits have been obtained from the
 1~$\sigma$ limits of Ref.~\cite{Beyer:2006hx}, $M_{\phi,\rho,f}=1.95,2.49,3.29$~TeV, which are 
based on $16 \pi^2 \Delta \alpha_4=0.5026,0.5671,0.6437$ for $\phi,\rho,f$ resonances. The latter have been doubled to 2~$\sigma$, which are then interpreted as 95\% CL limits for $M$ using Table~\ref{tab:restype}.}
\label{tab:mreslima4ilc}
\end{table}

\subsection{TGCs from a global fit to Higgs data}
\label{sec:electroweak-fit}


In this Section, we discuss the limits on higher-dimension operator
coefficients in EFT from measurements of the Higgs boson properties.

As discussed earlier in this report, effective Lagrangians can be used
to parametrize in a model-independent way the low--energy effects of
possible extensions of the standard model (SM)~\cite{effective,
Corbett:2012ja,Corbett:2012dm, Buchmuller:1986iq,59}.  There is a
freedom in the choice of the operator basis since operators connected
by the equations of motion lead to the same $S$--matrix
elements~\cite{eom}.  Taking advantage of the freedom in the choice of
the operator basis, it is convenient to include in the basis used to
analyze the Higgs couplings, operators that are directly related to
the existing data, in particular to triple gauge couplings (TGCs), as
well as, to the precision electroweak observables
~\cite{Corbett:2012ja, Corbett:2012dm, Corbett:2013pja}. Neglecting,
for the moment, modifications of the Higgs couplings to the first and
second families and CP violating interactions\footnote{{ We have
neglected in the analysis the effects of the operators ${\cal O}_{\phi
d}$ and ${\cal O}_{\phi B}$ which, strictly speaking, are not bounded
by the EWPT even after the use of equations of motion.  } }, a useful
basis is~\cite{Corbett:2012dm}
\begin{eqnarray}
{\cal L}_{eff} \!\!\!
&=& \!\!- \frac{\alpha_s v}{8 \pi} \frac{f_g}{\Lambda^2} 
{\cal O}_{GG}
+ \frac{f_{WW}}{\Lambda^2} {\cal O}_{WW}
+ \frac{f_{\rm bot}}{\Lambda^2} {\cal O}_{d\Phi,33} 
+ \frac{f_{\rm top}}{\Lambda^2} {\cal O}_{u\Phi,33} 
\label{ourleff}\\
&+& \frac{f_{\tau}}{\Lambda^2} {\cal O}_{e\Phi,33}
+ \frac{f_{W}}{\Lambda^2} {\cal O}_{W}
+ \frac{f_{B}}{\Lambda^2} {\cal O}_{B}
+ \frac{f_{WWW}}{\Lambda^2} {\cal O}_{WWW} 
\;\; \nonumber
\end{eqnarray}
with 
\begin{eqnarray}
\!\!\!\!&&{\cal O}_{GG} = \Phi^\dagger \Phi \; G_{\mu\nu}^a G^{a\mu\nu}  \;,
\;\;
{\cal O}_{WW} = \Phi^{\dagger} \hat{W}_{\mu \nu} 
 \hat{W}^{\mu \nu} \Phi  \; , \nonumber \\
&&{\cal O}_W  = (D_{\mu} \Phi)^{\dagger}  
 \hat{W}^{\mu \nu}  (D_{\nu} \Phi) \; , \;\;
 {\cal O}_B  =  (D_{\mu} \Phi)^{\dagger} 
  \hat{B}^{\mu \nu}  (D_{\nu} \Phi)  \; ,  \nonumber \\
&& {\cal O}_{WWW}=
\hbox{Tr}[\hat{W}_{\mu \nu}\hat{W}^{\nu\rho}\hat{W}_{\rho}^{\mu}]
\;, \;\;
{\cal O}_{u\Phi,ij}=(\Phi^\dagger\Phi)(\bar L_{i} \Phi u_{R_j}) 
\; , \\
&&{\cal O}_{e\Phi,ij}=(\Phi^\dagger\Phi)(\bar L_{i} \Phi e_{R_j}) 
\; ,  \;\;
{\cal O}_{d\Phi,ij}
=(\Phi^\dagger\Phi)(\bar Q_{i} \Phi d_{Rj})\; , \nonumber  \\
\label{ourope}
\end{eqnarray}
where $\Phi$ stands for the Higgs doublet with covariant derivative
$D_\mu\Phi= \left(\partial_\mu+i \frac{1}{2} g' B_\mu + i
g \frac{\sigma_a}{2} W^a_\mu \right)\Phi$ and $v=246$ GeV is its
vacuum expectation value. $\hat{B}_{\mu \nu} = i \frac{g'}{2}
B_{\mu \nu}$ and $\hat{W}_{\mu\nu} = i \frac{g}{2} \sigma^a
W^a_{\mu\nu}$ with $SU(2)_L$ ($U(1)_Y$) gauge coupling $g$
($g^\prime$) and Pauli matrices $\sigma^a$.  We also use the notation
of $L$ for the lepton doublet, $Q$ for the quark doublet and $f_R$ for
the $SU(2)$ singlet fermions, where $i, j$ are flavor indices.
${\cal O}_B$ and ${\cal O}_W$ contribute both to Higgs physics and
TGCs which means that some changes of the couplings of the Higgs field
to the vector gauge bosons are related to TGCs due to gauge invariance
in a model independent fashion~\cite{Corbett:2013pja}. In fact, the
TGCs $\gamma W^+ W^-$ and $Z W^+W^-$ can be parametrized
as~\cite{Hagiwara:1986vm}
\begin{equation}
{\cal L}_{WWV} = 
 -i g_{WWV} \Big\{ 
g_1^V \Big( W^+_{\mu\nu} W^{- \, \mu} V^{\nu} 
  - W^+_{\mu} V_{\nu} W^{- \, \mu\nu} \Big)
 + \kappa_V W_\mu^+ W_\nu^- V^{\mu\nu}
+ \frac{\lambda_V}{m_W^2} W^+_{\mu\nu} W^{- \, \nu\rho} V_\rho^{\; \mu}
 \Big\}
\;\;,
\label{eq:classical}
\end{equation}
where $g_{WW\gamma} = e=g s$ and $g_{WWZ} = g c$ 
with $s$($c$) being the sine (cosine) of the weak mixing angle.
In general these vertices involve six C and P conserving
couplings~\cite{Hagiwara:1986vm}.  Nevertheless, the electromagnetic
gauge invariance requires that $g_{1}^{\gamma} = 1$, while the five
remaining couplings are related to the dimension--six operators ${\cal
O}_B$, ${\cal O}_W$ and ${\cal O}_{WWW}$
\begin{equation}
\Delta \kappa_\gamma = 
 \frac{g^2 v^2}{8\Lambda^2}
\Big(f_W + f_B\Big)\,, \;
  \lambda_\gamma = \lambda_Z = 
\frac{3 g^2 M_W^2}{2 \Lambda^2} f_{WWW}\; ,
\Delta g_1^Z= \frac{g^2 v^2}{8 c^2\Lambda^2}f_W \, ,\; 
\Delta \kappa_Z =   \frac{g^2 v^2}{8 c^2\Lambda^2}
  \Big(c^2 f_W - s^2 f_B\Big)\, ,
\label{eq:wwv}
\end{equation}
where we wrote $\kappa^V=1+\Delta \kappa^V$ and $g^Z_1=1+\Delta g^Z_1$.

Here we assess the impact of Higgs physics on the TGC determination
at the LHC with a center--of-mass energy of 14 TeV and integrated
luminosities of 300 fb$^{-1}$ and 3000 fb$^{-1}$. We fit the ATLAS and CMS
expected sensitivities~\cite{proj_atlas, proj_cms} for the Higgs
signal strength using four independent parameters $\{ f_g~,~ f_W ~,~
f_B ~,~ f_{WW} \}$ and setting the Yukawa couplings to the fermions to
their SM values. This scenario captures most of the features of fits
using a larger set of free parameters since the addition of fermionic
operators has little impact on the Higgs couplings to gauge--boson
pairs and TGCs ~\cite{ Corbett:2012ja, Corbett:2012dm}.

\begin{figure}[t]
\begin{center}
\includegraphics[width=0.7\hsize]{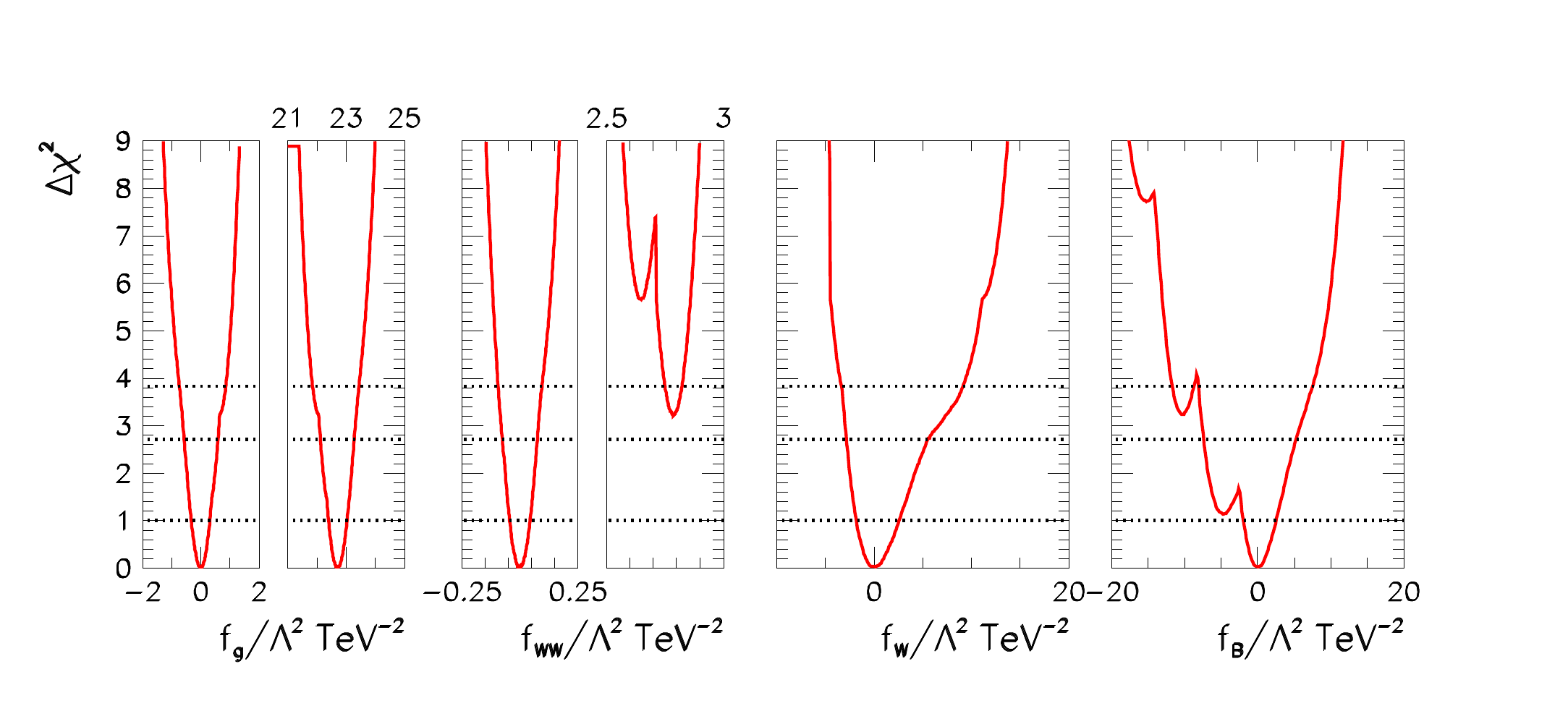}
\includegraphics[width=0.7\hsize]{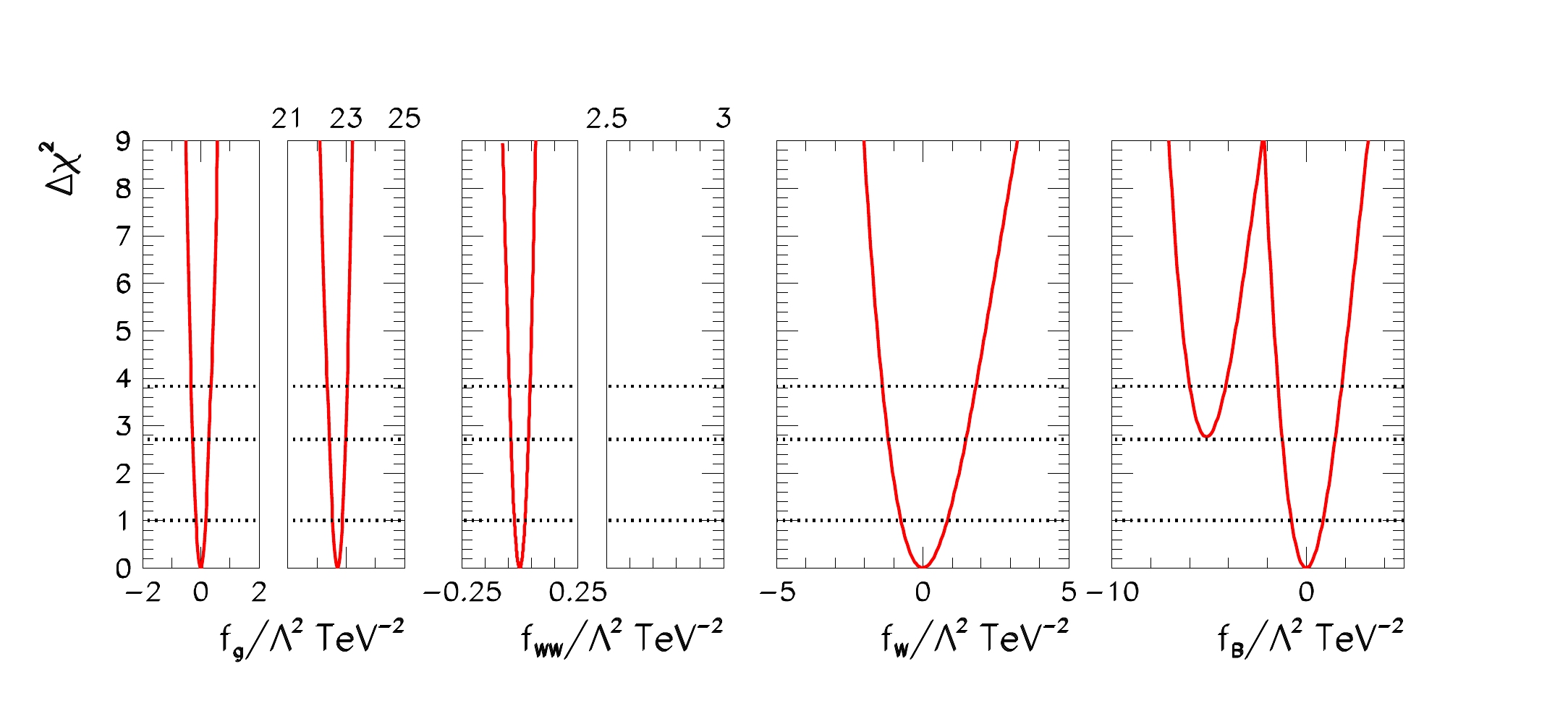}
\caption{$\Delta \chi^2 $ as a function of $f_{g}$, $f_{WW}$, $f_{W}$,
  and $f_B$ assuming $f_{\rm bot} = f_\tau = f_{\rm top}= 0$, after
  marginalizing over the three undisplayed ones.  The three horizontal
  dashed lines stand for the $\Delta \chi^2$ values associated to
  68\%, 90\% and 95\% from bottom to top respectively. The upper
  (lower) row was obtained for an integrated luminosity of 300 (3000)
  fb$^{-1}$. }
\label{fig:chi2_par}
\end{center}
\end{figure}

\begin{table}[htp]
\centering
\begin{tabular}{|c|p{2.5cm}|p{2.5cm}|p{2.5cm}|p{2.5cm}|}
 \cline{2-5}
 \multicolumn{1}{c|}{} &  \multicolumn{2}{c|}{68\% CL allowed range}
& \multicolumn{2}{c|}{95\% CL allowed range}
\\
\cline{2-5}
 \multicolumn{1}{c|}{} &  300 fb$^{-1}$ & 3000 fb$^{-1}$
& 300 fb$^{-1}$ & 3000 fb$^{-1}$\\
\hline
$f_g/\Lambda^2$ \footnotesize{(TeV$^{-2}$)} & $(-0.33,0.31)\cup(22.40,23.04)$ &
 $(-0.17,0.17)\cup(22.54,22.88)$ &
 $(-0.74,0.86)\cup(21.85,23.45)$ &  $(-0.33,0.34)\cup(22.36,23.04)$
\\
\hline
$f_{WW}/\Lambda^2$ \footnotesize{(TeV$^{-2}$)} & $(-0.043,0.044)$ & $(-0.023,0.022)$ & $(-0.093,0.096)\cup(2.75,2.82)$ & $(-0.045,0.044)$
\\
\hline
$f_{W}/\Lambda^2$ \footnotesize{(TeV$^{-2}$)} & $(-1.9,2.5)$ & $(-0.75,0.83)$ &
 $(-3.4,9.1)$ & $(-1.39,1.82)$
\\
\hline
$f_{B}/\Lambda^2$ \footnotesize{(TeV$^{-2}$)} & $(-2.0,2.5)$ & $(-0.78,0.85)$ &
$(-11.7,7.5)$ & $(-6.0,-4.1)\cup(-1.5,1.8)$
\\
\hline
\end{tabular}
\caption{68\% CL and 95\% expected allowed ranges for 300 and 3000
fb$^{-1}$ of integrated luminosity.}
\label{tab:anom}
\end{table}

Figure~\ref{fig:chi2_par} displays $\Delta\chi^2$ as a function of the
four fitting parameters \footnote{Details of the fitting procedure can
be seen in Refs.~\cite{ Corbett:2012ja, Corbett:2012dm}} for
integrated luminosities of 300 fb$^{-1}$ (upper row) and 3000
fb$^{-1}$ (lower row).  The corresponding 68\% CL and 95\% expected
allowed ranges can be found in Table~\ref{tab:anom}. We can observe in
the upper and lower left panels that the $\Delta\chi^2$ as a function
of $f_g$ exhibits two degenerate minima due to the interference
between SM and anomalous contributions to $g g \rightarrow H$
production . In the case of the $\chi^2$ dependence on $f_{WW}$ there
is also an interference between anomalous and SM contributions to
$H\rightarrow\gamma\gamma$ , however, the degeneracy of the minima is
lifted since the $f_{WW}$ coupling contributes also to Higgs decays
into $WW^*$, $ZZ^*$ and $\gamma Z$, as well as in $Vh$ associated and
vector boson fusion production mechanisms.  Clearly a larger
statistics helps eliminating the degeneracy in $f_{WW}$.  The
interference between $f_B$ and the SM contribution to
$H\rightarrow \gamma Z$ is responsible for the two local minima with
smaller $\Delta \chi^2$ while the additional minima in the upper right
panel originates from the marginalization of $f_{WW}$.  Comparing the
upper and lower rows, we can see that a larger integrated luminosity
also helps to significantly reduce the errors in the determination of
the anomalous couplings.

\begin{figure}[htp]
\begin{center}
  \includegraphics[width=0.7\hsize]{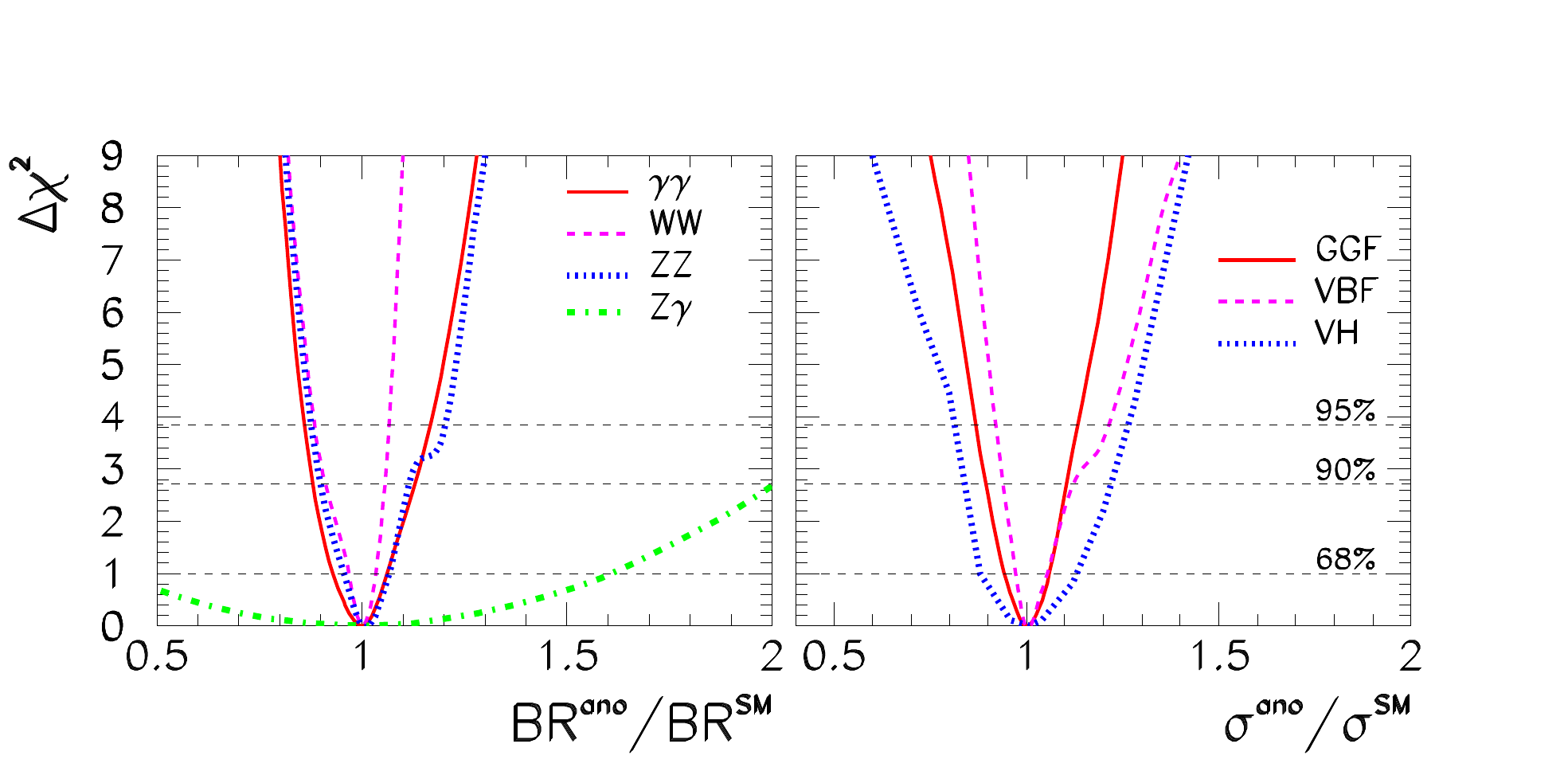}
  \includegraphics[width=0.7\hsize]{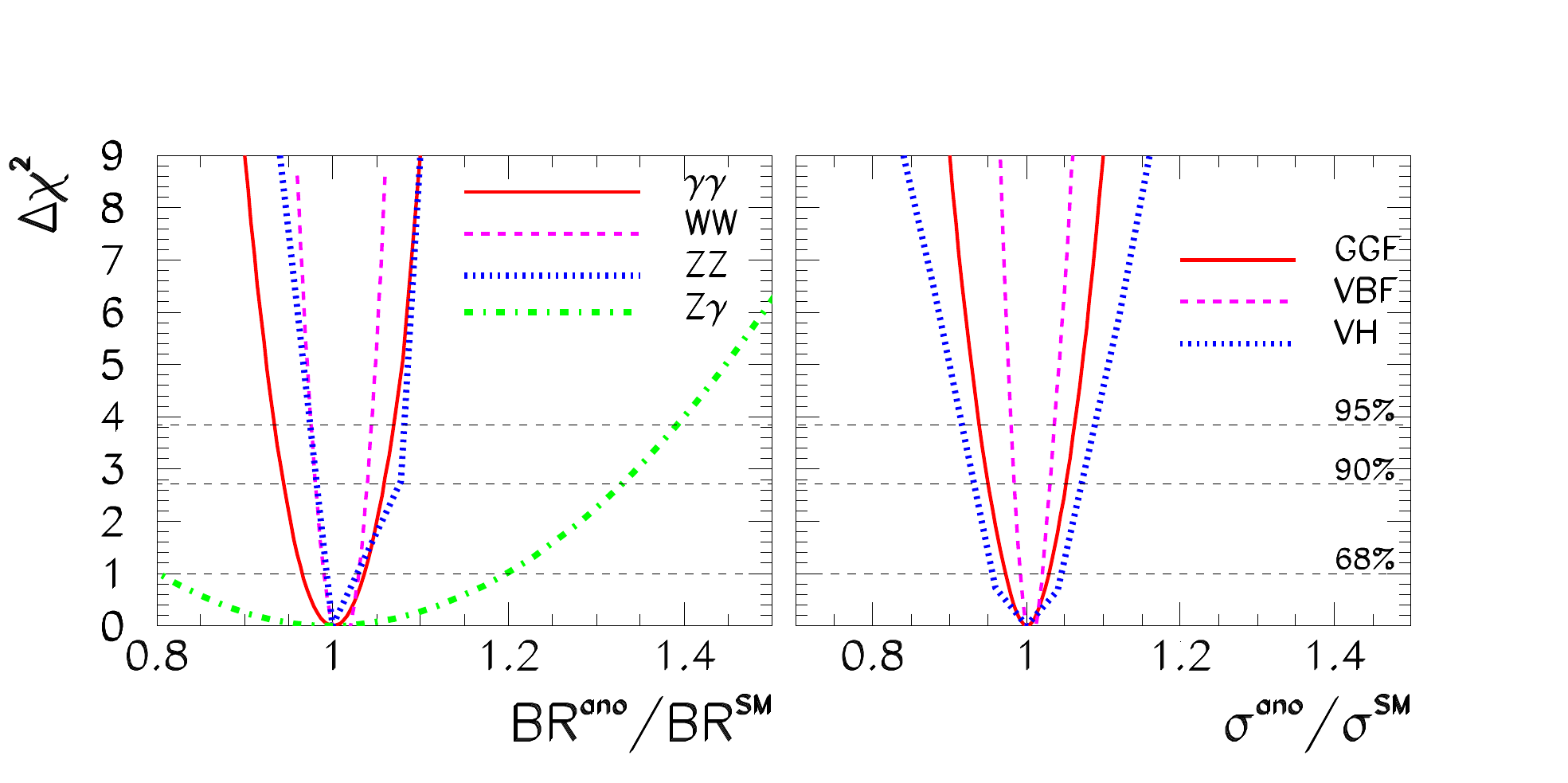}
  \caption{$\Delta \chi^2$ as a function of branching ratios (left panels)
    and production cross sections (right panels) when we use only the
    expected ATLAS and CMS sensitivity on the Higgs signal strengths
    for integrated luminosities of 300 fb$^{-1}$ (upper row) and 3000 fb$^{-1}$ (lower
    row).}
\label{fig:chi2_brsig}
\end{center}
\end{figure}

Figure~\ref{fig:chi2_brsig} depicts the $\chi^2$ dependence on
branching ratios and production cross sections for integrated
luminosities of 300 fb$^{-1}$ and 3000 fb$^{-1}$. As we can see these
quantities can be determined with a precision better than 20\% (5\%)
with 300 (3000) fb$^{-1}$. The only exception is the Higgs branching
ratio into $Z\gamma$ that can be measured within 20\% with 3000
fb$^{-1}$.  These results show the consistency of the extracted
accuracies in the production cross sections and branching ratios in
the dim-6 operator framework with those obtained by the experimental
collaborations in their simulations ~\cite{proj_atlas, proj_cms}
assuming a shift of the SM couplings.

\begin{figure}[htp]
\begin{center}
\includegraphics[width=0.75\hsize]{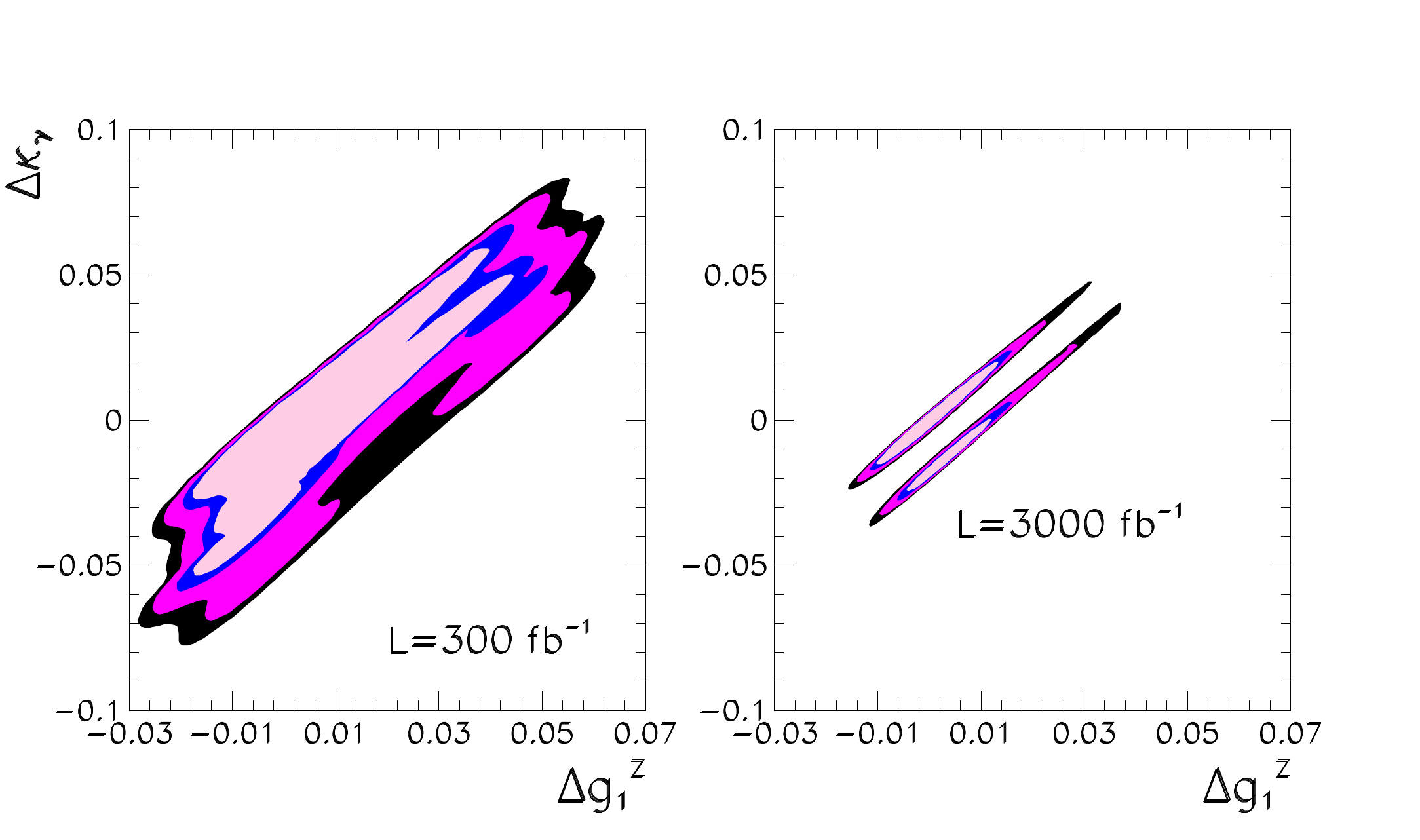}
\caption{ We present the expected 90\%, 95\%, 99\%, and 3$\sigma$
  allowed regions for the $\Delta\kappa_\gamma \otimes \Delta g^Z_1$
  plane from the analysis of the Higgs data from LHC at 14 TeV with
  integrated luminosities of 300 fb$^{-1}$ (left panel) and 3000
  fb$^{-1}$ (right panel).}
\label{fig:tgc_hbounds}
\end{center}
\end{figure}


Next we focus our attention to the expected TGC bounds derived from
this analysis of the Higgs data.  Eq.~\ref{eq:wwv} allows us to
translate the constraints on $f_W$ and $f_B$ coming from the Higgs
measurements to bounds on $\Delta\kappa_\gamma$, $\Delta\kappa_Z$ and
$\Delta g_1^Z$ of which only two are independent.
Fig.~\ref{fig:tgc_hbounds} displays the result of our fit to the Higgs
data where we plot the 90\%, 95\%, 99\%, and 3$\sigma$ CL allowed
region in the plane $\Delta\kappa_\gamma \otimes \Delta g_1^Z$ after
marginalizing over the other two parameters relevant to the Higgs
analysis, {\em i.e.}  $f_g$ and $f_{WW}$.
Notice that the two almost degenerate local minima in $f_B$ lead to
the appearance of two narrow disconnected regions due to the high
precision achieved with 3000 fb$^{-1}$.

Clearly the analysis of the Higgs data alone can improve the present
best bounds on TGC which are still coming from LEP. Further
improvement will come from combining the Higgs results with those from
direct study of the TGC couplings. Unfortunately the study of the
capabilities of the LHC14 runs on the constraints of the TGC couplings
from di-boson production in this scenario is still missing.


In summary, indirect new physics effects associated with extensions of
the electroweak symmetry breaking sector can be written in terms of an
effective Lagrangian whose lowest order operators are of dimension
six. The coefficients of these dimension--six operators parametrize
our ignorance of these effects.  In the above framework changes of the
couplings of the Higgs to electroweak gauge bosons are related to the
anomalous triple gauge--boson vertices~\cite{deCampos:1997ez}.
Therefore, the analysis of the Higgs boson data at LHC can be used to
constrain TGCs. Moreover, the combination of future TGC and Higgs
measurements have the potential to lead to the strongest constraints
on new physics effects associated with this sector.

\section{Conclusions}
\label{sec:electroweak-conclusions}

With the discovery of the Higgs boson and the measurement of its mass
at the LHC, the last missing component of the SM has been
determined. However, the big question related to the origin of the
Higgs "Mexican hat" potential remains to be answered. This question is
exacerbated by the instability of this potential under quantum
corrections.

The role of precision electroweak measurements is of increasing
importance in over-constraining the Higgs sector of the SM. The two
themes we have investigated in the arena of precision electroweak
measurements are (a) the electroweak precision observables (EWPOs)
that test the particle content and couplings in the SM and BSM
scenarios, and (b) the measurements involving multiple gauge bosons in
the final state which provide unique probes of the basic tenets of
electroweak symmetry breaking.

In the case of EWPOs, we have focused on the measurement of $M_W$ and
$\sin^2\theta_\eff^l$. Other EWPOs should be available, in particular
 at ILC/GigaZ and TLEP; for instance, 
 the $Z \rightarrow b\bar{b}$ partial width and the leptonic partial width are equally worthy of consideration. Our conclusions are as follows:

\begin{itemize}
\item The knowledge of the Higgs mass has sharpened the predictions of
  these EWPOs such that the predictions are a factor of 2-4 more
  precise than the experimental measurements.

\item In almost all extensions of the SM, which are associated with the
  electroweak symmetry-breaking sector, these EWPOs receive corrections
  due to quantum loops (due to e.g.  supersymmetric particles or
  techni-fermions), or due to effective operators (induced for example in
  strongly-interacting light Higgs models), or due to Kaluza-Klein
  modes in extra-dimensional models.
 
\item $M_W$ and $\sin^2\theta_\eff^l$ typically have different sensitivities
  to the sources of new physics. This may be demonstrated by the
  parametrization of new physics in the gauge boson self-energies in
  terms of the $S$, $T$ and $U$ ``oblique" corrections. Fixed values of
  $M_W$ and $\sin^2\theta_\eff^l$ correspond to lines in the $S-T$ plane
  with different slopes. Thus, improved measurements of both EWPOs can
  constrain all of the above sources of new physics in a relatively
  model-independent fashion.

\item The current world average $M_W$ has a precision of 15 MeV,
  dominated by the combined Tevatron measurement, which has a precision
  of 16 MeV based on the analysis of partial datasets. CDF and D\O\
  have projected that analyses of the full Tevatron statistics can
  yield a 10 MeV measurement, assuming a factor of two improvement in
  the uncertainty due to parton distribution functions, 
  improvement in the calculation of radiative corrections and improved
  understanding of the trackers and calorimeters.

\item Studies based on pseudo-data have demonstrated that measurements
  of boson distributions with the 2011-2012 LHC data may be able to
  improve the PDFs relevant for the $M_W$ measurement by a factor of
  two in the near future, enabling the Tevatron potential for $M_W$ to
  be realized.

\item Enormous statistics of $W$ bosons and control samples at the LHC
  offer the prospect of higher $M_W$ precision. Studies based on
  pseudo-data have shown that the PDF uncertainty in $M_W$ is about
  twice as big at the LHC as the Tevatron, due mainly to the larger
  fraction of sea quark-initiated production. Thus, further
  improvement by a factor of 2-3 in the PDFs will be required, beyond
  what is needed for the Tevatron.  Furthermore, additional
  improvements in the QED and electroweak radiative correction calculations and
  NNLO+NNLL generators for $W$ and $Z$ bosons will likely also be
  required. However, considering the 15-year time scale for the
  ultimate $M_W$ measurement from the LHC, we consider a target
  precision of 5 MeV to be appropriate for the LHC.

\item Studies of the $M_W$ measurement at the ILC using the threshold
  scan and final state reconstruction have been updated. It is
  projected that the ILC will be able to perform the $M_W$ measurement
  with a precision of 4-5 MeV with ${\cal L}=100\,{\rm fb}^{-1}$ and of 2.5 MeV with 
  ${\cal L}\approx 500 \, {\rm fb}^{-1}$.

\item The circular electron-positron TLEP machine, running at the $WW$
  threshold, can produce very high statistics for the $M_W$
  measurement, and is likely to achieve energy calibration at the
  level of a fraction of an MeV.  This potential motivates further
  studies of longitudinal beam polarization and control of other
  systematics achievable at TLEP. Given an integrated luminosity that
  can enable a statistical precision of $\sim 0.5$ MeV, further
  investigations of related issues are clearly warranted. Assuming a systematic uncertainty contribution of 1 MeV from missing higher-order corrections, TLEP could achieve an $M_W$ precision of $< 1.2$~MeV. 
 One of the goals of the ongoing TLEP study is to ensure that the $W$ mass can be measured with  a
precision better that 1 MeV; this includes a program of improvements in the theoretical
calculations.

\item The measurement of $\sin^2 \theta_\eff$ from LEP and SLC have
  averaged to a precision of $16 \times 10^{-5}$, albeit with a $\sim
  3 \sigma$ difference between them. Additional, especially improved,
  measurements will be valuable to shed light on this difference.

\item A measurement of $\sin^2\theta_\eff^l$ using the full Tevatron dataset
  is projected with a precision of $41 \times 10^{-5}$. This
  measurement will be interesting to compare with LEP and SLC.

\item Compared to the Tevatron, measurement of $\sin^2 \theta_\eff^l$ at the
  LHC is handicapped by a larger sensitivity to PDFs due to the
  dilution of the quark and antiquark directions. As with the $M_W$
  measurement, considerable control of the experimental and production
  model uncertainties will be required. Under the condition that a
  factor of 6-7 improvement on PDFs is achieved (a condition also
  required for the $M_W$ target for the LHC), a projected uncertainty
  on $\sin^2\theta_\eff^\ell$ of $21 \times 10^{-5}$ is obtained. This
  precision is similar to the current LEP and SLC measurements and is
  valuable before the advent of future lepton colliders.

\item Considerably more precise measurements of $\sin^2 \theta_\eff^\ell$ are
  highly desirable for taking the stringency of the SM tests to the
  next order of magnitude. Such measurements are possible at future
  lepton colliders running on the $Z-$pole such as ILC/GigaZ and TLEP. 

\item The ILC/GigaZ projection for the precision on $\sin^2 \theta_\eff^\ell$ is
  $1.3 \times 10^{-5}$, a factor of 10 improvement on the current
  world average.

\item TLEP has the potential to go beyond ILC/GigaZ in the
  precision on $\sin^2\theta_\eff^\ell$. This warrants a more detailed
  study. TLEP 
 can produce $10^{12}$ $Z$ bosons and could target improvements of a factor of 
 $\sim$100 in $\sin^2 \theta_\eff^\ell$ and a factor of 10-20 in other $Z$-pole
 observables beyond LEP and SLC. 
  The precision at TLEP would be high enough that
  all aspects of EWPOs, both theoretical and experimental, need to be
  revisited. Further studies on experimental and
 theory systematics,  and beam energy calibration and polarization are needed to 
 understand if this TLEP potential can be realized. 

\item Measurements of $M_W$ at the few MeV level, and $\sin^2
  \theta_\eff^\ell$ at the level of $10^{-5}$, require that the
  parametric uncertainties from $m_{top}, M_Z$, and $\Delta
  \alpha_{had}$ (the contribution to the running of $\alpha_{EM}$ from
  hadronic loops) as well as the missing higher order calculations be
  addressed.  Parametric uncertainties from $m_{top}$ and $\Delta
  \alpha_{had}$, if reduced by a factor of 2 compared to current
  uncertainties, will prevent them from exceeding the anticipated
  total precision on $M_W$ at the LHC. At the ILC and TLEP a factor of
  5 and 10 improvement, respectively, in the parametric uncertainties
  is needed, which is only achievable if the precision on $M_Z$ is
  considerably improved as well. TLEP can improve the $M_Z$ precision
  by a factor of at least 10. It is anticipated that calculations in
  the coming years will reduce the effect of missing higher-order
  calculations by a factor of 4 which is sufficient for the LHC and ILC
  target uncertainties, but further effort will be needed for  TLEP.

\end{itemize}

The second aspect of precision electroweak measurements we have
emphasized is vector boson scattering and the related process of
triboson production. Vector boson scattering addresses one of the 
crucial big questions that still needs to be tested experimentally, ie. the unitarization
of longitudinal vector boson scattering at high energy. In the SM, the
unitarization is achieved when Higgs boson exchange amplitudes are
included, and this mechanism relies on the longitudinal modes of the
massive gauge bosons being the would-be Goldstone modes of the
symmetry-breaking Higgs potential. A direct demonstration of this
mechanism is required, and is a prime motivation for the HL-LHC.

Models which explain the lightness of the discovered Higgs boson by
describing it as a pseudo-Goldstone boson associated with the breaking
of a larger symmetry, often introduce higher-dimension operators as an
effective field theory (EFT) approximation of the new
dynamics. Testing for these operators in vector boson scattering and
triboson production can answer one of the outstanding questions in the
Higgs sector: is the dynamics associated with the stabilization of the
Higgs potential under quantum corrections, weakly coupled (e.g. SUSY)
or strongly coupled (e.g. SILH models)? 

The EFT formulation is not limited to specific models; any high energy theory
 can be reduced to a low-energy EFT and the former will specify the values
 of operator coefficients in the latter. Therefore, EFT operators provide
 a general method of parameterizing the effects of new physics at a high scale. 

Some of these higher-dimension operators can alter the Higgs boson
couplings, some can affect the values of EWPOs while others have no
impact on these observables but still strongly affect multi-boson
production. The study of the latter processes can provide direct
evidence of new SILH dynamics through the energy-dependence of the
anomalous production. Further clarification of the new dynamics can be
provided by comparing final states involving different combinations of
$W$ and $Z$ bosons and photons, which can elucidate the group
structure of the new dynamics.

Our conclusions in the area of multi-boson production are as follows:

\begin{itemize}
\item Studies of vector boson scattering and triboson production have
  become possible, for the first time, at the LHC. These processes probe 
 quartic couplings which offer  a new and promising avenue of research into 
 electroweak symmetry breaking. 

\item For the next decade, the LHC will continue to be the facility to
  explore these processes at higher levels of precision. 

\item The LHC will improve the sensitivity to anomalous trilinear gauge couplings 
 by 1-2 orders of magnitude beyond LEP and the Tevatron. 

\item The HL-LHC is needed to demonstrate that the Higgs couplings to
  the electroweak vector bosons is an essential component of the unitarization
  mechanism for vector boson scattering. An integrated luminosity of
  300 fb$^{-1}$ is not enough.

\item The sensitivity to higher-dimension operators improves by a
  factor of 2-3 with the HL-LHC, in comparison with the 300 fb$^{-1}$
  at the LHC.

\item Triboson production and vector boson scattering are sensitive 
 and complementary probes of dimension-8 operators. These  
  processes  becomes rapidly more sensitive with increasing beam energy,
  providing strong motivation for a 100 TeV $pp$ collider.

\item  Anomalous trilinear gauge couplings, which are induced by
  dimension-6 operators, are significantly better probed by the 
 high-energy ILC options compared to the LHC. 
 On the other hand, anomalous quartic gauge couplings, which are 
 induced by dimension-8 operators, are significantly better probed 
 (by 1-2 orders of magnitude) by the LHC, due to the stronger growth of the anomalous
 cross section with energy. Interpreting the latter as being induced by electroweak resonances, 
 the LHC is sensitive       to resonance
 masses  that are higher    by more     than a factor of two, as compared to ILC1000.

\end{itemize}

\Acknowledgements

We are grateful to the many scientists who assisted us in various ways through the process. We have
 benefited from those who have given us technical advice, and from those who took time to help us with individual tasks.

We acknowledge the support of the U.S. Department of Energy and the National Science Foundation. 


C.~Degrande is supported in part by the U.~S. Department of Energy under Contract No. DE-FG02-13ER42001.

The work of the Gfitter collaboration is funded by the German Research Foundation (DFG) in the Collaborative Research Centre (SFB) 676 ``Particles, Strings and the Early Universe'' located in Hamburg.

J.~Reuter has been partially supported by the Strategic HGF Alliance "Physics at the Terascale".

O.~Eboli acknowledges support from Funda\c{c}\~ao de Amparo \`a Pesquisa do Estado de S\~ao Paulo (FAPESP)
and Conselho Nacional de Desenvolvimento Cient\'ifico e Tecnol\'ogico (CNPq).

The work of J.~Erler was supported by PAPIIT (DGAPA--UNAM) project IN106913 and CONACyT (Mexico) project 151234,
and grateful acknowledges the hospitality and support by the Mainz Institute for Theoretical Physics (MITP)
where part of his work was completed.

B.~Feigl and M.~Rauch acknowledge support by the
BMBF under Grant No. 05H09VKG (``Verbundprojekt HEP-Theorie'').

The work of S.~Heinemeyer was supported in part by CICYT (Grant No.\ FPA 2010--22163-C02-01), and by the Spanish MICINN's
Consolider-Ingenio 2010 Program under Grant MultiDark No.\ CSD2009-00064.

B.~Jaeger would like to thank the Research Center Elementary Forces and
Mathematical Foundations (EMG) of the Johannes Gutenberg University
Mainz. B.~Jaeger and D.~Wackeroth would also like to thank the hospitality of the Aspen Center for Physics where part of this 
work was done and supported in part by the National Science
Foundation under Grant No. PHYS-1066293.

O.~Mattelaer is a chercheur logistique post-doctoral of F.R.S-FNRS.

A.~Vicini is supported by an italian PRIN 2010 grant.

The work of D.~Wackeroth is supported in part by the U.S.~National
Science Foundation under grant no. PHY-1118138.



\begin{thebibliography}{999}

\bibitem{ALEPH:2010aa} 
ALEPH and CDF and D0 and DELPHI and L3 and OPAL and SLD and LEP Electroweak Working Group and Tevatron Electroweak Working Group and SLD Electroweak and Heavy Flavour Groups Collaborations,
  arXiv:1012.2367 [hep-ex].

\bibitem{Giudice:2007fh} 
  G.~F.~Giudice, C.~Grojean, A.~Pomarol and R.~Rattazzi,
  JHEP {\bf 0706}, 045 (2007)
  [hep-ph/0703164], and references therein.

\bibitem{zpole}
  The ALEPH, DELPHI, L3, OPAL, SLD Collaborations,
  the LEP Electroweak Working Group,
  the SLD Electroweak and Heavy Flavour Groups,
    Phys.\ Rept.\  {\bf 427} 257, (2006)
  [hep-ex/0509008].
  

\bibitem{sintev}
  T.~Aaltonen {\it et al.}  [CDF Collaboration],
    Phys.\ Rev.\ Lett.\  {\bf 106}, 241801 (2011)
  [arXiv:1103.5699 [hep-ex]];\\
    V.~M.~Abazov {\it et al.}  [D\O\ Collaboration],
    Phys.\ Rev.\ D {\bf 84}, 012007 (2011)
  [arXiv:1104.4590 [hep-ex]].
  

\bibitem{sinlhc}
  S.~Chatrchyan {\it et al.}  [CMS Collaboration],
    Phys.\ Lett.\ B {\bf 718}, 752 (2013)
  [arXiv:1207.3973 [hep-ex]].
    G.~Aad {\it et al.}  [ATLAS Collaboration],
  ATLAS-CONF-2013-043.

\bibitem{sineffl2}
  M.~Awramik, M.~Czakon, A.~Freitas, G.~Weiglein,
        Phys.\ Rev.\ Lett.\  {\bf 93}, 201805 (2004)
    [hep-ph/0407317];\\
  M.~Awramik, M.~Czakon and A.~Freitas,
    Phys.\ Lett.\  B {\bf 642}, 563 (2006)
  [hep-ph/0605339];\\
      W.~Hollik, U.~Meier and S.~Uccirati,
      Nucl.\ Phys.\ B {\bf 731}, 213 (2005)
  [hep-ph/0507158],
      Nucl.\ Phys.\ B {\bf 765}, 154 (2007)
    [hep-ph/0610312].
  

\bibitem{sineffl2a}
  M.~Awramik, M.~Czakon and A.~Freitas,
    JHEP {\bf 0611}, 048 (2006)
   [hep-ph/0608099].
  

\bibitem{sineffl3}
L.~Avdeev, J.~Fleischer, S.~Mikhailov and O.~Tarasov,
Phys.\ Lett.\ B {\bf 336}, 560 (1994)
[Erratum-ibid.\ B {\bf 349}, 597 (1994)]
[hep-ph/9406363];\\
K.~G.~Chetyrkin, J.~H.~K\"uhn and M.~Steinhauser,
Phys.\ Lett.\ B {\bf 351}, 331 (1995)
[hep-ph/9502291];\\
K.~G.~Chetyrkin, J.~H.~K\"uhn and M.~Steinhauser,
Phys.\ Rev.\ Lett.\  {\bf 75}, 3394 (1995)
[hep-ph/9504413];\\
K.~G.~Chetyrkin, J.~H.~K\"uhn and M.~Steinhauser,
Nucl.\ Phys.\ B {\bf 482}, 213 (1996)
[hep-ph/9606230].

\bibitem{sineffl4}
Y.~Schr\"oder and M.~Steinhauser,
Phys.\ Lett.\ B {\bf 622}, 124 (2005)
[hep-ph/0504055];\\
K.~G.~Chetyrkin, M.~Faisst, J.~H.~K\"uhn, P.~Maierhoefer and C.~Sturm,
  Phys.\ Rev.\ Lett.\  {\bf 97}, 102003 (2006)
  [hep-ph/0605201];\\
    R.~Boughezal and M.~Czakon,
    Nucl.\ Phys.\  B {\bf 755}, 221 (2006)
  [hep-ph/0606232].
  

\bibitem{sineffmt}
J.~J.~van der Bij, K.~G.~Chetyrkin, M.~Faisst, G.~Jikia and T.~Seidensticker,
Phys.\ Lett.\ B {\bf 498}, 156 (2001)
[hep-ph/0011373];\\
M.~Faisst, J.~H.~K\"uhn, T.~Seidensticker and O.~Veretin,
Nucl.\ Phys.\ B {\bf 665}, 649 (2003)
[hep-ph/0302275].

\bibitem{sineffmh}
 R.~Boughezal, J.~B.~Tausk and J.~J.~van der Bij,
      Nucl.\ Phys.\ B {\bf 713}, 278 (2005)
  [hep-ph/0410216],
        Nucl.\ Phys.\ B {\bf 725}, 3 (2005)
  [hep-ph/0504092].
  

\bibitem{zfitter}
D.~Y.~Bardin, P.~Christova, M.~Jack, L.~Kalinovskaya, A.~Olchevski, S.~Riemann
and T.~Riemann,
Comput.\ Phys.\ Commun.\  {\bf 133}, 229 (2001)
[hep-ph/9908433].

\bibitem{QEDLR}
 M.~B\"ohm and W.~Hollik,
      Nucl.\ Phys.\ B {\bf 204}, 45 (1982);\\
  S.~Jadach, J.~H.~K\"uhn, R.~G.~Stuart and Z.~W\c{a}s,
    Z.\ Phys.\ C {\bf 38}, 609 (1988)
  [Erratum-ibid.\ C {\bf 45}, 528 (1990)].
  

\bibitem{QEDFB}
  M.~Greco, G.~Pancheri-Srivastava and Y.~Srivastava,
    Nucl.\ Phys.\ B {\bf 171}, 118 (1980)
  [Erratum-ibid.\ B {\bf 197}, 543 (1982)];\\
    F.~A.~Berends, R.~Kleiss and S.~Jadach,
      Nucl.\ Phys.\ B {\bf 202}, 63 (1982).
  

\bibitem{hollikee}
W.~Hollik, {\it Predictions for $e^+e^-$ Processes}, in {\it Precision Tests of
the Standard Model}, ed. P.~Langacker (World Scientific, Singapur, 1993), p.~117.

\bibitem{Zqqaas}
  A.~Czarnecki and J.~H.~K\"uhn,
    Phys.\ Rev.\ Lett.\  {\bf 77}, 3955 (1996)
  [hep-ph/9608366];\\
    R.~Harlander, T.~Seidensticker and M.~Steinhauser,
    Phys.\ Lett.\ B {\bf 426}, 125 (1998)
  [hep-ph/9712228].
  

\bibitem{rb}
  A.~Freitas and Y.-C.~Huang,
    JHEP {\bf 1208}, 050 (2012)
  [Erratum arXiv:1205.0299v3 [hep-ph]].
  

\bibitem{gzmt}
R.~Barbieri, M.~Beccaria, P.~Ciafaloni, G.~Curci and A.~Vicere,
Phys.\ Lett.\ B {\bf 288}, 95 (1992)
[Erratum-ibid.\ B {\bf 312}, 511 (1993)]
[hep-ph/9205238];\\
R.~Barbieri, M.~Beccaria, P.~Ciafaloni, G.~Curci and A.~Vicere,
Nucl.\ Phys.\ B {\bf 409}, 105 (1993);\\
J.~Fleischer, O.~V.~Tarasov and F.~Jegerlehner,
Phys.\ Lett.\ B {\bf 319}, 249 (1993);\\
J.~Fleischer, O.~V.~Tarasov and F.~Jegerlehner,
Phys.\ Rev.\ D {\bf 51}, 3820 (1995);\\
  G.~Degrassi and P.~Gambino,
    Nucl.\ Phys.\  B {\bf 567}, 3 (2000)
  [hep-ph/9905472].
  

\bibitem{muqed}
  T.~van Ritbergen and R.~G.~Stuart,
    Phys.\ Rev.\ Lett.\  {\bf 82}, 488 (1999)
  [hep-ph/9808283];\\
    M.~Steinhauser and T.~Seidensticker,
    Phys.\ Lett.\ B {\bf 467}, 271 (1999)
  [hep-ph/9909436].
  

\bibitem{mw2}
  A.~Freitas, W.~Hollik, W.~Walter and G.~Weiglein,
    Phys.\ Lett.\ B {\bf 495}, 338 (2000)
  [Erratum-ibid.\ B {\bf 570}, 260 (2003)]
  [hep-ph/0007091];\\
    M.~Awramik and M.~Czakon,
    Phys.\ Rev.\ Lett.\  {\bf 89}, 241801 (2002)
  [hep-ph/0208113];\\
    A.~Onishchenko and O.~Veretin,
    Phys.\ Lett.\ B {\bf 551}, 111 (2003)
  [hep-ph/0209010];\\
      M.~Awramik and M.~Czakon,
    Phys.\ Lett.\ B {\bf 568}, 48 (2003)
  [hep-ph/0305248].
    

\bibitem{Awramik:2003rn} 
  M.~Awramik, M.~Czakon, A.~Freitas and G.~Weiglein,
    Phys.\ Rev.\ D {\bf 69}, 053006 (2004)
  [hep-ph/0311148].
  

\bibitem{zfitter2}
    A.~B.~Arbuzov, M.~Awramik, M.~Czakon, A.~Freitas, M.~W.~Gr\"unewald,
    K.~M\"onig, S.~Riemann,~T. Riemann,
      Comput.\ Phys.\ Commun.\  {\bf 174}, 728 (2006)
  [hep-ph/0507146].
  

\bibitem{gfitter}
  H.~Fl\"acher, M.~Goebel, J.~Haller, A.~H\"ocker, K.~M\"onig and J.~Stelzer,
      Eur.\ Phys.\ J.\  C {\bf 60}, 543 (2009)
  [Erratum-ibid.\  C {\bf 71}, 1718 (2011)]
  [arXiv:0811.0009 [hep-ph]].
  

\bibitem{pole}
R.~G.~Stuart,
Phys.\ Lett.\ B {\bf 262}, 113 (1991);\\
H.~Veltman,
Z.\ Phys.\ C {\bf 62}, 35 (1994).

\bibitem{Davier:2010nc} 
  M.~Davier, A.~Hoecker, B.~Malaescu and Z.~Zhang,
    Eur.\ Phys.\ J.\ C {\bf 71}, 1515 (2011)
  [Erratum-ibid.\ C {\bf 72}, 1874 (2012)]
  [arXiv:1010.4180 [hep-ph]].
  

\bibitem{Hagiwara:2011af} 
  K.~Hagiwara, R.~Liao, A.~D.~Martin, D.~Nomura and T.~Teubner,
    J.\ Phys.\ G {\bf 38}, 085003 (2011)
  [arXiv:1105.3149 [hep-ph]].
    

\bibitem{mtop}
  The Tevatron Electroweak Working Group and the CDF and D\O\
                  Collaborations,
      arXiv:1107.5255 [hep-ex].
  

\bibitem{pdg}
  J.~Beringer {\it et al.}  [Particle Data Group Collaboration],
    Phys.\ Rev.\ D {\bf 86}, 010001 (2012).
  

\bibitem{Peskin:1991sw} 
  M.~E.~Peskin and T.~Takeuchi,
    Phys.\ Rev.\ D {\bf 46}, 381 (1992).
    

\bibitem{Kotwal:2008zz}
  A.~V.~Kotwal and J.~Stark,
    Ann.\ Rev.\ Nucl.\ Part.\ Sci.\  {\bf 58}, 147 (2008).
  

\bibitem{Balazs:1997xd}
  C.~Balazs and C.~P.~Yuan,
    Phys.\ Rev.\  D {\bf 56}, 5558 (1997)
  [arXiv:hep-ph/9704258];
    R.~K.~Ellis and S.~Veseli,
      Nucl.\ Phys.\  B {\bf 511}, 649 (1998)
  [arXiv:hep-ph/9706526].
  

\bibitem{Landry:1999an}
  F.~Landry, R.~Brock, G.~Ladinsky and C.~P.~Yuan,
      Phys.\ Rev.\  D {\bf 63}, 013004 (2001)
  [arXiv:hep-ph/9905391].
  

\bibitem{Frixione:2002ik}
  S.~Frixione and B.~R.~Webber,
    JHEP {\bf 0206}, 029 (2002)
  [arXiv:hep-ph/0204244];
    S.~Frixione, F.~Stoeckli, P.~Torrielli, B.~R.~Webber and C.~D.~White,
    [arXiv:hep-ph/1010.0819].
  

\bibitem{Alioli:2008gx}
  S.~Alioli, P.~Nason, C.~Oleari and E.~Re,
    JHEP {\bf 0807}, 060 (2008)
  [arXiv:hep-ph/0805.4802].
  

\bibitem{Hamilton:2008pd}
  K.~Hamilton, P.~Richardson and J.~Tully,
      JHEP {\bf 0810}, 015 (2008)
  [arXiv:0806.0290 [hep-ph]]

\bibitem{Anastasiou:2003ds}
  C.~Anastasiou, L.~J.~Dixon, K.~Melnikov and F.~Petriello,
      Phys.\ Rev.\  D {\bf 69}, 094008 (2004)
  [arXiv:hep-ph/0312266].
  

\bibitem{Melnikov:2006di}
  K.~Melnikov and F.~Petriello,
    Phys.\ Rev.\ Lett.\  {\bf 96}, 231803 (2006)
  [arXiv:hep-ph/0603182].
  

\bibitem{Catani:2009sm}
  S.~Catani, L.~Cieri, G.~Ferrera, D.~de Florian and M.~Grazzini,
      Phys.\ Rev.\ Lett.\  {\bf 103}, 082001 (2009)
  [arXiv:0903.2120 [hep-ph]].
 
\bibitem{Aaltonen:2007ps} 
  T.~Aaltonen {\it et al.}  [CDF Collaboration],
  Phys.\ Rev.\ D {\bf 77}, 112001 (2008)
  [arXiv:0708.3642 [hep-ex]].

\bibitem{Abazov:2009cp} 
  V.~M.~Abazov {\it et al.}  [D0 Collaboration],
  Phys.\ Rev.\ Lett.\  {\bf 103}, 141801 (2009)
  [arXiv:0908.0766 [hep-ex]].

\bibitem{cdfwmass}
F.~Abe {\it et al.} [CDF Collaboration], Phys. Rev. Lett. {\bf 75}, 11
(1995) and Phys. Rev. D {\bf 52}, 4784 (1995); T.~Affolder {\it et al.} 
[CDF Collaboration], Phys. Rev. D {\bf 64}, 052001 (2001).

\bibitem{d0wmass}
S.~Abachi {\it et al.} [D0 Collaboration], Phys. Rev. Lett.
{\bf 77}, 3309 (1996), B.~Abbott {\it et al.} [D0 Collaboration],
Phys. Rev. D {\bf 58}, 012002 (1998); 
Phys. Rev. D {\bf 58}, 092003 (1998); Phys. Rev. Lett. {\bf 80}, 3008
(1998); Phys. Rev. Lett. {\bf 84}, 222 (2000); Phys. Rev. D {\bf 62} 092006
(2000); V.~M.~Abazov {\it et al.} [D0 Collaboration], 
Phys.\ Rev. D {\bf 66}, 012001 (2002). 

\bibitem{unknown:2003sv}
W.~Ashmanskas {\it et al.}  [TEVEWWG],
arXiv:hep-ex/0311039 and references therein.

\bibitem{Abe:1994qn}
F.~Abe {\it et al.}  [CDF Collaboration],
Phys.\ Rev.\ Lett.\  {\bf 74}, 341 (1995).

\bibitem{Affolder:2000mt}
T.~Affolder {\it et al.}  [CDF Collaboration],
 Phys.\ Rev.\ Lett.\  {\bf 85}, 3347 (2000).

\bibitem{Abazov:2002xj}
V.~M.~Abazov {\it et al.}  [D0 Collaboration],
Phys.\ Rev. D {\bf 66}, 032008 (2002).

\bibitem{Wackeroth:1996hz}
  D.~Wackeroth and W.~Hollik,
      Phys.\ Rev.\  D {\bf 55}, 6788 (1997)
  [arXiv:hep-ph/9606398].
  

\bibitem{Baur:1998kt}
  U.~Baur, S.~Keller and D.~Wackeroth,
      Phys.\ Rev.\  D {\bf 59}, 013002 (1999)
  [arXiv:hep-ph/9807417].
  

\bibitem{Dittmaier:2001ay}
  S.~Dittmaier and M.~Kr\"amer,
      Phys.\ Rev.\  D {\bf 65}, 073007 (2002)
  [arXiv:hep-ph/0109062].
  

\bibitem{Baur:2004ig}
  U.~Baur and D.~Wackeroth,
      Phys.\ Rev.\  D {\bf 70}, 073015 (2004)
  [arXiv:hep-ph/0405191].
  

\bibitem{Arbuzov:2005dd}
  A.~Arbuzov, D.~Bardin, S.~Bondarenko, P.~Christova, L.~Kalinovskaya, G.~Nanava and R.~Sadykov,
      Eur.\ Phys.\ J.\  C {\bf 46}, 407 (2006)
  [Erratum-ibid.\  C {\bf 50}, 505 (2007)]
  [arXiv:hep-ph/0506110].
  

\bibitem{CarloniCalame:2006zq}
  C.~M.~Carloni Calame, G.~Montagna, O.~Nicrosini and A.~Vicini,
      JHEP {\bf 0612}, 016 (2006)
  [arXiv:hep-ph/0609170].
  

\bibitem{Zykunov:2008zz}
  V.~A.~Zykunov,
      Phys.\ Atom.\ Nucl.\  {\bf 71}, 732 (2008);
        Eur.\ Phys.\ J.\ direct C {\bf 3}, 9 (2001)
  [arXiv:hep-ph/0107059].
  

\bibitem{Buttar:2006zd}
  C.~Buttar {\it et al.},
      [arXiv:hep-ph/0604120.]
  

\bibitem{Gerber:2007xk}
  C.~E.~Gerber {\it et al.}  [TeV4LHC-Top and Electroweak Working Group],
    [arXiv:hep-ph/0705.3251].
  

\bibitem{CarloniCalame:2003ux}
C.~M.~Carloni Calame, G.~Montagna, O.~Nicrosini and M.~Treccani,
 Phys.\ Rev. D {\bf 69}, 037301 (2004).

\bibitem{Placzek:2003zg}
W.~Placzek and S.~Jadach,
Eur.\ Phys.\ J. C {\bf 29}, 325 (2003).

\bibitem{Golonka:2005pn}
  P.~Golonka and Z.~Was,
      Eur.\ Phys.\ J.\  C {\bf 45}, 97 (2006)
  [arXiv:hep-ph/0506026].
  

\bibitem{Hamilton:2006xz}
  K.~Hamilton and P.~Richardson,
    JHEP {\bf 0607}, 010 (2006)
  [arXiv:hep-ph/0603034].
  

\bibitem{Brensing:2007qm}
  S.~Brensing, S.~Dittmaier, M.~Kr\"amer and A.~M\"uck,
      Phys.\ Rev.\  D {\bf 77}, 073006 (2008)
  [arXiv:0710.3309 [hep-ph]].
  

\bibitem{Laenen:2009zz}
        E.~Laenen and D.~Wackeroth,
                Ann.\ Rev.\ Nucl.\ Part.\ Sci.\  {\bf 59}, 367 (2009).
        

\bibitem{Cao:2004yy}
  Q.~H.~Cao and C.~P.~Yuan,
      Phys.\ Rev.\ Lett.\  {\bf 93}, 042001 (2004)
  [arXiv:hep-ph/0401026].
  

\bibitem{Balossini:2009sa}
  G.~Balossini {\it et al.},
      JHEP {\bf 1001}, 013 (2010)
  [arXiv:hep-ph/0907.0276].
  

\bibitem{Corcella:2000bw}
  G.~Corcella {\it et al.},
    JHEP {\bf 0101 } (2001)  010.
  [hep-ph/0011363].

\bibitem{Arbuzov:2007db}
  A.~Arbuzov, D.~Bardin, S.~Bondarenko, P.~Christova, L.~Kalinovskaya, G.~Nanava, R.~Sadykov,
    Eur.\ Phys.\ J.\ C {\bf 54}, 451 (2008)
  [arXiv:0711.0625 [hep-ph]].
  

\bibitem{Sjostrand:2006za}
  T.~Sjostrand, S.~Mrenna, P.~Z.~Skands,
    JHEP {\bf 0605 } (2006)  026.
  [hep-ph/0603175].


\bibitem{Richardson:2010gz}
  P.~Richardson, R.~R.~Sadykov, A.~A.~Sapronov, M.~H.~Seymour and P.~Z.~Skands,
    arXiv:1011.5444 [hep-ph].
  

\bibitem{Bernaciak:2012hj} 
  C.~Bernaciak and D.~Wackeroth,
    Phys.\ Rev.\ D {\bf 85}, 093003 (2012)
  [arXiv:1201.4804 [hep-ph]].
  

\bibitem{Barze:2012tt} 
  L.~Barze, G.~Montagna, P.~Nason, O.~Nicrosini and F.~Piccinini,
    JHEP {\bf 1204}, 037 (2012)
  [arXiv:1202.0465 [hep-ph]].
    

\bibitem{Barze:2013yca} 
  L.~Barze, G.~Montagna, P.~Nason, O.~Nicrosini, F.~Piccinini and A.~Vicini,
    Eur.\ Phys.\ J.\ C {\bf 73}, 2474 (2013)
  [arXiv:1302.4606 [hep-ph]].
    
\bibitem{robertprivate}
Andreas von Manteuffel and Robert Schabinger, private communication.

\bibitem{Kilgore:2011pa}
  W.~B.~Kilgore and C.~Sturm,
  Phys.\ Rev.\ D {\bf 85}, 033005 (2012)
  [arXiv:1107.4798 [hep-ph]].

\bibitem{stefanradcorr}
S.~Dittmaier, talk given at RADCOR 2013, www.ippp.dur.ac.uk/Workshops/13/RADCOR2013.

\bibitem{Bozzi:2011ww}
  G.~Bozzi, J.~Rojo and A.~Vicini,
    Phys.\ Rev.\ D {\bf 83} (2011) 113008
  [arXiv:1104.2056 [hep-ph]].
    

\bibitem{Ball:2012cx}
  R.~D.~Ball, V.~Bertone, S.~Carrazza, C.~S.~Deans, L.~Del Debbio, S.~Forte, A.~Guffanti and N.~P.~Hartland {\it et al.},
    Nucl.\ Phys.\ B {\bf 867} (2013) 244
  [arXiv:1207.1303 [hep-ph]].
    

\bibitem{Rojo:2013nia} 
  J.~Rojo and A.~Vicini,
  arXiv:1309.1311 [hep-ph].

\bibitem{Martin:2009iq}
  A.~D.~Martin, W.~J.~Stirling, R.~S.~Thorne, G.~Watt,
    Eur.\ Phys.\ J.\  {\bf C63 } (2009)  189-285.
  [arXiv:0901.0002 [hep-ph]].

\bibitem{Lai:2010vv} 
  H.~-L.~Lai, M.~Guzzi, J.~Huston, Z.~Li, P.~M.~Nadolsky, J.~Pumplin and C.~-P.~Yuan,
  Phys.\ Rev.\ D {\bf 82}, 074024 (2010)
  [arXiv:1007.2241 [hep-ph]].

\bibitem{Demartin:2010er} 
  F.~Demartin, S.~Forte, E.~Mariani, J.~Rojo and A.~Vicini,
    Phys.\ Rev.\ D {\bf 82}, 014002 (2010)
  [arXiv:1004.0962 [hep-ph]].
    

\bibitem{Ball:2011mu} 
  R.~D.~Ball, V.~Bertone, F.~Cerutti, L.~Del Debbio, S.~Forte, A.~Guffanti, J.~I.~Latorre and J.~Rojo {\it et al.},
    Nucl.\ Phys.\ B {\bf 849}, 296 (2011)
  [arXiv:1101.1300 [hep-ph]].
    

\bibitem{Martin:2004dh} 
  A.~D.~Martin, R.~G.~Roberts, W.~J.~Stirling and R.~S.~Thorne,
  Eur.\ Phys.\ J.\ C {\bf 39}, 155 (2005)
  [hep-ph/0411040].

\bibitem{Ball:2013hta} 
  R.~D.~Ball {\it et al.}  [ The NNPDF Collaboration],
    arXiv:1308.0598 [hep-ph].
  
\bibitem{Aaltonen:2012bp} 
  T.~Aaltonen {\it et al.}  [CDF Collaboration],
  Phys.\ Rev.\ Lett.\  {\bf 108}, 151803 (2012)
  [arXiv:1203.0275 [hep-ex]].

\bibitem{Abazov:2012bv} 
  V.~M.~Abazov {\it et al.}  [D0 Collaboration],
  Phys.\ Rev.\ Lett.\  {\bf 108}, 151804 (2012)
  [arXiv:1203.0293 [hep-ex]].

\bibitem{Aaltonen:2013iut} 
  T.~A.~Aaltonen {\it et al.}  [CDF and D0 Collaborations],
  Phys.\ Rev.\ D {\bf 88}, 052018 (2013)
  [arXiv:1307.7627 [hep-ex]].

\bibitem{Bodek:2010}
A. Bodek, Eur.Phys.J.C67:321 (2010),  arXiv:0911.2850 [hep-ex]

\bibitem{Bodek:2012}
A. Bodek et. al. , Eur.Phys.J.C72: 2194 (2012),  arXiv:1208.3710 [hep-ex]

\bibitem{Aaltonen:2013wcp} 
  T.~Aaltonen {\it et al.}  [CDF Collaboration],
    arXiv:1307.0770 [hep-ex].
  
\bibitem{Abazov:2011ws} 
  V.~M.~Abazov {\it et al.}  [D0 Collaboration],
  Phys.\ Rev.\ D {\bf 84}, 012007 (2011)
  [arXiv:1104.4590 [hep-ex]].

    

\bibitem{atlassin2theta}
The ATLAS collaboration, ATLAS-CONF-2013-043

\bibitem{Denner:1999kn} 
  A.~Denner, S.~Dittmaier, M.~Roth and D.~Wackeroth,
  Phys.\ Lett.\ B {\bf 475}, 127 (2000)
  [hep-ph/9912261];
          Nucl.\ Phys.\ B {\bf 587}, 67 (2000)
  [hep-ph/0006307].
    

\bibitem{Jadach:2000kw} 
  S.~Jadach, W.~Placzek, M.~Skrzypek, B.~F.~L.~Ward and Z.~Was,
    Phys.\ Rev.\ D {\bf 65}, 093010 (2002)
  [hep-ph/0007012].
    

\bibitem{Denner:2005es} 
  A.~Denner, S.~Dittmaier, M.~Roth and L.~H.~Wieders,
    Phys.\ Lett.\ B {\bf 612}, 223 (2005)
  [Erratum-ibid.\ B {\bf 704}, 667 (2011)]
  [hep-ph/0502063];
         Nucl.\ Phys.\ B {\bf 724}, 247 (2005)
  [Erratum-ibid.\ B {\bf 854}, 504 (2012)]
  [hep-ph/0505042].
    

\bibitem{Fadin:1999bq} 
  V.~S.~Fadin, L.~N.~Lipatov, A.~D.~Martin and M.~Melles,
    Phys.\ Rev.\ D {\bf 61}, 094002 (2000)
  [hep-ph/9910338];
      A.~Denner, M.~Melles and S.~Pozzorini,
    Nucl.\ Phys.\ B {\bf 662}, 299 (2003)
  [hep-ph/0301241];
      M.~Beccaria, F.~M.~Renard and C.~Verzegnassi,
    Nucl.\ Phys.\ B {\bf 663}, 394 (2003)
  [hep-ph/0304175].
    

\bibitem{Kuhn:2007ca} 
  J.~H.~K\"uhn, F.~Metzler and A.~A.~Penin,
    Nucl.\ Phys.\ B {\bf 795}, 277 (2008)
  [Erratum-ibid.\  {\bf 818}, 135 (2009)]
  [arXiv:0709.4055 [hep-ph]].
    

\bibitem{Fadin:1995fp}
V.~S. Fadin, V.~A. Khoze, A.~D. Martin and W.~J. Stirling,
\newblock Phys. Lett. {\bf B363}, 112 (1995).

\bibitem{Actis:2008rb} 
  S.~Actis, M.~Beneke, P.~Falgari and C.~Schwinn,
    Nucl.\ Phys.\ B {\bf 807}, 1 (2009)
  [arXiv:0807.0102 [hep-ph]].
    

\bibitem{Beneke:2007zg} 
  M.~Beneke, P.~Falgari, C.~Schwinn, A.~Signer and G.~Zanderighi,
    Nucl.\ Phys.\ B {\bf 792}, 89 (2008)
  [arXiv:0707.0773 [hep-ph]].
    

\bibitem{Skrzypek:1992vk}
M.~Skrzypek,
\newblock Acta Phys. Polon. {\bf B23}, 135 (1992).

\bibitem{Wilson_2001aw}
``Precision measurement of the $W$ 
mass with a polarized threshold scan at a linear collider.''
{}G.~W.~Wilson.
{}In *2nd ECFA/DESY Study 1998-2001* 1498-1505 {}LC-PHSM-2001-09 and in Proceedings of LCWS 1999, Sitges, Spain.

\bibitem{Wilson_LC2013}
``Investigating In-Situ Center-of-Mass Energy Determination with Di-Muon Events'', 
G.~W.~Wilson, ECFA LC2013 Workshop, Hamburg, May2013.

\bibitem{Schael:2013ita} 
  S.~Schael {\it et al.}  [ALEPH and DELPHI and L3 and OPAL and LEP Electroweak Working Group Collaborations],
  ``Electroweak Measurements in Electron-Positron Collisions at W-Boson-Pair Energies at LEP,''
  arXiv:1302.3415 [hep-ex].
    

\bibitem{Beckmann:2010ib} 
  M.~Beckmann, B.~List and J.~List,
  ``Treatment of Photon Radiation in Kinematic Fits at Future e+ e- Colliders,''
  Nucl.\ Instrum.\ Meth.\ A {\bf 624}, 184 (2010)
  [arXiv:1006.0436 [hep-ex]].
    

\bibitem{Abramowicz:2013tzc} 
  H.~Abramowicz {\it et al.}  [CLIC Detector and Physics Study Collaboration],
    arXiv:1307.5288 [hep-ex].
    
\bibitem{TLEPwhitepaper}
  M.~Bicer, H.~Duran Yildiz, I.~Yildiz, G.~Coignet, M.~Delmastro, T.~Alexopoulos, C.~Grojean and S.~Antusch {\it et al.},
  arXiv:1308.6176 [hep-ex].


\bibitem{kmogigaz} R.~Hawkings and K.~Moenig, 
  Eur.\ Phys.\ J.\ direct C {\bf 1}, 8 (1999), 
                      hep-ex/9910022.
                       

\bibitem{peter_mike}
                    P.~C.~Rowson and M.~Woods,
                    {\it Experimental issues for precision electroweak 
                    physics at a high-luminosity Z factory}, 
                    hep-ex/0012055,
                    Proceedings of the ``Linear Collider Workshop 2000'',
                    Fermilab Natl.\ Lab., October 2000.
                    

\bibitem{Erler:2002su} 
  J.~Erler, K.~Flottmann, S.~Heinemeyer, K.~Moenig, G.~A.~Moortgat-Pick, P.~C.~Rowson, E.~Torrence and G.~Weiglein {\it et al.},
    eConf C {\bf 010630}, E3004 (2001)
  [hep-ph/0112070].

\bibitem{alrsld}  
  SLD Collaboration, K. Abe et al., Phys. Rev. Lett. {\bf 84} (2000) 5945.

\bibitem{Boogert:2009ir} 
  S.~Boogert, M.~Hildreth, D.~Kafer, J.~List, K.~Moenig, K.~C.~Moffeit, G.~Moortgat-Pick and S.~Riemann {\it et al.},
    JINST {\bf 4}, P10015 (2009)
  [arXiv:0904.0122 [physics.ins-det]].

\bibitem{Blondel:1987wr} 
  A.~Blondel,
    Phys.\ Lett.\ B {\bf 202}, 145 (1988)
  [Erratum-ibid.\  {\bf 208}, 531 (1988)].

\bibitem{Group:2012gb} 
  Tevatron Electroweak Working Group [CDF and D0 Collaborations],
  arXiv:1204.0042 [hep-ex].

\bibitem{Assman:1995tb} 
  R.~Assman, A.~Blondel, B.~Dehning, A.~Drees, P.~Grosse-Wiesmann, H.~Grote, R.~Jacobsen and J.~P.~Koutchouk {\it et al.},
    CERN-SL-95-21.
    
\bibitem{Assmann:2004gc} 
  R.~Assmann {\it et al.}  [LEP Energy Working Group Collaboration],
    Eur.\ Phys.\ J.\ C {\bf 39}, 253 (2005)
  [hep-ex/0410026].

\bibitem{Flacher:2008zq} 
  H.~Flacher, M.~Goebel, J.~Haller, A.~Hocker, K.~Moenig and J.~Stelzer,
    Eur.\ Phys.\ J.\ C {\bf 60}, 543 (2009)
  [Erratum-ibid.\ C {\bf 71}, 1718 (2011)]
  [arXiv:0811.0009 [hep-ph]].
    

\bibitem{Baak:2011ze} 
  M.~Baak, M.~Goebel, J.~Haller, A.~Hoecker, D.~Ludwig, K.~Moenig, M.~Schott and J.~Stelzer,
    Eur.\ Phys.\ J.\ C {\bf 72}, 2003 (2012)
  [arXiv:1107.0975 [hep-ph]].
    
\bibitem{Baak:2012kk} 
  M.~Baak, M.~Goebel, J.~Haller, A.~Hoecker, D.~Kennedy, R.~Kogler, K.~Moenig and M.~Schott {\it et al.},
    Eur.\ Phys.\ J.\ C {\bf 72}, 2205 (2012)
  [arXiv:1209.2716 [hep-ph]].


\bibitem{Peskin:1990zt} 
  M.~E.~Peskin and T.~Takeuchi,
    Phys.\ Rev.\ Lett.\  {\bf 65}, 964 (1990).
    

\bibitem{Heinemeyer:2006px}
  S.~Heinemeyer, W.~Hollik, D.~Stockinger, A.~M.~Weber and G.~Weiglein,
  {\em JHEP} {\bf 0608} (2006) 052
  [arXiv:hep-ph/0604147].
  

\bibitem{MWlisa} S.~Heinemeyer, G.~Weiglein and L.~Zeune,
                 DESY 13--015,
                 {\em in preparation}.

\bibitem{Frank:2006yh}
  M.~Frank, T.~Hahn, S.~Heinemeyer, W.~Hollik, H.~Rzehak and G.~Weiglein,
  JHEP {\bf 0702} (2007) 047
  [hep-ph/0611326].
  

\bibitem{Degrassi:2002fi}
  G.~Degrassi, S.~Heinemeyer, W.~Hollik, P.~Slavich and G.~Weiglein,
  Eur.\ Phys.\ J.\ C {\bf 28} (2003) 133
  [hep-ph/0212020].
  

\bibitem{Heinemeyer:1998np}
  S.~Heinemeyer, W.~Hollik and G.~Weiglein,
  Eur.\ Phys.\ J.\ C {\bf 9} (1999) 343
  [hep-ph/9812472].
  

\bibitem{Heinemeyer:1998yj}
  S.~Heinemeyer, W.~Hollik and G.~Weiglein,
  Comput.\ Phys.\ Commun.\  {\bf 124} (2000) 76
  [hep-ph/9812320].
  

\bibitem{Hahn:2009zz}
  T.~Hahn, S.~Heinemeyer, W.~Hollik, H.~Rzehak and G.~Weiglein,
  Comput.\ Phys.\ Commun.\  {\bf 180} (2009) 1426.
  

\bibitem{Bechtle:2008jh}
  P.~Bechtle, O.~Brein, S.~Heinemeyer, G.~Weiglein and K.~E.~Williams,
  Comput.\ Phys.\ Commun.\  {\bf 181} (2010) 138
  [arXiv:0811.4169 [hep-ph]].
  
 

\bibitem{Bechtle:2011sb}
  P.~Bechtle, O.~Brein, S.~Heinemeyer, G.~Weiglein and K.~E.~Williams,
  Comput.\ Phys.\ Commun.\  {\bf 182} (2011) 2605
  [arXiv:1102.1898 [hep-ph]].
  

\bibitem{Heinemeyer:2011aa}
  S.~Heinemeyer, O.~St{\aa}l and G.~Weiglein, 
  Phys.\ Lett.\ B {\bf 710} (2012) 201
                [arXiv:1112.3026 [hep-ph]]; 
  

\bibitem{Heinemeyer:2007bw}
  S.~Heinemeyer, W.~Hollik, A.~M.~Weber and G.~Weiglein,
  JHEP {\bf 0804} (2008) 039
  [arXiv:0710.2972 [hep-ph]].
  

\bibitem{Allanach:2002nj}
  B.~C.~Allanach, M.~Battaglia, G.~A.~Blair, M.~S.~Carena, A.~De Roeck, A.~Dedes, A.~Djouadi and D.~Gerdes {\it et al.},
  Eur.\ Phys.\ J.\ C {\bf 25} (2002) 113
  [hep-ph/0202233].
  

\bibitem{Barate:2003sz}
  R.~Barate {\it et al.}  [LEP Working Group for Higgs boson searches and ALEPH and DELPHI and L3 and OPAL Collaborations],
  Phys.\ Lett.\ B {\bf 565} (2003) 61
  [hep-ex/0306033].

\bibitem{Schael:2006cr}
  S.~Schael {\it et al.}  [ALEPH and DELPHI and L3 and OPAL and LEP Working Group for Higgs Boson Searches Collaborations],
  Eur.\ Phys.\ J.\ C {\bf 47} (2006) 547
  [hep-ex/0602042].
  
\bibitem{Erler:2009jh} 
  J.~Erler, P.~Langacker, S.~Munir and E.~Rojas,
  JHEP {\bf 0908}, 017 (2009)
  [arXiv:0906.2435 [hep-ph]].

\bibitem{Reuter:2013gla} 
  J.~Reuter, W.~Kilian and M.~Sekulla,
  arXiv:1307.8170 [hep-ph].

\bibitem{Degrande:2013rea} 
  C.~Degrande, O.~Eboli, B.~Feigl, B.~Jaeger, W.~Kilian, O.~Mattelaer, M.~Rauch and Jür.~Reuter {\it et al.},
  arXiv:1309.7890 [hep-ph].

\bibitem{Grzadkowski:2010es} 
  B.~Grzadkowski, M.~Iskrzynski, M.~Misiak and J.~Rosiek,
    JHEP {\bf 1010}, 085 (2010)
  [arXiv:1008.4884 [hep-ph]].
      

\bibitem{Arzt:1994gp} 
  C.~Arzt, M.~B.~Einhorn and J.~Wudka,
    Nucl.\ Phys.\ B {\bf 433}, 41 (1995)
  [hep-ph/9405214].
    

\bibitem{Eboli:2006wa} O.~J.~P.~Eboli, M.~C.~Gonzalez-Garcia and
  J.~K.~Mizukoshi, 
Phys.\ Rev.\ D {\bf
    74}, 073005 (2006) [hep-ph/0606118].
  

    

\bibitem{Belyaev:1998ih} 
  A.~S.~Belyaev, O.~J.~P.~Eboli, M.~C.~Gonzalez-Garcia, J.~K.~Mizukoshi, S.~F.~Novaes and I.~Zacharov,
    Phys.\ Rev.\ D {\bf 59}, 015022 (1999)
  [hep-ph/9805229];
      O.~J.~P.~Eboli, M.~C.~Gonzalez-Garcia and S.~M.~Lietti,
    Phys.\ Rev.\ D {\bf 69}, 095005 (2004)
  [hep-ph/0310141];
      O.~J.~P.~Eboli, M.~C.~Gonzalez-Garcia, S.~M.~Lietti and S.~F.~Novaes,
    Phys.\ Rev.\ D {\bf 63}, 075008 (2001)
  [hep-ph/0009262].

\bibitem{Eboli:2000ad} 
  O.~J.~P.~Eboli, M.~C.~Gonzalez-Garcia, S.~M.~Lietti and S.~F.~Novaes,
  Phys.\ Rev.\ D {\bf 63}, 075008 (2001)
  [hep-ph/0009262].

\bibitem{Eboli:2003nq} 
  O.~J.~P.~Eboli, M.~C.~Gonzalez-Garcia and S.~M.~Lietti,
  Phys.\ Rev.\ D {\bf 69}, 095005 (2004)
  [hep-ph/0310141].

\bibitem{Hagiwara:1986vm} 
  K.~Hagiwara, R.~D.~Peccei, D.~Zeppenfeld and K.~Hikasa,
    Nucl.\ Phys.\ B {\bf 282}, 253 (1987).
    
    
\bibitem{Hagiwara:1993ck} 
  K.~Hagiwara, S.~Ishihara, R.~Szalapski and D.~Zeppenfeld,
    Phys.\ Rev.\ D {\bf 48}, 2182 (1993).
  

\bibitem{Wudka:1994ny} 
  J.~Wudka,
    Int.\ J.\ Mod.\ Phys.\ A {\bf 9}, 2301 (1994)
  [hep-ph/9406205].
    

\bibitem{quarticATlep} 
  P.~S.~Wells,
  Eur.\ Phys.\ J.\ C {\bf 33}, S5 (2004); J. Beringer et al. (Particle Data Group), Phys. Rev. D86, 010001 (2012).

\bibitem{Stirling:1999ek} 
  W.~J.~Stirling and A.~Werthenbach,
    Eur.\ Phys.\ J.\ C {\bf 14}, 103 (2000)
  [hep-ph/9903315].
    

\bibitem{Belanger:1992qh} 
  G.~Belanger and F.~Boudjema,
    Phys.\ Lett.\ B {\bf 288}, 201 (1992).
    

\bibitem{Eboli:1993wg} 
  O.~J.~P.~Eboli, M.~C.~Gonzalez-Garcia and S.~F.~Novaes,
    Nucl.\ Phys.\ B {\bf 411}, 381 (1994)
  [hep-ph/9306306].


\bibitem{Eboli:anom4}
  \href{http://feynrules.irmp.ucl.ac.be/wiki/AnomalousGaugeCoupling}{http://feynrules.irmp.ucl.ac.be/wiki/AnomalousGaugeCoupling}

\bibitem{Alboteanu:2008my}
  A.~Alboteanu, W.~Kilian and J.~Reuter,
    JHEP {\bf 0811} (2008) 010
  [arXiv:0806.4145 [hep-ph]].
  

\bibitem{Degrande:2012wf} 
  C.~Degrande, N.~Greiner, W.~Kilian, O.~Mattelaer, H.~Mebane, T.~Stelzer, S.~Willenbrock and C.~Zhang,
    arXiv:1205.4231 [hep-ph].
  

\bibitem{Jager:2011ms}
  B.~J\"ager and G.~Zanderighi,
    JHEP {\bf 1111} (2011) 055
  [arXiv:1108.0864 [hep-ph]].

\bibitem{Jager:2006zc}
  B.~J\"ager, C.~Oleari, D.~Zeppenfeld, 
  JHEP {\bf 07 } (2006)  015.
  [arXiv:hep-ph/0603177].

\bibitem{Jager:2006cp} 
  B.~Jager, C.~Oleari and D.~Zeppenfeld,
    Phys.\ Rev.\ D {\bf 73}, 113006 (2006)
  [hep-ph/0604200].
  

\bibitem{Bozzi:2007ur} 
  G.~Bozzi, B.~Jager, C.~Oleari and D.~Zeppenfeld,
    Phys.\ Rev.\ D {\bf 75}, 073004 (2007)
  [hep-ph/0701105].
  

\bibitem{Jager:2009xx}
  B.~J\"ager, C.~Oleari, D.~Zeppenfeld, 
  Phys.\ Rev.\  {\bf D80 } (2009)  034022.
  [arXiv:0907.0580 [hep-ph]].

\bibitem{Denner:2012dz}
  A.~Denner, L.~Hosekova and S.~Kallweit,
    Phys.\ Rev.\ D {\bf 86} (2012) 114014
  [arXiv:1209.2389 [hep-ph]].
  

\bibitem{Nason:2004rx}
  P.~Nason,
    JHEP {\bf 0411 } (2004)  040.
  [hep-ph/0409146].

\bibitem{Frixione:2007vw}
  S.~Frixione, P.~Nason, C.~Oleari,
    JHEP {\bf 0711 } (2007)  070.
  [arXiv:0709.2092 [hep-ph]].

\bibitem{Alioli:2010xd}
  S.~Alioli, P.~Nason, C.~Oleari, E. Re,
    JHEP {\bf 1006 } (2010)  043.
  [arXiv:1002.2581 [hep-ph]].

\bibitem{Figy:2003nv}
  T.~Figy, C.~Oleari and D.~Zeppenfeld,
    Phys.\ Rev.\ D {\bf 68} (2003) 073005
  [hep-ph/0306109].

\bibitem{Oleari:2003tc}
  C.~Oleari and D.~Zeppenfeld,
    Phys.\ Rev.\ D {\bf 69} (2004) 093004
  [hep-ph/0310156].

\bibitem{Nason:2009ai}
  P.~Nason and C.~Oleari,
    JHEP {\bf 1002} (2010) 037
  [arXiv:0911.5299 [hep-ph]].

\bibitem{Jager:2012xk}
  B.~J\"ager, S.~Schneider and G.~Zanderighi,
    JHEP {\bf 1209} (2012) 083
  [arXiv:1207.2626 [hep-ph]].

\bibitem{Schissler:2013nga}
  F.~Schissler and D.~Zeppenfeld,
    JHEP {\bf 04} (2013) 057
   [JHEP {\bf 1304} (2013) 057]
  [arXiv:1302.2884 [hep-ph]].

\bibitem{Jager:2013mu}
  B.~J\"ager and G.~Zanderighi,
    JHEP {\bf 1304} (2013) 024
  [arXiv:1301.1695 [hep-ph]].



\bibitem{Cacciari:2005hq}
M.~Cacciari, G.~P.~Salam, 
Phys.\ Lett.\ {\bf B641} (2006) 57. 
[hep-ph/0512210]. 


\bibitem{Lazopoulos:2007ix} 
  A.~Lazopoulos, K.~Melnikov and F.~Petriello,
    Phys.\ Rev.\ D {\bf 76}, 014001 (2007)
  [hep-ph/0703273].
  

\bibitem{Hankele:2007sb} 
  V.~Hankele and D.~Zeppenfeld,
    Phys.\ Lett.\ B {\bf 661}, 103 (2008)
  [arXiv:0712.3544 [hep-ph]].
  

\bibitem{Campanario:2008yg} 
  F.~Campanario, V.~Hankele, C.~Oleari, S.~Prestel and D.~Zeppenfeld,
    Phys.\ Rev.\ D {\bf 78}, 094012 (2008)
  [arXiv:0809.0790 [hep-ph]].
  

\bibitem{Binoth:2008kt} 
  T.~Binoth, G.~Ossola, C.~G.~Papadopoulos and R.~Pittau,
    JHEP {\bf 0806}, 082 (2008)
  [arXiv:0804.0350 [hep-ph]].
  

\bibitem{Bozzi:2009ig} 
  G.~Bozzi, F.~Campanario, V.~Hankele and D.~Zeppenfeld,
    Phys.\ Rev.\ D {\bf 81}, 094030 (2010)
  [arXiv:0911.0438 [hep-ph]].
  

\bibitem{Bozzi:2010sj} 
  G.~Bozzi, F.~Campanario, M.~Rauch, H.~Rzehak and D.~Zeppenfeld,
    Phys.\ Lett.\ B {\bf 696}, 380 (2011)
  [arXiv:1011.2206 [hep-ph]].
  

\bibitem{Baur:2010zf} 
  U.~Baur, D.~Wackeroth and M.~M.~Weber,
    PoS RADCOR {\bf 2009}, 067 (2010)
  [arXiv:1001.2688 [hep-ph]].
  

\bibitem{Bozzi:2011wwa} 
  G.~Bozzi, F.~Campanario, M.~Rauch and D.~Zeppenfeld,
    Phys.\ Rev.\ D {\bf 83}, 114035 (2011)
  [arXiv:1103.4613 [hep-ph]].
  

\bibitem{Bozzi:2011en} 
  G.~Bozzi, F.~Campanario, M.~Rauch and D.~Zeppenfeld,
    Phys.\ Rev.\ D {\bf 84}, 074028 (2011)
  [arXiv:1107.3149 [hep-ph]].
  

\bibitem{Bozzi:2012mh} 
  G.~Bozzi, F.~Campanario, C.~NEnglert, M.~Rauch, M.~Spannoswky and D.~Zeppenfeld,
    arXiv:1205.2506 [hep-ph].
  

\bibitem{Campbell:2012ft} 
  J.~M.~Campbell, H.~B.~Hartanto and C.~Williams,
    JHEP {\bf 1211}, 162 (2012)
  [arXiv:1208.0566 [hep-ph]].
  

\bibitem{Brown:1982xx} 
  R.~W.~Brown, K.~L.~Kowalski and S.~J.~Brodsky,
    Phys.\ Rev.\ D {\bf 28}, 624 (1983).
  

\bibitem{Baur:1993ir} 
  U.~Baur, T.~Han and J.~Ohnemus,
    Phys.\ Rev.\ D {\bf 48}, 5140 (1993)
  [hep-ph/9305314].
  

\bibitem{Baur:1997bn} 
  U.~Baur, T.~Han, N.~Kauer, R.~Sobey and D.~Zeppenfeld,
    Phys.\ Rev.\ D {\bf 56}, 140 (1997)
  [hep-ph/9702364].

\bibitem{Nhung:2013jta} 
  D.~T.~Nhung, L.~D.~Ninh and M.~M.~Weber,
  arXiv:1307.7403 [hep-ph].
  
\bibitem{Aihara:1995iq} 
  H.~Aihara, T.~Barklow, U.~Baur, J.~Busenitz, S.~Errede, T.~A.~Fuess, T.~Han and D.~London {\it et al.},
  In *Barklow, T.L. (ed.) et al.: Electroweak symmetry breaking and new physics at the TeV scale* 488-546
  [hep-ph/9503425].

\bibitem{Lombardo:2013daa} 
  V.~Lombardo [on behalf of the ATLAS and CMS Collaboration],
    arXiv:1305.3773 [hep-ex].
  
\bibitem{Aad:2012awa} 
  G.~Aad {\it et al.}  [ATLAS Collaboration],
  JHEP {\bf 1303}, 128 (2013)
  [arXiv:1211.6096 [hep-ex]].

\bibitem{Aad:2012mr} 
  G.~Aad {\it et al.}  [ATLAS Collaboration],
  Phys.\ Lett.\ B {\bf 717}, 49 (2012)
  [arXiv:1205.2531 [hep-ex]].

\bibitem{Chatrchyan:2013foa} 
  S.~Chatrchyan {\it et al.}  [CMS Collaboration],
  JHEP {\bf 1307}, 116 (2013)
  [arXiv:1305.5596 [hep-ex]].

\bibitem{ATLAS-Collaboration:1496527}
      ATLAS Collaboration,
CERN preprint ATL-PHYS-PUB-2012-005,
     {\it http://cds.cern.ch/record/1496527}. 

\bibitem{ATLAS-Collaboration:1558703}
     ATLAS Collaboration,
    CERN Preprint ATL-PHYS-PUB-2013-006,
{\it http://cds.cern.ch/record/1558703}. 

\bibitem{Truong:1988zp}
 T.~N.~Truong,
 Phys.\ Rev.\ Lett.\  {\bf 61}, 2526 (1988).

\bibitem{Ballestrero:2011pe} 
  A.~Ballestrero, D.~B.~Franzosi, L.~Oggero and E.~Maina,
  JHEP {\bf 1203}, 031 (2012)
  [arXiv:1112.1171 [hep-ph]], and references therein.

\bibitem{snowmass-multiboson-whitepaper}
  C.~Degrande, J.~L.~Holzbauer, S.~-C.~Hsu, A.~V.~Kotwal, S.~Li, M.~Marx, O.~Mattelaer and J.~Metcalfe {\it et al.},
  SNOW13-00189,
  arXiv:1309.7452 [physics.comp-ph].

\bibitem{ilc:rdr} 
  G.~Aarons {\it et al.}  [ILC Collaboration],
  arXiv:0709.1893 [hep-ph].

\bibitem{Freitas:2013xga} 
  A.~Freitas, K.~Hagiwara, S.~Heinemeyer, P.~Langacker, K.~Moenig, M.~Tanabashi and G.~W.~Wilson,
    arXiv:1307.3962 [hep-ph].
    

\bibitem{Accomando:2004sz}
E.~Accomando \textit{et~al.}, Physics at the CLIC multi-TeV linear collider,
  2004, hep-ph/0412251.


\bibitem{Beyer:2006hx} 
  M.~Beyer, W.~Kilian, P.~Krstonosic, K.~Moenig, J.~Reuter,
  E.~Schmidt and H.~Schroder, 
      Eur.\ Phys.\ J.\ C {\bf 48}, 353 (2006)
  [hep-ph/0604048].
  

\bibitem{Baer:2013cma} 
  H.~Baer, T.~Barklow, K.~Fujii, Y.~Gao, A.~Hoang, S.~Kanemura, J.~List and H.~E.~Logan {\it et al.},
    arXiv:1306.6352 [hep-ph].
    

\bibitem{EWLag}
  S.~Weinberg,
    Phys.\ Rev.\  {\bf 166}, 1568 (1968);
    M.~S.~Chanowitz and M.~K.~Gaillard,
    Nucl.\ Phys.\ B {\bf 261}, 379 (1985);
    W.~Kilian,
    Springer Tracts Mod.\ Phys.\  {\bf 198}, 1 (2003).
  


  

\bibitem{Kilian:2007gr} 
  W.~Kilian, T.~Ohl and J.~Reuter,
    Eur.\ Phys.\ J.\ C {\bf 71}, 1742 (2011)
  [arXiv:0708.4233 [hep-ph]].
    
\bibitem{KRS}
  W.~Kilian, J.~Reuter, M.~Sekulla,
  {\em Search for New Physics in Vector boson scattering},
  in preparation. 


  

  

  

  

\bibitem{effective} 
S.~Weinberg,
    Physica A {\bf 96}, 327 (1979).
    

\bibitem{Corbett:2012ja} 
  T.~Corbett, O.~J.~P.~Eboli, J.~Gonzalez-Fraile and M.~C.~Gonzalez-Garcia,
    Phys.\ Rev.\ D {\bf 87}, 015022 (2013)
  [arXiv:1211.4580 [hep-ph]].
    

\bibitem{Corbett:2012dm} 
  T.~Corbett, O.~J.~P.~Eboli, J.~Gonzalez-Fraile and M.~C.~Gonzalez-Garcia,
    Phys.\ Rev.\ D {\bf 86}, 075013 (2012)
  [arXiv:1207.1344 [hep-ph]].
    

\bibitem{Buchmuller:1986iq} 
  W.~Buchm\"uller and D.~Wyler,
    Phys.\ Lett.\ B {\bf 177}, 377 (1986).

\bibitem{59}
B.~Grzadkowski, M.~Iskrzynski, M.~Misiak and J.~Rosiek,
    JHEP {\bf 1010}, 085 (2010)
  [arXiv:1008.4884 [hep-ph]].
    


\bibitem{eom}
H.~D.~Politzer,
    Nucl.\ Phys.\ B {\bf 172}, 349 (1980);
  H.~Georgi,
    Nucl.\ Phys.\ B {\bf 361}, 339 (1991);
  C.~Arzt,
    Phys.\ Lett.\ B {\bf 342}, 189 (1995);
    H.~Simma,
    Z.\ Phys.\ C {\bf 61}, 67 (1994).
  

\bibitem{Corbett:2013pja} 
  T.~Corbett, O.~J.~P.~Eboli, J.~Gonzalez-Fraile and M.~C.~Gonzalez-Garcia,
    arXiv:1304.1151 [hep-ph].
    

\bibitem{proj_atlas} 
  ATLAS collaboration,
    ATL-PHYS-PUB-2012-004.
  

\bibitem{proj_cms}
CMS collaboration,
\url{https://twiki.cern.ch/twiki/bin/view/CMSPublic/HigProjectionEsg2012TWiki}

\bibitem{deCampos:1997ez} For an early application of this framework
  to the Tevatron data see F.~de Campos, M.~C.~Gonzalez-Garcia and
  S.~F.~Novaes,
    Phys.\ Rev.\ Lett.\  {\bf 79}, 5210 (1997)
  [hep-ph/9707511].
    

\end{thebibliography}
\end{document}